 \documentclass[review,12pt]{elsarticle}

 \usepackage{graphicx}
\usepackage{amsmath}
\usepackage{amssymb}
 \usepackage{amsthm}
 \usepackage{epstopdf}

 \usepackage{lineno}

\usepackage[colorlinks=true, pdfstartview=FitV, linkcolor=blue, citecolor=blue, urlcolor=blue]{hyperref}

\usepackage{relsize}
\usepackage{url}
\usepackage{hyperref}
\journal{Physics Reports C}
\def\babar{\mbox{\slshape B\kern-0.1em{\smaller A}\kern-0.1em
    B\kern-0.1em{\smaller A\kern-0.2em R}}}
\bibpunct{[}{]}{,}{a}{,}{,}
\begin{document}

\begin{frontmatter}



\title{Charged Lepton Flavor Violation:\\ An Experimenter's Guide}

\author[rhb]{R.H.\ Bernstein}
\ead{rhbob@fnal.gov}
\author[psc]{Peter S.\ Cooper}
\ead{pcooper@fnal.gov}
\address{Fermi National Accelerator Laboratory, Batavia IL 60510. 
}

\begin{abstract}
Charged lepton flavor violation (CLFV) is a clear signal of new physics; it directly addresses the physics of flavor and of generations. The search for CLFV has continued from the early 1940's, when the muon was identified as a separate particle, until today.    Certainly in the LHC era the motivations for continued searches are clear and have been covered in many reviews.  This review is focused on the experimental history with a view toward how these searches might progress.  We examine of the status of searches for charged lepton flavor violation in the muon, tau, and other channels,  and then examine the prospects for new efforts  over the next decade.   Finally, we examine what paths might be taken after the conclusion of upcoming experiments and what facilities might be required.
\end{abstract}


\end{frontmatter}


\section{Introduction \label{sec:intro}}

Isidor Isaac Rabi's famous question about the muon's existence, ``Who ordered that?",  was prescient and deep.\begin{footnote}{We have tried to track down the provenance of that quote.  John Rigden, Rabi's biographer, thinks it would be apocryphal if not that it sounds so much like something Rabi would have said. (priv.\ comm.) According to Jon Rosner (priv.\ comm), it springs from a Columbia tradition.  T.D. Lee would take the group to lunch at a Chinese restaurant and often so much would be ordered that no one would know who had ordered individual items.  Rabi compared the muon to one of the mystery dishes --- ``who ordered that?"  Although the date of the quote remains a mystery it would likely have come from Rabi's time at Columbia. In the absence of a more fitting story, we choose this one.}\end{footnote} His question, in modern terms, asked why are there flavors and generations?  Why are there muons and taus in addition to the electron?  The same question applies to the quark and neutrino sectors.  We believe there are three generations in each sector, and that the number in each sector must be the same. We see quarks changing generations, as codified in the CKM matrix, and neutrinos changing from muon to electron to tau neutrinos according to the PMNS matrix.   Lepton Flavor Violation (LFV) is an established fact, but only in the neutral neutrinos. What about their charged partners?  Is there Charged Lepton Flavor Violation (CLFV)?  

This article reviews the experimental history of searches for CLFV.  It concentrates on a subset of the experiments with a focus on the most sensitive ones, and attempts to guide the reader through the development of experimental techniques and their current status.  The purpose is to collect a fraction of the knowledge we have on these searches as an explanation of ongoing and planned experiments with a view toward how to develop these experiments in the future.  There are many  review articles on the theory and phenomenology: the encyclopedic review of \cite{Kuno:1991} is an excellent point-of-departure, and more recent reviews by  \cite { deGouvea:2013zba}, \cite{Marciano:2008}, \cite{Raidal:2008} and \cite{deGouvea:2010} update the subject.  There seems little point in repeating the contents of those articles, and no chance of surpassing them; therefore the reader is referred to those articles for an overview of the underlying physics.  However, there is a dearth of articles on the development of the experimental methods and the present article was written to address that absence.

The most powerful searches have used the muon state or the $\tau$ state with additional contributions from the kaon system.  The $\tau$ has a ``per-particle" advantage since, as we will see, the GIM suppressions are smaller than in muons but given the high statistics available in muon beams, the muon searches have been the most powerful.    The best limits have been set in the muon sector at the Paul Scherrer Institute (PSI) in Zurich, primarily $\mu \rightarrow e \gamma$ and $\mu N \rightarrow e  N$ (muon-to-electron conversion) along with a number of other muon processes. \babar\  and BELLE have made significant measurements with taus, and elegant kaon experiments at Brookhaven and Fermilab have produced important limits as well. In the future, the flavor factories (and possibly an electron-ion collider) can be competitive.  Each of J-PARC and Fermilab are planning a new muon-to-electron  conversion experiment, COMET and Mu2e respectively, to reach four orders-of-magnitude beyond current limits.  PSI  is discussing an innovative $\mu \rightarrow 3e$ search.  It is possible to envisage another two orders-of-magnitude beyond Mu2e and COMET with upgrades to muon flux and new beams.  J-PARC could build on COMET using innovative muon beam technology in PRISM/PRIME.  Fermilab's Project X has the potential to make intense muon and kaon beams that could push the limits of currently planned experiments another two orders-of-magnitude or study a signal by varying the $Z$ of the target. High-$Z$ studies could illuminate the underlying physics of a signal (as explained in \cite{Okada:2009}), and must be pursued in the future despite the experimental difficulties we will discuss.

In closing this introduction we want to stress the tremendous difficulty of these experiments.   The TWIST experiment at TRIUMF was designed to perform a precision measurement of some of the parameters of muon decay.  The TRIUMF Experimental Evaluation committee for TWIST put it nicely in July 1990:

\begin{quote}
In order to thoroughly understand the problems that may be encountered, we recommend that the proponents start serious discussions with the authors of the LAMPF experiment who were less ambitious by factors of two for $\rho$ and $\delta$ and by a factor of five for $P_{\mu}\xi$ compared to this proposal and nevertheless failed completely.(\cite{Marshall:2012})
\end{quote}

We point out that TWIST met its goals, no doubt thanks to,  in part,  an appreciation of the challenges that lay ahead.

\section{Theory Overview \label{sec:theoryOverview}}

This review will focus on the experimental methods, history, and prospects for charged lepton flavor violation experiments. For context and completeness we devote this Section to theoretical considerations.  The interested reader should consult the reviews mentioned in Section~\ref{sec:intro} for details. This Section is heavily indebted to the reviews by \cite{Marciano:2008} and \cite{deGouvea:2013zba}.

The discovery of neutrino mass and neutrino oscillations guarantees that Standard Model charged lepton flavor violation must occur through oscillations in loops.  Such transitions are suppressed by sums over $\left( \Delta m_{ij}/M_W \right)^4$.  Now that $\theta_{13}$ has been measured by \cite{DayaBay:2012} and \cite {RENO:2012} we can calculate, for the $\mu \rightarrow e \gamma$ decay:
\begin{eqnarray}
{\cal B}(\mu \to e \gamma )=\frac{3\alpha}{32 \pi}\left|\sum_{i=2,3} 
U_{\mu i}^*U_{ei} \frac{\Delta m_{i1}^2}{M_W^2}\right|^2 \sim 10^{-54}\,,
\label{meg_sm}
\end{eqnarray}
and as calculated in \cite{Marciano:2008}
other muon processes we will discuss are suppressed to similar unmeasurable levels.  

Therefore any detection of charged lepton flavor violation is an unambiguous signal of physics beyond the Standard Model.  It is often speculated that the rates for charged lepton flavor violation are ``just around the corner" from existing experimental limits.  The reason is that the physics of electroweak symmetry breaking is expected to have mass scales ${\cal O}(1)$ TeV/$c^2$.  If one assumes  large couplings , as is typical in SUSY models, then the next generation of experiments should see a signal.  There is no dearth of models and it would be convenient to have some generic parameterization.  de Gouv\^{e}a, in \cite {deGouvea:2010} or \cite{Appel:2008aa} and  most recently in \cite{deGouvea:2013zba}, has written:
\begin{eqnarray}
{\cal L}_{\rm CLFV}=\frac{m_{\mu}}{(\kappa+1)\Lambda^2}\bar{\mu}_R\sigma_{\mu\nu}e_LF^{\mu\nu}+ {\rm ~h.c.} \nonumber \\ 
 + \frac{\kappa}{(1+\kappa)\Lambda^2}\bar{\mu}_L\gamma_{\mu}e_L\left(\bar{u}_L\gamma^{\mu}u_L+\bar{d}_L\gamma^{\mu}d_L\right)+ {\rm ~h.c.}\,.
\label{eqn:leptonquark}
\end{eqnarray}
Very roughly one can characterize this a sum of ``loop" and ``contact" terms.  Supersymmetry belongs with the first term; particle exchange is reflected in the second.  The coefficients of the two types of operators are parameterized by two independent constants: $\Lambda$, the mass scale of the new physics,  and $\kappa$, a dimensionless parameter that mediates between the two terms. $L$ and $R$ indicate the chirality of the different Standard Model fermion fields, $F^{\mu\nu}$ is the photon field strength and $m_{\mu}$ is the muon mass.   This Lagrangian coupling quarks to leptons will govern $\mu N \rightarrow e N$, $\mu \rightarrow e \gamma$ and $\mu \rightarrow 3e$ in many models.

There is of course a similar expression for a ``lepton only" Lagrangian:
\begin{eqnarray}
{\cal L}_{\rm CLFV}= \frac{m_{\mu}}{(\kappa+1)\Lambda^2}\bar{\mu}_R\sigma_{\mu\nu}e_LF^{\mu\nu}+ {\rm ~h.c.} \nonumber \\
 + \frac{\kappa}{(1+\kappa)\Lambda^2}\bar{\mu}_L\gamma_{\mu}e_L\left(\bar{e}\gamma^{\mu}e\right) + {\rm ~h.c.}\,.
\label{eqn:leptons}
\end{eqnarray}

\begin{figure}[h]
\begin{center}
\includegraphics[scale=0.8]{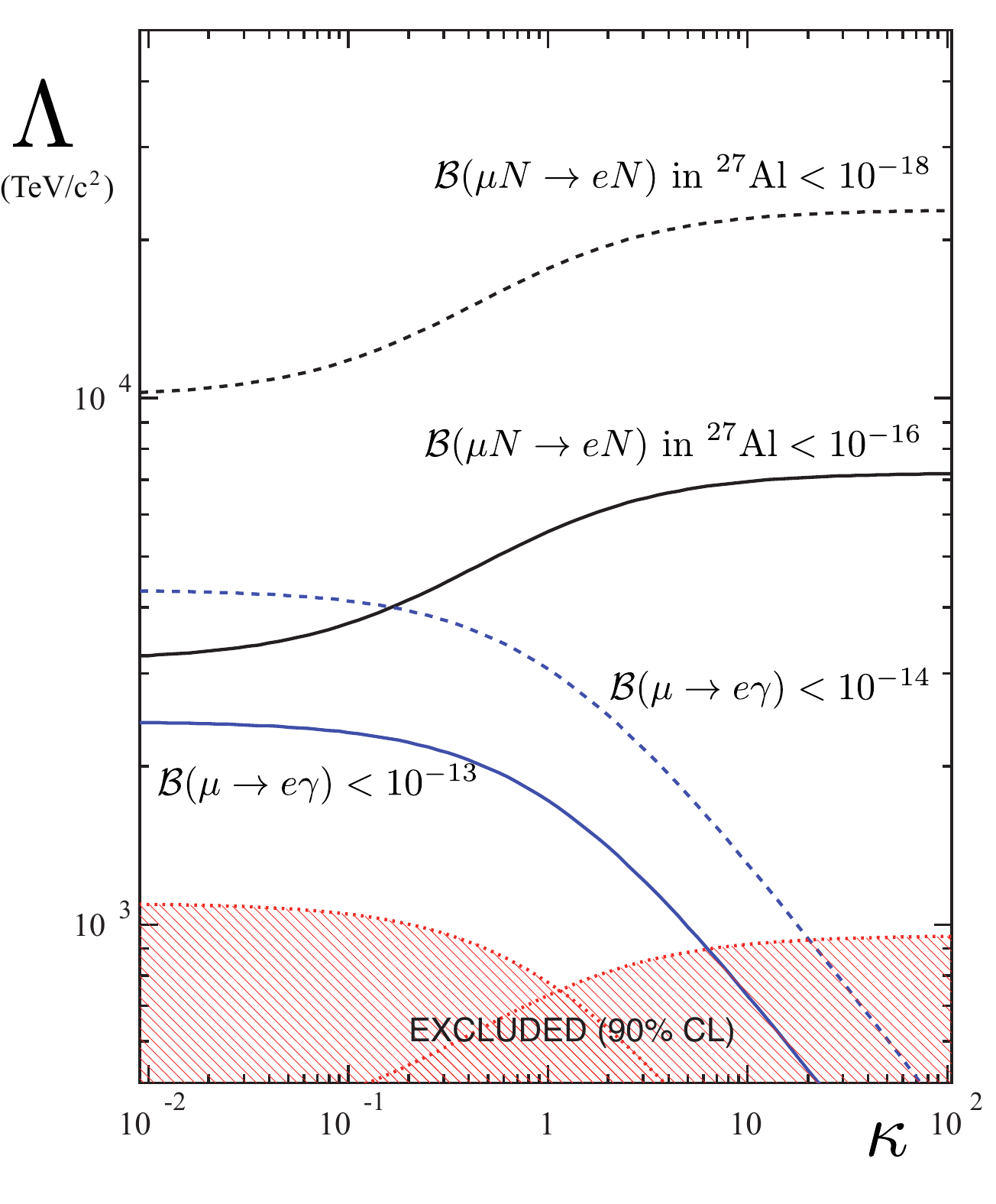}
\end{center}
\caption{Sensitivity of a $\mu\to e$~conversion in  $^{27}$Al  that can probe a normalized capture rate of $10^{-16}$ and $10^{-18}$, and of a $\mu\to e\gamma$ search that is sensitive to a branching ratio of $10^{-13}$ and $10^{-14}$, to the new physics scale $\Lambda$ as a function of $\kappa$, as defined in Eqn.~(\ref{eqn:leptonquark}). These correspond roughly to the discovery limits for the Mu2e experiment at the FNAL Booster, currently approved, and an ``ultimate experiment."  The $\mu \rightarrow e \gamma$ values are indicative of the signals-event sensitivity for MEG and its approved upgrade.  Also depicted are the currently excluded regions of this parameter space from the MEG and SINDRUM-II experiments.  See Sec~\ref{sec:muonoverview} for references and explanations.  Figure and caption adapted from \cite{deGouvea:2013zba}.\label{fig:lvsk1}}
\end{figure}

\begin{figure}[h]
\begin{center}
\includegraphics[scale=0.8]{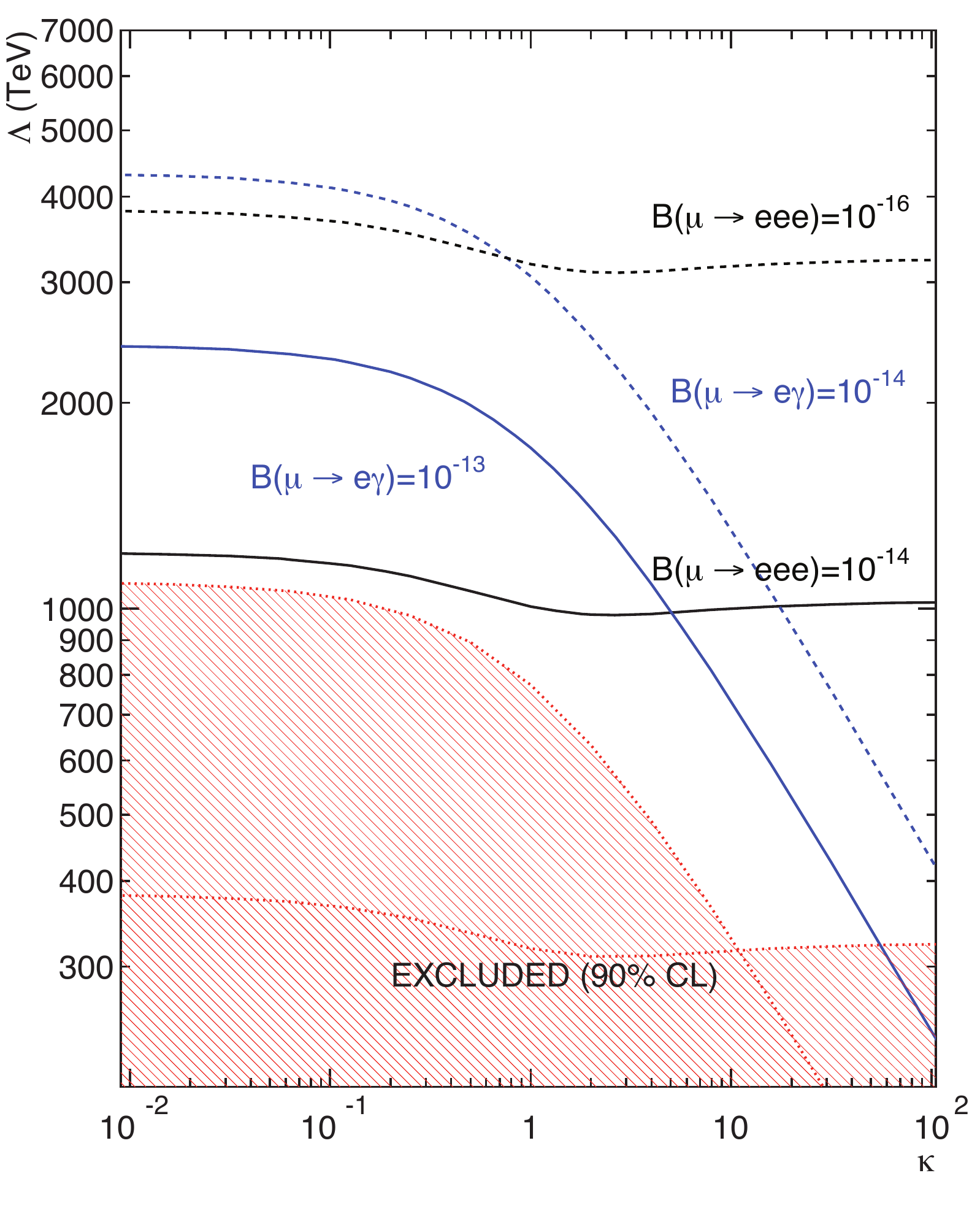}
\end{center}
\caption{Sensitivity of a $\mu\to eee$ experiment that is sensitive to branching ratios $10^{-14}$ and $10^{-16}$, and of a $\mu\to e\gamma$ search that is sensitive to a branching ratio of $10^{-13}$ and $10^{-14}$, to the new physics scale $\Lambda$ as a function of $\kappa$ Eqn.~(\ref{eqn:leptons}). These correspond roughly to the discovery limits for the Mu2e experiment at the FNAL Booster, currently approved, and an ``ultimate experiment". The $\mu \rightarrow e \gamma$ values are indicative of the signals-event sensitivity for MEG and its approved upgrade.  Also depicted are the currently excluded regions of this parameter space from the MEG and SINDRUM-II experiments.  See Sec~\ref{sec:muonoverview} for references and explanations.  Figure and caption adapted from \cite{deGouvea:2013zba}.\label{fig:lvsk2}}

\end{figure}

It has become commonplace in discussions to use this form\begin{footnote}{ It should be noted, as in \cite{Kuno:1991},  that there are several other terms and the possibility for constructive or destructive interference among terms that are ignored here.}\end{footnote} to plot $\Lambda$ vs.\ $\kappa$.  We borrow the plots from \cite{deGouvea:2013zba} in Figures~\ref{fig:lvsk1} and \ref{fig:lvsk2}.  One sees that mass scales up to 1000 TeV/$c^2$ have already been excluded if the assumptions behind this Lagrangian are valid.  

One pitfall of these plots, as convenient as they are, is that the casual reader often sees the leptonic and lepton-quark plots shown and immediately compares them as if they represented the same physics.  The underlying diagrams are of course different and that is a strength of performing a suite of such experiments, especially if $\tau$ charged lepton flavor violating modes are discovered as well.

Finally, with the first run of the LHC and the apparent discovery of the Higgs boson, we should ask whether such experiments are still relevant.  The answer is a resounding ``yes."  Charged lepton flavor violation may be related to the physics behind neutrino mass, and hence the seesaw mechanism, with possible ramifications for grand unified theories and the matter-antimatter asymmetry.  If new physics is found at the LHC, these experiments are required to discriminate among models.  If not, charged lepton flavor violation experiments can either severely constrain physics inaccessible at foreseeable colliders.  The discovery potential of these experiments, reaching mass scales at nearly $10^4$ TeV/$c^2$, is enormous.      We conclude with a recently written quotation from  \citet{Glashow:2013via}:
\begin{quotation}
 Because their standard-model branching ratios are far
too tiny  for possible detection,  observation of any
 mode  would be certain evidence of new physics.  That's
what makes such sensitive searches potentially
transformative.  
\end{quotation}

\section{Searches for Charged Lepton Flavor Violation with Muons \label{sec:muonoverview}}

\begin{figure}[h]
\begin{center}
\includegraphics[scale=0.6]{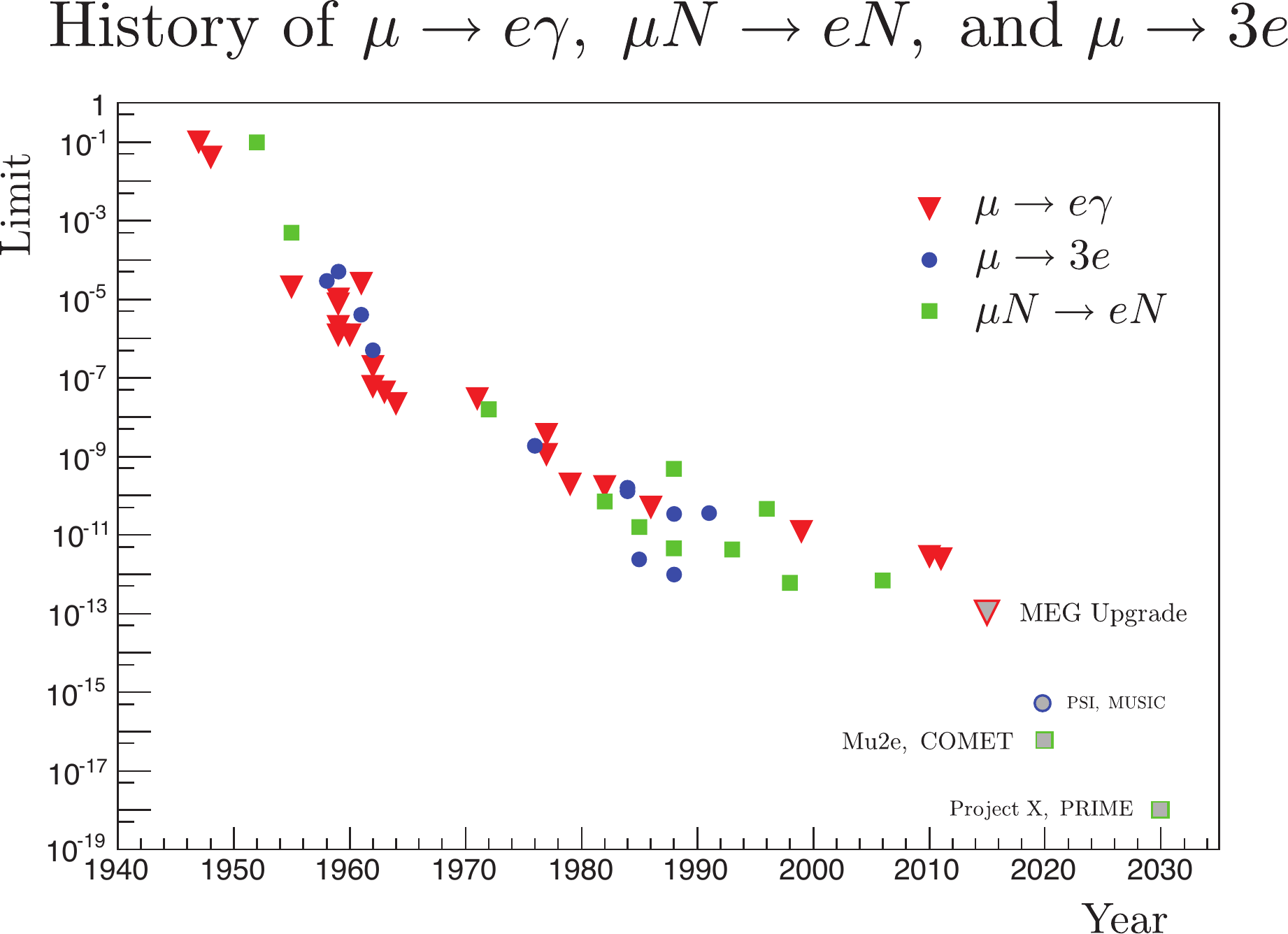}
\end{center}
\caption{The history of CLFV searches in muons (not including muonium.)  One sees a steady improvement in all modes and then a flattening of the rate improvement throughout the 1990s.  MEG has upgrade plans for the $\mu \rightarrow e \gamma$ search. The two next generations of $\mu N \rightarrow e N$, Mu2e/COMET at FNAL and J-PARC are labeled, and possible extensions at Project X and PRIME are shown.  Letters-of-intent are in process for $\mu \rightarrow 3e$ experiments at PSI and Osaka's MUSIC facility.   Individual experiments are discussed in the text.\label{fig:clfvHistory}}
\end{figure}
\clearpage
\subsection{$\mu^+ \rightarrow e^+ \gamma$}

The first search for the process $\mu \rightarrow e \gamma$, ``Search for Gamma-Radiation in the 2.2-
Microsecond Meson Decay Process"  was performed by \citet*{Pontecorvo:1947} at Chalk River and confirmed shortly thereafter by \citet*{Sard:1948}.  The search was motivated by the results from an experiment of  \citet*{Conversi:1947} .   The Conversi et al.\ experiment showed that in a heavy element (Fe) only positive stopped muons decay, while in a light element (C) both positive and negative muons decayed. The theoretical situation had been described  by \citet*{Fermi:1947} and the interest was intense.   At that time, it was expected that the muon would be captured in the process  $p + \mu \rightarrow n + h\nu$, where ``$h \nu$" was some light quantum.  Surprisingly, the experiment showed that the interaction between the muon and the nucleus was twelve orders of magnitude less than that required by a Yukawa particle.  Pontecorvo suggested that there might be no neutrino at all in the decay, and the decay of the muon may be simply $\mu \rightarrow e \gamma$.   The paper concludes ``that each decay electron is not accompanied by a photon of about 50 MeV".  In contrast, the \citet*{Sard:1948} paper explicitly quotes having observed nine events with a background of five and is in that sense a more reliable first measurement.\begin{footnote}{This paper has escaped mention in a number of reviews and the authors thank G.~Signorelli for pointing it out to us.}\end{footnote}    We now know the two-neutrino hypothesis is required to make sense of the situation, and although it is out of the scope of this article, it is fascinating to trace the development of these ideas through the demonstration of the existence of two neutrino species in the Nobel Prize-winning experiment of \citet*{Danby:1962}.

\begin{table}[h]

\begin{tabular}{|l|c|c|c|}\hline
Year&90\% CL on ${\cal B}(\mu \rightarrow e \gamma)$&Collaboration/Lab&Reference\\\hline
1947	&$1.0 	\times 10^{-1}$	&Chalk River&		\cite{Pontecorvo:1947}	\\
1948 &     $.04$ & Washington University&\cite{Sard:1948}\\
1955&$	2.0 	\times 10^{-5}$&	Nevis	&	\cite{Steinberger:1955a}	\\
1959&$	7.5 	\times 10^{-6}$	&Liverpool	&	\cite{Okeefe:1959}	\\
1959&$	2.0 	\times 10^{-6}$	&Nevis	&	\cite{Berley:1959}	\\
1959&$	1.0 	\times 10^{-5}$&	Rochester	&	\cite{Davis:1959}	\\
1959&$	1.2 	\times 10^{-6}$	&CERN	&	\cite{Ashkin:1959}\\
1960&$	1.2 	\times 10^{-6}$&	LBL	&	\cite{Frankel:1960}	\\
1961&$	2.5 	\times 10^{-5}$&	Carnegie	&	\cite{Crittenden:1961}\\
1962&$	1.9 	\times 10^{-7}$&	LBL	&	\cite{Frankel:1962}	\\
1962&$	6.0 	\times 10^{-8}$&	Nevis	&	\cite{Bartlett:1962}	\\
1963&$	4.3 	\times 10^{-8}$&	LBL	&	\cite{Frankel:1963}	\\
1964&$	2.2 	\times 10^{-8}$&	Chicago	&	\cite{Parker:1964}	\\
1971&$	2.9 \times 10^{-8}$&Dubna		& \cite{Korenchenko:1971}\\
1977&$	3.6 	\times 10^{-9}$&	TRIUMF	&	\cite{Depommier:1977}\\
1977&$	1.1 	\times 10^{-9}$&	SIN		&\cite{Povel:1977}\\
1979&$	1.9 	\times 10^{-10}$&	LAMPF	&	\cite{Bowman:1979}	\\
1982&$	1.7 	\times 10^{-10}$&	LAMPF	&	\cite{Kinnison:1982}\\
1986&$	4.9 	\times 10^{-11}$&	LAMPF/Crystal Box&  	\cite{Bolton:1986,Bolton:1988}	\\
1999&$	1.2 	\times 10^{-11}$&	LAMPF/MEGA &	\cite{Brooks:1999}	\\
2010&$	2.8 	\times 10^{-11}$	&PSI/MEG  &	\cite{meg:2010}	\\
2011&$	2.4 	\times 10^{-12}$&	PSI/MEG	&\cite{Adam:2011}\\\hline
\end{tabular}

\caption{History of $\mu \rightarrow e \gamma$ experiments.  \citet*{Pontecorvo:1947} does not set a limit; the limit usually quoted is actually a number of counts/hour and it is difficult to set a limit from the paper. \label{tab:muegamma}}
\end{table}

Before turning to the experimental status and prospects, we look at the process and intrinsic backgrounds in order to understand the design of the experiments and the problems they face.  First, we note that in $\mu \rightarrow e \gamma$ the electron energy is 52.8 MeV and the electron and photon  have equal but opposite momenta.  The experiments use stopped $\mu^+$ rather than $\mu^-$ and bring the muons to rest in a thin target.  Why $\mu^+$ rather than $\mu^-$?  First (and less important) is that one gets more $\pi^+$ than $\pi^-$ from proton collisions and so the final data sample is somewhat larger.  The more important reason is muon capture.  Muons captured on the nucleus typically cause the nucleus to eject protons, neutrons, and photons, which produce accidental rates in the detector, and this problem will recur in the muon-electron conversion experiments of Sec.~\ref{sec:mu2e}.)  Further, if one uses a surface beam of $\pi^-$ to make the muons, one has to deal with all the $\pi^-$ capture products as well.  Hence $\mu^+$ searches are preferable to $\mu^-$-based ones.

There are no Standard Model backgrounds of importance.\begin{footnote}{Assuming that in the Standard Model the neutrino mass is zero; the phrase $\nu$SM, indicating non-zero neutrino masses, is beginning to be used to make the distinction.}\end{footnote} The simplest way to see this is to write the branching ratio as a function of the lepton masses, as in \citet*{Marciano:1977} and \citet*{Lee:1977} (assuming the partial width to $\mu \rightarrow e \nu \nu = 1$):
\begin{eqnarray}
{\cal B}(\mu  \rightarrow e \gamma) &=& \frac{3 \alpha}{32\pi} \left | \sum_{i=2,3} U^*_{\mu i}U_{e i} \frac{\Delta m^2_{1i}} {M_W^2} \right |^2 \label{eq:BRSM}
\end{eqnarray}
which for $\mu \rightarrow e \gamma$ becomes (following the assumptions in \cite{Marciano:1977}):
\begin{eqnarray}
{\cal B}(\mu  \rightarrow e \gamma) &=& \frac{3 \alpha}{32\pi}\left (\frac{1}{4}\right) \sin^2 2\theta_{13} \sin^2 \theta_{23}  \left| \frac{\Delta m^2_{13}} {M_W^2} \right|^2 
\end{eqnarray}
We use  neutrino masses and mixings from the PDG, \cite{Nakamura:2010}), noting  the recent observations of  \cite{DayaBay:2012} and \cite{RENO:2012}, along with indications from \cite{MINOS:2011}, and \cite{T2K:2011} give  $\sin^2 \theta_{13} \approx 0.1$.  Combining these values, we find ${\cal B}(\mu  \rightarrow e \gamma)= {\cal O}(10^{-54})$, an effectively unmeasurable value.
We can thus ignore any  Standard Model background.  Similar levels are obtained from analogous calculations in the other muon processes we will examine.  This is an important advantage of these searches since any signal is clear evidence for physics beyond the Standard Model.

\begin{figure}[h]
\begin{center}
\includegraphics{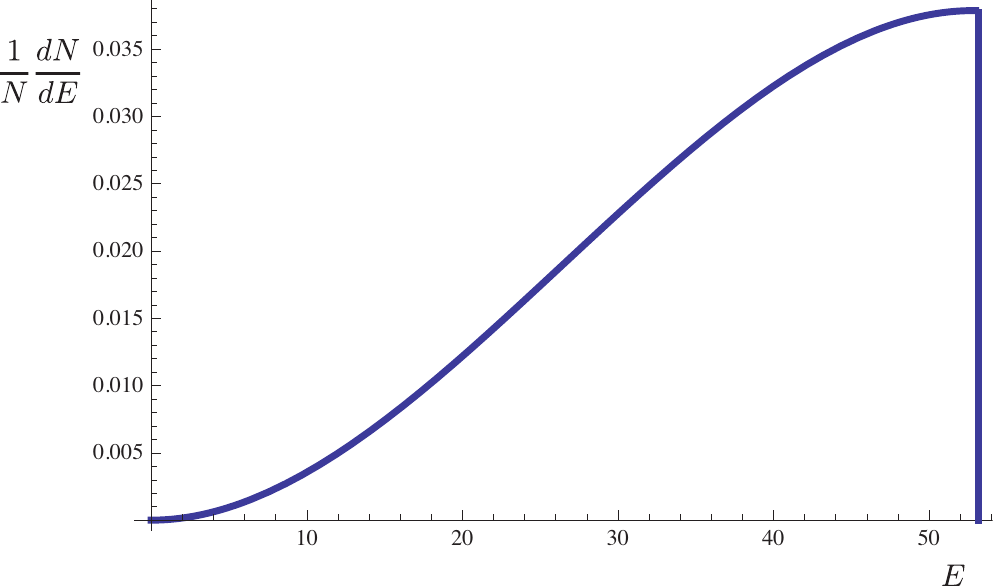}
\end{center}
\caption{The spectrum of $\mu^+ \rightarrow e^+ \nu_e \bar{\nu}_{\mu}$ decays, commonly known as the Michel spectrum, for the free decay of a muon at rest.(\cite{Michel:1950})  This calculation does not include radiative corrections.\label{fig:Michel}}
\end{figure}

 There are two important backgrounds: the first is an intrinsic, ``in-time"  physics background from the inner bremsstrahlung Radiative Muon Decay (RMD) process $\mu^+ \rightarrow e^-\gamma{\nu}_e  \bar{\nu}_{\mu}$, where the neutrinos carry off small momenta.  The second set of backgrounds is ``accidentals."  The search for $\mu \rightarrow e \gamma$ takes place in a sea of normal Michel decays: $\mu^+ \rightarrow e^+ \nu_e \bar{\nu}_{\mu}$.  The Michel spectrum is given in Fig.~\ref{fig:Michel} and is  derived in \citet*{Michel:1950}, \citet*{Kinoshita:1957}, and \citet*{ Commins:1973}.

Michel decays can combine with the following  processes to produce background if Michel electron  and a photon overlap within the time resolution:  
\begin{enumerate}
\item Radiative muon decay, $\mu^+ \rightarrow e^+ \gamma  \nu_e \bar{\nu}_{\mu}$ where the neutrino momenta are small.
\item Annihilation in flight of positrons (from, for example, another muon decay):  $e^+ e^- \rightarrow \gamma \gamma$, with a photon  of appropriate momentum and direction to combine with a regular Michel decay.
\item$eN \rightarrow eN \gamma$ from scattering off a nucleus.
 \end{enumerate}

 The ``accidental" processes, where an electron born from one stopped muon combines with a photon from another, dominate (the intrinsic RMD background is only about 10\% of the accidental background in modern experiments.) The size of the accidental backgrounds are tied to the detector resolutions: as one searches for smaller and smaller signals, the resolution requirements on energy, angle, and timing  become progressively more stringent.  The dependence of the background on the various factors is given by the convenient form of Eqn.~\ref{eqn:muegback}.  ${\cal B}$ is the ``single-event sensitivity" for  one background event; another way to understand the Equation is by examining $1/{\cal B}$: $1/{\cal B}$, up to acceptances and the statistical factor for a 90\% CL, is the number of muons that need to be examined to expect one background event.
 \begin{eqnarray}
 {\cal B} & \propto & ( \frac{R_{\mu}}{D}  )(\Delta t_{e \gamma})\frac{ \Delta E_e}{m_{\mu}/2}  \left (\frac{\Delta E_{\gamma}}{15 m_{\mu}/2} \right )^2  \left (\frac{ \Delta \theta_{e \gamma}}{2}\right )^2 \label{eqn:muegback}
 \end{eqnarray}
 The terms are the muon stop rate divided by the beam duty factor multiplied by the detector time resolution, the positron energy resolution, the photon energy resolution, and the angular resolution factors.  The sources of the terms are not difficult to understand:
 \begin{enumerate}
 \item The time difference between any two stops is essentially random, hence the $\Delta t_{e \gamma}$ term and the ${R_{\mu}}/{D}$ dependences.
 \item The \cite{Michel:1950} spectrum is (as derived in \cite{Commins:1973}) $\Gamma(\epsilon) \, d\epsilon \propto (3-2 \epsilon) \epsilon^2 \, d\epsilon$, where $\epsilon = 2E_e/m_{\mu}$.  Near $\epsilon=1$ at the maximum the derivative is zero.  Hence the $\Delta E_e /({m_{\mu}/2} )$ dependence.
 \item As derived in \cite{Kuno:1991} the radiative decay $\mu \rightarrow e \nu \nu \gamma$ near the zero-energy neutrino edge is a  bremsstrahlung term that behaves as $(1-y)\, dy$ where $y = 2E_{\gamma}/m_{\mu}$.  Hence the background under the $\mu \rightarrow e \gamma$ peak is proportional to the integral over the resolution window of width $\Delta$: $\int_ {(1 - \Delta)}^1 (1- y) \, dy $ which is just proportional to   $ \Delta^2$.
\item The angular term is simple as well.  Since  the direction of the photon in a $\mu \rightarrow e \gamma$ decay is opposite to the direction of the electron, the area of the angular phase space is  a small patch of area $\Delta \theta_{e \gamma} \Delta \phi_{e \gamma}$, yielding a quadratic dependence in angular resolution.   The precise form will depend on whether the photon is converted and details of the apparatus.
\end{enumerate}
 One sees that both good photon energy and angle measurements are especially important since the terms appear quadratically.  For the photon angle, this translates into a requirement of excellent position resolution and a well-known target location.  We will return to Eqn.~\ref{eqn:muegback} in our discussion of MEG and graphically illustrate how the data reveal the resolution terms.
 
This analysis then leads to a set of design choices on how to deal with the photon from $\mu \rightarrow e \gamma$: one can either (a) convert the photon and track the outgoing $e^+e^-$ pair in a magnetic field,  (b) use a calorimeter, or (c) use internal conversion of the photon and no converter.  The tracking solutions have much better resolution, but then one must pay a price in rate.  Converting the photon requires material, and using internal conversion suffer by at least ${\cal O}(\alpha)$; too much material spoils the resolution, but too little limits the size of the data sample.  However, determining the photon trajectory (and $\Delta \theta_{e \gamma}$) without a conversion then demands using the electron information, which leads to a different set of experimental difficulties.  We will look at three generations of experiments, the Crystal Box, MEGA, and MEG, and see how each has dealt with this dilemma.

\subsubsection{Experimental Status}

The Crystal Box at LAMPF (\cite{Bolton:1984,Bolton:1988}) was  arguably the first ``modern" $\mu \rightarrow e \gamma$ search.  The LAMPF linear accelerator produced a 300-$\mu$A, 800-MeV proton beam at 120 pulses per second, with duration 530 $\mu$sec. The average duty factor for the experiment was 6.4\%.  This small duty factor ultimately limited the experiment by causing pile-up in the tracking chambers, frequently making track reconstruction unsuccessful and reducing the acceptance.  A significant part of the history of $\mu \rightarrow e \gamma$ searches is the negotiation of the tradeoffs between rate, duty factor, running time, and sensitivity.  

\begin{figure}[htbp]
\begin{center}
\includegraphics[scale=0.5]{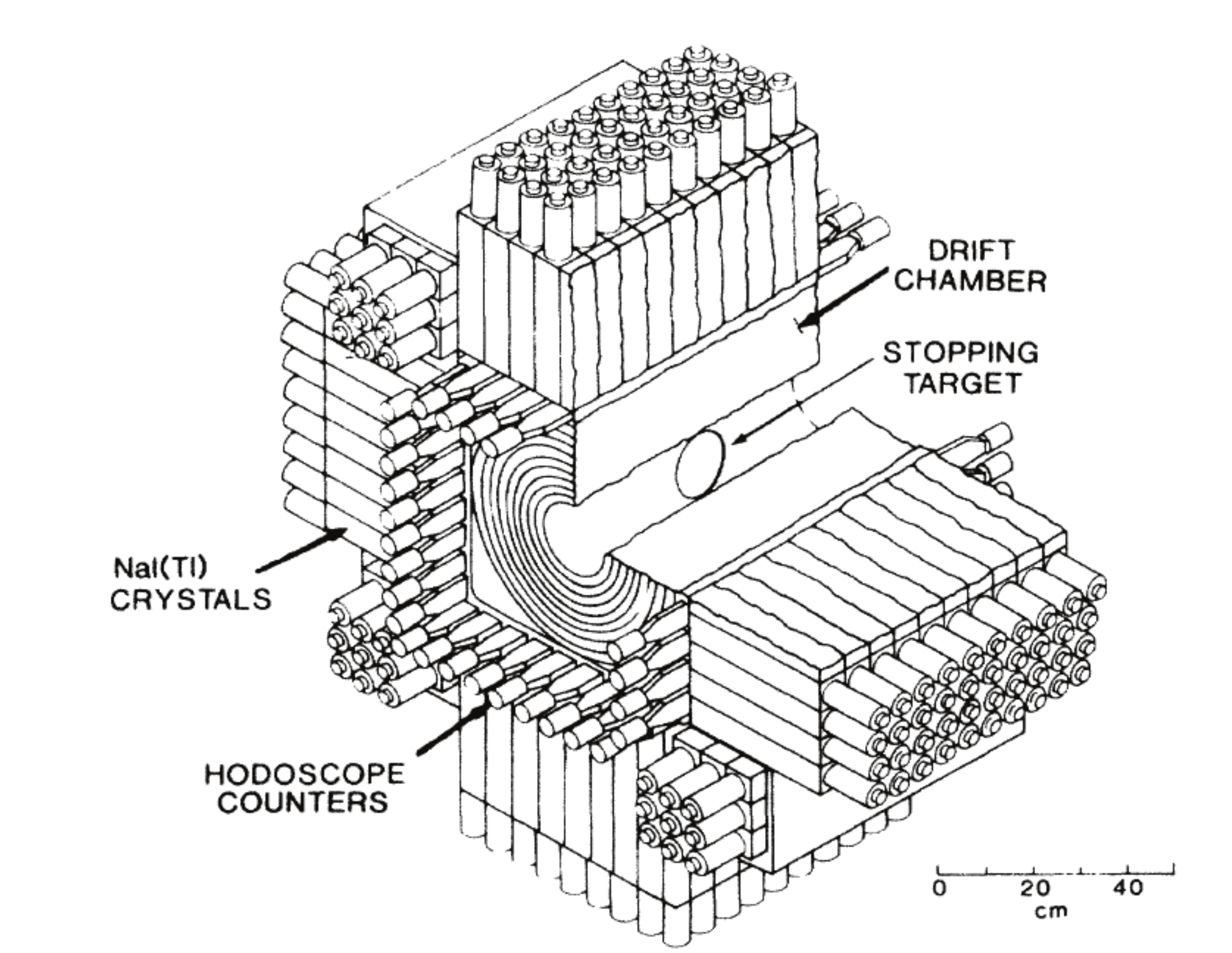}
\end{center}
\caption{The Crystal Box detector, Figure and Caption from \cite{Bolton:1986}.\label{fig:crystalbox}}
\end{figure}

The experiment used a surface pion beam:  pions brought to rest decayed near the surface of a target.   The daughter muons therefore come from a well-defined source.  Unfortunately the beam also transported a large contamination of positrons created in the same target, which were  then separated from the muons with a degrader.  The experiment examined $3 \times 10^{12}$  muons stopped in a thin polystyrene stopping target. The experiment then chose to detect the electron from $\mu \rightarrow e \gamma$ with tracking, and the photon with an NaI(Tl) calorimeter, employing choice (b) above.   There were 396  NaI(Tl) crystals surrounding a cylindrical drift chamber and plastic scintillation counters. There was no magnetic field.  The apparatus is displayed  in Fig.~\ref{fig:crystalbox}.  

\cite{Bolton:1986} tells us the positron energy resolution
averaged over all data was 8.8\% FWHM, and the $\gamma$-ray energy resolution was 8\% FWHM, both at 52.8 MeV. The angle between the photon and electron is determined by taking the vector from the reconstructed photon position in the crystals, and then using the extrapolated positron trajectory to the stopping target.  The RMS uncertainty in the measurement of the angle between the positron and photon momentum vectors was then 37 mrad, dominated by the knowledge of the photon position resolution in the NaI.  Since there was no magnetic field, the energy of the electron and photon were both measured in the crystals and so their resolutions are approximately equal.  The time resolution given in a later analysis by \cite{Bolton:1988} was determined to be 1.27 nsec for the photons, and 290 psec for the positrons.   A maximum-likelihood analysis established a
90\% CL upper limit for the branching ratio $\Gamma (\mu^+ \rightarrow e^+ \gamma)/\Gamma(\mu^+ \rightarrow e^+ \nu \bar{\nu}$) of
$4.9 \times 10^{-11}$.

\begin{figure}[htbp]
\begin{center}
\includegraphics[scale=0.6]{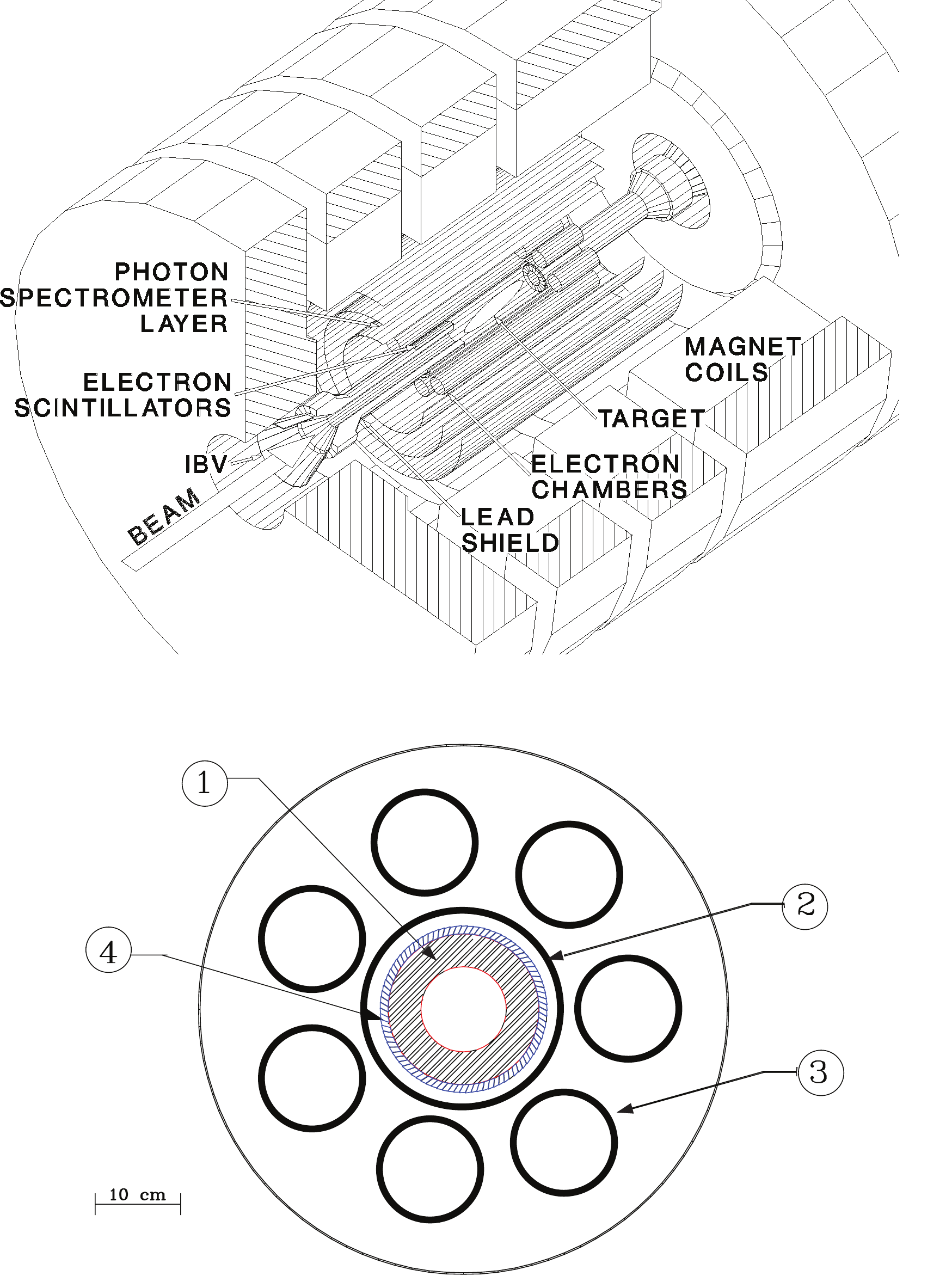}
\end{center}
\caption{The MEGA detector, from \cite{Ahmed:2002}.\label{fig:MEGA}}
\end{figure}

The next-generation experiment, MEGA, described in (\cite{Ahmed:2002}), was also performed at Los Alamos. MEGA also converted the photon, and surrounded the stopped muons with a cylindrical detector. The inner chamber, ``Snow White", surrounded the stopping target.  The ``Seven Dwarves" were smaller cylindrical chambers surrounding central Snow White.  The apparatus is shown in Fig.~\ref{fig:MEGA}.

 MEGA measured $\mu \rightarrow e \gamma < 1.2 \times 10^{-11}$ at 90\% CL.  The reader might reasonably wonder why, with all the evident improvements from the Crystal Box to MEGA, there was no concomitant improvement in the MEGA limit: only 4.9/1.2 = $\times 4.1$.  The reason is revealed by examining the acceptance.  Pile-up of hits in the chambers made it difficult to reconstruct tracks, greatly reducing the acceptance.  The 6.4\% duty factor limited the experiment with a vengeance: if one follows through the equation for ${\cal B}$ and applies the acceptance factors and total muon stops, one finds that MEGA should have done about only about five times better than the Crystal Box despite all the advantages and we see only the $\times 4.1$ above.  The acceptance effect, driven by the duty factor and resultant instantaneous rates, overwhelmed the other advantages.  This is a precautionary tale for other experiments trying to push the rate in any of the CLFV experiments.  

\begin{table}[h]
\begin{center}
\begin{tabular}{|l|c|c|c|}\hline
Experiment&Crystal Box&MEGA&MEG\\\hline
Date& 1986  &1999 & 2011 \\\hline
Rate (stops/sec) &$4 \times 10^5$ &$1.5\times 10^7$ & $2.9 \times 10^7$ \\
Duty Factor&5--10\% &3\% &  $\approx 50$\% \\
$\Delta E_{\gamma}$&8.0\% &1.7 or 3.0\%  & 4.5\% \\
$\Delta \theta_{e \gamma} $(mrad)& 87 & 33 & 50 \\
$\Delta E_e$ (at $\approx 53$ MeV) &8.0\% & 1.0\%&  1.5\% \\
$\Delta t_{e \gamma} $(nsec)& 1.2 & 1.6 & 0.305 \\
Acceptance& 0.17& $4 \times 10^{-3}$&0.18 \\
Muon Stops& $1.35 \times 10^{12}$& $1.2 \times 10^{14}$& $1.8 \times 10^{14}$ \\\hline
90\% CL Limit&$4.9 \times 10^{-11}$ &$1.2 \times 10^{-11}$ & $2.4 \times 10^{-12}$  \\\hline
\end{tabular}
\end{center}
\caption{Comparison of Modern $\mu \rightarrow e \gamma$ experiments. Recall the background is proportional to  $(R_{\mu}/{D} ) (\Delta E_{\gamma})^2( \Delta \theta_{e \gamma})^2(\Delta t_{e \gamma}) (\Delta E_e) $ (Eqn.~\ref{eqn:muegback}, with definitions provided there.)  All resolutions are FWHM (MEG reports $\sigma$ and we multiply by 2.35.)
\label{table:muegcompare}}
\end{table}

The state-of-the-art in $\mu \rightarrow e \gamma$ is MEG. (\cite{Adam:2013})   MEG	covers	a	10\%	solid angle, centered around a thin muon stopping target (205 $\mu$-thick polyethylene) and is composed of a positron spectrometer and a photon detector in search of back-to-back, monoenergetic, time coincident photons and positrons from the two-body $\mu  \rightarrow  e \gamma$ decays.   There is a positron arm with a electromagnetic calorimeter on the other side.  A key difference between MEG and its predecessors is that MEG chose not to convert the photon and then accept the consequent loss in rate; instead it relies on electromagnetic calorimetry and an innovative spectrometer.  This method  avoids  the pileup pattern recognition problems that limited MEGA.

\begin{figure}[h]
\begin{center}
\includegraphics[scale=2.5]{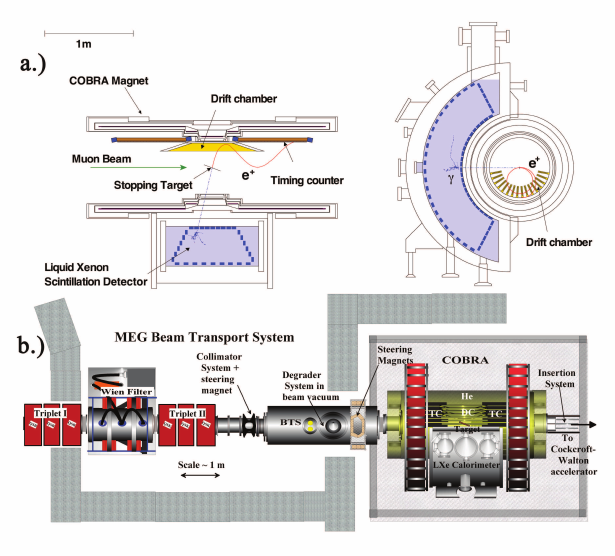}
\end{center}
\caption{The MEG detector and beam transport system, taken from \cite{meg:2010}.\label{fig:meg_apparatus}}
\end{figure}

 The apparatus is shown in Fig.~\ref{fig:meg_apparatus}.  It is preceded by degraders and collimators, and the detector surrounds a stopping target in the usual form.  In this sense nothing has changed, but in fact MEG made a major shift in technique: MEGA converted the photon in $\mu \rightarrow e  \gamma$ and MEG does not.  MEG uses a state-of-the-art liquid Xenon calorimeter.  Furthermore, MEG has avoided many of the pattern recognition problems that limited MEGA with an innovative spectrometer.  

Solenoidal fields have the advantage of confining low momentum tracks, which is useful for keeping Michel positrons out of the detector.  However, a simple solenoidal field has two disadvantages: (1) positrons emitted close to $90^{\circ}$ to the field curl many times, yielding large numbers of hits and potential problems in pattern recognition and momentum resolution, and (2) the bending radius depends on the angle, which makes it difficult to select the desired high-momentum tracks.  Therefore MEG adopted a gradient field near 1.1 T at $z=0$ that slowly decreased as $|z|$ increased.  This gradient quickly sweeps out the positrons of case (1). The precise gradient is set so that monochromatic positrons follow a (co)nstant projected (b)ending (ra)dius independent of emission angle.  The bending radius is thus set by the absolute momentum, not the transverse component --- hence the name COBRA.(\cite{Nishiguchi:2008})

The calorimeter must detect the photons with good efficiency and resolution, and relies on liquid Xenon.  The material was chosen after balancing a number of requirements: (1) light output, (2) fast decay time to avoid pileup, (3) high-$Z$ and density in order to make a compact device, (4) uniform response, and (5) radiation hardness.  Photons pass through the spectrometer and the thin wall of the   superconducting coil with about 80\% probability.  The photons then enter a 800 liter liquid Xe calorimeter surrounded by 846 photomultipliers.  The entrance window is a thin structure made of aluminum and carbon fiber plates.   The photomultipliers are immersed in the liquid Xe in order to directly observe the scintillation light. (\cite{Sawada:2008, Nishimura:2010, Signorelli:2004})

\begin{figure}[h]
\begin{center}
\includegraphics[scale=1.25]{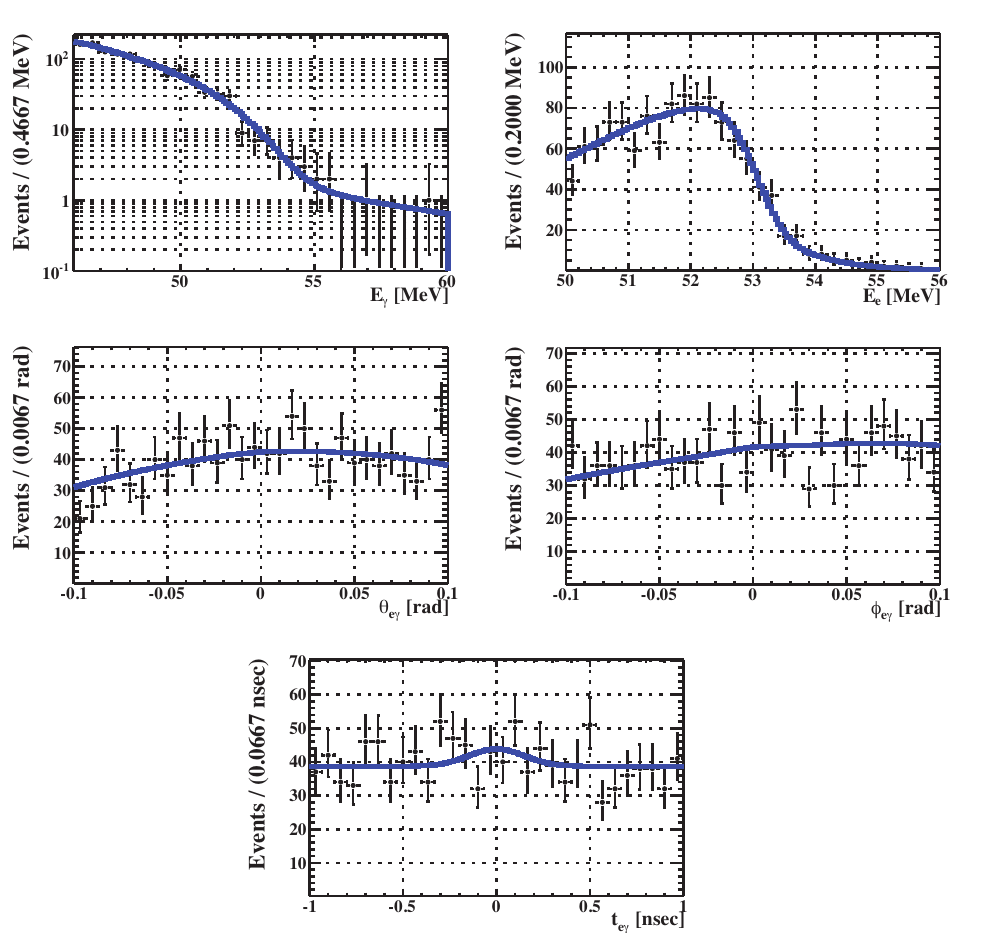}
\end{center}
\caption{Kinematic and time distributions from \cite{meg:2010} demonstrating the background dependences.  The ``bump" in $\Delta t_{e \gamma}$ near $t=0$ is from radiative muon decay.  The reader will also note the shift in the $\Delta \theta_{e \gamma}$ and $\Delta \phi_{e \gamma}$ distributions.  These exist because there is a slight correlation between angle and energy in the apparatus; since the experiment is performed at the kinematic edge one tends to have an average underestimate of the momentum (seen in the $E_e$ distribution), which then causes the correlation.  The MEG simulations correctly account for these effects.   \label{fig:megresolv}}
\end{figure}

 The experiment has now set a limit of ${\cal B} (\mu^+ \rightarrow e^+ \gamma ) < 5.7 \times 10^{-13}$ at 90\% CL.(\cite{Adam:2013,Adam:2011})  The limitations are not statistical.  One can see from Table~\ref{table:muegcompare} that the calorimeter resolutions yield energy and angle resolutions worse than MEGA, pointing to the difficulties intrinsic to using electromagnetic calorimetry for the photon instead of converting it and using tracking.  One does not get the photon momentum vector but has to rely on the extrapolation of the track to the stopping target.  Although a number of careful calibrations and studies were performed,  the observed light in 2007 was approximately one-third of the expected level, and the original electronics suffered from noise and instability.  Improvements to both the electronics and a set of sophisticated calibration runs with a Cockroft-Walton accelerator and charge-exchange were made after the 2009 run. Returning to Eqn.~\ref{eqn:muegback},    Fig.~\ref{fig:megresolv} shows dependences from \cite{meg:2010} in the MEG experiment, discussed below.  One can see how the various terms reveal themselves in the distributions.
 
 \begin{table}[h]
\begin{center}
\begin{tabular}{|l|c|c|}\hline
Variable&Foreseen&Obtained\\\hline
$\Delta E_{\gamma}$ (\%) &1.2&1.9\\
$\Delta t_{\gamma}$ (psec)&43&67\\
$\gamma$ position (mm)&4 (u,v), 6(w)&5(u,v),6(w)\\
$\gamma$ efficiency&$>40$&60\\
$\Delta p_e$ (keV/$c)$& 200&380\\
$e^+$ angle (mrad)&$5(\phi_e),5(\theta_e)$&$11(\phi_e),9(\theta_e)$\\
$\Delta t_{e^+}$ (psec) &50 & 107\\
$e^+$ efficiency (\%) &90&40\\
$\Delta t_{e \gamma}$ (psec) & 65&120\\\hline
\end{tabular}
\end{center}
\caption{Foreseen and obtained resolutions in the MEG experiment,  from \cite{Baldini:2012}.\label{table:megcompare}}
\end{table}
 
\clearpage

\subsubsection{Prospects}

 An upgrade proposal has recently been approved by PSI to reach ${\cal B} < 6 \times 10^{-14}$ at 90\% CL.(\cite{Baldini:2013}) The essential upgrades here are to improve the granularity of the photomultiplier system and a new tracking system with cluster timing.   More intense beams at future accelerators (FNAL's  Project X, J-PARC, or PSI upgrades) could provide more stopped muons if the resolution errors and backgrounds can be made sufficiently low to take advantage of higher statistics.

\subsection{$\mu$---$e$ Conversion \label{sec:mu2e}}

The conversion of a muon captured by a nucleus into an electron has been one of the most powerful methods to  search for CLFV.  The process can be written as
\begin{eqnarray}
\mu^-  + N \rightarrow e^-  + N
\end{eqnarray}
where $N$ is a nucleus of atomic mass  $A$ and atomic number $Z$.  The core advantage of this mode is that the outgoing electron is monoenergetic at an energy far above the normal Michel endpoint:
\begin{eqnarray}
E_{\mu e} &=& m_{\mu} - E_{b} - \frac{E_{\mu}^2}{2m_N}
\end{eqnarray}
where $m_{\mu}$ is the muon mass,  $E_{b}\approx Z^2 \alpha^2 m_{\mu}/2$ is the muonic binding energy, and the last term is from nuclear recoil energy up to terms of order $1/m_N^2$ and neglects variations of the weak-interaction matrix element with energy.  For Al ($Z=13$), a currently favored candidate nucleus,  the outgoing electron has energy $E_{\mu e} \approx 104.96$ MeV.\begin{footnote}{In fact, not all $\mu N \rightarrow e N$ conversions are coherent; the coherent process is enhanced by $Z$ since all nucleons participate in the coherent conversion. This ``enhancement by $Z$" is, like most statements, only an approximation.  The reader is referred to \cite{Okada:2002} for details. We ignore this effect in this discussion and wait for a precision measurement of the conversion ratio before considering those terms.}\end{footnote}

The quantity one measures is:\begin{footnote}{$R_{\mu e}$ is commonly used; some authors use $C$ for conversion instead. }\end{footnote}
\begin{eqnarray}
R_{\mu e} &=& \frac{ \Gamma(\mu^- + N \rightarrow e^- + N) }{\Gamma(\mu^- + N \rightarrow {\rm ~all~captures}) }
\end{eqnarray}

The normalization to captures has a calculational advantages since many details of the nuclear wavefunction cancel in the ratio.  Detailed calculations have been performed by \cite{Okada:2002}, \cite{Okada:2002e}, and \cite{Okada:2009}.  There is another link in the chain.  Experiments do not observe captures on the nucleus, but instead see the signature of a ``stopped" muon, one that comes to rest from energy loss.  A muon that stops falls into a 1s state of some target nucleus; in so doing, X-rays are emitted and their characteristic spectrum serves as the signal of a ``stopped" muon.  The muon either then (a) is captured by the nucleus, (b) decays by $\mu \rightarrow e \bar{\nu}_{e} \nu_{\mu}$ while in the 1s state, or (c) converts into an electron.  The lifetime on Al is 864 nsec (as summarized in the classic, strongly-recommended reading of \cite{Measday:2001}.)  The X-ray spectrum is well-known and one can use a variety of methods to detect the characteristic X-rays (of course one needs to know the acceptance, and  if one were measuring $R_{\mu e}$ one would need to know the uncertainties on the acceptance.)  The lifetime of the muonic atom is known and the stopped muon either decays or is captured (or converts, which occurs at an unfortunately negligible rate for this calculation.)  Both the decay lifetime of the free muon and the total lifetime in Aluminum are known, and therefore using
\begin{eqnarray}
\frac{1}{\Gamma} &=& \frac{1}{\Gamma_{ {\rm decay}} }+ \frac{1}{\Gamma{{\rm capture}}}
\end{eqnarray}
 by measuring the number of stops one can infer the number of captures.  Hence experiments count the number of stops, infer the number of captures, and use the calculated $R_{\mu e}$ when reporting a result.   One could, in fact, normalize to the accepted portion of the decay-in-orbit spectrum if the stopping rate in modern high-rate experiments were to exceed the capability of the Ge detectors commonly used for detection of the X-rays.

Experiments using negative  muons captured by a nucleus (such as muon-to-electron conversion) are intrinsically less clean than positive muon experiments (such as $\mu \rightarrow e \gamma$) using stopped muons.  When a muon is captured, neutrons, and photons are produced in the $\mu^- N \rightarrow \nu N^{\prime}$ transition.  These particles  can then travel into and produce extra activity in the detector that can obscure a signal track or potentially create backgrounds. For rates and spectra of ejected protons and neutrons, see \cite{Measday:2001} and \cite{Mukhopadhyay:1977}.

 The experiments require shielding against cosmic rays --- for example, one potential background arises from a cosmic ray muon that produces a $\delta$-ray in the muon stopping target.  If the electron is in the momentum signal window and the parent muon is unseen, the electron then fakes a signal.   Therefore, in addition to shielding to lower the rate, a cosmic ray veto is required to detect the parent muon.  
 
The cosmic ray veto, a required feature, then presents a new problem.  Captures can ``self-veto" the event in the cosmic ray veto system.  Approximately one neutron is produced for every captured muon.(\cite{Measday:2001})The neutron has a kinetic energy of order a few MeV.   As the neutron thermalizes, any time information associated with the parent muon capture is lost and the resultant neutron background then has a uniform time distribution.  If a neutron from muon capture stops in the veto counter and is then captured on hydrogen, as would happen in a scintillator-based veto system, the outgoing 2.2 MeV gamma can convert and ``self-veto" the event.  
 
In the geometries for modern experiments, one typically uses solenoids, which do not charge select. Protons produced in the  capture can enter tracking chambers and since they are highly ionizing,  can deaden a detector element.  They can also induce cross-talk between channels.  They can be removed by looking for their large ionization signal, but such removals are never perfect and the cross-talk issue remains.  
 
Nonetheless the single-particle electron signal is relatively clean. Normally one does not want to search for a single-particle final state  since it can be prone to accidental backgrounds but this conversion process is an exception.  In this case, the electron stands out from the background:  the Michel spectrum for free muon decay peaks and ends at $52.8$ MeV. Typical experimental resolutions on the momentum of a 100 MeV electron are a few hundred keV or less, so there would effectively be no background if the muon were free.     Hence muon-electron conversion does not suffer from accidental coincidences in the same manner as does $\mu \rightarrow e  \gamma$ or $\mu \rightarrow 3e$ where one is searching for electrons near the peak of the Michel spectrum. 

As usual, things are not that simple.  Experiments searching for muon-electron conversion require that the muon be bound in orbit around the nucleus.  The muon can then be captured by the nucleus (and possibly then convert into an electron through some interaction.) The muon can also decay.   The outgoing electron from a decaying muon can exchange a photon with the nucleus, which then distorts the Michel spectrum.   The tail of the muon decay spectrum produces background called  DIO (for ``decay-in-orbit") or MIO (``muon decay-in-orbit") in the literature. The form of the DIO  spectrum near the endpoint is approximately given by:
\begin{eqnarray}
N(E_e)dE_e &=& CE^2_e (\frac{\delta_1}{m_{\mu}})^5 \, dE_e
\end{eqnarray}
with 
\begin{eqnarray}
\delta_1 &=& m_{\mu} - E_e - \frac{E^2_e}{2M_N}
\end{eqnarray}

The recoil energy is:
\begin{eqnarray}
\vec{p}_N &=& - (\vec{p}_e+ \vec{p}_{\bar{\nu}_e} + \vec{p}_{\nu_{\mu}}) \nonumber \\
E_{{\rm recoil}} &=& \frac {|p_N|^2}{2m_N} \nonumber \\
 ~&=& \frac{(\vec{p}_{e} + \vec{p}_{ \bar{\nu}_{e} } + \vec{p}_{\nu_{\mu}})^2 } {2m_N} \: \approx \:    \frac{ |p_e|^2}{2m_N} \: = \:  \frac{E_e^2}{2m_N}\label{eqn:recoil}
 \end{eqnarray}
 and  the addition of the recoil term $E_e^2/(2m_N)$ reduces the rate  by $\times 2$ or more near the endpoint.   

\begin{figure}[h]
\begin{center}
\includegraphics[scale=0.8]{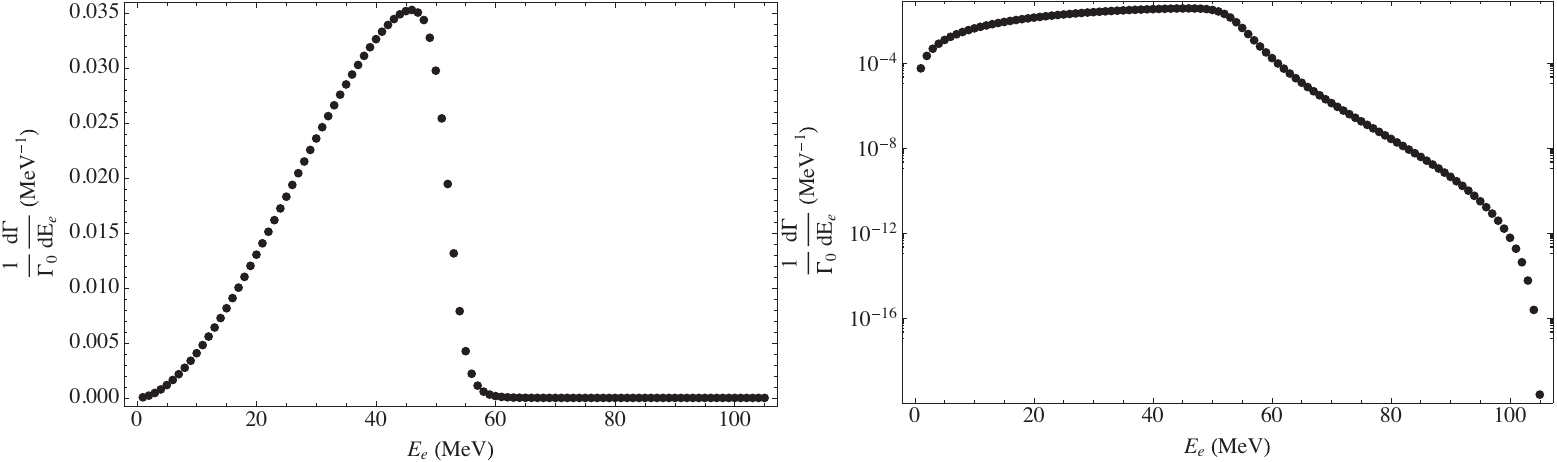}
\end{center}
\caption{Electron spectrum for aluminum on a linear and logarithmic scale. \cite{Czarnecki:2011}.\label{fig:czarnecki}}
\end{figure}

\begin{figure}[h]
\begin{center}
\includegraphics[scale=0.8]{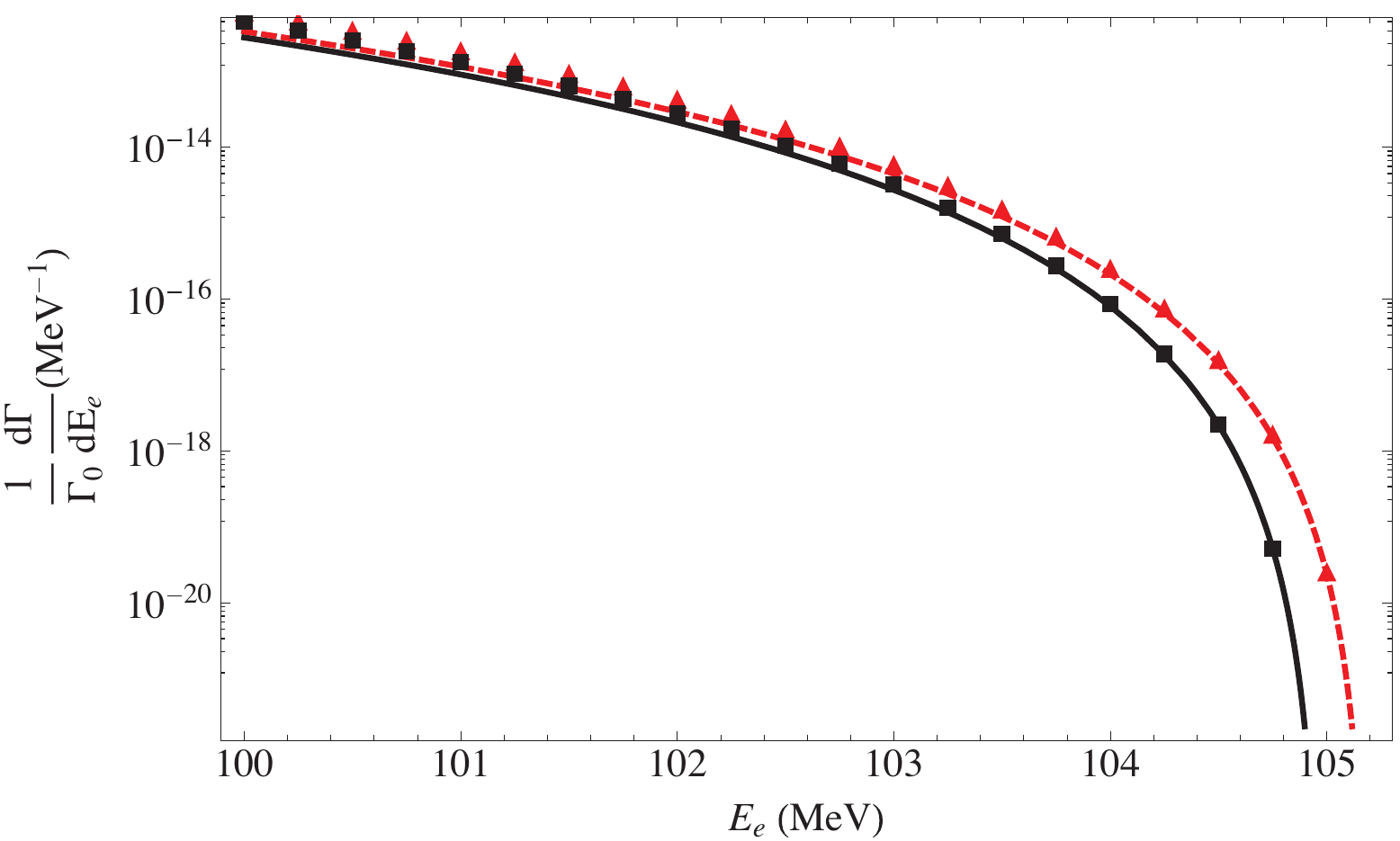}
\end{center}
\caption{Endpoint region of the electron spectrum for aluminum. The squares correspond to the spectrum with recoil effects,  the triangles neglect the recoil effects. The solid (dashed) lines correspond to a Taylor expansion with and without recoil. Figure and modified caption from \cite{Czarnecki:2011}.\label{fig:czarnecki2}}
\end{figure}

One sees that overall the DIO spectrum is falling as $(E_{\mu} - E_e )^5$ before recoil; this is the usual three-body form as in Sargent's rule  as discussed in \cite{Perkins:1999}.  Corrections to these formulae from (a) the relativistic wave function for the electron ($E_e \approx m_{\mu} >> m_e$),  (b) the finite size of the nucleus, (c) screening, and (d) radiative corrections must also be included.    A series of papers calculated the DIO spectrum; until recently, the most complete calculation was from \cite{Shanker:1982} and \cite{Shanker:1997}. \cite{Czarnecki:2011}    have recently performed a new calculation.  Figs.~\ref{fig:czarnecki} and \ref{fig:czarnecki2} show the new results.

What should we learn from this plot? First, only $\sim 10^{-17}$ of the spectrum is within the last MeV from the endpoint; second, the spectrum is falling rapidly.  Broadly speaking, the less the signal is smeared by measurement resolution or experimental effects, the more powerful the search.  Therefore to reach the goals of current experiments, ${\cal O}(10^{-17})$,  the experimental resolution of the detector should be well below an MeV.   An experimental effect, such as energy loss of the conversion electron from any source in the apparatus, or from the capture material itself is also problematic --- a converted electron must pass through the material used to stop and capture the muon before it can be detected, and therefore the signal has an unavoidable energy loss.  Although both conversion electrons and DIOs near the endpoint will be equally shifted, since energy loss is stochastic it will widen the conversion signal, automatically forcing the experiment to integrate over a wider region, adding more DIO background.  Thus minimizing the energy loss and the detector resolution are important to the design.  The upcoming generation of experiments, Mu2e (see the Conceptual Design Report in \cite{Abrams:2012}, and much of the information about Mu2e or these experiments appears there) and COMET, both expect about 1 MeV FWHM for the signal peak, dominated by energy loss.  Hence, from DIOs alone, one will not do better than ${\cal O}(10^{-17})$ without narrowing the signal peak.  Both experiments are studying ways to reduce the smearing from energy loss.  This will be especially important at next-generation experiments at FNAL's Project X or at J-PARC; at either laboratory one could hope for $\times 100$ in the number of muon stops.  To take full advantage of those statistics the DIO background must be reduced, along with other improvements to be discussed.  Since the DIO background is an intrinsic physics background --- the endpoint of the spectrum is the conversion energy, up to neutrino mass --- minimizing the detector resolution and energy loss have to be the focus of improvements in detector design.   A ``first-pass" at future Mu2e-style experiments is discussed by \cite{Knoepfel:2013ouy} in the context of \cite{Kronfeld:2013uoa}.

The second main background to muon-to-electron conversion searches comes from radiative pion capture (RPC), the process $\pi N \rightarrow \gamma N^*$, with a subsequent conversion $\gamma \rightarrow e^+e^-$. One must also include the internal conversion process $\pi^- N \rightarrow e^+e^- N^*$.  Depending on the details of the experimental arrangement this can be as large as the direct process if the probability of conversion is as small as the ratio of internal conversions to direct photon production, approximately 0.007, calculated in \cite{Kroll:1955}.
  
\begin{figure}[h]
\begin{center}
\includegraphics{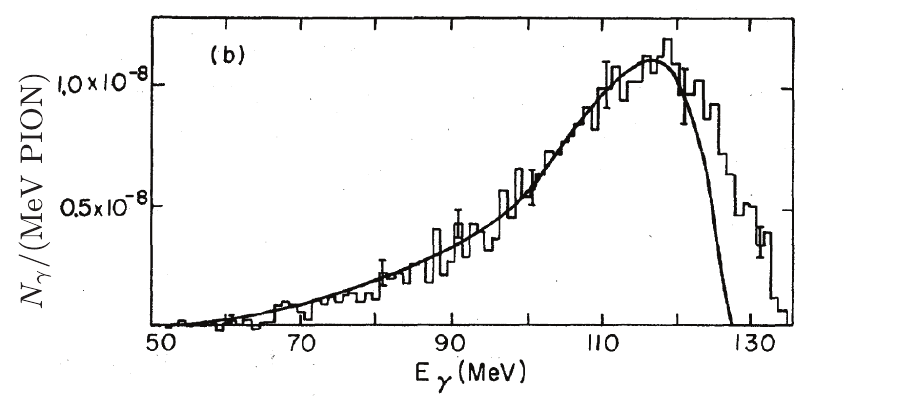}
\end{center}
\caption{\label{fig:rpc} Energy spectrum of photons from radiative pion capture in magnesium; solid line: pole model predictions.(Figure reproduced with permission from Physical Review, caption adapted from \cite{Bistirlich:1972}.)}
\end{figure}

  Normally the experiments make muons by striking a production target with protons; pions are produced and the resultant muons from $\pi$ decay are used for the measurement.  However, not all the pions decay and whatever fraction strike the ``stopping target" can be captured and undergo RPC.    Fig.~\ref{fig:rpc} shows the RPC photon spectrum on Mg (recall Al is a typical target material.)  We see the peak is in the 120 MeV range and asymmetric conversions can produce an electron in the same energy regime as conversion electrons.

How does one reduce the RPC background?  It depends on the intensity of the beam and the time structure, and as the experiments have progressed the choice has changed.  The simplest method is to use a veto counter.     Since RPC occurs at time scales shorter than a nanosecond, one  can veto beam particles immediately after the proton pulse (which also reduces the number of scattered electrons.)  This becomes impractical at sufficiently high intensities.  One can next use a passive degrader to reduce the pion content of the muon beam---the pions have a mean range about half of that for muons at relevant energies.  With sufficient numbers of stopped muons, one can use a third method: simply use the pion lifetime of $\gamma \times 26.$ nsec, which is short compared to the 864 nsec lifetime in (e.g.) aluminum, by waiting for the pions to decay.   This method has a limitation of its own.  When the protons strike the target they create a ``flash" of electrons and other particles headed down the muon beam line to the stopping target.  The apparatus requires a clean separation of this flash from the measurement period so that (a) the conversion signal is not hidden in extra activity, (b) that extra hits do not produce misreconstructed tracks, and (c) that the detectors themselves are operational having had time to recover from the high-intensity flash.  Note that as the lifetime of the bound muon decreases, this method then becomes problematic.  The muon lifetime in gold, for example, is only 79 nsec.  With a typical proton pulse of order 100 nsec.\ RMS, the beam flash would overwhelm the detector.  We will see in Sec.~\ref{sec:currentSolenoids} with the  solenoids as in current proposals not only would any conceivable detector would be overwhelmed by the rate from electrons transported through the system but the radiative pion capture background would be ${\cal O}(10^{11})$ higher.

There is an additional source of RPC background.  Antiprotons produced in the production target drift slowly (since they have small kinetic energies) and can annihilate on the stopping target, producing pions.  Those pions are then an additional source of RPC background, and this source of RPCs evades the extinction requirement since the antiprotons move so slowly---there is effectively no time period associated with a pulse that one can wait out. Both Mu2e and COMET use beams at around 8 GeV (8 GeV kinetic energy at FNAL),  with a threshold for $p + p \rightarrow  p + p +p+ \bar{p}$ at 5.6 GeV.  Fermilab's Project X upgrades to Mu2e will use a lower energy (nominally 1--3 GeV) beam with details in \cite{Kronfeld:2013uoa} and this source of background will not be important there since the beam energy will be below threshold.  

There are other processes that produce background but at a smaller level.  Radiative Muon Capture, analogous to Radiative Pion Capture, has a kinematic endpoint on aluminum of 102.5 MeV but \cite{Bergsbusch:1999} indicate the spectrum extends only to  $\sim 90$ MeV, a much lower energy;  therefore at most the electrons produced in this process distort the measured DIO spectrum away from the endpoint.  It is worth noting that this process then contributes a background to the $\Delta L = 2$ process $\mu^- + {\rm (A,Z)} \rightarrow e^+ +{\rm (A, Z-2)}$ described in Sec.~\ref{sec:deltaL2}. The are other ``prompt",  beam-related processes such as muon decay-in-flight (for muons with momentum $>$ 76.5 MeV/$c$ can yield an electron at the conversion energy of 105 MeV), or remnant electrons in the beam.   These backgrounds must be controlled, but the design of the experiment is driven by RPCs and DIOs.  Simply put, RPCs drive the beam structure and design; DIOs drive the detector and resolution issues.

\subsubsection{Experimental History and Status\label{sec:mueconv}}

There is a long history of muon-to-electron conversion experiments, starting in 1952  with \cite{Lagarrigue:1952} in cosmic rays and then moving to accelerators.  A  list is given in Table~\ref{tab:mueconv}.

\begin{table}[h]
\begin{center}
\begin{tabular}{|c|c|c|c|c|c|}\hline
Year&90\% Limit&Lab/Collaboration&Reference&Material\\\hline
1952&	$1.0 	\times 10^{-1}$	&Cosmic Ray	&	\cite{Lagarrigue:1952}	&Sn, Sb\\
1955&	$5.0 	\times 10^{-4}$	&Nevis	&	\cite{Steinberger:1955}	&Cu\\
1961&	$4.0 	\times 10^{-6}$	&LBL	&	\cite{Sard:1961}	&Cu\\
1961&	$5.9 	\times 10^{-6}$	&CERN	&	\cite{Conversi:1961}&	Cu\\
1962&	$2.2 	\times 10^{-7}$	&CERN&		\cite{Conforto:1962}&	Cu\\
1964&	$2.2 	\times 10^{-7}$	&Liverpool	&	\cite{Bartley:1964}&	Cu\\
1972&	$1.6	\times 10^{-8}$& SREL	&	\cite{Bryman:1972}	&Cu\\
1977&	$4.0 	\times 10^{-10}$	&SIN	&	\cite{Badertscher:1977}&	S\\
1982&	$7.0 	\times 10^{-11}$	&SIN	&	\cite{Badertscher:1982}&	S\\
1988&	$4.6 	\times 10^{-12}$	&TRIUMF&		\cite{Ahmad:1988}&	Ti\\
1993& $	4.3 \times 10^{-12}$& 	SINDRUM II&	\cite{Dohmen:1993}&	Ti\\
1996& $	4.6 \times 10^{-11}$	&SINDRUM II&	\cite{Honecker:1996}&	Pb\\
2006	&$7.0 \times 10^{-13}$ &	SINDRUM II&	\cite{Bertl:2006}&	Au\\\hline
\end{tabular}
\end{center}
\caption{History of $\mu^- N \rightarrow e^- N$ conversion experiments.  \cite{Lagarrigue:1952} saw $\approx 1 \sigma$ signals for Sn and Sb; we have averaged their results  and set an approximate limit.  We thank E.~Craig Dukes for help in the preparation of this Table. \label{tab:mueconv}}
\end{table}

The most recent series were the SINDRUM and SINDRUM-II  experiments at the Paul Scherrer Institute (PSI).  We will describe the final SINDRUM-II series  in more detail.  Although it is instructive to follow the upgrades, the essential ideas can be covered in the later experiment. 

 There was a strong desire to measure heavy targets in SINDRUM-II  There were two reasons, one of which is still valid.  As described in \cite{Okada:2002}, the dependence of $R_{\mu e}$ on $Z$ can reveal the nature of the interaction responsible for CLFV.  A second reason was an early calculation of that rate vs. $Z$ predating~\cite{Okada:2002} that made it seem that heavy nuclei would produce a large effect from \cite{Kosmas:1998} and \cite{Kosmas:1990}.  The combination of these reasons explains the historical emphasis on heavy nuclei despite the considerable experimental difficulties described below.

The $\pi$E5 experimental area at PSI of SINDRUM-II used a proton beam with kinetic energy of 590 MeV and a time structure of 0.3 nsec bursts every 19.75 nsec.     It is therefore impractical to use the pulse structure and wait for the pions to decay since the separation between pulses is shorter than the pion lifetime.  The intensity is too high for a veto counter and so the experimenters chose an 8 mm thick CH$_2$ degrader to reduce the RPC (and other prompt) contamination, requiring fewer than $10^4$ pion stops during the total measurement time. Cosmic ray backgrounds using a combination of passive shielding, veto counters, and reconstruction cuts.   The typical muon energy arriving at the end of the transport channel, before the moderator, was  $52 \pm 1 $ MeV/$c$, ideal for the experiment since  one can stop the muons in a well-defined volume.\begin{footnote}{The beam line is tunable; it can select momenta between 20 and 100 MeV/c with resolution of a few percent and choose either positive or negative particles.  As used by MEG, it is used to select positive 28 MeV/c muons  
thereby making a surface muon beam.}\end{footnote}

\begin{figure}[h]
\begin{center}
\includegraphics{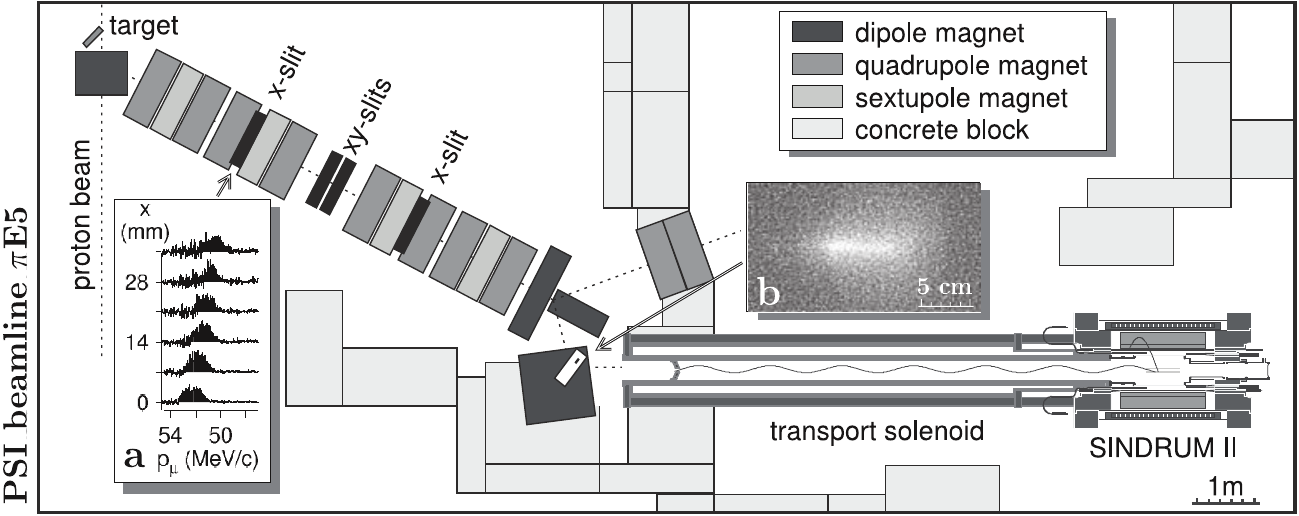}
\end{center}
\caption{\label{fig:sindrum_plan} Plan view of the SINDRUM-II experiment. The 1MW 590 MeV proton beam extracted from the PSI ring cyclotron hits the 40 mm carbon
production target (top left of \ref{fig:sindrum_plan}.) The $\pi$E5 beam line transports secondary particles ($\pi, \mu, e$ ) emitted in the backward
direction to a degrader situated at the entrance of a transport solenoid connected axially to the SINDRUM II spectrometer.   The CH$_2$  degrader preferentially removes pions relative to muons ---  pions have half the range of muons in the degrader.   Inset
a) shows the momentum dispersion measured at the position of the first slit system. The momentum was calculated from the flight
time through the channel and the distributions show the increase when opening one side of the slit. Inset b) shows a cross section
of the beam observed at the position of the beam focus. Caption taken from ~\cite{Bertl:2006}.}
\end{figure}

\begin{figure}[h]
\begin{center}
\includegraphics{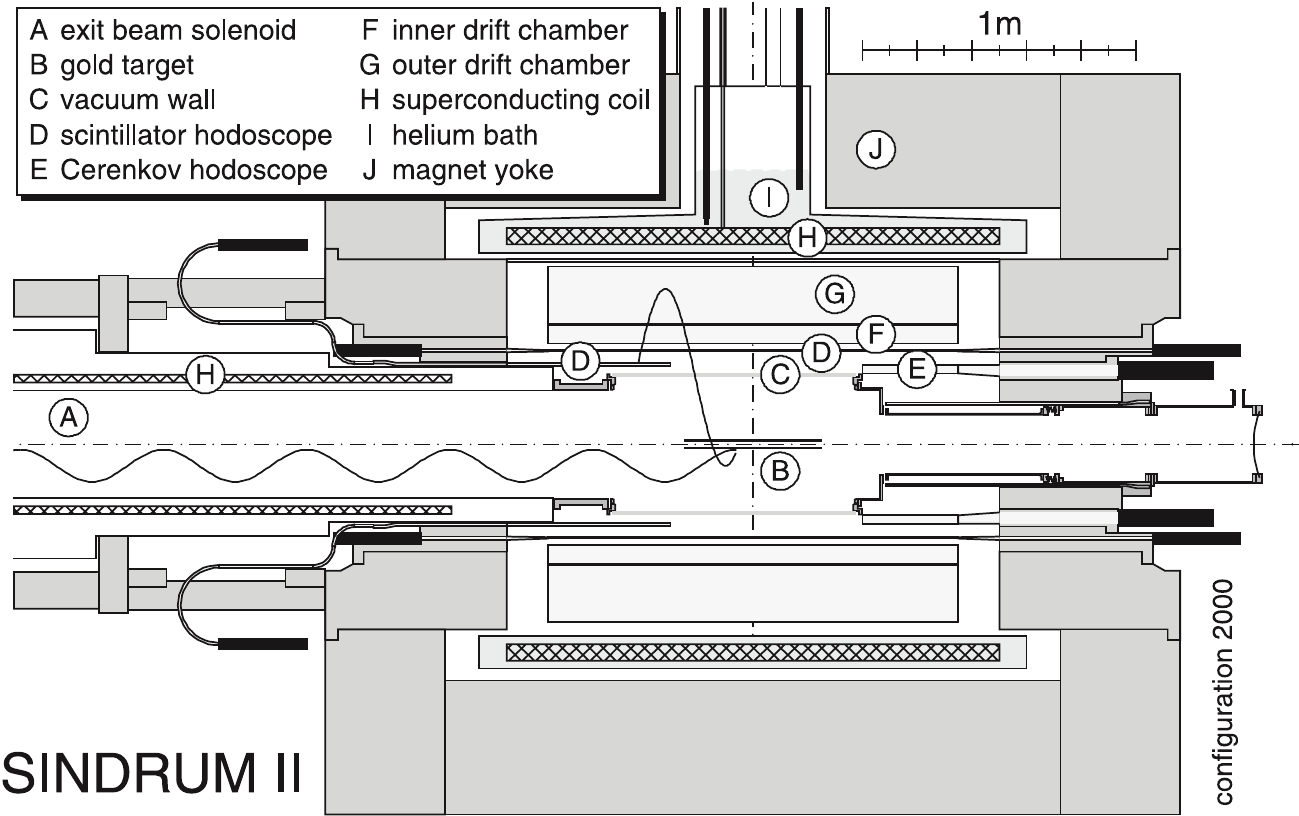}
\end{center}
\caption{\label{fig:sindrum_detector}  The SINDRUM II spectrometer.
Typical trajectories of a beam muon and a hypothetical conversion electron are indicated  (Figure and  caption taken from \cite{Bertl:2006}.)}
\end{figure}

The spectrometer employed a superconducting solenoid, scintillation counters, and drift chambers to track the helical trajectory of conversion electrons.  We see the target was centered in the detector (this will change in the next generation of experiments.)  More specifically, the SINDRUM-II detector consisted of radial drift chambers and a cylindrical array of 64 scintillation counters viewing a hollow double-cone target.  The entire apparatus was in a 0.33 T field with axis parallel to the beam direction.   The series of experiments reached 90\% confidence limits in the 6--7 $\times 10^{-13}$ range, a considerable accomplishment.  

We next examine the results, asking what the limitations were and how subsequent experiments might improve on this impressive series of experiments.

   The beam structure of the $\pi$E5 beam (300 psec bursts every 19.75 nsec) allowed the authors to define two sets of events based on two cuts:
\begin{itemize}
\item A  cut on $\cos \theta$ where $ \theta$ is  the  polar angle of the reconstructed helix .  Small $\cos \theta$ (forward) events are associated with (a) RPCs produced in the degrader itself and (b) pion decay in flight ($\pi^- \rightarrow e^- \bar{\nu}_e$)  in the region just before the degrader. 
\item A cut on $|t _{\rm rf}| < 4.5$ nsec.  $t_{\rm rf}$ is the time of the beam burst. This essentially divides the data sample in two time groups, those near the beam burst and those ``far" from the burst.  This cut preferentially removes RPCs arising from pions striking the target.  
\end{itemize}

\begin{figure}[h]
\begin{center}
\includegraphics{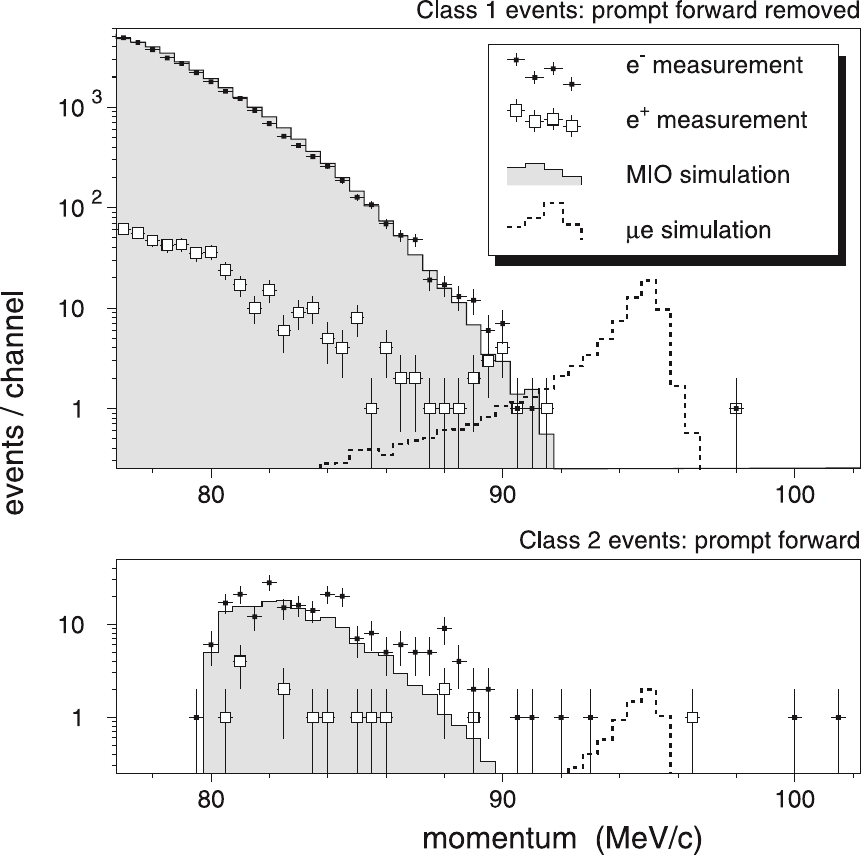}
\end{center}
\caption{\label{fig:sindrum_result}  Momentum distributions of electrons and positrons
for the two event classes. Measured distributions are compared
with the results of simulations of muon decay in orbit and $\mu$ -- $e$
conversion.  ``MIO" is the decay-in-orbit background.   Figure and Caption taken from \cite{Bertl:2006}.}
\end{figure}

Based on these cuts, the authors divided the key data sample into two classes:
\begin{enumerate}[a.]
\item Class 1:  events with $\cos \theta < 0.4$ or $ |t _{\rm rf} - 10  {\rm ~nsec}| > 4.5$ nsec.    These should be muon-based DIO or conversion events and are practically free of pion induced background.
\item Class 2:  events with $\cos \theta > 0.4$ and $ |t _{\rm rf} - 10  {\rm ~nsec}| < 4.5$ nsec.   This is the class that is more likely to arise from pion contamination.
\end{enumerate}

Fig.~\ref{fig:sindrum_result} shows the effect clearly.  However, we note that an event past the signal region still remains even in the Class 1 events.  It is therefore unlikely this technique can be used for an experiment probing significantly smaller values of $R_{\mu e}$ and a new method is required.  Nevertheless, study of these experiments informs us of the problems with radiative pion capture and decays of either pions (or muons, a smaller source of problems) in flight.  Both the next generation and subsequent generations of experiments must design muon beams as free of pion contamination as possible and find clean ways to allow the pions to decay before looking for a conversion signal.  

\subsubsection{Prospects: Mu2e and COMET \label{sec:currentSolenoids}}

Lobashev and collaborators(\cite*{Lobashev:1989} and \cite{Dzhilkibaev:1989zb}) first suggested the basic idea for the next generation of $\mu-e$ conversion experiments at MELC (the Moscow Muon Factory); this idea was then used to develop the MECO (\cite{Bachman:1997}) experiment at BNL, which was canceled because of budget constraints.  The two modern experiments, Mu2e (\cite{Carey:2008},\cite{Abrams:2012} at Fermilab and COMET (\cite{Bryman:2007})at J-PARC, follow Lobashev's initial idea with upgrades and modifications.  COMET and Mu2e  are quite similar in broad outlines.

The first step is to increase the muon intensity.  Fermilab's current accelerator complex can be re-used in the post-Tevatron era.  It can generate $\approx 23$ kW of power from 8 GeV kinetic energy protons, or $2.4 \times 10^{13}$ protons/spill with a spill every 1.33 sec.  One could then have $3.6 \times 10^{20}$ protons/year. The FNAL antiproton rings, not needed after the colliders shut down, are about 1.7 $\mu$sec in circumference. At J-PARC,  the beam power is 56 kW and the proton beam has 8 GeV total energy.    Both experiments will have bunch lengths of approximately 100 nsec.  The J-PARC bunch time separation is between approximately 1 and 1.3 $\mu$sec, depending on the precise beam delivery scheme.  Both FNAL and J-PARC  are well suited for searches in aluminum or, for example, titanium with a lifetime of about 338 nsec.

\begin{figure}[h]
\begin{center}
\includegraphics[scale=0.6]{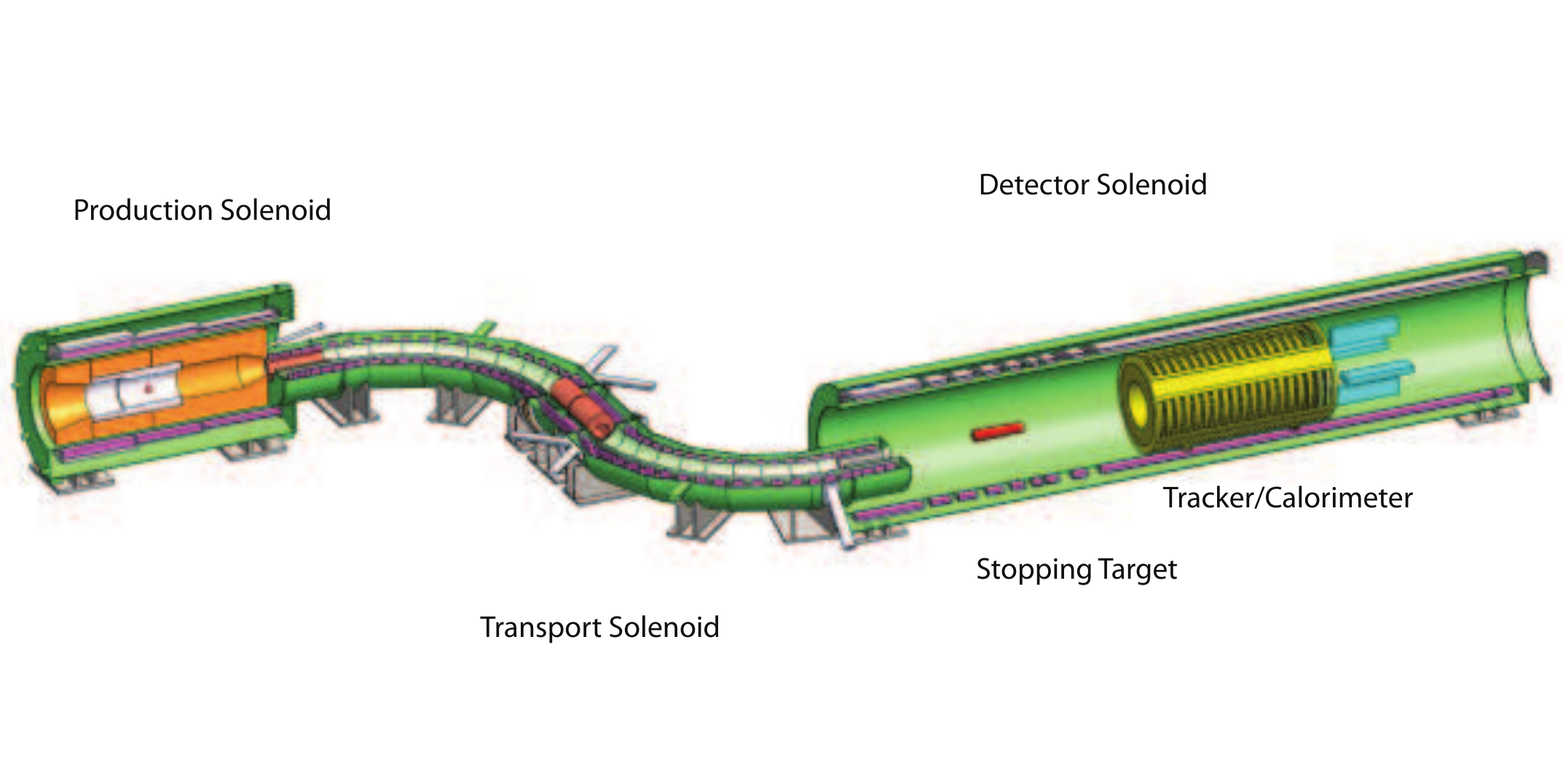}
\end{center}
\caption{\label{fig:mu2eOverview} Overview of the Mu2e muon-electron conversion experiment.}
\end{figure}

\begin{figure}[h]
\begin{center}
\includegraphics[scale=0.6]{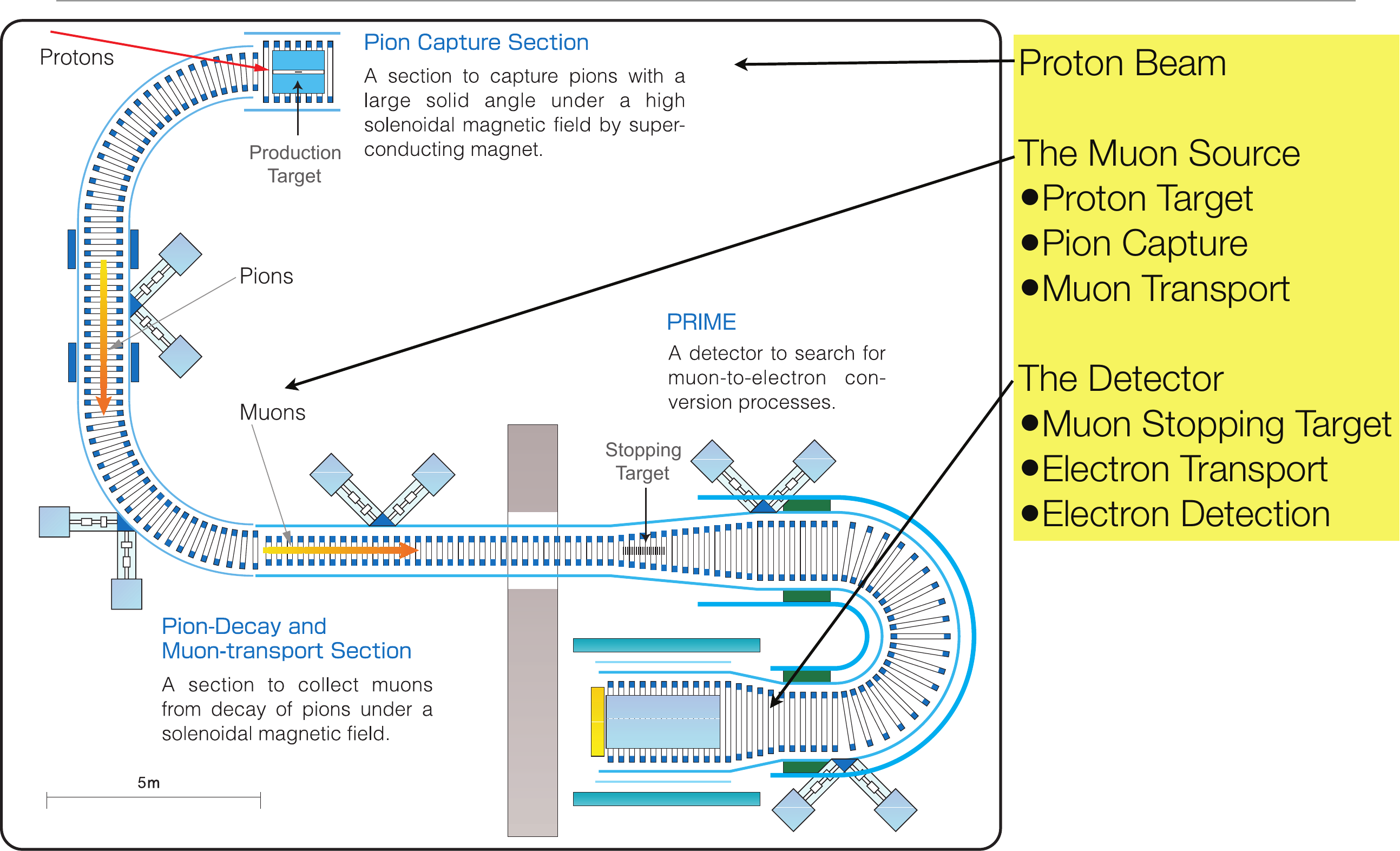}
\end{center}
\caption{\label{fig:cometOverview} Overview of the COMET muon-electron conversion experiment. Figure taken from \cite{Kuno:2008zz}. }
\end{figure}

The experimental design of both Mu2e and COMET looks fundamentally different from the SINDRUM series.  At PSI, SINDRUM and SINDRUM-II brought muons to rest and surrounded the stopping target with the detector, like the $\mu \rightarrow e \gamma$ experiments.  The next generation resembles  a ``fixed-target" experiment with a stopping target and a spectrometer downstream.   The experiments are based on the following concepts:
\begin{enumerate}
\item A pulsed proton beam with the time between pulses approximately the muon lifetime in the stopping target.   Recall muons have a total lifetime of 864 nsec in aluminum.(\cite{Measday:2001} )  Thus one can send in a short pulse of beam and wait for prompt backgrounds such as RPCs to die away before beginning a ``measurement period."  
\item A graded field solenoid to collect pions and allow them to decay into muons.  The field is graded almost opposite to the direction of the incoming proton direction.  Therefore the experiments capture primarily backwards-going muons but reflections in the graded field add $\approx 15$\% to the rate.  This method yields about a $10^3$ increase in the number of muons/second relative to SINDRUM-II, up to about $10^{11}$.  The muons need to be low energy (typically 40 MeV kinetic energy) so that they can be stopped and subsequently captured by a target nucleus in a stopping target.  
\item A curved solenoid then``transports" the muons to a final solenoid containing the detector.  This Transport Solenoid has a curved shape that eliminates line-of-sight neutrals and similar backgrounds.  Mu2e and COMET use different designs for this section.  As shown in Figs.~\ref{fig:mu2eOverview} and  \ref{fig:cometOverview}, Mu2e uses a ``S"-shape, and COMET an ``C".  The S-shape provides somewhat more rate ($\approx 30$\%) according to the Mu2e simulations, but the COMET C tends to produce a tighter time and momentum distribution.  The curved solenoid serves to momentum and charge-select.  As derived in \cite{Jackson:1975}, charged particles following a curved solenoid are deflected with positives and negatives receiving opposite deflections.  Mu2e uses a rotating central collimator in the central straight section of the ``S" to select negative muons and eliminate other particles.
\item Finally, there is a Detector Solenoid containing the stopping target and the detectors.  The muons stop in the stopping target and any electrons from conversion are then identified.  Again, the Mu2e and COMET designs diverge.  COMET has a final bend after the stopping target, largely eliminating particles ejected when the muons stop at the loss of some acceptance for electrons.  
\end{enumerate}

The prompt backgrounds will be suppressed because the pions will decay as they travel from the production target, where they are born, to the stopping target and detector.  In Mu2e or COMET, they spiral in helical paths through solenoids over $\sim 12$ m and the suppression is ${\cal O} (10^{-11})$ or more.   Protons in between pulses can evade or greatly lessen this suppression depending on when they arrive --- if protons arrive ``late" they can produce prompt background inside the measurement period.   The experiments define an 	``extinction" as the ratio of out-of-pulse to in-pulse protons.  The required extinction is obviously a function of time relative to the pulse and the beginning of the measurement period, but crudely modeling the extinction factor as flat over the out-of-pulse period.  both Mu2e and COMET find they need an extinction factor at the level of $10^{-9}$--$10^{-10}$.

A tracking system, probably made from straw tubes, then tracks the outgoing conversion electrons and provides their momentum.  A downstream calorimeter may aid in background rejection from catastrophic misreconstruction.  One important point (again see \cite{Measday:2001} and references therein) is that   roughly 0.1 $p$, 2 $n$, and 0.1 $\gamma$'s are produced per captured muon stop and these particles, especially the slow, highly-ionizing protons,  can overwhelm the tracking detector. The two experiments cope with this problem in different ways.  The two experiments also cope with the decay-in-orbit backgrounds differently.

The two essential differences between the designs are:
\begin{enumerate}
\item Mu2e has an ``S-shaped" solenoid for the curved Transport Solenoid and COMET uses a ``C"-shape.  The C yields somewhat less flux but a smaller momentum spread.  The smaller momentum spread aids in the design of the stopping target and could reduce the spread in energy loss, yielding a cleaner signal.  The time distribution of arriving muons is also  tighter than in the S-shape.
\item COMET has a curved solenoid after the stopping target. The particles associated with muon capture do not pass through the curve in the Detector Solenoid.  In addition,  most of the decay-in-orbit events (including those up to the Michel peak and beyond) are stopped in the curve.  
\end{enumerate}

Mu2e solves the first problem with a ``proton absorber" that filters the protons associated with muon capture.  Instead of using a second ``C" curve to filter out low energy decay-in-orbit events, Mu2e has a central hole sized so that only a small number ($\sim 100$K in the proposed run) have sufficient $p_{\perp}$ to be seen by the tracker or calorimeter.  COMET's second ``C" entails a loss of acceptance and does not perfectly reject decay-in-orbit events so that a ``DIO-blocker" must be employed.  Optimizations of both experiments are being performed at this writing.

There are two other ideas which should be included in this article even though they are in an early stage.  The first, DeeMe, is a proposal at J-PARC to use conversions {\em in the production target} to search for muon-to-electron conversion in a silicon-carbide target.\cite{Aoki:2012}  This experiments hopes to improve on the SINDRUM-II sensitivity by up to two orders-of-magnitude.  COMET has suggested a phased approach for their effort where the construct the first half of their C transport solenoid and possibly build a SINDRUM-style detector.  After a brief run of 12 days they hope to obtain a similar sensitivity to DeeMe.  

\subsubsection{Prospects for Future Muon-to-Electron Conversion Experiments}

 It is perhaps an excellent example of {\it hubris}, while still planning experiments $10^4$ times better than existing ones, to  consider experiments an additional two or more orders-of-magnitude beyond those.  Nonetheless we can set out some general ideas and what the limitations might be.  
 
The experimental goals for a search post-Mu2e or COMET depend on what is seen by those experiments.  In the case of a signal, the new physics must be pinned down and explored.   Perhaps the most powerful discriminant is to change the $Z$ of the capturing nucleus as explained by \cite{Okada:2009}.  Looking at higher-Z nuclei presents challenges.  The lifetime of the muonic atom  shrinks as $Z$ grows (e.g. $\tau_{{\rm Al}} = 864$ nsec and $\tau_{ {\rm Au}} =  72.6$ nsec.)  Since a typical proton pulse is between tens and a couple of hundred nsec,  the ability to wait until after the beam flash, ensuring a quiet detector, disappears for a 72.6 nsec Au target.  Therefore the solenoid system in Mu2e or COMET probably will not work.  One could lengthen the central ``C" or ``S" solenoid, or perhaps add another bend, but then of course more muons would decay and this would not guarantee that no electrons were transmitted down the muon beamline.  

Two more technical experimental issues should be covered in this discussion.  First, cosmic rays are a potentially fatal background.  There are a number of ways in which cosmic rays can produce background, but perhaps the most insidious is in the stopping target itself.  A cosmic ray muon can pass through the stopping target and eject an electron in the signal region, as discussed earlier in Sec.~\ref{sec:mu2e}.  Such an electron is indistinguishable from the signal.  \cite{Abrams:2012} tell us that in the Mu2e experiment one would find about one such electron per day of running.  This requires a cosmic ray veto system surrounding the detector at some appropriate degree of efficiency.

The cosmic ray veto is made more complicated from neutron associated backgrounds.  The most copious source of neutrons is the primary production target.  Such neutrons would overwhelm any detector with raw rate and require considerable shielding.  The neutrons can also stop in the cosmic ray veto material and produce a 2.2 MeV photon which then fires the veto.  Furthermore, \cite{Abrams:2012} point out that photo-sensensors such as APDs can fail after being exposed to ${\cal O} (10^{10})$ neutrons/cm$^2$.  One could go to ``neutron-blind" technologies such as cathode-strip chambers. A second source of neutrons is the stopping target itself as discussed earlier.  While Mu2e and COMET believe they have solved the problems, going to higher power systems or longer runs or more demanding more stringent limits on $R_{\mu e}$ will put great demands on the cosmic ray veto system and managing the neutron flux.  

Here, technologies of muon cooling or FFAGs may be the right next step.  In either case, one can capture and hold the muons until the pions have decayed and the beam flash has ended before directing the muons to the stopping target and place the production target, with its flood of neutrons, far from the apparatus.  Designs for muon beams for ``next-generation" muon-to-electron conversion experiments are naturally part of the planning for neutrino factories and research is ongoing and the literature on the subject is vast; see, for example, \cite{Johnson:2005}.  Some ideas for a Mu2e-like system are discussed by \cite{Knoepfel:2013ouy}.

Assuming one could successfully design a muon beam, a number of detector problems then present themselves.  Energy loss in the stopping target smears out the conversion peak, and since the process is stochastic, DIO events with small losses can fall under the smeared-out signal peak.  An obvious way to solve this problem is to cut on the reconstructed momentum hard enough that a negligible number of DIOs remain, and since the spectrum falls as $(E - E_{\rm conv})^5$ this can be a productive strategy.  Unfortunately one then loses acceptance for the conversion events with ``large" energy loss.  A more productive strategy could be to minimize the energy loss with the stopping target design.  For example, targets shaped along Archimedean screws following the trajectory of the electron would minimize its interactions.  Unfortunately such a scheme then {\em increases} the energy loss for positively charged particles.  The problems are then: (a) one may want to use the two-body decay $\pi^+ \rightarrow e \nu$, yielding monochromatic electrons, as a calibration, and (b) a search for outgoing positrons in $\mu^- N \rightarrow e^+N$ (discussed later in Sec.~\ref{sec:deltaL2}) will be compromised.

Another notion is to phase-rotate the beam so that the muons entering the stopping target are at a well-defined energy and all stop in the same place: the current Mu2e and COMET designs stop their muons over just under a meter of spaced foils.  One would employ a single, thin target (this does not work in the upcoming generation simply because the smaller amount of material means fewer stops and less sensitivity to conversions.)

The detectors for such experiments present their own set of difficulties.  The muon capture process produces photons, neutrons, and protons, as we have discussed.  This is an intrinsic source of extra hits that scales with stopped muons.  The problem is compounded if the experiment runs at even higher instantaneous intensities at next-generation experiments.

We can therefore see the following dilemma: suppose we reduce the number of foils and increase the number of muons/sec so that we can have a well-defined vertex with minimum, constant energy loss but maintain the statistics.  Then the detector can easily by overwhelmed by the increase in instantaneous rate from the stopping process, such that the delta-rays and occupancy becomes intolerable.  

If we switch to a higher-$Z$ target then there are two competing effects: the capture fraction increases relative to decays, but the fraction of DIOs near the signal peak increases.  \cite{Czarnecki:2011} calculate that even though the total number of DIO events in Ti, for example, is six times lower than in Al, the problematic part of the spectrum near the conversion peak is six times {\em higher} in Ti than in Al. 

We have thus shown there are  difficulties with the beam flash, but even if that problem is solved by improved muon beam technology, the intrinsic backgrounds are quite dangerous.  One could attack this in a variety of ways: for example, one could place a bend or some channel after the stopping target, as in COMET.  Unfortunately all of them tend to lower the stopping rate.  The challenges in detector technology in order to avoid lowering the rate while still maintaining (even-better) resolution are considerable.  Only by performing the current generation will we have the necessary information on how to proceed.

\subsection{Calibration Issues in Future Muon-Electron Conversion Experiments}

\begin{figure}[h]
\begin{center}
\includegraphics[scale=0.8]{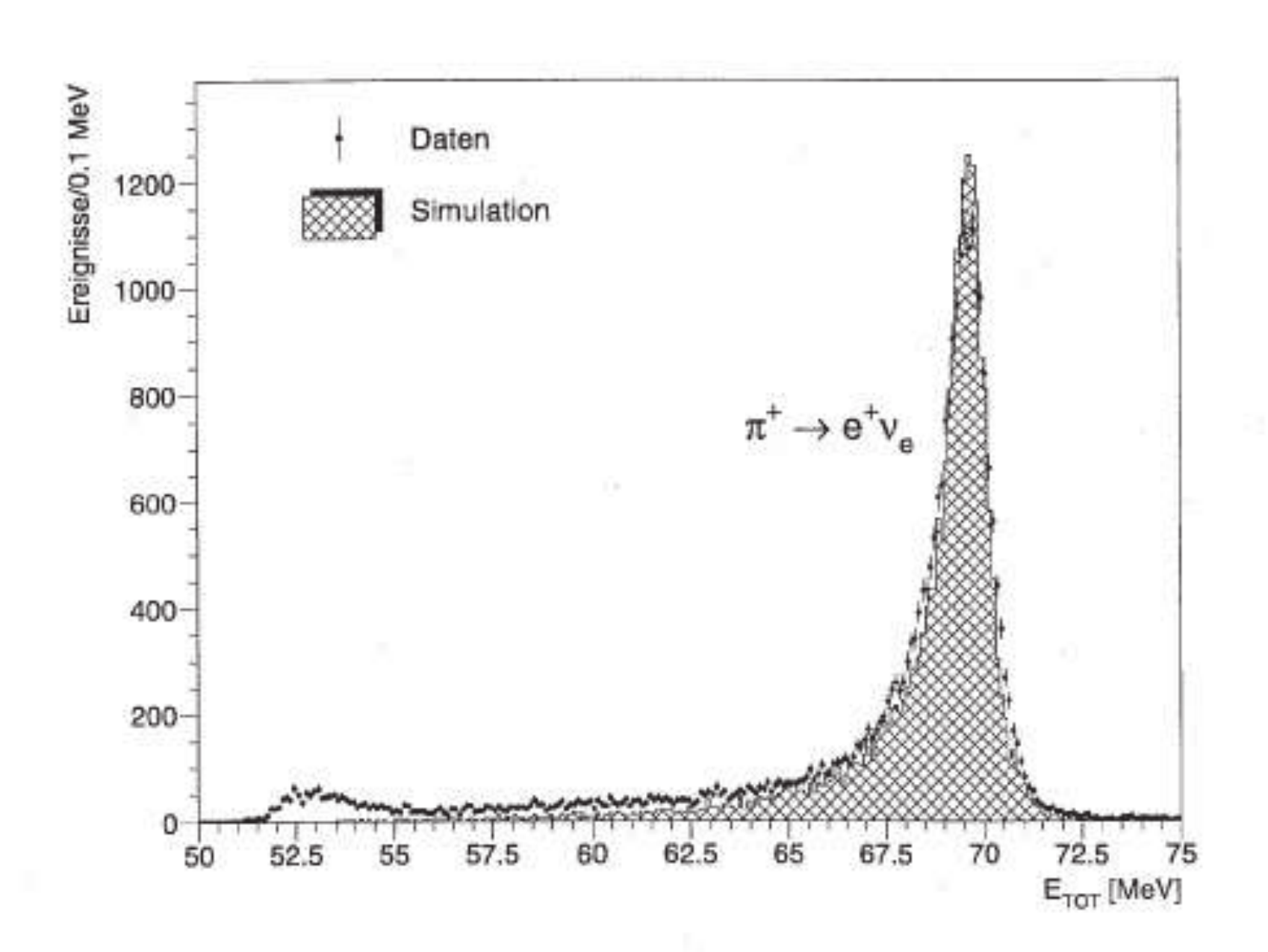}
\end{center}
\caption{Calibration of the SINDRUM-II Spectrometer from $\pi^+ \rightarrow e \nu_e$.  Peaks at 69.8 MeV/$c$ from the $\pi^+ \rightarrow e \nu$ two-body decay and the lower-energy Michel peak are evident.  The original caption tells us the spectrum was shifted by $+0.1$ MeV. Figure taken from \cite{Dohmen:1993}. \label{fig:dohmen} }
\end{figure}

The absolute calibration of the momentum scale will be important in future muon-to-electron conversion experiments.  Prior experiments at SINDRUM-II used $\pi^+ \rightarrow e \nu_e$ decays.  The resultant electrons are monoenergetic at 69.8 MeV/$c$.  
The quality of the calibration is shown in Fig.~\ref{fig:dohmen}.  It is clear that a shift of $\approx$ 100-200 keV is certainly possible.  

How important is this calibration? One needs to set cuts to define a range of accepted momenta (this is effectively true if one uses a shape analysis or ``cut and count", so we imagine we are in the latter situation.)  The radiative pion capture background yields a relatively flat electron spectrum, so the precise cuts are unimportant.  However, the decay-in-orbit background is rising rapidly.  Both COMET and Mu2e use signal regions close to $103.5 < E_e < 105$ MeV/$c$.  Using the decay-in-orbit spectrum from \cite{Czarnecki:2011} and a toy simulation of the Mu2e apparatus, one finds  a -200 keV calibration error can produce abut one background event.   This problem is being addressed in Mu2e using the S-shaped section of the solenoid system.  Particles of negative charge are deflected perpendicular to the axis of the solenoid for a curved solenoid and particles of positive charge are deflected in the opposite direction.  Mu2e employs a rotating collimator to filter out positives for normal running; by reversing this collimator one can selected positives, which will yield a $\pi^+$ sample for the $\pi e 2$ calibration.  This calibration, as we saw in Fig.~\ref{fig:dohmen}, is quite difficult.  Simple extrapolations indicate it should work for the Mu2e/COMET required sensitivities, but a next generation experiment could easily be limited by the absolute knowledge of the field and momentum scale.  A natural idea might be to use an ``electron gun" from a small accelerator to fire electrons of known momentum into the solenoid and perform an {\it in situ} calibration, but at the current time this idea is no more than fanciful (although surely expensive.)  Experiments that go beyond the planned generation must squarely face this problem if they hope to push the limit on muon-electron conversion.  The calibration problem is not generally part of the discussion about future facilities, since it requires a fairly deep understanding of the experiments, and we hope this article calls this problem to the attention of the community.

\subsection{$\mu^{\pm} \rightarrow e^{\pm}e^+e^-$}

The decay $\mu \rightarrow 3e$ is of great interest; it is sensitive to supersymmetry, littlest Higgs scenarios, leptoquarks, and other physics models and is complementary to the other modes.  The decay mode has signatures in a wide variety of BSM physics models: see \cite{Blondel:2012} for references.  The mode has been examined in Littlest Higgs scenarios by \cite{Blanke:2007db}.  An investigation of $\mu \rightarrow 3e$ in polarized muons beams can be found in \cite{Okada:1997fz}. The case is nicely made for a leptoquark model in \cite{Babu:2010} (and then see Sec.~\ref{sec:eic} for further discussion.)

A new measurement should strive to set a limit $< ~{\cal O}(10^{-16})$ to be competitive with existing limits and other planned measurements.  The current limit from \cite{Bellgardt:1988} in SINDRUM is $B(\mu \rightarrow 3e) < 1.0 \times 10^{-12}$ at 90\% CL. Therefore a factor of $10^4$ improvement is required.  
With a $10^7$ second run, one then requires $10^{9-10}$ decays/sec before acceptances, etc.\ are included.  The current $\pi e5$ (MEG) beamline yields about $10^9$ muons/sec, barely enough.  A proposed spallation neutron source at PSI (SINQ, \url{http://www.psi.ch/sinq/} ) could provide $5 \times 10^{10}$ muons/sec, probably an effective minimum requirement.

Existing experiments have used stopped muons and muon decay-at-rest.  In that case the outgoing electron and positrons can be tracked and the kinematic constraints $\left | \,  \sum \vec{p} \, \right |= 0$ and $\sum E = m_e$, along with timing, can then be used to identify the rare decay. 

Unfortunately, this mode suffers from many of the same problems as $\mu \rightarrow e \gamma$. Because it is a decay, unlike muon-to-electron conversion, $\mu \rightarrow 3e$ electrons are in the same momentum range as ordinary Michel decays.  Therefore there are accidental backgrounds from Michel positrons that coincide with $e^+e^-$ pairs from $\gamma$ conversions or from other Michel positrons that undergo Bhabha scattering.  ( One could cut on the opening angle between the positrons and each of the electrons, since conversions tend to have a small opening angle, but  if the $\mu \rightarrow 3e$ process occurs through processes with a photon, one then loses acceptance.)

\begin{table}[h]
\begin{center}
\begin{tabular}{|l|c|c|c|}\hline
Year&90\% CL Limit &Collaboration/Lab&Reference\\\hline
1958	&$3.0 \times 10^{-5}$&	Nevis	&	\cite{Lynch:1958}\\
1959	&$5.0 \times 10^{-5}$&	Nevis	&	\citet*{Lee:1959}\\
1961	&$4.0 \times 10^{-6}$	&Carnegie	&	\cite{Crittenden:1961}\\
1962	&$5.0 \times 10^{-7}$	&Chicago	&\cite{Parker:1962}	\\
1976&	$1.9 \times 10^{-9}$	&Dubna		&\cite{Korenchenko:1976}\\
1984&	$1.3 \times 10^{-10}$	&LAMPF/Crystal Box &	\cite{Bolton:1984}\\
1984&	$1.6 \times 10^{-10}$	&SIN/SINDRUM&	\cite{Bertl:1984}\\
1985&	$2.4 \times 10^{-12}$	&SIN/SINDRUM&	\cite{Bertl:1985}\\
1988&	$3.5 \times 10^{-11}$	&LAMPF/Crystal Box &\cite{Bolton:1988}\\
1988&	$1.0 \times 10^{-12}$	&SIN/SINDRUM	&\cite{Bellgardt:1988}\\
1990&	$3.6 \times 10^{-11}$	&JINR	&	\cite{Baranov:1991}\\\hline
\end{tabular}
\caption{History of $\mu \rightarrow 3e$ results.}
\end{center}
\end{table}

 This leads to a requirement for a high duty-factor muon beam as employed by SINDRUM or MEG.     A second class of background comes from $\mu \rightarrow 3e \nu\nu$ radiative decays, requiring excellent momentum resolution to eliminate low energy neutrinos.  The radiative process has a branching fraction ${\cal B} = 3.4 \times 10^{-5}$, large compared to the $10^{-16}$ requirement.  The precise form is derived in \citet*{Kuno:1991}.  A more useful way to look at the problem is to consider the background by examining  $m_{\mu} - E_{\rm tot}$, where $E_{\rm tot}$ is the observed total energy of the three electrons.  \cite{Djilkibaev:2009} derive for the branching fraction:
\begin{eqnarray}
R&=& 2.99 \times 10^{-19} \left ( \frac{m_{\mu} - E_{\rm tot}}{m_e} \right)^6 \label{eqn:mu3eback}
\end{eqnarray}
near the endpoint of the spectrum where there is small missing energy.  The differential spectrum that gives Eqn.~\ref{eqn:mu3eback}, convoluted with the detector resolution and energy loss, then yields the background.   Because of the rapid rise of the background,  $(m_{\mu} - E_{\rm tot})^6$,  a $\mu \rightarrow 3e$ search requires  excellent tracking and momentum resolution along with minimum energy loss in the tracking material.  MEG certainly surpassed SINDRUM in these regards, but the COBRA spectrometer of MEG is optimized for the high end of the Michel spectrum and could not be adapted.  In order to suppress the rapidly rising background to the level of $10^{-16}$ an energy resolution in $E_{ {\rm tot}}$ of better than 1 MeV is required, with corresponding resolution for the individual tracks.

We show results from \cite{Bellgardt:1988}, performed with the SINDRUM-II apparatus, to illustrate how the analysis proceeds.  Recall the SINDRUM-II detector consists of five concentric MWPCs, a stopping target, and a solenoidal field as described in Sec.~\ref{sec:mueconv}.  The tracks were examined to check for a common vertex, an obvious requirement.  

 Kinematic constraints were then applied.  For a $\mu \rightarrow 3e$ decay, since 
\begin{eqnarray}
{ \textstyle\sum_i} E_i &=& m_{\mu}c^2\\ \nonumber
\left| \, {\textstyle \sum_i} \vec{p_i} \,  \right| &=& 0 \label{eqn:mu3evecsum}
\end{eqnarray}
the analysis defined a kinematically allowed region:
\begin{eqnarray}
{\textstyle \sum_i} E_i + \left| \, { \textstyle \sum_i} \vec{p_i} \, \right| c &\leq& m_{\mu}c^2 \label{eqn:mu3ekin}
\end{eqnarray}

The next step defined a $\Delta t$ in that allowed region between the $e^+e^-$ pair with the smallest invariant mass compared to the time of the second positron (as obtained from the scintillation counters.)   One sees a peak near $\Delta t = 0$ with a flat background.  The peak was interpreted as $\mu^+ \rightarrow e^+e^-e^+ 2\nu$ decays with a potential $\mu \rightarrow 3e$ signal.   

After an event had passed vertex and timing cuts, the final selection was made on the basis of Eqn.~\ref{eqn:mu3evecsum}.  The analysis actually chose to examine
\begin{eqnarray}
\hat{p}^2 &=& \left( p_{\perp}/\sigma_{p_{\perp}} \right)^2 + \left( p_{\parallel}/\sigma_{p_{\parallel}} \right)^2
\end{eqnarray}
 since the uncertainties on the components perpendicular and parallel to the field axis differed significantly: 0.7 MeV/$c$ and 1.8 MeV/$c$ respectively.  The final selection was made in a two-dimensional distribution of $\hat{p}$ vs.\ ${\textstyle \sum_i} E_i$, as shown in Fig.~\ref{fig:bellgardtkine}.

  \begin{figure}[h]
\begin{center}
\includegraphics[scale=0.7]{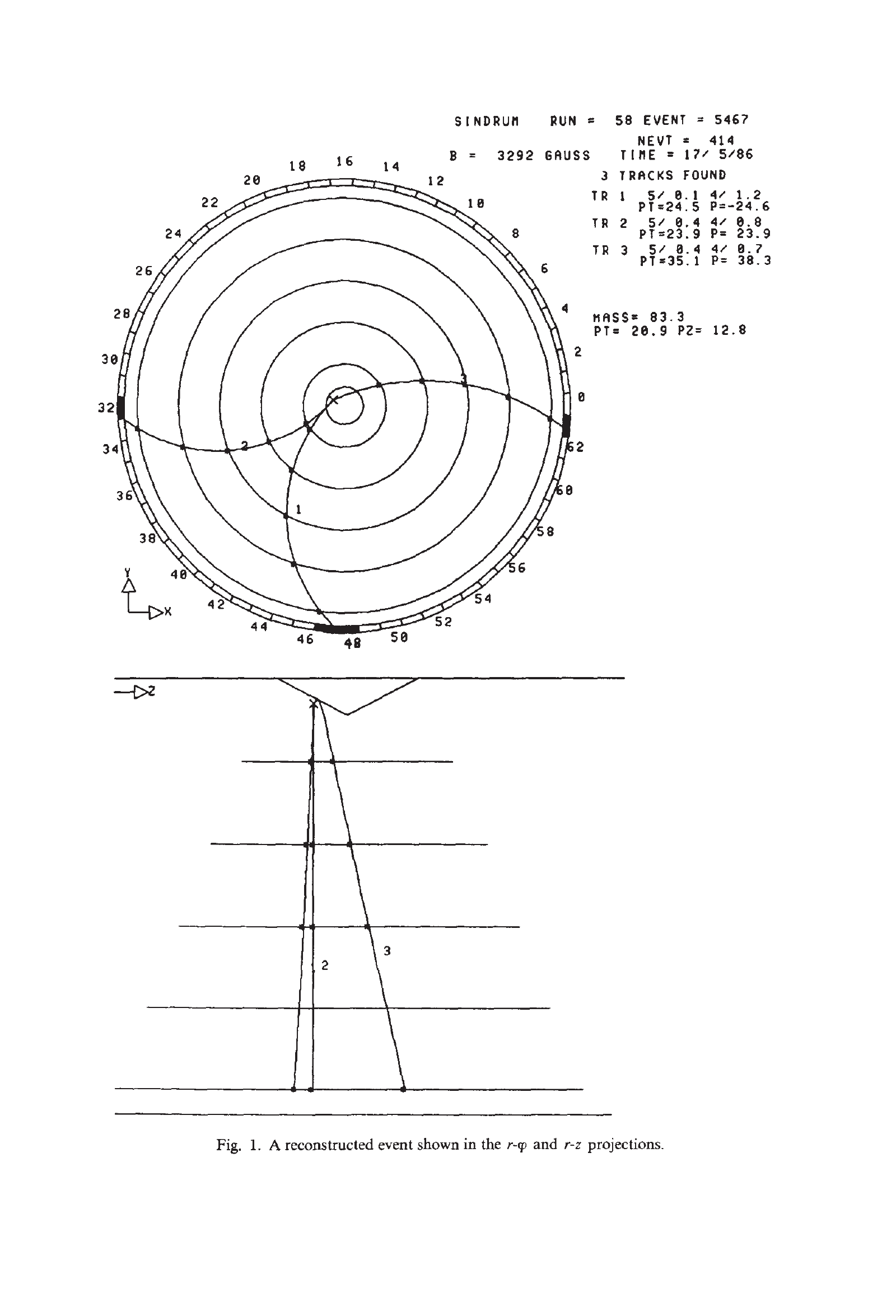}
\end{center}
\caption{\label{fig:bellgardttracker} Tracking for a typical candidate event in the  $\mu^+ \rightarrow e^+e^-e^-$ search at  SINDRUM-II. Figure taken from \cite{Bellgardt:1988}.}
\end{figure}

  \begin{figure}[h]
\begin{center}
\includegraphics[scale=0.8]{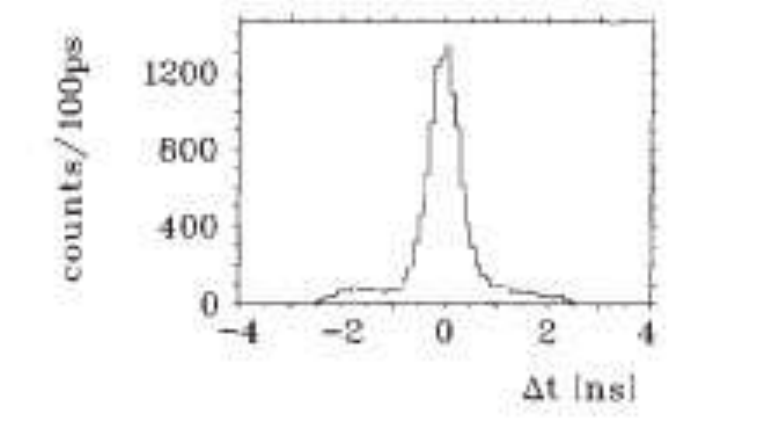}
\end{center}
\caption{\label{fig:bellgardttiming} The timing distribution for the kinematically allowed region for the $\mu^+ \rightarrow e^+e^-e^-$ search at SINDRUM-II. Figure taken from \cite{Bellgardt:1988}.}
\end{figure}

  \begin{figure}[h]
\begin{center}
\includegraphics[scale=0.8]{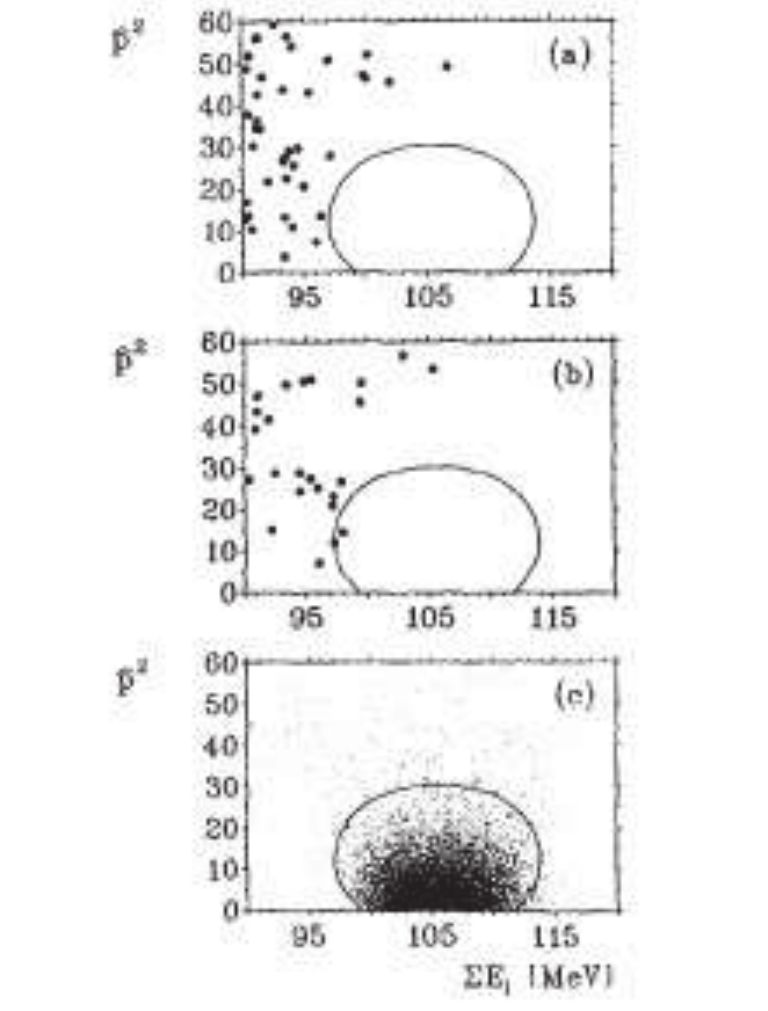}
\end{center}
\caption{\label{fig:bellgardtkine} Final kinematic selection for $\mu^+ \rightarrow e^+e^-e^-$ search at SINDRUM-II. The top distribution is for ``prompt" events within timing cuts, the center plot is for accidentals, and the bottom plot is for a simulated $\mu \rightarrow 3e$ signal, where the contours define a region containing 95\% of the signal. Figure taken from \cite{Bellgardt:1988}.}
\end{figure}

\begin{figure}[h]
\begin{center}
\includegraphics[scale=0.8]{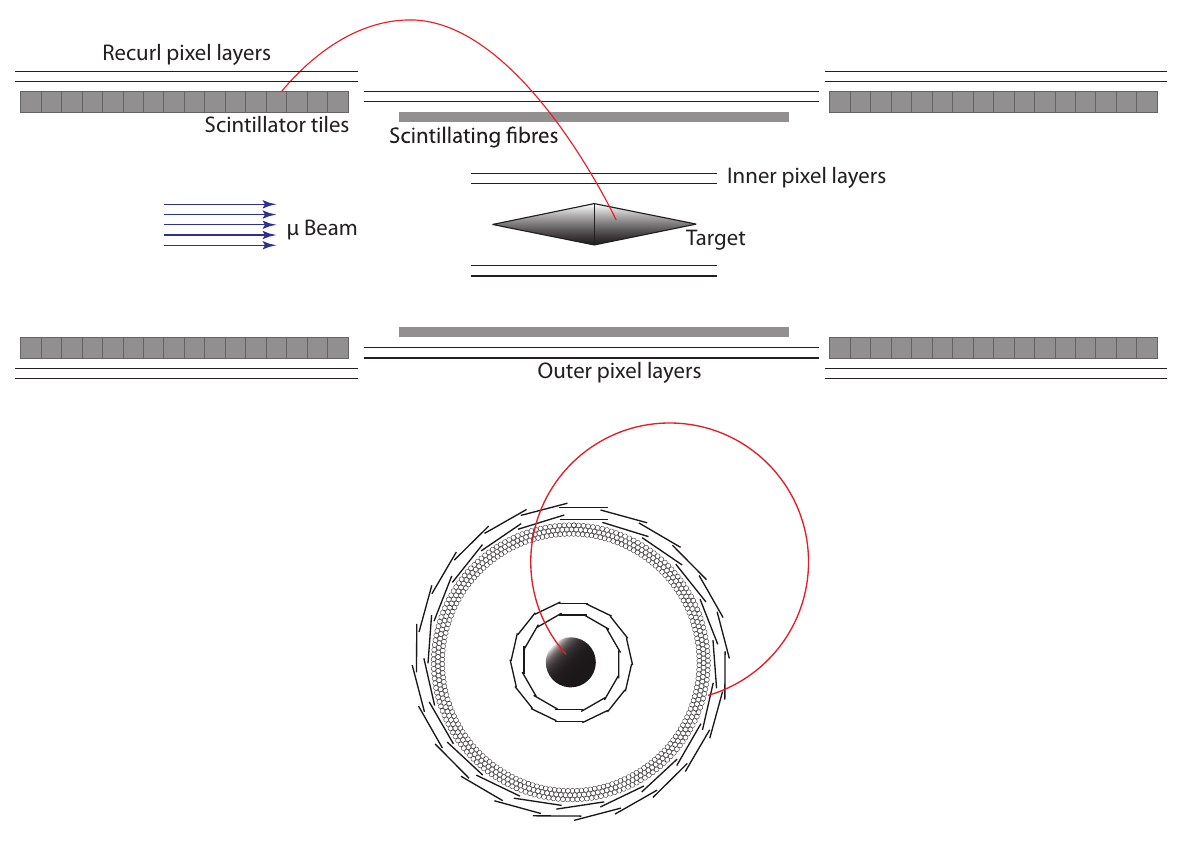}
\end{center}
\caption{\label{fig:mu3ePSI} Proposed $\mu \rightarrow 3e$ apparatus at PSI. Figure taken from \cite{Blondel:2012}.}
\end{figure}

\cite{Berger:2011}  are investigating a new experiment using monolithic active pixel sensors;  the experiment has just received preliminary approval at PSI.  As detailed in \cite{Blondel:2012} the proponents plan to overcome the difficulties above by making the tracking material so thin that multiple scattering is small and backgrounds from radiative muon decay  are negligible. The apparatus is depicted in Fig.~\ref{fig:mu3ePSI}.  Variations are also being considered.  The location of the experiment is a matter of logistics, time-sharing with MEG, etc.  A first-round would achieve $10^{-15}$ with eventual improvements in the beam (possibly moving  to a spallation neutron source at PSI) and the detector yielding a potential limit of $10^{-16}$.  The phase space for accepting the radiative decays and their being indistinguishable from a $\mu \rightarrow 3e$ signal may be the ultimate limitation of these experiments.

\subsection{$|\Delta L = 2|$ Processes}

\subsubsection{$\Delta L = 2$ Transitions and Muonic Atoms \label{sec:deltaL2}}

 The $\Delta L=2$ process
 \begin{eqnarray}
 \mu^- + (A,Z) \rightarrow e^+ + (A,Z-2)
 \end{eqnarray}
 is of interest as well.  As described in \citet{Littenberg:2000} this mode searches for $\left| \Delta L \right| = 2$ transitions with $\left| \Delta L_e \right| \left| \Delta L_{\mu} \right|= \pm 1$.  The decay is intimately related to $K^+  \rightarrow \pi^- \l^+ (l^{\prime})^+$ transitions and  neutrinoless double $\beta$ decay as covered in Sec.~\ref{sec:chargedK}.
 
Experiments have been performed and continue to be proposed as described in \cite{Kuno:2010}.  It is also possible to use the next generation of $\mu e$ conversion experiments to search for these as well if they can be configured to look for both electrons and positrons.  It would be best to have a charge-symmetric detector; the overhead with reconfiguration and the likely time needed to run the main search successfully make ``add-on" experiments unlikely unless the experiment is designed for both at once.  
 
  The $\Delta L=2$ process is in many ways similar to muon-electron conversion.  A single positron is produced at 
 \begin{eqnarray}
 E_e &=& m_{\mu} - B_{\mu} - E_{\rm recoil} - \Delta_{Z-2}
 \end{eqnarray}
 where $\Delta_{Z-2}$ is the difference in nuclear binding energy between the final and initial nuclear states (the other terms are as in muon-electron conversion.)
 
However, this mode suffers from experimental difficulties not present in muon-electron conversion.  First, since the initial and final nuclear states are different:  it is not a coherent process; therefore it is not amplified by $Z$.  Therefore the ``intrinsic" rate is lower. Next, the enormous advantage of the monoenergetic electron of $\mu e$ conversion does not apply.  Since the initial and final states are different, the final nucleus can be in either the ground or excited states. If the excited state is a giant dipole resonance, the width of the final state is $\approx 20$ MeV (together with a downward shift of about 20 MeV) and so the positron is far from monoenergetic.  In this case, both radiative pion capture and radiative muon capture become  backgrounds: emitted photons that convert can produce positrons in the signal region and the RMC rate in the relevant region is not negligible.  In the case of an Al target, the final state is Na; for Ti, as discussed below, the final state is Ca.  A list of GDR cross-sections, widths, and other properties can be found in \cite{Varlamov:1999}.

 There have been a series of experiments searching for $\mu^- \rightarrow e^+$ transitions summarized in Table~\ref{tab:delta2}.  The last, \cite{Kaulard:1998} in SINDRUM-II on a Ti stopping target, set limits for transitions to the ground and GDR states separately at  $1.7 \times 10^{-12}$ and $3.6 \times 10^{-11}$ respectively.  There were two main limitations.  First, there were  backgrounds from scattered electrons in a final collimator.  These were identified by timing relative to a beam counter.  However, there was a  background component outside the timing window that was never understood.  The maximum RMC electron energy is $91.4 \pm 2.0$ MeV (based on an argument in the text) but events were observed  up to 3 MeV beyond the endpoint, well outside the resolution.  The experimenters interpreted this as due to an additional component with a 93 MeV endpoint, corresponding to the maximum photon energy in the reaction $^{48}$Ti($\mu^-,\nu_{\mu} \gamma)\rightarrow \,^{48}$Sc($0^+$, 6.68 MeV) but  no calculation of the size of the contribution was supplied and the authors state ``Such a weak transition to a discrete final state could not have been resolved in the available RMC data."  This lack of clear understanding remains troubling and future experiments should be aware of it.  
 

  \begin{table}[h]
\begin{center}
\begin{tabular}{|l|c|c|c|c|}\hline
Year&90\% CL Limit &Material&Collaboration/Lab&Reference\\\hline
1972&	$2.6\times 10^{-8	}$&Cu&SREL	& \cite{Bryman:1972}\\
1978&$1.5 \times 10^{-9}$ &S& SIN&\cite{Badertscher:1978}\\
1980&$9.0 \times 10^{-10}$ &S&SIN&\cite{Badertscher:1980}\\
1980&$3 \times 10^{-10}$&$^{127}$I & --- &\cite{Abela:1980}\\
1988&$1.7 \times 10^{-10}$&Ti&TRIUMF/TPC&\cite{Ahmad:1988}\\
1993&$8.9 \times 10^{-11}$&Ti&SINDRUM II&\cite{Dohmen:1993}\\
1993&$4.3 \times 10^{-12}$&Ti&SINDRUM II&\cite{Dohmen:1993}\\
1998&$1.7 \times 10^{-12}$&Ti&SINDRUM II&\cite{Kaulard:1998}\\
1998&$3.6 \times 10^{-11}$&Ti&SINDRUM II&\cite{Kaulard:1998}\\\hline
\end{tabular}
\caption{History of $\mu^-  N\rightarrow e^+ N$ results. Limits are normalized to captures.  Note \cite{Abela:1980} used a radiochemical technique to detect particle-stable states of $^{127}$Sb.  The two results in \cite{Dohmen:1993} and \cite{Kaulard:1998} refer  to the assumption the final state is a giant dipole resonance excitation or that the daughter nucleus is left in the ground state.  As described in the text, the relative probabilities of these two final states are unknown.\label{tab:delta2}}
\end{center}
\end{table}

\begin{figure}[h]
\begin{center}
\includegraphics[scale=0.8]{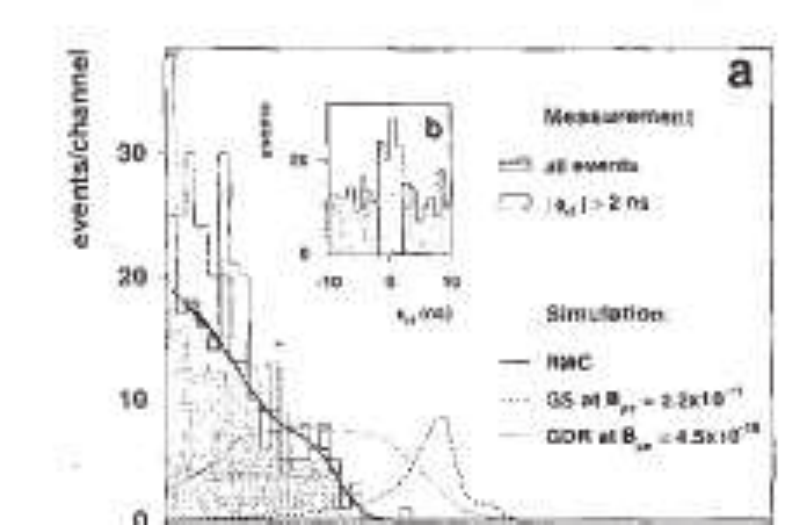}
\end{center}
\caption{\label{fig:deltal2} Fig.~2 from \cite{Kaulard:1998}.  The ground state and GDR transitions for the quoted $B_{\mu e}$ are shown against the observed spectrum.}
\end{figure}

 The experiment then set two limits based on simulations of the expected signal for the ground state transition and for the excited transition.  Fig.~\ref{fig:deltal2} shows the positron momentum spectrum with the two potential signals overlaid.  The grey histogram is for events outside the timing peak associated with the beam; these events are classified and fit to a RMC spectrum. One can see the effect of the additionally modeled Sc reaction in the behavior of the spectrum at 90 MeV/$c$.

$\mu^- N \rightarrow e^+N$ is perhaps the only experiment that can be run simultaneously, in the same detector, with another measurement in CLFV.  \citet{Rashid:2010}  have suggested a ``three-in-one" arrangement but the idea is only a sketch. The authors proposed keeping an ``S"-shaped solenoid (although the ``C" would work just as well, or poorly) but then rearranging the detector elements inside the final solenoid.  The scheme presented would not work simply because of the neutron flux in the calorimeter elements surrounding the stopping target. A scintillating crystal such as LYSO would be able to handle the radiation, energy deposit from the ambient neutron background would likely mask any signal.(\cite{Zhu:2006cn})  The extinction factor quoted has been demonstrated in Mu2e (\cite{Abrams:2012}) to be too low, and the authors claim a reach better (it is not clear whether an SES or 90\% CL is intended) than Mu2e with an apparatus that has never been carefully simulated.  Although it would be a great coup to design such an experiment, it looks as though the experimental requirements on beam structure and the difference between stopped muon and captured muon experiments are just too great; it seems more likely that increasing specialization is the future of the field and that a single experiment capable of studying all three modes is not workable.

\subsubsection{Muonic Atoms: $\mu^+e^- \rightarrow \mu^-e^+$}

Hydrogenic bound states of $\mu^+e^-$ (muonium, or ``Mu") can convert to $\mu^-e^+$ (``$\overline{{\rm Mu}}$"), violating individual electron and muon number by two units.  This process is analogous to $K^o\bar{K}^o$ mixing; \cite{Pontecorvo:1958} suggested the process could proceed through an intermediate state of two neutrinos.  Part of the calculation is performed in \cite{Willmann:1998}.   One typically states the result of a search as an upper limit on  an effective coupling analogous to $G_F$:  $G_{ {{\rm Mu}}  \overline{ {\rm Mu} } }$, where the exchange is mediated by such particles as a doubly charged Higgs, dileptonic gauge bosons,  a  heavy Majorana neutrino, or a supersymmetric R-parity violating $\tau$-sneutrino. (\cite{Hou:1996,Horikawa:1996,Cvetic:2005, Liu:2009}) The new interaction leads to a splitting of the otherwise degenerate energy levels (recall the coupling is $V-A$.)  Such a new interaction would break the degeneracy by an amount
\begin{eqnarray}
\frac{\delta}{2} &=& \frac {8 \, G_F}{\sqrt{2} n^2 \pi a_o^3} \left (  \frac{G_{ {{\rm Mu}}  \overline{ {\rm Mu} } }}{G_F} \right )
\end{eqnarray}
where $n$ is the principal quantum number and $a_o$ is the Bohr radius of the muonium atom.  For $n=1$,  
\begin{eqnarray} 
\delta &=& 2.16 \times 10^{-12} \,  \frac{G_{ {{\rm Mu}}  \overline{ {\rm Mu} } }}{G_F} \,\, {\rm eV}
\end{eqnarray}
Assuming an initially pure $\mu^+e^-$ state, the probability of transition is given by:
\begin{eqnarray}
{\cal P}(t) &=& \sin^2 \left ( \frac{\delta t}{2 \hbar} \right ) \,\,\lambda_{\mu}e^{-\lambda_{\mu} t}
\end{eqnarray}
where $\lambda_{\mu}$ is the muon lifetime.  Modulating the oscillation probability against the muon lifetime tells us the maximum probability of decay as anti-muonium  occurs at $t_{\rm max}= 2 \tau_{\mu}$.  The overall probability of transition is
\begin{eqnarray}
P_{\rm total}&=& 2.5 \times 10^{-3} \left (  \frac{G_{ {{\rm Mu}}  \overline{ {\rm Mu} } }}{G_F} \right )\label{eq:totalprobmuonium}
\end{eqnarray}

Normally the experiments quote a limit on $G_{ {{\rm Mu}}  \overline{ {\rm Mu} } } $.  Experimentally, of course, no such thing is measured; one measures a probability of transition.  The limit is set assuming an interaction of (V$\pm$A)$\times$(V$\pm$A) although one can also set limits on masses of, for example, dileptonic gauge bosons.  We follow the practice of quoting a limit on the ratio of coupling constants.  

\begin{figure}[h]
\begin{center}
\includegraphics[scale=0.8]{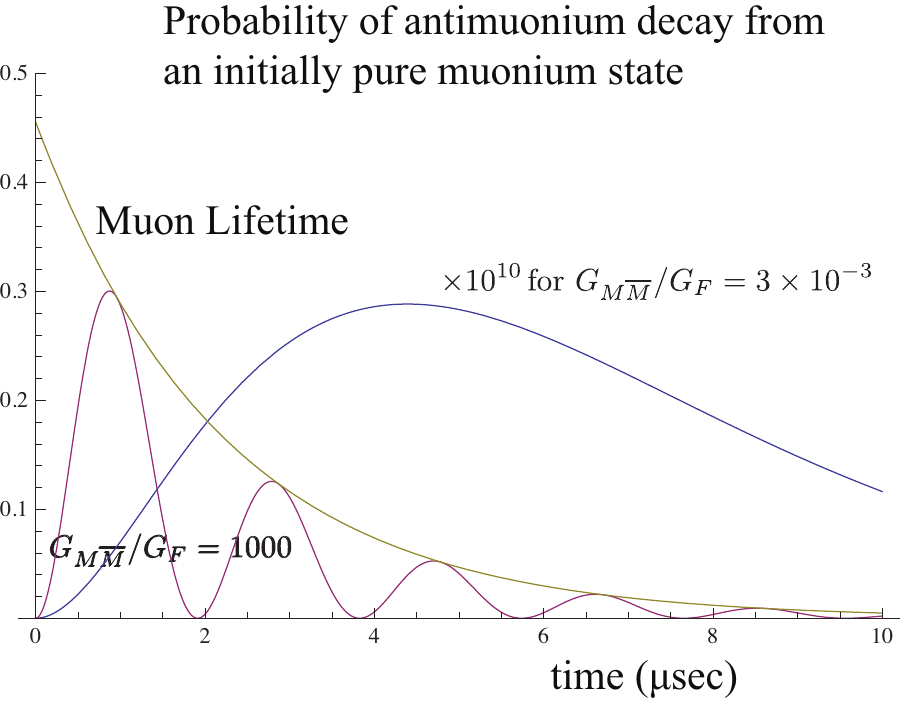}
\end{center}
\caption{\label{fig:muonium} Time dependence of the probability of observing antimuonium decay . The value of $3 \times 10^{-3}\,G_{ {{\rm Mu}}  \overline{ {\rm Mu}}}/{G_F}$  is the limit from the best experiment, \cite{Willmann:1999}; Figure updated.}
\end{figure}

It is interesting to consider placing the muonium system in a magnetic field, since the muonium energy levels will be split (see \cite{Matthias:1991}.)   We refer the reader to \citet*{Kuno:1991} and \cite{Feinberg:1961} for a fuller discussion of the physics.  Because the spectrometers used to detect and measure electron momenta require a magnetic field, this effect must be included in the calculation of the transition rate.  In this more general case, $\delta \rightarrow \sqrt{\delta^2  + \Delta^2}$.  The effect is significant even for a weak ($\sim$ 0.1T) field because of the Zeeman splitting of the energy levels.  The reduction factor for fields of about 0.1 Gauss to  0.1 Tesla is nearly flat at a factor of two, but \cite{Hou:1995} show the reduction becomes rapidly more suppressed at higher fields.

We now examine the experimental history.  It is clear that the current limits can be significantly improved with modern technology and the advent of new, intense muon sources.

In general, one wants the muonium to be in vacuum as much as possible before detection.   Losing the  negative muon in antimuonium to an atom is energetically favored over that negative muon remaining bound in the  antimuonium atom.  The trick, then, is to capture as many positive muons as possible while having no material in which they can interact.  Obviously the goals are mutually exclusive and the experiments have progressed in balancing the requirements.  Furthermore, the time dependence of the system has not yet been exploited.  

The first experiments made muonium by exposing a $\mu^+$ beam to 1 ATM of Ar.  If the conversion then occurs, the $\overline{\rm Mu}$ atom would likely collide with an Ar atom and form the argon muonic atom.  The capture rate is much higher than the muon lifetime at this pressure; hence capture dominates in separating the anti-muonium component from the oscillating system.    The signal would have been the $2P$--$1S$ muonic X-ray at 643 keV, and the experiment set a limit of $G_{ {{\rm Mu}}  \overline{ {\rm Mu} } } \leq 5680 \, G_F$ at 95\% CL.\begin{footnote}{The technique is the same used by \cite{Hughes:1960} and collaborators to discover muonium approximately eight years earlier.}\end{footnote}(\cite{Amato:1968} and \cite{Feinberg:1961})

The next series of experiments moved away from gas to silica-powder (after attempts with metal foils.)  Muons that come to rest in silica powder form muonium, which can then react through spin exchange with oxygen in the SiO$_2$.  The muonium moves thermally and has a mean free path of order $3 \times 10^{-7}$ m in the silica powders used in the early experiments.  This relatively short distance still greatly suppresses muonium conversion, but an atom can migrate between voids for $~ 0.1$ mm before decay.  Hence muonium formed near a surface can escape.  Any $\mu^-$ from antimuonium would have been detected through observation of a Ca $2P$--$1S$ X-ray from calcium oxide layers adjacent to the drift regions. The first such experiment at $1.7 \times 10^5 \mu^+$/sec using silica powder, at TRIUMF, set a limit at $42 \, G_F$ at 95\% CL, a huge improvement over the existing $5680 \, G_F$.  This technique, with refinements, has been used by \cite{Marshall:1982}.  Note that neither the muon nor the positron is directly  detected.  Studies at J-PARC have investigated muonium formation from hot tungsten wires but as of this writing the efficiencies are too small. (see \cite{Tungsten:2012})

The next experiment in the sequence (\cite{Huber:1989}) , four years later, was performed at a lower  muon intensity of $2 \times 10^4$ $\mu^+$/sec,  but had a significant experimental improvement.  It  detected the positron, measuring the time and position of muon decay and confirming the thermal emission hypothesis.  \cite{Beer:1986} halved the limit to $20 \, G_F$.  The group then  improved the measurement with a radiochemical technique.  It relied on a antimuonium signature with a $\mu^-$ creating $^{184}$Ta in an  W foil.  The surface layer was chemically extracted.  Tantalum is then observed by the triple coincidence of $\beta$-decay (8.7 hr lifetime), a 414 keV $\gamma$ decay, and a delayed $\gamma$ cascade decay (mostly 921 keV.)  These were counted in a low-background germanium spectrometer.   The result improved the limit  to $0.29 \,G_F$ at 90\% CL.

In 1991 the field turned back to the coincident detection of the muon and positron in an LANL experiment in \cite{Matthias:1991}.  A subsurface $\mu^+$ at $\approx$ 20 MeV/$c$ was passed into the by-now canonical SiO$_2$ powder.  The apparatus could detect the decay of both muonium and antimuonium.  Decay positrons or electrons were observed in a spectrometer at right angles to the beam and after passing through a pair of MWPCs were detected in CsI.  Atomic electrons (or positrons) were electrostatically collected, focused, and accelerated to 5.7 keV. A dipole then charge- and momentum-selected the particles, which were finally detected by an MCP.  The advantages of observing the thermal muonium are obvious: one can verify the experimental method and calibrate the detectors, study acceptances with reversed polarities, etc.  The experiment examined $9.8 \times 10^{11}$ incident muons and set a limit of $< 0.16 \, G_F$ at 90\% CL.  

\begin{figure}[h]
\begin{center}
\includegraphics[scale=0.6]{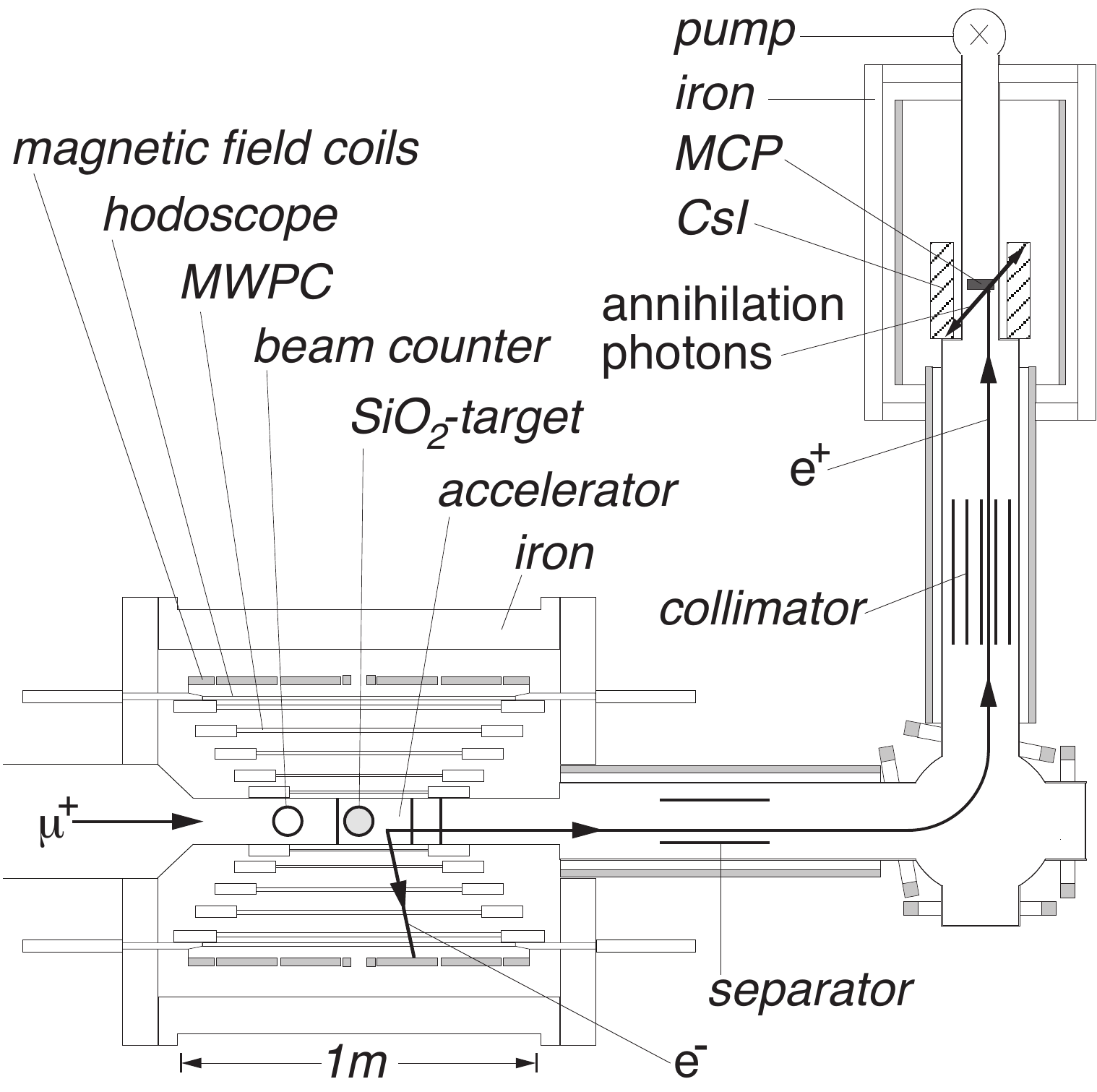}
\end{center}
\caption{\label{fig:muonium_apparatus} MACS apparatus at PSI.  The signature requires the energetic $e^-$ from the $\mu^-$ decay of $\overline{\rm Mu}$ in a magnetic spectrometer, in coincidence with the atomic shell $e^+$, which is accelerated and magnetically guided onto a microchannel plate; at least one annihilation photon is then detected in a CsI calorimeter.  Figure and caption taken from \cite{Willmann:1999}. }
\end{figure}

The most recent experimental series, performed at PSI (\cite{Willmann:1999}) used an upgraded version of the LANL technique, setting a limit $< 3.0 \times 10^{-3} \,G_F$ at 90\% CL.  This paper nicely covers the dominant background of the method, which must be understood in planning subsequent generations of searches.  The apparatus is shown in Fig.~\ref{fig:muonium_apparatus}.

The experimental signature of anti-muonium decay is an energetic electron from normal muon decay in coincidence with an approximately  13.5 eV kinetic energy positron (the Rydberg energy in the $1s$ state.)  Because the negative muon can be captured, the signal rate is suppressed by the capture fraction (depending on $Z$,  the suppression is $\approx \times 2$ for (V$\mp$A)$\times$(V$\pm$A) processes.)  This measurement suffers rate-dependent backgrounds not dissimilar to those found in $\mu \rightarrow e \gamma$ and $\mu \rightarrow 3e$, from accidentals and radiative decay processes:
\begin{enumerate}
\item The rare decay mode $\mu^+ \rightarrow e^+ e^+e^- \nu_{e}\bar{\nu}_{\mu}$ with a branching ratio of $3.4 \pm 0.4 \times 10^{-5}$ (value from \cite{RPP}.)  If one of the positrons has low kinetic energy and the electron is detected, this channel can fake a signal.  
\item The system starts as muonium, hence $\mu^+ \rightarrow e^+ \nu_e \bar{\nu}_{\mu}$ yields a positron.  If the $e^+$ undergoes Bhabha scattering, an energetic electron can be produced.  Background results from the coincidence of that scattering with a scattered $e^+$.   The positron's time-of-flight is is used to reject background.  
\end{enumerate}

Could the radiochemical experiments be improved? \cite{Aoki:2003} has argued that because there is no active device, with modern intensities this method could surpass the counter/chamber techniques and the associated physics background limits.   One significant background would be cosmic ray production of   $^{184}$Ta.   His estimate is that the limit would be $ \leq 10^{-4} \, G_F$, about $\times 30$ better than the current limit.  $^{184}$Ta production from cosmic rays could be handled by using the fact that such production from cosmic rays occurs uniformly throughout the target, whereas anti-muonium only affects a layer of about 28 nm.  However, no estimate of the relative rate is given; obviously if the fluctuations in the CR rate are sufficiently large then any signal would be masked.  Therefore detailed geometry calculations are required.  But perhaps more fatal, at least in the near term,  is the  $\mu^-$ beam contamination. Using Eqn.~\ref{eq:totalprobmuonium}, assuming we want to reach $\leq 10^{-4} \, G_F$, and assuming an detection/reconstruction efficiency of $10^{-5}/\mu$, one can calculate the beam  
$\mu^-$ contamination must be  $\leq {\cal O}(10^{-14}) $.  This is extraordinarily difficult without the technology of a fixed-field alternating gradient (FFAG) accelerator or some functional equivalent as described in \cite{Symon:1956}.

For the time being we return to the standard chamber/counter techniques.  \cite{Willmann:1999} employed a MCP-based TOF system with a FWHM of 3.3 nsec.  Modern TOF systems with MCPs can do at least $\times 10$ better.  This improvement affects the coincidence background; it seems straightforward to reduce that background by $\times 100$.  This process yielded an expected background of 1.7  events.  The size of the radiative decay background is again a function of resolution.  In \cite{Willmann:1999} the resolution of the electron at 50 MeV/$c$ was given as 54\% as determined by a cathode strip hodoscope in a 0.1 T field, limited by the 2 mm wire spacing.  The positron energy was measured by a CsI crystal calorimeter with 350 keV (FWHM) resolution.  The positron was accelerated to 7 keV, and struck the calorimeter, leaving a signature of at least one annihilation photon.  

\begin{figure}[h]
\begin{center}
\includegraphics[scale=0.6]{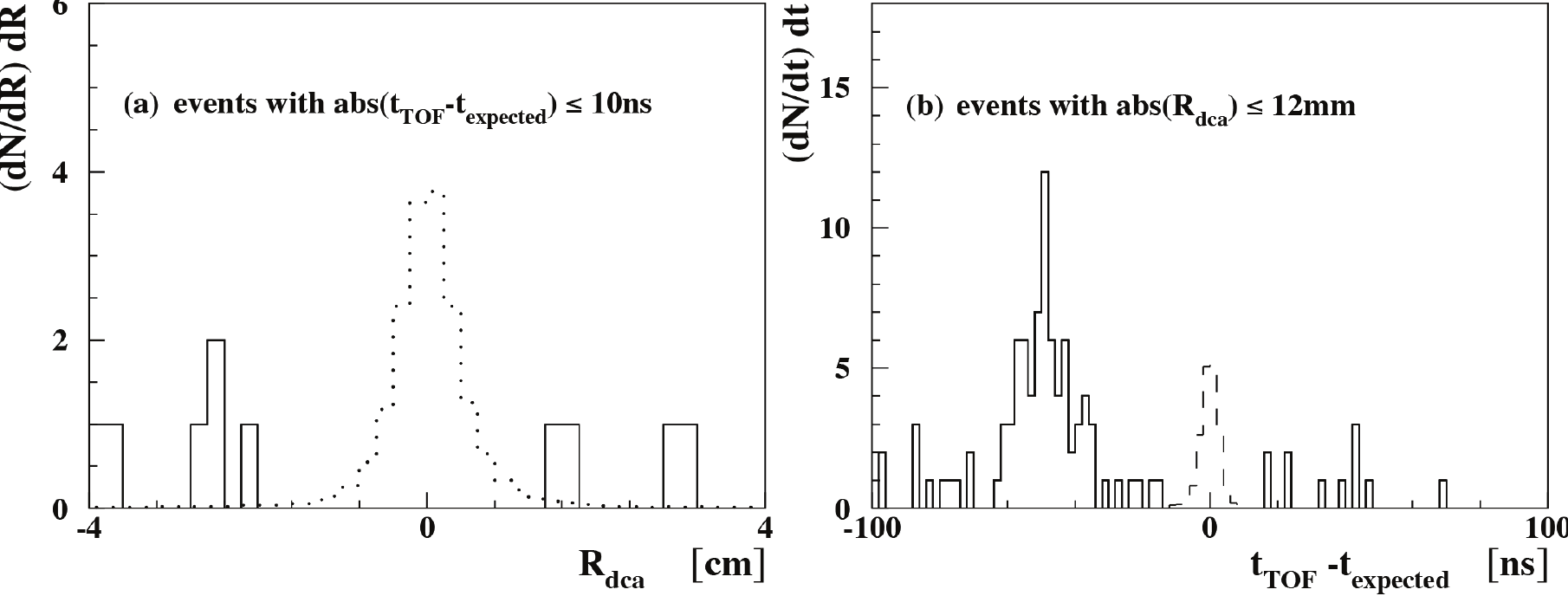}
\end{center}
\caption{\label{fig:muonium_resolution} Number of events with identified energetic electron and slow positron as a function of (a) the distance of closest approach between the electron track in the magnetic spectrometer and the back projection of the positron measured at the MCP and (b) the difference of the positron's time-of-flight and the expected arrival time. Figure and caption taken from \cite{Willmann:1998}. }
\end{figure}

It is clear from Fig.~\ref{fig:muonium_resolution} that better resolution is required for a significantly better measurement.   \cite{Bertl:1985} point out if one were to use a pulsed muon source one could use the muon lifetime to suppress the radiative decay, which unfortunately peaks near the Michel endpoint.  However, \cite{Willmann:1999} have already achieved $\leq 3 \times 10^{-3} \, G_F$; by waiting five muon lifetimes, one would suppress the radiative decay an additional factor of 150.  Therefore it seems the same technique with modest improvements at current pulsed beam intensities could reduce the limit two orders-of-magnitude or more. \cite{Willmann:1998} explain that that final state interactions in muonium decay could kick out a positron through internal Bhabha scattering.  However, the energy spectrum can be used to reduce this background to a negligible level.

  \section{Searches for Charged Lepton Flavor Violation with Taus}
\begin{figure}[h]
\begin{center}
\includegraphics[scale=0.6]{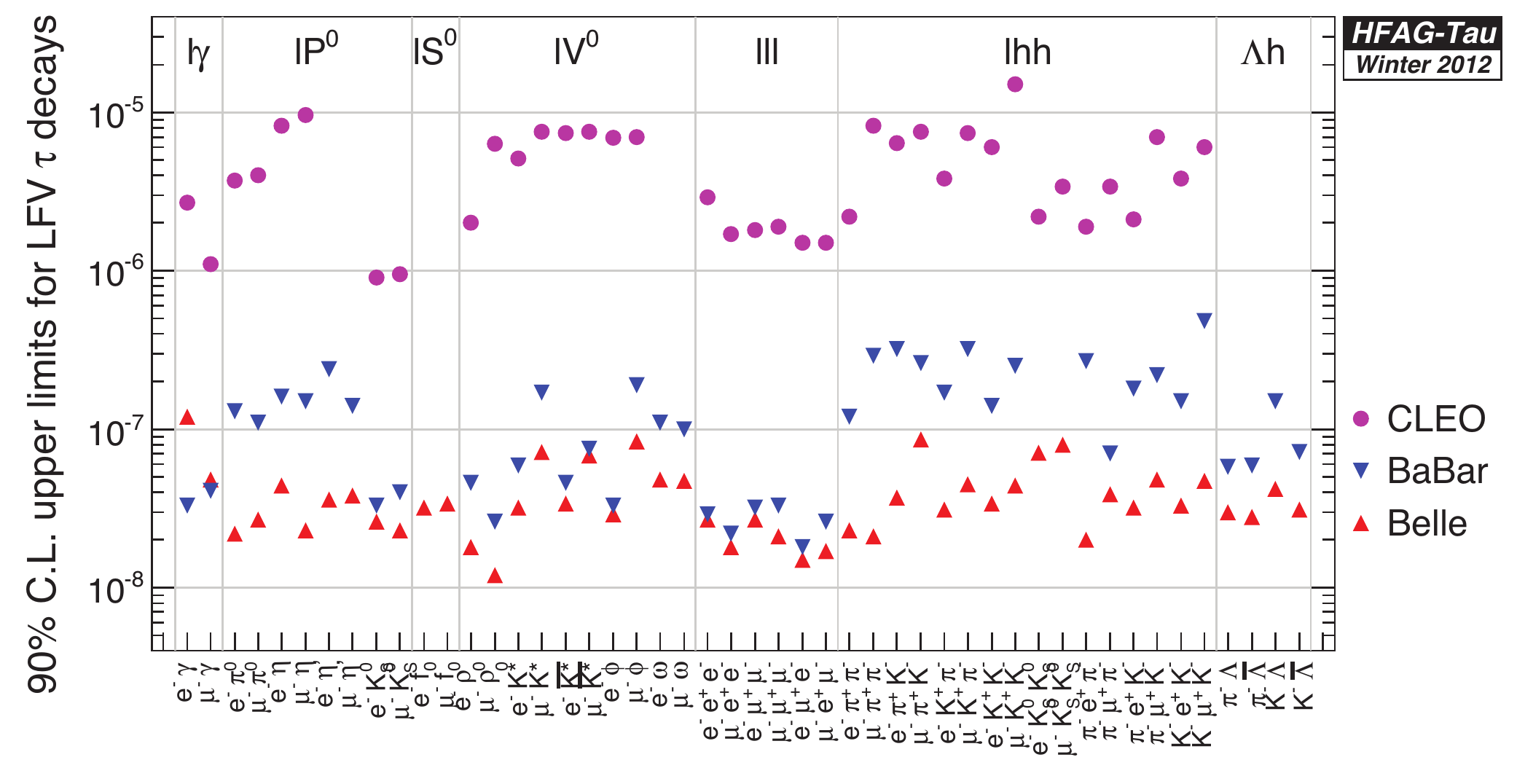}
\end{center}
\caption{\label{fig:currentTauLimits} Current limits for $\tau$-based CLFV processes.  Taken from the HFAG Working Group, \cite{HFAG:2012}.}
\end{figure}

Searches with the tau lepton have both advantages and disadvantages. Recalling Eq.~\ref{eq:BRSM} from Sec.~\ref{sec:muonoverview}, Standard Model  $\tau \rightarrow \mu \gamma$ backgrounds are negligible but many models predict rates smaller than the ${\cal O}(10^{-8})$ limits seen in Fig.~\ref{fig:currentTauLimits}.(\cite{Antusch:2006, Cvetic:2002}.)  

However, the $\tau \rightarrow   3l$ modes or $\tau \rightarrow h l$ can produce much larger effects.  In the $\tau \rightarrow   3l$ case, there are additional diagrams involving only logarithmic suppressions.(\cite{Pham:1998}) One can expect rates as large as ${\cal O}(10^{-8})$.   Higgs-mediated CLFV has also been studied and seems a promising mode.(\cite{Paradisi:2006, Babu:2002})

Although the predictions of  supersymmetric models for relevant $\tau$ decays are beyond the current luminosities of  \babar\ or BELLE, the next generation of $B$ factories may be able to see these modes. The flavor factories can access CLFV decay rates as much as $\times 100$ smaller than existing limits in the  cleanest channels, such as $\tau \rightarrow 3e$, and over $\times 10$ smaller for other modes such as $\tau  \rightarrow l \gamma$ that have irreducible backgrounds (from the analogous radiative decay processes to $\mu \rightarrow 3e$, i.e. $l \rightarrow 3l^{\prime} l  \nu \bar{\nu}$. A polarized electron beam provides additional advantages to determine the properties of the LFV interaction from the polarization-dependent angular distribution of the $\tau$ decay products, and to improve the selection for specific NP models. Polarization improves the sensitivity by $\approx 2.6$ (with an assumption that the coupling is of the appropriate handedness, since if one polarizes the beam one loses cross-section.) 

We begin with $\tau \rightarrow \mu \gamma$ in the \babar\ experiment as described in \cite{Aubert:2010} or \cite{Aubert:2010a}.  The signal is extremely clean.  One imagines $e^+ e^- \rightarrow \tau^+ \tau^-$.  The event can be divided into two hemispheres, each containing one $\tau$ decay.  Each candidate decay must have the $\tau$ mass, and the combined energy must be $\sqrt{s}/2$ . A two-dimensional analysis of $E_{lX}$--$M_{lX}$ then provides a clean region to investigate:  the  missing momentum must be near zero and the $\tau$'s are back-to-back. 

\begin{figure}[h]
\begin{center}
\includegraphics[scale=0.4]{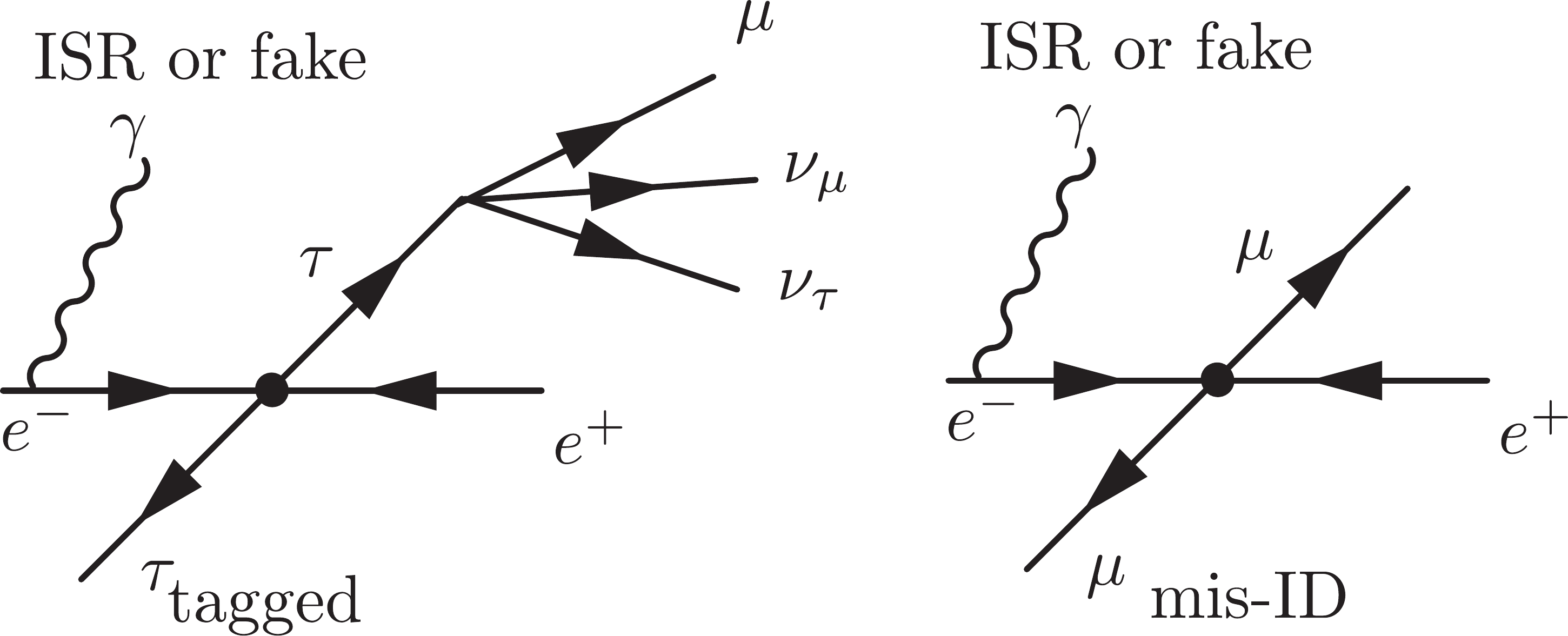}
\end{center}
\caption{\label{fig:tauback} A depiction of major sources of background in the $\tau \rightarrow \mu \gamma$ search at $e^+e^- $ colliders.  The particle-antiparticle distinctions are arbitrary.  On the left, an electron undergoes ISR or an unassociated photon combines with a tagged $\tau$ decay.  This process can produce background if the reconstructed quantities fall in the signal window.  On the right, we see a $e^+ e^- \rightarrow \mu^+ \mu^-$ event where a muon is misidentified and the event is tagged as as containing a $\tau$, from, for example, a hadronic $\tau$ decay $\tau \rightarrow \pi X$ process.  In this case the same ISR/fake photon situation can produce background exactly as on the left.}
\end{figure}

There are many  features and problems analogous to the $\mu \rightarrow e \gamma$ search.  For example, $\tau \rightarrow \mu \nu \nu$ decays with an accompanying photon from, for example, initial state radiation (ISR) yield a muon and photon.  If the opposite hemisphere provided a mistag then the event may fall into the signal region.   One could have a mistag in $e^+e^- \rightarrow \mu^+ \mu^-$ where one $\mu$ is identified as an electron, or from $e^+e^- \rightarrow \tau^+ \tau^-$ where one of the $\tau$ leptons decays hadronically and the outgoing $\pi$ is misidentified as the electron or the muon.  Radiative processes such as $e^+e^- \rightarrow e^+e^- \gamma$ or $\rightarrow \mu^+ \mu^- \gamma$ also contribute.  Exploiting the statistical power of the flavor factories  will require improvements in the photon background and will require  better granularity than \babar\ or BELLE.  Fig.~\ref{fig:tauback} is a graphical and perhaps easier to follow description of these background sources.

In $\tau \rightarrow 3 l$ and $\tau \rightarrow lh$ one does not have the ISR problem, and  the mass resolution is much better than for modes using a photon and calorimeter since one has the vertex and superior momentum measurement.     \cite{SuperKEKB:2004} at $50^{-1}$ ab  projects an improvement on the $\tau \rightarrow \mu \eta$ limits of $\times 100$.

 \subsubsection{Other Modes}
 
 One can also search for hadronic modes such as $B^{\pm} \rightarrow h \tau l$, where $ h = \pi, K$ and $l = e, \mu$.  One divides the event into a signal $B$ and a tagged $B$.  The tagged $B$ can come from hadronic decays such as $B^{\pm} \rightarrow D^{(0)*}X$.  The signal $B$ can then be three charged tracks ($\pi,K$ along with a $\tau$ and $l$.)  \cite{Lees:2012} set limits on mass scales for models such as \cite{Sher:1991} of $\Lambda_{bd} > 11$ TeV/$c^2$ and $\Lambda_{bs} > 15$ TeV/$c^2$.
 
 \subsection{Searches at  LHC,  LHCb, and other Colliders}
 
 THe LHC and LHCb can certainly contribute to the search for CLFV although such searches are just beginning.  We will not cover possible CLFV in SUSY particles (for example, CDF and D0 have published limits in \cite{Aaltonen:2010fv} and  \cite{Abazov:2006nw}), but will simply briefly discuss the $\tau$ modes.

The LHC can search for a number of modes: $\tau \rightarrow \mu  \gamma$, $Z \rightarrow e \mu$, and $\tau \rightarrow 3\mu$.  First results can be found in \cite{Aad:2012yw} with a nice summary of the possibilities in \cite{Morrissey:2012}.  Of course if supersymmetric particles are discovered there will be immediate searches for their charged lepton flavor violating decays.  It should be realized that this is a rapidly developing avenue of investigation and we are just at the beginning of learning what can and will be done.

 LHCb has already searched for both lepton and baryon number violation through the $\tau$ modes.  The inclusive $\tau^-$ production cross-section is so large such that LHCb can collect per year of nominal running more than $\approx \times 100$ the total samples collected by \babar\ and BELLE.  The $1.0^{-1}$ fb 2011 data has set the upper limits of Table~\ref{tab:LHCb}.  
 
The first results are reported in \cite{Harrison:2012} and \cite{Aaij:2011ex}.   The $\tau \rightarrow 3 \mu$ search used two likelihood functions and a cut on the $\tau$ mass.   The first ${\cal M}_{\rm 3body}$ classifier distinguishes three-body decays  from combinations of tracks from different vertices.  The second identifier is a PID cut using the LHCb RICH, calorimeters, and muon stations.  The classifiers are trained on $D_S^- \rightarrow \phi (\mu^+ \mu^-) \pi^-$ and $J/\psi \rightarrow \mu^+ \mu^-$.  These limits are currently about $\times 3$ worse than the existing limits, but it is early in the data-taking.  For the latter $B^+$ modes at $36^{-1}$ pb and $\sqrt{s} = 7$ TeV the search is based on the selection of $B^+ \rightarrow h^{\pm} \mu^+ \mu^{\mp}$ candidates using particle identification and the $B^+ \rightarrow J/\psi K^+$ mode as a proxy for the signal.
 
\begin{table}[h]
\begin{center}
\begin{tabular}{|l|c|}\hline
Mode&Limit\\\hline
${\cal B}(\tau^- \rightarrow \mu^+ \mu^- \mu^-)$ & $6.3 \times 10^{-8}$ (90\% CL)\\
${\cal B}(\tau^- \rightarrow  {p} \mu^- \mu^-)$ & $4.6 \times 10^{-7}$(90\% CL)\\
${\cal B}(\tau^- \rightarrow \bar{p} \mu^+ \mu^-)$ & $3.4 \times 10^{-7}$(90\% CL)\\
${\cal B}(B^+ \rightarrow \pi^- \mu^+ \mu^+)$ & $5.8 \times 10^{-7}$(95\% CL)\\
${\cal B}(B^+ \rightarrow K^- \mu^+ \mu^+)$ & $5.4 \times 10^{-8}$(95\% CL)\\
\hline
\end{tabular}
\end{center}
\caption{Limits from LHCb on CLFV and Baryon Violating Modes from \cite{Harrison:2012} and \cite{Aaij:2011ex} at $\sqrt{s} = 7$ TeV.\label{tab:LHCb}}
\end{table}
 
  The experiment also has performed new searches for $\tau \rightarrow p \mu^+\mu^-,$ and $ \tau \rightarrow \bar{p} \mu^+ \mu^-$, using the three-body classifier and the invariant mass cut.  No measurements of these modes currently exist, although the results of searches for $\tau \rightarrow \Lambda h$ ($h = \pi, K$) can be seen in Fig.~\ref{fig:currentTauLimits}.

\subsection{Electron-Ion Collider and Electron-to-Tau conversion \label{sec:eic}}

A high energy, high luminosity electron-proton/ion collider (EIC), as described in \cite{Boer:2011}, 
 is being considered by the US nuclear science community with a variable center-of-mass
energy of $50 \rightarrow 160$ GeV and with $100-1000$ times the
accumulated luminosity of HERA over a comparable operation time.  Recently,  \cite{Deshpande:2012} building on an earlier study in  \cite{Gonderinger:2010yn} argue  that an e-p collider 
could set a limit on leptoquark coupling-over-mass ratios that would surpass the current best limits from HERA experiments, summarized in \cite{Chekanov:2005au} and \cite{Aktas:2007ji}.  HERA's must powerful CLFV results were in leptoquarks and EIC proponents  have concentrated on those as well. The study also
shows that the proposed EIC could compete or surpass the updated leptoquark limits from $\tau \rightarrow e\gamma$ for a subset of 
quark flavor diagonal couplings. Finally, \cite{Gonderinger:2010yn} found that although $e\rightarrow \tau$ LFV is indeed severely suppressed,  $e \rightarrow \tau$ transitions could still exist within the reach of the EIC, under certain situations. Furthermore, depending on the models which give rise to the effective operators discussed above, there may be large log enhancements in the charge radius contribution to photon exchange $e\rightarrow\tau$ which could overcome the limits on the four lepton operators.

\subsubsection{Experimental Issues and Measurements at the Electron-Ion Collider}

The DIS process $e p \rightarrow \tau X$ can be used to search for $\tau$ leptons through $\tau \rightarrow e \nu_e \nu_{\tau}$, $\tau \rightarrow \mu \nu_{\mu} \nu_{\tau}$ and $\tau \rightarrow$ hadrons, and is the most studied by the proponents.  The proposed searches concentrate on differentiating Standard Model $\tau$'s created in the DIS interaction (background) from $\tau$'s created from the decay of a leptoquark.  

What variables separate Standard Model  processes (such as $D_s \rightarrow \tau \nu_{\tau}$)
 from leptoquark creation of a $\tau$?  The leptoquark signature for leptonic tau decays  is an isolated high $p_T$ muon or electron back-to-back to the hadronic system in the plane transverse to the incoming momentum vectors.  If the $\tau$ decays hadronically it typically decays through three-prong processes which constitute about 15\% of the total, 10\% of that to $3\pi$ (values from the PDG in \cite{Nakamura:2010}.)  One would then see a ``narrow" jet of the three prongs recoiling against the more diffuse jet of the proton breakup.  Finally, one can examine the angular distribution through the inelasticity $y = (1/2)(1 - \cos \theta)$, where $\theta$ is the angle between the lepton relative to the proton.  The $y$ distribution  of different leptoquarks differ from each other and from the $y$ distribution of DIS.  

The studies for the EIC do not yet have realistic detector simulations and assume 100\% reconstruction efficiency, where studies from BELLE, \babar\ or the HERA experiments suggest 10--20\% efficiencies.  However, the uncertainty in the luminosity, center-of-mass energy, and running time are large and so the possibilities for  CLFV studies at an EIC are intriguing and deserving of further study.

\begin{figure}[h]
\begin{center}
\includegraphics[scale=0.8]{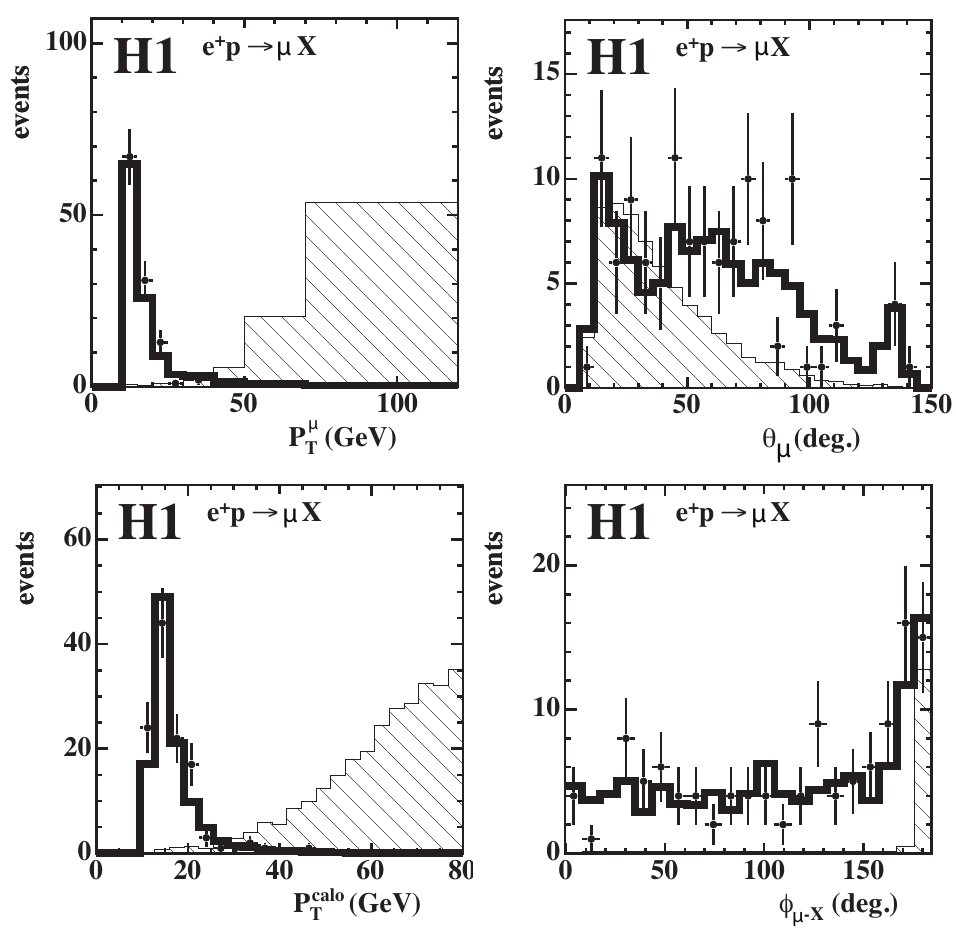}
\end{center}
\caption{\label{fig:heralfv} Distributions in muon transverse momentum, polar angle, transverse momentum, and acoplanarity between the muon and hadronic final state in muonic CLFV from H1.  The hatched histogram is the LFV signal MC sample of a
leptoquark $\tilde{S}L_{1/2}$ with $m_{ {\rm LQ}} = 200$ GeV/$c^2$ and $\lambda_{eq} =  \lambda_{\mu q} = 0.3$  with arbitrary  normalization in each plot.  The ``data" are SM background distributions. The distributions illustrate the kinematic differences between leptoquarks and SM backgrounds.  The Figure and this caption are adapted from \cite{Aktas:2007ji}. The same ideas would be used at the EIC.}
\end{figure}

\section{Searches for Charged Lepton Flavor Violation with Kaons and Other Mesons}

We concentrate here on the searches for CLFV in pseudoscalar mesons, in particular $K_L \rightarrow \mu e$, which may or may not be accompanied by a $\pi^o$.  Fermilab E791 has performed a number of searches in the $D$ system;  \babar\ , CDF, and CLEO have concentrated on the $D$ system.  The single most sensitive search was by \cite{Ambrose:1998} in BNL E871, which measured ${\cal B}(K_L \rightarrow \mu^{\pm} e^{\mp}) < 4.7 \times 10^{-12}$ at 90\% CL.  Such experiments have many experimental challenges, some of which are (a) making enough kaons, (b) having good acceptance, and (c) adequate rate-handling capacity.  The high rates and acceptance needed to achieve the $10^{-12}$ level then make it difficult to achieve sufficient background rejection.    We will concentrate on these two experiments to examine the designs and determine what might be done in the future.  

\begin{table}[h]
\begin{center}
\begin{tabular}{|l|c|c|c|}\hline
Year&90\% CL &Collaboration/Lab&Reference\\\hline
1966& $	1.0\times 10^{-4}	$&BNL	&\cite{Carpenter:1966}	 \\
1967& $8.0\times 10^{-6}	 $&BNL		&\cite{Fitch:1967} \\
1967&$	9.0\times 10^{-6}	$&CERN&		\cite{Bott:1967} \\
1988&$	1.1\times 10^{-8}	$&BNL	&	\cite{Cousins:1988}  \\
1988&$	6.7\times 10^{-9}	$&BNL	&\cite{Greenlee:1988}\\
1989&	$1.9\times 10^{-9}	$&BNL	&	\cite{Schaffner:1989} \\
1989&	$2.2\times 10^{-10}	$&BNL/E791&	\cite{Mathiazhagan:1989} \\
1989&	$4.3\times 10^{-10}	$&KEK	&	\cite{Inagaki:1989}\\
1993&	$3.3\times 10^{-11}	$&BNL/E791&	\cite{Arisaka:1993}\\
1995&	$9.4\times 10^{-11}	$&KEK/E137&	\cite{Akagi:1995} \\
1998&	$4.7\times 10^{-12}	$&BNL/E871&\cite{Ambrose:1998}\\\hline
\end{tabular}
\end{center}
\caption{History of $K_L \rightarrow \mu e$ experiments. }
\end{table}

\begin{figure}[h]
\begin{center}
\includegraphics[scale=0.6]{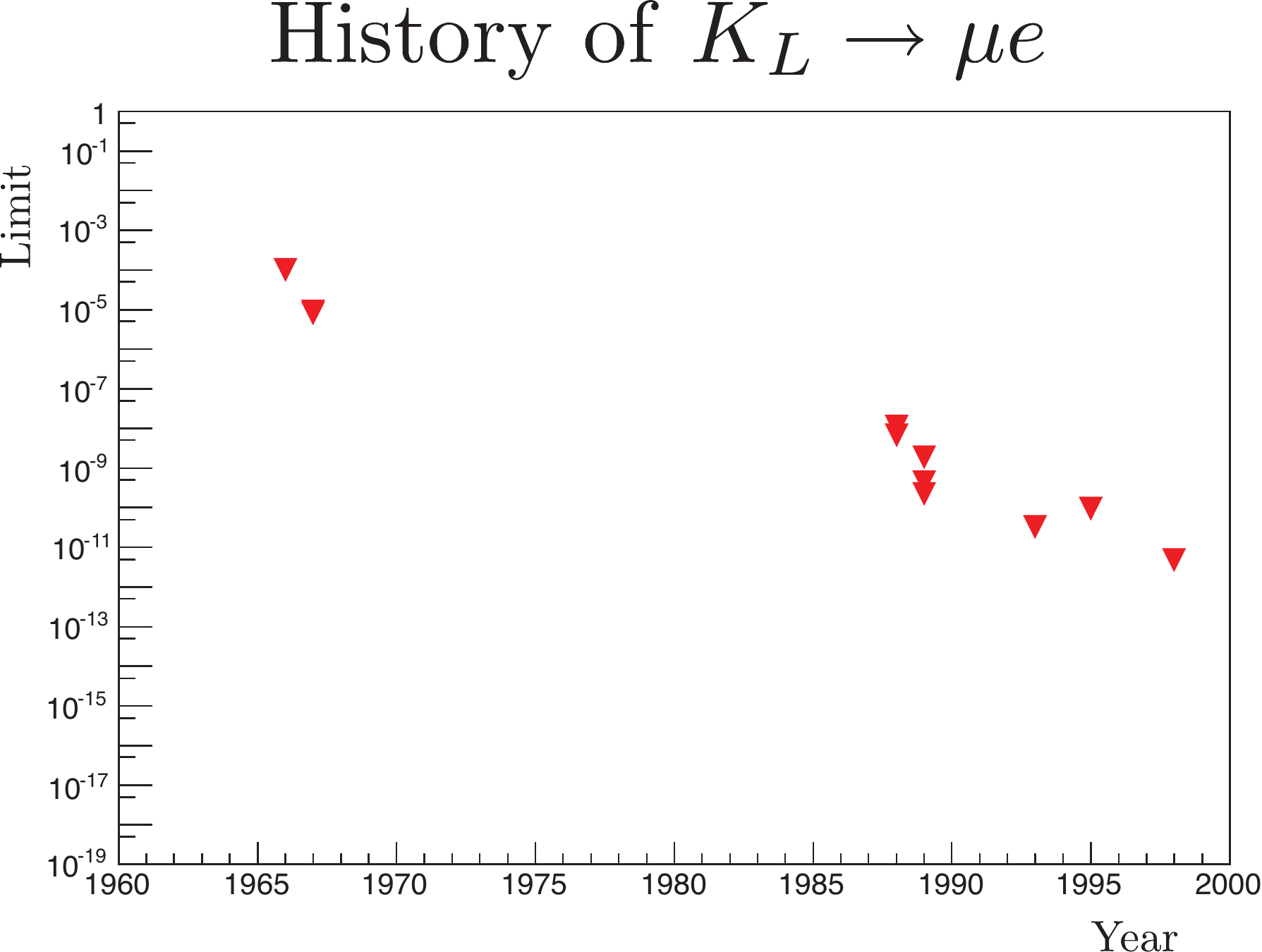}
\end{center}
\caption{The history of $K_L \rightarrow \mu e$ searches.  \label{fig:klclfv}}
\end{figure}

\begin{figure}[h]
\includegraphics[scale=0.4
]{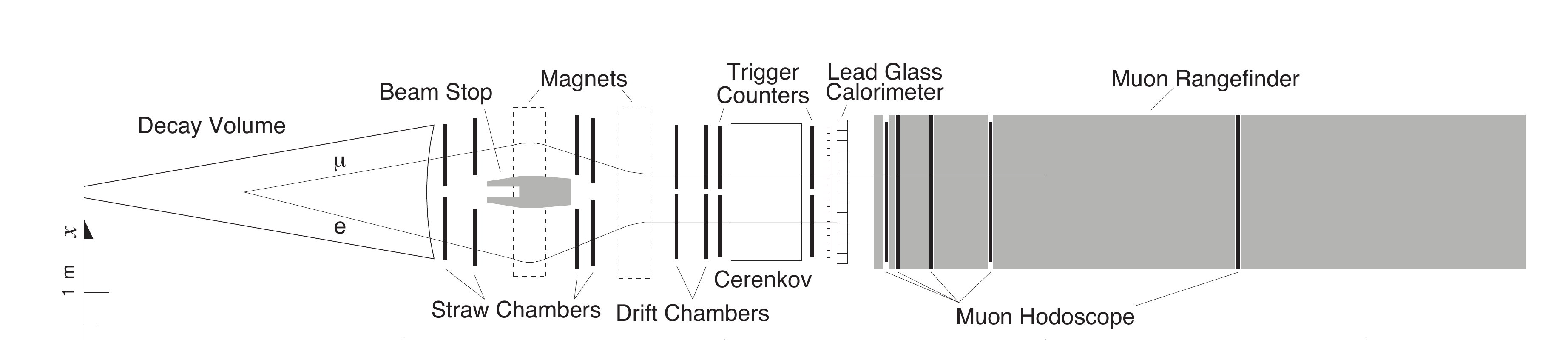}
\caption{The BNL871 apparatus; the Figure is taken from \cite{Ambrose:1998}.     \label{fig:bnl871}}
\end{figure}

The BNL871 experiment, which searched  for $K_L \rightarrow \mu e$, is shown in Fig.~\ref{fig:bnl871}. Neutral beams in $K_L$ experiments have large numbers of neutrons and photons.  One typically uses thin lead foils to convert photons and then sweeps away the decay products.  The $n/K_L$ ratio in BNL871 was about 10:1, controlled by setting the targeting angle at 3.75$^o$.  In BNL871, this implied about $2 \times 10^8$ $K_L$/(1.2--1.6) sec spill, with ten times that number of neutrons.  Many experiments have used a central hole to allow the neutrons to pass through without interacting. Rather than having a central hole, BNL871 used a compact beam dump described in \cite{Belz:1999} to absorb the neutral beam.  

 The dominant background was $K_L \rightarrow \pi e \nu$ decays with a subsequent $\pi \rightarrow \mu \nu$. Scattering in the vacuum window and first straw chamber was the leading background source. The spectrometer consisted of tracking chambers and two dipole magnets.  The two magnets were critical to the success of the experiment.  First,  the topology of two-body decays was imposed in the trigger, reducing rates to $\approx 70$ kHz for the lowest-level trigger.  The analysis then exploited the double-magnet design.   Two different algorithms fitted the final tracks.  The first made a selection based  on the overall $\chi^2$ and the second fit the front and back halves of the spectrometer separately.  Scattering in the vacuum window and first straw chamber was the leading background source: a hard scatter of a electron in $K_L \rightarrow \pi e 3$ followed by $\pi$ decay can cause the vertex to be misreconstructed and fake a $K_L \rightarrow \mu e$ decay.  This particular background arises from electron scattering in the vacuum window and the $\pi \rightarrow \mu$ decay upstream of they window;  both effects tend to increase the invariant mass toward the $K_L$ mass.  A fuller description of the backgrounds can be found in \cite{Ambrose:1998}.

 The KTeV experiment (\cite{AlaviHarati:2003}) was designed to study CP-violation.   Because of the $K_L \rightarrow \pi^0 \pi^0$ mode relevant to $\epsilon^{\prime}$, the experiment was optimized for $\pi^0$ detection and measurement.  Table~\ref{tab:ktev} from \cite{Abouzaid:2008} gives the results.  
 
Only axial and pseudoscalar hadron currents can contribute to $K_L \rightarrow \mu e$ since the K-meson is a pseudoscalar particle (as discussed, for example in \cite{Landsberg:2005}.)  KTeV, which searched for modes with a $\pi^o$, was sensitive to scalar, vector, and tensor hadron currents as well (because the $\pi$ is also a pseudoscalar.)  Therefore $K_L$ decays with a $\pi^o$ are interesting on their own.  \cite{Abouzaid:2008}, like \cite{Ambrose:1998}, believed the experiments could be improved:
 \begin{quote}
 Given that we find negligible backgrounds, our techniques could clearly be extended to higher intensity neutral kaon beams.
 \end{quote}
 
   \begin{table}[h]
 \begin{center}
 \begin{tabular}{|c|c|}\hline
Mode& 90\% CL\\\hline
 ${\cal B}(K_L \rightarrow \pi^0 \mu^{\pm}e^{\mp} )$&$< 7.56\times 10^{-11}$\\
 ${\cal B}(K_L \rightarrow \pi^o \pi^o \mu{\pm} e^{\mp}) $&$< 1.64 \times 10^{-10}$\\
${\cal B}(\pi^o \rightarrow \mu^{\pm} e^{\mp} ) $&$ < 3.59 \times 10^{-10}$\\\hline
  \end{tabular}
  \caption{CLFV Limits set by the KTeV experiment, from \cite{Abouzaid:2008}\label{tab:ktev} }
  \end{center}
  \end{table}

FNAL's proposed Project X, described in \cite{Kronfeld:2013uoa},  has already stimulated discussion about rare decay searches in the $K \rightarrow \pi \nu \bar{\nu}$ modes (both charged and neutral.)  The intensities to go beyond the BNL experiments would certainly exist.  For the $K_L \rightarrow \mu e $ mode,  future experiments probably need to have the absolute minimum amount of material (tracking in a vacuum, like the Mu2e/COMET muon experiments.)  One could even drop the classic neutral kaon spectrometer and think about transporting the neutral beam {\it in vacuo} along with a magnetic field and then observing the outgoing leptons.  The challenges presented by rate, resolution, and background rejection are daunting.

\subsection{Charged Kaon Searches\label{sec:chargedK}}

\begin{figure}[h]
\begin{center}
\includegraphics[scale=0.6]{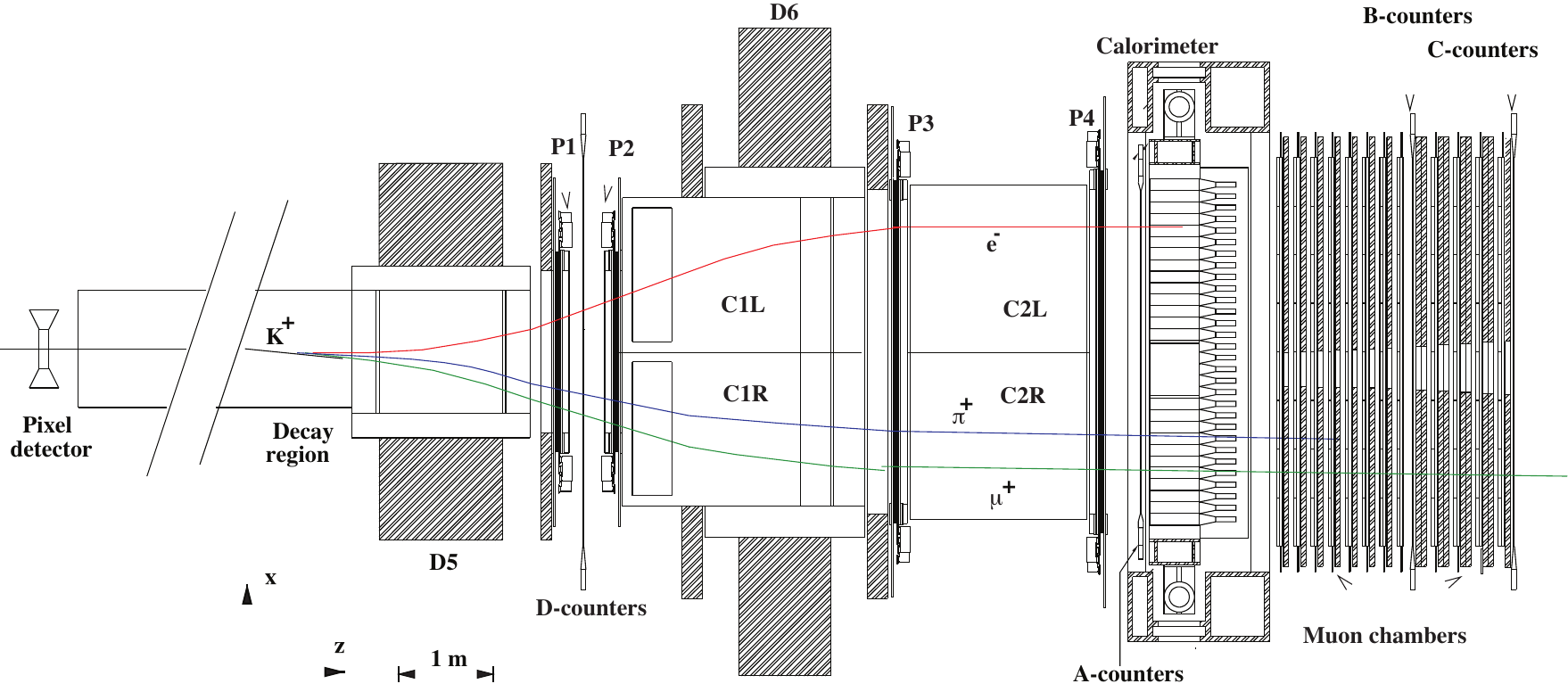}
\end{center}
\caption{Plan view of the BNL-865 detector.  A $K^+ \rightarrow \pi \mu e$ event is superimposed.  C1 and C2 are gas \c{C}erernkov counters; P1-4 are proportional chambers; D5 and D6 are dipole magnets. A--D are scintillation counter trigger hodoscopes.  The calorimeter was an early use of the Shaslyk design as described in \cite{atoyan:1992}.\label{fig:bnl865}}
\end{figure}

The decays $K^+ \rightarrow \pi^+ l^+ l^-$ and $K^+ \rightarrow \pi^+ l^+ (l^{\prime})^-$ have been studied in both the electron and muon modes by the BNL-865 experiment, the Hyper{\it CP} experiment at Fermilab, the NA48 collaboration at CERN, and at TRIUMF. (\cite{Adler:1997, Appel:1999, Appel:2000, Park:2002, Batley:2009,Sher:2005}.)  The charged-lepton flavor violating modes  were examined by \cite{Appel:2000} with the previous searches being 25 years earlier, and final results for  $K^+ \rightarrow \pi^+ \mu^+ e^-$ were presented in \cite{Sher:2005}.  We summarize the BNL-865 limits in Table~\ref{tab:bnl865}.  The experiment, performed at the BNL AGS, used a 6 GeV/$c$ $K^+$ beam with $1.5 \times 10^7$ $K^+$ every 1.6 sec AGS pulse.  The apparatus is shown in Fig.~\ref{fig:bnl865}.

The modes $K^+ \rightarrow \mu^+ \mu^+ \pi^- $, $K^+ \rightarrow e^+ e^+ \pi^- $, and $K^+ \rightarrow \mu^+ e^+ \pi^-  $ violate {\em both} generation number and lepton flavor.  They are of special theoretical interest because they are sensitive to Majorana neutrinos, a second generation analog to neutrinoless double $\beta$ decays) as first described in Sec.~\ref{sec:deltaL2}.( see \cite{Littenberg:2000})  The $K^+ \rightarrow \pi^+ \mu^+e^-$ mode is not sensitive to heavy neutrinos (to lowest order) but the techniques are the same. We focus on the final paper of \cite{Sher:2005} for concreteness.

Events were required to have three charged tracks from a common vertex in the decay volume, and a timing spread among the tracks of about 0.5 nsec.  The experiment also used the phase space of the charged beam: the five-dimensional phase space was broken up into three two-dimensional distributions: $x$ vs.\ $\theta_x$, $y$ vs.\ $\theta_y$, and $P_{K^+}$ vs.\ $x$.  A likelihood function was then constructed from the vertex quality, kaon momentum vector, beam phase space, and track $\chi^2$.  
The primary background was from the decay chain $K_{\tau}$, or $K^+ \rightarrow \pi^+ \pi^+ \pi^-$, where $\pi^+ \rightarrow \mu^+ \nu$, $\pi^- \rightarrow  e^- \bar{\nu}$.     Another background for the $K^+ \rightarrow e^+ \pi^{\pm}\mu^{\mp}$ searches was $K^+ \rightarrow \pi^+ \pi^-  e^+ \nu_e$ ($K_{e4}$) where  one of the pions was misidentified as a muon (about 5\% from pion decays and punchthrough.)   Other backgrounds were less significant.  These two, especially the first, drove the detector design.  Particle ID was used to reduce the backgrounds.  Accidental combinations of $\pi^+$, $\mu^+$ and $e^-$ from separate kaon decays tended to have poor track timing and kinematic quality and thus had poor likelihood. A signal window was then chosen in a plot of the likelihood function vs.\ the reconstructed mass, with a resolution of about 4 MeV/$c^2$ for simulated $K_{\pi \mu e}$ events.  The final results from the experiment (combined with earlier results from predecessor \cite{Lee:1990})are given in Table~\ref{tab:bnl865}.

\begin{table}[h]
\begin{center}
\begin{tabular}{|l|c|}\hline
Mode& 90\% CL\\\hline
$K^+ \rightarrow \mu^+ \mu^+ \pi^- $& $3.0 \times 10^{-9}$\\
$K^+ \rightarrow e^+ e^+ \pi^- $& $3.0 \times 10^{-9}$\\
$K^+ \rightarrow \mu^+ e^+ \pi^-  $& $5.0 \times 10^{-10}$\\
$K^+ \rightarrow \pi^+ \mu^+ e^-$& $2.1 \times 10^{-11}$\\
$\pi^o \rightarrow \mu^- e^+   $& $3.4 \times 10^{-9}$\\\hline
\end{tabular}
\end{center}
\caption{Limits on lepton flavor violation from kaon and pion decays in BNL-865.\label{tab:bnl865}}
\end{table}

How could the experiments be improved? In order to increase the statistics while maintaining an acceptable accidental background  one needs to separate pions from kaons.  A high-flux separated kaon beam is one method.  Such beams have been discussed by \cite{tschirhart:2010} but no carefully thought-out proposal exists as of this writing.

The existing generation of $K^+ \rightarrow \pi^+ \nu \bar{\nu}$ experiments such as the CERN NA62 search, as described in \cite{na62:2012} has already published limits on lepton universality and we look forward to their lepton flavor violation analyses.

\subsection{$Z$ Decays \label{sec:zdecays}}

DELPHI, L3, and OPAL have searched for modes involving the $Z$, and such searches could be performed at other colliders.  The current best  bounds are given in Table~\ref{tab:zbounds}.   There are indirect constraints on these processes from other CLFV processes, as discussed in \cite{Marciano:2008}.

\begin{table}[h]
\begin{center}
\begin{tabular}{|l|c|c|}\hline
Mode& 90\% CL&Reference\\\hline
$Z \rightarrow \mu e$& $1.7\times 10^{-6}$&\cite{Akers:1995gz}\\
$Z \rightarrow \tau \mu$& $1.2 \times 10^{-5}$&\cite{Akers:1995gz} and \cite{Abreu:1996mj}\\
$Z \rightarrow \tau e $& $9.8 \times 10^{-6}$&\cite{Akers:1995gz} and \cite{Adriani:1993sy}\\\hline
\end{tabular}
\end{center}
\caption{Limits on lepton flavor violation from $Z$ decays.\label{tab:zbounds}}
\end{table}

\section{Summary}

Sixty five years later, Rabi's ``Who ordered that?" is still both  profound and unanswered.  We do not understand the flavor puzzle and the generation problem.  Mixing among the generations will provide key information on solving the problem, but we are only partway there.  Why the neutral leptons (neutrinos) mix  while the charged  leptons, so far, do not is an undoubtedly important piece of this puzzle.  The models predicting charged lepton flavor violation are either already seriously constrained by present measurements and present new fine-tuning problems,  or so diverse and un-compelling as to  provide little clear guidance to experiment.  The lack of guidance from theory, the enormous power of CLFV to rule out models, and the importance of a discovery lead to one conclusion: we have to keep looking as far as we can see.   

But this is an experimental review and in that spirit (and in the current world of funding),  grandiose statements about needing better experiments in all channels are na\"{\i}ve and intellectually insufficient.  How  many experiments? In which channels?  This article is not intended to replace program committees by answering those questions, but to serve as a resource for future plans.  We have attempted to delineate the difficulty of such measurements, set forth the ``state of the art," and present the current ideas for going beyond it, in order to make difficult decisions among the range of options.

The measurements that come along with existing experiments, $B$ meson 
and tau decays from collider experiments, or searches for $K \rightarrow \pi \mu e$ in high 
sensitivity experiments already looking for $K \rightarrow \pi \nu \bar{\nu}$, for example,
can provide significant 
improvements on present limits for small marginal cost.   Similarly, dedicated searches in the $\tau$ sector will are natural measurements to perform at flavor factories studying $B$ decays.

Muons will play a central role in upcoming experiments, and so we summarize the prospects in MEG, Mu2e/COMET, and $\mu \rightarrow 3e$. For MEG, there is already a clear plan for upgrading to ${\cal B}\sim 5 \times 10^{-14}$.  There are no extant plans for going beyond that.  The flux for a more sensitive search would exist at Fermilab's Project X but it is far from clear that the resolution could be made good enough to justify the flux.  One critical question is whether to convert the photon and use an all-tracking solution.  Of course if MEG sees a signal measuring the size will be a necessary next step and will influence the other experiments.

For the muon-to-electron conversion experiments, 
\begin{enumerate}
\item  If there is a signal, we need to advance the beam technology required to measure the conversion rate in higher-$Z$ atoms.  The essential problem is the shorter lifetime of muons at higher-$Z$.  The radiative pion background increases, the rate associated with the beam flash goes up, and the time of the  flash and the signal region overlap, a significant challenge to the detector technology.  Therefore the most essential ingredient would be new muon beams that avoid the flash and the radiation pion capture background.  The focus must then be on storage rings, similar to those suggested for the neutrino factories, so that pions can decay and the beam flash does not propagate to the detector.
\item If there is no signal, more statistics are required which will involve new facilities capable of producing the requisite flux.  But with the intrinsic decay-in-orbit background in Mu2e or COMET at the $\leq 0.25$ event level, (for detailed background estimates see the Mu2e CDR, \cite{Abrams:2012}) it is hard to see how to improve the limits beyond approximately an order-of-magnitude.  The radiative pion capture backgrounds, currently at $\sim 0.1$ event will become more significant.   requiring a suppression of pions by some combination of shorter beam pulses, better extinction, and muon transport.  Backgrounds from cosmic rays will require deeper detectors with more efficient and hermetic vetoes. To be explicit, if nothing were done other than to increase the number of stopped muons by an order of magnitude by running time and power, the background from DIO would be $\approx 2.5$  events and the background from RPC would be $\approx 1$ event.  Tighter cuts on momentum and a narrower beam pulse, or further delay in beginning the measurement period could reduce those backgrounds but there would be a loss of acceptance.  The tradeoffs and ultimate sensitivity would likely place the experiment  in the background dominated region of $1/\sqrt{N}$ improvement.  Finally, the proposed, and audacious, $10^4$  improvement in the limit at Mu2e and COMET will undoubtedly present unforeseen problems.  
\end{enumerate}

We are also beginning to hit limits of computation and limits on knowledge of secondary processes.  Monte Carlo simulations for Mu2e are already occupying weeks of grid time; the design of neutron shielding is limited by the knowledge of neutron production rates in captures and by our ability to model neutron transport; the measurements of radiative pion and muon capture are barely adequate as they are (and given the unlikelihood of new, dedicated experiments, will have to be measured {\it in situ}.)  The entire issue of calibration will have to be thought afresh for going beyond level the required for COMET or Mu2e.

The $\mu \rightarrow 3e$ channel has always been performed as an additional measurement in an experiment designed for something else, although it provides a powerful complement to the other searches. It requires a new home because the $\mu \rightarrow e \gamma $ and muon-to-electron conversion experiments have become so specialized. An add-on to the $\mu \rightarrow e \gamma$ is probably not feasible: the rates required overwhelm photon detection required in $\mu \rightarrow e \gamma$, and reaching the intrinsic background limits requires an overwhelming flux.  No one has systematically explored a modern experiment in the style of MEGA: converting the photon in $\mu \rightarrow e \gamma $ and performing an experiment that could search for $\mu \rightarrow e \gamma$ and $\mu \rightarrow 3e$ simultaneously.   It is hard to imagine doing these experiments at a modern muon-to-electron conversion experiment like Mu2e or COMET, since their design requires eliminating the vast majority of the Michel spectrum and, in particular, the region required for a $\mu \rightarrow 3e$ search.   The dedicated proposal at PSI discussed in this Article seems the only way to progress in the near-to-medium term.  We hope the brutal realities of flux, real estate, budgets, and human resources will not require postponement until some future facility can handle both experiments.

Having been realistic about the challenges and difficulties of these experiments, one might be dissuaded from further investigations. However, one should note from history that the first searches for CLFV in the muon sector  started at $10^{-4}$ and have now progressed to $10^{-13}$ with $10^{-17}$ seemingly within reach, as demonstrated in Fig.~\ref{fig:clfvHistory}. Presently CLFV is an active and exciting area of research and given that history one should not be too pessimistic.   Perhaps it is fitting to end this article with a metaphor on generations:  by studying the parents and grandparents of the current generation, we can perhaps glimpse and plan for what will develop next.  

Both authors acknowledge the support of Fermi National Accelerator Laboratory, operated by Fermi Research Alliance, LLC under Contract No.\ DE-AC02-07CH11359 with the United States Department of Energy.

The authors would like to thank Martin Cooper, David G. Hitlin, Klaus Jungmann,  Andries van der Schaaf, Lawrence Littenberg,  Jack Ritchie, and Giovanni Signorelli  for careful readings of parts or all of the manuscript and many useful comments; remaining errors, omissions,  infelicities of prose,  and plain bad writing are entirely ours and we regret them.




    \def\citeapos#1{\citeauthor{#1}:\citeyear{#1}}


\bibliography{Biblio-RHB}

\begin{thebibliography}{184}%
\makeatletter
\providecommand \@ifxundefined [1]{%
 \@ifx{#1\undefined}
}%
\providecommand \@ifnum [1]{%
 \ifnum #1\expandafter \@firstoftwo
 \else \expandafter \@secondoftwo
 \fi
}%
\providecommand \@ifx [1]{%
 \ifx #1\expandafter \@firstoftwo
 \else \expandafter \@secondoftwo
 \fi
}%
\providecommand \natexlab [1]{#1}%
\providecommand \enquote  [1]{``#1''}%
\providecommand \bibnamefont  [1]{#1}%
\providecommand \bibfnamefont [1]{#1}%
\providecommand \citenamefont [1]{#1}%
\providecommand \href@noop [0]{\@secondoftwo}%
\providecommand \href [0]{\begingroup \@sanitize@url \@href}%
\providecommand \@href[1]{\@@startlink{#1}\@@href}%
\providecommand \@@href[1]{\endgroup#1\@@endlink}%
\providecommand \@sanitize@url [0]{\catcode `\\12\catcode `\$12\catcode
  `\&12\catcode `\#12\catcode `\^12\catcode `\_12\catcode `\%12\relax}%
\providecommand \@@startlink[1]{}%
\providecommand \@@endlink[0]{}%
\providecommand \url  [0]{\begingroup\@sanitize@url \@url }%
\providecommand \@url [1]{\endgroup\@href {#1}{\urlprefix }}%
\providecommand \urlprefix  [0]{URL }%
\providecommand \Eprint [0]{\href }%
\providecommand \doibase [0]{http://dx.doi.org/}%
\providecommand \selectlanguage [0]{\@gobble}%
\providecommand \bibinfo  [0]{\@secondoftwo}%
\providecommand \bibfield  [0]{\@secondoftwo}%
\providecommand \translation [1]{[#1]}%
\providecommand \BibitemOpen [0]{}%
\providecommand \bibitemStop [0]{}%
\providecommand \bibitemNoStop [0]{.\EOS\space}%
\providecommand \EOS [0]{\spacefactor3000\relax}%
\providecommand \BibitemShut  [1]{\csname bibitem#1\endcsname}%
\let\auto@bib@innerbib\@empty
\bibitem [{\citenamefont {Aad}\ \emph {et~al.}(2012)\citenamefont {Aad} \emph
  {et~al.}}]{Aad:2012yw}%
  \BibitemOpen
  \bibfield  {author} {\bibinfo {author} {\bibnamefont {Aad}, \bibfnamefont
  {G.}},  \emph {et~al.} (\bibinfo {collaboration} {ATLAS Collaboration})}
  (\bibinfo {year} {2012}),\ \href {\doibase 10.1140/epjc/s10052-012-2040-z}
  {\bibfield  {journal} {\bibinfo  {journal} {Eur.Phys.J.}\ }\textbf {\bibinfo
  {volume} {C72}},\ \bibinfo {pages} {2040}},\ \Eprint
  {http://arxiv.org/abs/1205.0725} {arXiv:1205.0725 [hep-ex]} \BibitemShut
  {NoStop}%
\bibitem [{\citenamefont {Aaij}\ \emph {et~al.}(2012)\citenamefont {Aaij} \emph
  {et~al.}}]{Aaij:2011ex}%
  \BibitemOpen
  \bibfield  {author} {\bibinfo {author} {\bibnamefont {Aaij}, \bibfnamefont
  {R.}},  \emph {et~al.} (\bibinfo {collaboration} {LHCb Collaboration})}
  (\bibinfo {year} {2012}),\ \href {\doibase 10.1103/PhysRevLett.108.101601}
  {\bibfield  {journal} {\bibinfo  {journal} {Phys.Rev.Lett.}\ }\textbf
  {\bibinfo {volume} {108}},\ \bibinfo {pages} {101601}},\ \Eprint
  {http://arxiv.org/abs/1110.0730} {arXiv:1110.0730 [hep-ex]} \BibitemShut
  {NoStop}%
\bibitem [{\citenamefont {Aaltonen}\ \emph {et~al.}(2010)\citenamefont
  {Aaltonen} \emph {et~al.}}]{Aaltonen:2010fv}%
  \BibitemOpen
  \bibfield  {author} {\bibinfo {author} {\bibnamefont {Aaltonen},
  \bibfnamefont {T.}},  \emph {et~al.} (\bibinfo {collaboration} {CDF
  Collaboration})} (\bibinfo {year} {2010}),\ \href {\doibase
  10.1103/PhysRevLett.105.191801} {\bibfield  {journal} {\bibinfo  {journal}
  {Phys.Rev.Lett.}\ }\textbf {\bibinfo {volume} {105}},\ \bibinfo {pages}
  {191801}},\ \Eprint {http://arxiv.org/abs/1004.3042} {arXiv:1004.3042
  [hep-ex]} \BibitemShut {NoStop}%
\bibitem [{\citenamefont {Abadjev}\ \emph {et~al.}(1992)\citenamefont {Abadjev}
  \emph {et~al.}}]{Lobashev:1989}%
  \BibitemOpen
  \bibfield  {author} {\bibinfo {author} {\bibnamefont {Abadjev}, \bibfnamefont
  {V.~S.}},  \emph {et~al.}} (\bibinfo {year} {1992}),\ \href@noop {} {\emph
  {\bibinfo {title} {{MELC Experiment to Search for the} $\mu${A} $\rightarrow$
  e{A Process}}}},\ \bibinfo {type} {Tech. Rep.}\ (\bibinfo  {institution}
  {{INR--786/92}})\ \bibinfo {note} {\url
  {https://mu2e-docdb.fnal.gov:440/cgi-bin/RetrieveFile?docid=76;filename=meco002.pdf;version=1}}\BibitemShut
  {NoStop}%
\bibitem [{\citenamefont {Abazov}\ \emph {et~al.}(2006)\citenamefont {Abazov}
  \emph {et~al.}}]{Abazov:2006nw}%
  \BibitemOpen
  \bibfield  {author} {\bibinfo {author} {\bibnamefont {Abazov}, \bibfnamefont
  {V.}},  \emph {et~al.} (\bibinfo {collaboration} {D0 Collaboration})}
  (\bibinfo {year} {2006}),\ \href {\doibase 10.1016/j.physletb.2006.05.077}
  {\bibfield  {journal} {\bibinfo  {journal} {Phys.Lett.}\ }\textbf {\bibinfo
  {volume} {B638}},\ \bibinfo {pages} {441}},\ \Eprint
  {http://arxiv.org/abs/hep-ex/0605005} {arXiv:hep-ex/0605005 [hep-ex]}
  \BibitemShut {NoStop}%
\bibitem [{\citenamefont {Abe}\ \emph {et~al.}(2011)\citenamefont {Abe} \emph
  {et~al.}}]{T2K:2011}%
  \BibitemOpen
  \bibfield  {author} {\bibinfo {author} {\bibnamefont {Abe}, \bibfnamefont
  {K.}},  \emph {et~al.} (\bibinfo {collaboration} {T2K Collaboration})}
  (\bibinfo {year} {2011}),\ \href {\doibase 10.1103/PhysRevLett.107.041801}
  {\bibfield  {journal} {\bibinfo  {journal} {Phys. Rev. Lett.}\ }\textbf
  {\bibinfo {volume} {107}},\ \bibinfo {pages} {041801}}\BibitemShut {NoStop}%
\bibitem [{\citenamefont {Abela}\ \emph {et~al.}(1980)\citenamefont {Abela},
  \citenamefont {Backenstoss}, \citenamefont {Kowald}, \citenamefont {Wuest},
  \citenamefont {Seiler} \emph {et~al.}}]{Abela:1980}%
  \BibitemOpen
  \bibfield  {author} {\bibinfo {author} {\bibnamefont {Abela}, \bibfnamefont
  {R.}}, \bibinfo {author} {\bibfnamefont {G.}~\bibnamefont {Backenstoss}},
  \bibinfo {author} {\bibfnamefont {W.}~\bibnamefont {Kowald}}, \bibinfo
  {author} {\bibfnamefont {J.}~\bibnamefont {Wuest}}, \bibinfo {author}
  {\bibfnamefont {H.}~\bibnamefont {Seiler}},  \emph {et~al.}} (\bibinfo {year}
  {1980}),\ \href {\doibase 10.1016/0370-2693(80)90495-5} {\bibfield  {journal}
  {\bibinfo  {journal} {Phys.Lett.}\ }\textbf {\bibinfo {volume} {B95}},\
  \bibinfo {pages} {318}}\BibitemShut {NoStop}%
\bibitem [{\citenamefont {Abouzaid}\ \emph {et~al.}(2008)\citenamefont
  {Abouzaid} \emph {et~al.}}]{Abouzaid:2008}%
  \BibitemOpen
  \bibfield  {author} {\bibinfo {author} {\bibnamefont {Abouzaid},
  \bibfnamefont {E.}},  \emph {et~al.} (\bibinfo {collaboration} {KTeV
  Collaboration})} (\bibinfo {year} {2008}),\ \href {\doibase
  10.1103/PhysRevLett.100.131803} {\bibfield  {journal} {\bibinfo  {journal}
  {Phys. Rev. Lett.}\ }\textbf {\bibinfo {volume} {100}},\ \bibinfo {pages}
  {131803}}\BibitemShut {NoStop}%
\bibitem [{\citenamefont {Abrams}\ \emph {et~al.}(2012)\citenamefont {Abrams}
  \emph {et~al.}}]{Abrams:2012}%
  \BibitemOpen
  \bibfield  {author} {\bibinfo {author} {\bibnamefont {Abrams}, \bibfnamefont
  {R.}},  \emph {et~al.} (\bibinfo {collaboration} {Mu2e Collaboration})}
  (\bibinfo {year} {2012}),\ \href@noop {} {\ }\Eprint
  {http://arxiv.org/abs/1211.7019} {arXiv:1211.7019 [physics.ins-det]}
  \BibitemShut {NoStop}%
\bibitem [{\citenamefont {Abreu}\ \emph {et~al.}(1997)\citenamefont {Abreu}
  \emph {et~al.}}]{Abreu:1996mj}%
  \BibitemOpen
  \bibfield  {author} {\bibinfo {author} {\bibnamefont {Abreu}, \bibfnamefont
  {P.}},  \emph {et~al.} (\bibinfo {collaboration} {DELPHI Collaboration})}
  (\bibinfo {year} {1997}),\ \href {\doibase 10.1007/s002880050313} {\bibfield
  {journal} {\bibinfo  {journal} {Z.Phys.}\ }\textbf {\bibinfo {volume}
  {C73}},\ \bibinfo {pages} {243}}\BibitemShut {NoStop}%
\bibitem [{\citenamefont {Adam}\ \emph {et~al.}(2010)\citenamefont {Adam} \emph
  {et~al.}}]{meg:2010}%
  \BibitemOpen
  \bibfield  {author} {\bibinfo {author} {\bibnamefont {Adam}, \bibfnamefont
  {J.}},  \emph {et~al.} (\bibinfo {collaboration} {MEG collaboration})}
  (\bibinfo {year} {2010}),\ \href {\doibase 10.1016/j.nuclphysb.2010.03.030}
  {\bibfield  {journal} {\bibinfo  {journal} {Nucl.Phys.}\ }\textbf {\bibinfo
  {volume} {B834}},\ \bibinfo {pages} {1}},\ \Eprint
  {http://arxiv.org/abs/0908.2594} {arXiv:0908.2594 [hep-ex]} \BibitemShut
  {NoStop}%
\bibitem [{\citenamefont {Adam}\ \emph {et~al.}(2011)\citenamefont {Adam} \emph
  {et~al.}}]{Adam:2011}%
  \BibitemOpen
  \bibfield  {author} {\bibinfo {author} {\bibnamefont {Adam}, \bibfnamefont
  {J.}},  \emph {et~al.} (\bibinfo {collaboration} {MEG Collaboration})}
  (\bibinfo {year} {2011}),\ \href {\doibase 10.1103/PhysRevLett.107.171801}
  {\bibfield  {journal} {\bibinfo  {journal} {Phys. Rev. Lett.}\ }\textbf
  {\bibinfo {volume} {107}},\ \bibinfo {pages} {171801}},\ \Eprint
  {http://arxiv.org/abs/arXiv:hep-ex/1107.5547} {arXiv:hep-ex/1107.5547}
  \BibitemShut {NoStop}%
\bibitem [{\citenamefont {Adam}\ \emph {et~al.}(2013)\citenamefont {Adam} \emph
  {et~al.}}]{Adam:2013}%
  \BibitemOpen
  \bibfield  {author} {\bibinfo {author} {\bibnamefont {Adam}, \bibfnamefont
  {J.}},  \emph {et~al.} (\bibinfo {collaboration} {MEG Collaboration})}
  (\bibinfo {year} {2013}),\ \href@noop {} {\ }\Eprint
  {http://arxiv.org/abs/1303.0754} {arXiv:1303.0754 [hep-ex]} \BibitemShut
  {NoStop}%
\bibitem [{\citenamefont {Adamson}\ \emph {et~al.}(2011)\citenamefont {Adamson}
  \emph {et~al.}}]{MINOS:2011}%
  \BibitemOpen
  \bibfield  {author} {\bibinfo {author} {\bibnamefont {Adamson}, \bibfnamefont
  {P.}},  \emph {et~al.} (\bibinfo {collaboration} {MINOS Collaboration})}
  (\bibinfo {year} {2011}),\ \href {\doibase 10.1103/PhysRevLett.107.181802}
  {\bibfield  {journal} {\bibinfo  {journal} {Phys. Rev. Lett.}\ }\textbf
  {\bibinfo {volume} {107}},\ \bibinfo {pages} {181802}}\BibitemShut {NoStop}%
\bibitem [{\citenamefont {Adler}\ \emph {et~al.}(1997)\citenamefont {Adler},
  \citenamefont {Atiya}, \citenamefont {Chiang}, \citenamefont {Frank},
  \citenamefont {Haggerty}, \citenamefont {Kycia}, \citenamefont {Li},
  \citenamefont {Littenberg}, \citenamefont {Sambamurti}, \citenamefont
  {Stevens}, \citenamefont {Strand}, \citenamefont {Witzig}, \citenamefont
  {Louis}, \citenamefont {Akerib}, \citenamefont {Ardebili}, \citenamefont
  {Convery}, \citenamefont {Ito}, \citenamefont {Marlow}, \citenamefont
  {McPherson}, \citenamefont {Meyers}, \citenamefont {Selen}, \citenamefont
  {Shoemaker}, \citenamefont {Smith}, \citenamefont {Blackmore}, \citenamefont
  {Bryman}, \citenamefont {Felawka}, \citenamefont {Konaka}, \citenamefont
  {Kuno}, \citenamefont {Macdonald}, \citenamefont {Numao}, \citenamefont
  {Padley}, \citenamefont {Poutissou}, \citenamefont {Poutissou}, \citenamefont
  {Roy}, \citenamefont {Turcot}, \citenamefont {Kitching}, \citenamefont
  {Nakano}, \citenamefont {Rozon},\ and\ \citenamefont {Soluk}}]{Adler:1997}%
  \BibitemOpen
  \bibfield  {author} {\bibinfo {author} {\bibnamefont {Adler}, \bibfnamefont
  {S.}}, \bibinfo {author} {\bibfnamefont {M.~S.}\ \bibnamefont {Atiya}},
  \bibinfo {author} {\bibfnamefont {I.-H.}\ \bibnamefont {Chiang}}, \bibinfo
  {author} {\bibfnamefont {J.~S.}\ \bibnamefont {Frank}}, \bibinfo {author}
  {\bibfnamefont {J.~S.}\ \bibnamefont {Haggerty}}, \bibinfo {author}
  {\bibfnamefont {T.~F.}\ \bibnamefont {Kycia}}, \bibinfo {author}
  {\bibfnamefont {K.~K.}\ \bibnamefont {Li}}, \bibinfo {author} {\bibfnamefont
  {L.~S.}\ \bibnamefont {Littenberg}}, \bibinfo {author} {\bibfnamefont
  {A.}~\bibnamefont {Sambamurti}}, \bibinfo {author} {\bibfnamefont
  {A.}~\bibnamefont {Stevens}}, \bibinfo {author} {\bibfnamefont {R.~C.}\
  \bibnamefont {Strand}}, \bibinfo {author} {\bibfnamefont {C.}~\bibnamefont
  {Witzig}}, \bibinfo {author} {\bibfnamefont {W.~C.}\ \bibnamefont {Louis}},
  \bibinfo {author} {\bibfnamefont {D.~S.}\ \bibnamefont {Akerib}}, \bibinfo
  {author} {\bibfnamefont {M.}~\bibnamefont {Ardebili}}, \bibinfo {author}
  {\bibfnamefont {M.}~\bibnamefont {Convery}}, \bibinfo {author} {\bibfnamefont
  {M.~M.}\ \bibnamefont {Ito}}, \bibinfo {author} {\bibfnamefont {D.~R.}\
  \bibnamefont {Marlow}}, \bibinfo {author} {\bibfnamefont {R.}~\bibnamefont
  {McPherson}}, \bibinfo {author} {\bibfnamefont {P.~D.}\ \bibnamefont
  {Meyers}}, \bibinfo {author} {\bibfnamefont {M.~A.}\ \bibnamefont {Selen}},
  \bibinfo {author} {\bibfnamefont {F.~C.}\ \bibnamefont {Shoemaker}}, \bibinfo
  {author} {\bibfnamefont {A.~J.~S.}\ \bibnamefont {Smith}}, \bibinfo {author}
  {\bibfnamefont {E.~W.}\ \bibnamefont {Blackmore}}, \bibinfo {author}
  {\bibfnamefont {D.~A.}\ \bibnamefont {Bryman}}, \bibinfo {author}
  {\bibfnamefont {L.}~\bibnamefont {Felawka}}, \bibinfo {author} {\bibfnamefont
  {A.}~\bibnamefont {Konaka}}, \bibinfo {author} {\bibfnamefont
  {Y.}~\bibnamefont {Kuno}}, \bibinfo {author} {\bibfnamefont {J.~A.}\
  \bibnamefont {Macdonald}}, \bibinfo {author} {\bibfnamefont {T.}~\bibnamefont
  {Numao}}, \bibinfo {author} {\bibfnamefont {P.}~\bibnamefont {Padley}},
  \bibinfo {author} {\bibfnamefont {J.-M.}\ \bibnamefont {Poutissou}}, \bibinfo
  {author} {\bibfnamefont {R.}~\bibnamefont {Poutissou}}, \bibinfo {author}
  {\bibfnamefont {J.}~\bibnamefont {Roy}}, \bibinfo {author} {\bibfnamefont
  {A.~S.}\ \bibnamefont {Turcot}}, \bibinfo {author} {\bibfnamefont
  {P.}~\bibnamefont {Kitching}}, \bibinfo {author} {\bibfnamefont
  {T.}~\bibnamefont {Nakano}}, \bibinfo {author} {\bibfnamefont
  {M.}~\bibnamefont {Rozon}}, \ and\ \bibinfo {author} {\bibfnamefont
  {R.}~\bibnamefont {Soluk}}} (\bibinfo {year} {1997}),\ \href {\doibase
  10.1103/PhysRevLett.79.4756} {\bibfield  {journal} {\bibinfo  {journal}
  {Phys. Rev. Lett.}\ }\textbf {\bibinfo {volume} {79}},\ \bibinfo {pages}
  {4756}}\BibitemShut {NoStop}%
\bibitem [{\citenamefont {Adriani}\ \emph {et~al.}(1993)\citenamefont {Adriani}
  \emph {et~al.}}]{Adriani:1993sy}%
  \BibitemOpen
  \bibfield  {author} {\bibinfo {author} {\bibnamefont {Adriani}, \bibfnamefont
  {O.}},  \emph {et~al.} (\bibinfo {collaboration} {L3 Collaboration})}
  (\bibinfo {year} {1993}),\ \href {\doibase 10.1016/0370-2693(93)90348-L}
  {\bibfield  {journal} {\bibinfo  {journal} {Phys.Lett.}\ }\textbf {\bibinfo
  {volume} {B316}},\ \bibinfo {pages} {427}}\BibitemShut {NoStop}%
\bibitem [{\citenamefont {Ahmad}\ \emph {et~al.}(1988)\citenamefont {Ahmad}
  \emph {et~al.}}]{Ahmad:1988}%
  \BibitemOpen
  \bibfield  {author} {\bibinfo {author} {\bibnamefont {Ahmad}, \bibfnamefont
  {S.}},  \emph {et~al.}} (\bibinfo {year} {1988}),\ \href {\doibase
  10.1103/PhysRevD.38.2102} {\bibfield  {journal} {\bibinfo  {journal} {Phys.
  Rev. D}\ }\textbf {\bibinfo {volume} {38}},\ \bibinfo {pages}
  {2102}}\BibitemShut {NoStop}%
\bibitem [{\citenamefont {Ahmed}\ \emph {et~al.}(2002)\citenamefont {Ahmed}
  \emph {et~al.}}]{Ahmed:2002}%
  \BibitemOpen
  \bibfield  {author} {\bibinfo {author} {\bibnamefont {Ahmed}, \bibfnamefont
  {M.}},  \emph {et~al.} (\bibinfo {collaboration} {MEGA Collaboration})}
  (\bibinfo {year} {2002}),\ \href {\doibase 10.1103/PhysRevD.65.112002}
  {\bibfield  {journal} {\bibinfo  {journal} {Phys. Rev. D}\ }\textbf {\bibinfo
  {volume} {65}},\ \bibinfo {pages} {112002}},\ \Eprint
  {http://arxiv.org/abs/arXiv:hep-ex/0111030v1} {arXiv:hep-ex/0111030v1}
  \BibitemShut {NoStop}%
\bibitem [{\citenamefont {Ahn}\ \emph {et~al.}(2012)\citenamefont {Ahn} \emph
  {et~al.}}]{RENO:2012}%
  \BibitemOpen
  \bibfield  {author} {\bibinfo {author} {\bibnamefont {Ahn}, \bibfnamefont
  {J.~K.}},  \emph {et~al.} (\bibinfo {collaboration} {RENO Collaboration})}
  (\bibinfo {year} {2012}),\ \href {\doibase 10.1103/PhysRevLett.108.191802}
  {\bibfield  {journal} {\bibinfo  {journal} {Phys. Rev. Lett.}\ }\textbf
  {\bibinfo {volume} {108}},\ \bibinfo {pages} {191802}}\BibitemShut {NoStop}%
\bibitem [{\citenamefont {Akagi}\ \emph {et~al.}(1995)\citenamefont {Akagi},
  \citenamefont {Fukuhisa}, \citenamefont {Hemmi}, \citenamefont {Inagaki},
  \citenamefont {Ishikawa}, \citenamefont {Kishida}, \citenamefont {Kobayashi},
  \citenamefont {Komatsubara}, \citenamefont {Kuze}, \citenamefont {Sai},
  \citenamefont {Sato}, \citenamefont {Shinkawa}, \citenamefont {Suekane},
  \citenamefont {Takamatsu}, \citenamefont {Toyoura}, \citenamefont
  {Yamamoto},\ and\ \citenamefont {Yoshimura}}]{Akagi:1995}%
  \BibitemOpen
  \bibfield  {author} {\bibinfo {author} {\bibnamefont {Akagi}, \bibfnamefont
  {T.}}, \bibinfo {author} {\bibfnamefont {R.}~\bibnamefont {Fukuhisa}},
  \bibinfo {author} {\bibfnamefont {Y.}~\bibnamefont {Hemmi}}, \bibinfo
  {author} {\bibfnamefont {T.}~\bibnamefont {Inagaki}}, \bibinfo {author}
  {\bibfnamefont {K.}~\bibnamefont {Ishikawa}}, \bibinfo {author}
  {\bibfnamefont {T.}~\bibnamefont {Kishida}}, \bibinfo {author} {\bibfnamefont
  {M.}~\bibnamefont {Kobayashi}}, \bibinfo {author} {\bibfnamefont {T.~K.}\
  \bibnamefont {Komatsubara}}, \bibinfo {author} {\bibfnamefont
  {M.}~\bibnamefont {Kuze}}, \bibinfo {author} {\bibfnamefont {F.}~\bibnamefont
  {Sai}}, \bibinfo {author} {\bibfnamefont {T.}~\bibnamefont {Sato}}, \bibinfo
  {author} {\bibfnamefont {T.}~\bibnamefont {Shinkawa}}, \bibinfo {author}
  {\bibfnamefont {F.}~\bibnamefont {Suekane}}, \bibinfo {author} {\bibfnamefont
  {K.}~\bibnamefont {Takamatsu}}, \bibinfo {author} {\bibfnamefont
  {J.}~\bibnamefont {Toyoura}}, \bibinfo {author} {\bibfnamefont {S.~S.}\
  \bibnamefont {Yamamoto}}, \ and\ \bibinfo {author} {\bibfnamefont
  {Y.}~\bibnamefont {Yoshimura}}} (\bibinfo {year} {1995}),\ \href {\doibase
  10.1103/PhysRevD.51.2061} {\bibfield  {journal} {\bibinfo  {journal} {Phys.
  Rev. D}\ }\textbf {\bibinfo {volume} {51}},\ \bibinfo {pages}
  {2061}}\BibitemShut {NoStop}%
\bibitem [{\citenamefont {Akeroyd}\ \emph {et~al.}(2004)\citenamefont {Akeroyd}
  \emph {et~al.}}]{SuperKEKB:2004}%
  \BibitemOpen
  \bibfield  {author} {\bibinfo {author} {\bibnamefont {Akeroyd}, \bibfnamefont
  {A.}},  \emph {et~al.} (\bibinfo {collaboration} {SuperKEKB Physics Working
  Group})} (\bibinfo {year} {2004}),\ \href@noop {} {\ }\Eprint
  {http://arxiv.org/abs/hep-ex/0406071} {arXiv:hep-ex/0406071 [hep-ex]}
  \BibitemShut {NoStop}%
\bibitem [{\citenamefont {Akers}\ \emph {et~al.}(1995)\citenamefont {Akers}
  \emph {et~al.}}]{Akers:1995gz}%
  \BibitemOpen
  \bibfield  {author} {\bibinfo {author} {\bibnamefont {Akers}, \bibfnamefont
  {R.}},  \emph {et~al.} (\bibinfo {collaboration} {OPAL Collaboration})}
  (\bibinfo {year} {1995}),\ \href {\doibase 10.1007/BF01553981} {\bibfield
  {journal} {\bibinfo  {journal} {Z.Phys.}\ }\textbf {\bibinfo {volume}
  {C67}},\ \bibinfo {pages} {555}}\BibitemShut {NoStop}%
\bibitem [{\citenamefont {Aktas}\ \emph {et~al.}(2007)\citenamefont {Aktas}
  \emph {et~al.}}]{Aktas:2007ji}%
  \BibitemOpen
  \bibfield  {author} {\bibinfo {author} {\bibnamefont {Aktas}, \bibfnamefont
  {A.}},  \emph {et~al.} (\bibinfo {collaboration} {H1 Collaboration})}
  (\bibinfo {year} {2007}),\ \href {\doibase 10.1140/epjc/s10052-007-0440-2}
  {\bibfield  {journal} {\bibinfo  {journal} {The European Physical Journal C -
  Particles and Fields}\ }\textbf {\bibinfo {volume} {52}},\ \bibinfo {pages}
  {833}}\BibitemShut {NoStop}%
\bibitem [{\citenamefont {Alavi-Harati}\ \emph {et~al.}(2003)\citenamefont
  {Alavi-Harati} \emph {et~al.}}]{AlaviHarati:2003}%
  \BibitemOpen
  \bibfield  {author} {\bibinfo {author} {\bibnamefont {Alavi-Harati},
  \bibfnamefont {A.}},  \emph {et~al.} (\bibinfo {collaboration} {KTeV
  Collaboration})} (\bibinfo {year} {2003}),\ \href {\doibase
  10.1103/PhysRevD.67.012005} {\bibfield  {journal} {\bibinfo  {journal} {Phys.
  Rev. D}\ }\textbf {\bibinfo {volume} {67}},\ \bibinfo {pages}
  {012005}}\BibitemShut {NoStop}%
\bibitem [{\citenamefont {Amato}\ \emph {et~al.}(1968)\citenamefont {Amato},
  \citenamefont {Crane}, \citenamefont {Hughes}, \citenamefont {Rothberg},\
  and\ \citenamefont {Thompson}}]{Amato:1968}%
  \BibitemOpen
  \bibfield  {author} {\bibinfo {author} {\bibnamefont {Amato}, \bibfnamefont
  {J.~J.}}, \bibinfo {author} {\bibfnamefont {P.}~\bibnamefont {Crane}},
  \bibinfo {author} {\bibfnamefont {V.~W.}\ \bibnamefont {Hughes}}, \bibinfo
  {author} {\bibfnamefont {J.~E.}\ \bibnamefont {Rothberg}}, \ and\ \bibinfo
  {author} {\bibfnamefont {P.~A.}\ \bibnamefont {Thompson}}} (\bibinfo {year}
  {1968}),\ \href {\doibase 10.1103/PhysRevLett.21.1709} {\bibfield  {journal}
  {\bibinfo  {journal} {Phys. Rev. Lett.}\ }\textbf {\bibinfo {volume} {21}},\
  \bibinfo {pages} {1709}}\BibitemShut {NoStop}%
\bibitem [{\citenamefont {Ambrose}\ \emph {et~al.}(1998)\citenamefont {Ambrose}
  \emph {et~al.}}]{Ambrose:1998}%
  \BibitemOpen
  \bibfield  {author} {\bibinfo {author} {\bibnamefont {Ambrose}, \bibfnamefont
  {D.}},  \emph {et~al.} (\bibinfo {collaboration} {BNL E871 Collaboration})}
  (\bibinfo {year} {1998}),\ \href {\doibase 10.1103/PhysRevLett.81.5734}
  {\bibfield  {journal} {\bibinfo  {journal} {Phys. Rev. Lett.}\ }\textbf
  {\bibinfo {volume} {81}},\ \bibinfo {pages} {5734}}\BibitemShut {NoStop}%
\bibitem [{\citenamefont {Amhis}\ \emph {et~al.}(2012)\citenamefont {Amhis}
  \emph {et~al.}}]{HFAG:2012}%
  \BibitemOpen
  \bibfield  {author} {\bibinfo {author} {\bibnamefont {Amhis}, \bibfnamefont
  {Y.}},  \emph {et~al.} (\bibinfo {collaboration} {Heavy Flavor Averaging
  Group})} (\bibinfo {year} {2012}),\ \href@noop {} {\enquote {\bibinfo {title}
  {{Averages of b-hadron, c-hadron, and $\tau$-lepton properties as of early
  2012}},}\ }\Eprint {http://arxiv.org/abs/1207.1158} {arXiv:1207.1158
  [hep-ex]} \BibitemShut {NoStop}%
\bibitem [{\citenamefont {An}\ \emph {et~al.}(2012)\citenamefont {An} \emph
  {et~al.}}]{DayaBay:2012}%
  \BibitemOpen
  \bibfield  {author} {\bibinfo {author} {\bibnamefont {An}, \bibfnamefont
  {F.~P.}},  \emph {et~al.}} (\bibinfo {year} {2012}),\ \href {\doibase
  10.1103/PhysRevLett.108.171803} {\bibfield  {journal} {\bibinfo  {journal}
  {Phys. Rev. Lett.}\ }\textbf {\bibinfo {volume} {108}},\ \bibinfo {pages}
  {171803}},\ \Eprint {http://arxiv.org/abs/1203.1669} {arXiv:1203.1669
  [hep-ex]} \BibitemShut {NoStop}%
\bibitem [{\citenamefont {Antusch}\ \emph {et~al.}(2007)\citenamefont {Antusch}
  \emph {et~al.}}]{Antusch:2006}%
  \BibitemOpen
  \bibfield  {author} {\bibinfo {author} {\bibnamefont {Antusch}, \bibfnamefont
  {S.}},  \emph {et~al.}} (\bibinfo {year} {2007}),\ \href {\doibase
  10.1016/j.nuclphysbps.2007.02.102} {\bibfield  {journal} {\bibinfo  {journal}
  {Nucl.Phys.Proc.Suppl.}\ }\textbf {\bibinfo {volume} {169}},\ \bibinfo
  {pages} {155}},\ \Eprint {http://arxiv.org/abs/hep-ph/0610439}
  {arXiv:hep-ph/0610439 [hep-ph]} \BibitemShut {NoStop}%
\bibitem [{\citenamefont {Aoki}(2003)}]{Aoki:2003}%
  \BibitemOpen
  \bibfield  {author} {\bibinfo {author} {\bibnamefont {Aoki}, \bibfnamefont
  {M.}}} (\bibinfo {year} {2003}),\ \href {\doibase
  10.1016/S0168-9002(03)00689-2} {\bibfield  {journal} {\bibinfo  {journal}
  {Nuclear Instruments and Methods in Physics Research Section A: Accelerators,
  Spectrometers, Detectors and Associated Equipment}\ }\textbf {\bibinfo
  {volume} {503}},\ \bibinfo {pages} {258 }},\ \bibinfo {note} {proceedings of
  the 3rd International Workshop on Neutrino Factories based on Muon Storage
  Rings}\BibitemShut {NoStop}%
\bibitem [{\citenamefont {Aoki}\ \emph {et~al.}(2012)\citenamefont {Aoki} \emph
  {et~al.}}]{Aoki:2012}%
  \BibitemOpen
  \bibfield  {author} {\bibinfo {author} {\bibnamefont {Aoki}, \bibfnamefont
  {M.}},  \emph {et~al.} (\bibinfo {collaboration} {DeeMe})} (\bibinfo {year}
  {2012}),\ \href@noop {} {\enquote {\bibinfo {title} {Expression of interest
  for an experiment searching for $\mu$-e conversion at j-parc muon
  facility},}\ }\bibinfo {note}
  {\url{http://deeme.hep.sci.osaka-u.ac.jp/documents}}\BibitemShut {NoStop}%
\bibitem [{\citenamefont {Appel}\ \emph {et~al.}(2008)\citenamefont {Appel},
  \citenamefont {Asner}, \citenamefont {Bigi}, \citenamefont {Bryman},
  \citenamefont {Buras} \emph {et~al.}}]{Appel:2008aa}%
  \BibitemOpen
  \bibfield  {author} {\bibinfo {author} {\bibnamefont {Appel}, \bibfnamefont
  {J.}}, \bibinfo {author} {\bibfnamefont {D.}~\bibnamefont {Asner}}, \bibinfo
  {author} {\bibfnamefont {I.}~\bibnamefont {Bigi}}, \bibinfo {author}
  {\bibfnamefont {D.}~\bibnamefont {Bryman}}, \bibinfo {author} {\bibfnamefont
  {A.}~\bibnamefont {Buras}},  \emph {et~al.}} (\bibinfo {year} {2008}),\
  \href@noop {} {\ }\BibitemShut {NoStop}%
\bibitem [{\citenamefont {Appel}\ \emph {et~al.}(1999)\citenamefont {Appel},
  \citenamefont {Atoyan}, \citenamefont {Bassalleck}, \citenamefont {Bergman},
  \citenamefont {Brown}, \citenamefont {Cheung}, \citenamefont {Dhawan},
  \citenamefont {Do}, \citenamefont {Egger}, \citenamefont {Eilerts},
  \citenamefont {Felder}, \citenamefont {Fischer}, \citenamefont {Gach},
  \citenamefont {Herold}, \citenamefont {Issakov}, \citenamefont {Kaspar},
  \citenamefont {Kraus}, \citenamefont {Lazarus}, \citenamefont {Leipuner},
  \citenamefont {Lichard}, \citenamefont {Lowe}, \citenamefont {Lozano},
  \citenamefont {Ma}, \citenamefont {Majid}, \citenamefont {Menzel},
  \citenamefont {Pislak}, \citenamefont {Poblaguev}, \citenamefont {Postoev},
  \citenamefont {Proskurjakov}, \citenamefont {Rehak}, \citenamefont {Robmann},
  \citenamefont {Sher}, \citenamefont {Thomas}, \citenamefont {Thompson},
  \citenamefont {Tru\"ol}, \citenamefont {Weyer},\ and\ \citenamefont
  {Zeller}}]{Appel:1999}%
  \BibitemOpen
  \bibfield  {author} {\bibinfo {author} {\bibnamefont {Appel}, \bibfnamefont
  {R.}}, \bibinfo {author} {\bibfnamefont {G.~S.}\ \bibnamefont {Atoyan}},
  \bibinfo {author} {\bibfnamefont {B.}~\bibnamefont {Bassalleck}}, \bibinfo
  {author} {\bibfnamefont {D.~R.}\ \bibnamefont {Bergman}}, \bibinfo {author}
  {\bibfnamefont {D.~N.}\ \bibnamefont {Brown}}, \bibinfo {author}
  {\bibfnamefont {N.}~\bibnamefont {Cheung}}, \bibinfo {author} {\bibfnamefont
  {S.}~\bibnamefont {Dhawan}}, \bibinfo {author} {\bibfnamefont
  {H.}~\bibnamefont {Do}}, \bibinfo {author} {\bibfnamefont {J.}~\bibnamefont
  {Egger}}, \bibinfo {author} {\bibfnamefont {S.}~\bibnamefont {Eilerts}},
  \bibinfo {author} {\bibfnamefont {C.}~\bibnamefont {Felder}}, \bibinfo
  {author} {\bibfnamefont {H.}~\bibnamefont {Fischer}}, \bibinfo {author}
  {\bibfnamefont {M.}~\bibnamefont {Gach}}, \bibinfo {author} {\bibfnamefont
  {W.}~\bibnamefont {Herold}}, \bibinfo {author} {\bibfnamefont {V.~V.}\
  \bibnamefont {Issakov}}, \bibinfo {author} {\bibfnamefont {H.}~\bibnamefont
  {Kaspar}}, \bibinfo {author} {\bibfnamefont {D.~E.}\ \bibnamefont {Kraus}},
  \bibinfo {author} {\bibfnamefont {D.~M.}\ \bibnamefont {Lazarus}}, \bibinfo
  {author} {\bibfnamefont {L.}~\bibnamefont {Leipuner}}, \bibinfo {author}
  {\bibfnamefont {P.}~\bibnamefont {Lichard}}, \bibinfo {author} {\bibfnamefont
  {J.}~\bibnamefont {Lowe}}, \bibinfo {author} {\bibfnamefont {J.}~\bibnamefont
  {Lozano}}, \bibinfo {author} {\bibfnamefont {H.}~\bibnamefont {Ma}}, \bibinfo
  {author} {\bibfnamefont {W.}~\bibnamefont {Majid}}, \bibinfo {author}
  {\bibfnamefont {W.}~\bibnamefont {Menzel}}, \bibinfo {author} {\bibfnamefont
  {S.}~\bibnamefont {Pislak}}, \bibinfo {author} {\bibfnamefont {A.~A.}\
  \bibnamefont {Poblaguev}}, \bibinfo {author} {\bibfnamefont {V.~E.}\
  \bibnamefont {Postoev}}, \bibinfo {author} {\bibfnamefont {A.~L.}\
  \bibnamefont {Proskurjakov}}, \bibinfo {author} {\bibfnamefont
  {P.}~\bibnamefont {Rehak}}, \bibinfo {author} {\bibfnamefont
  {P.}~\bibnamefont {Robmann}}, \bibinfo {author} {\bibfnamefont
  {A.}~\bibnamefont {Sher}}, \bibinfo {author} {\bibfnamefont {T.~L.}\
  \bibnamefont {Thomas}}, \bibinfo {author} {\bibfnamefont {J.~A.}\
  \bibnamefont {Thompson}}, \bibinfo {author} {\bibfnamefont {P.}~\bibnamefont
  {Tru\"ol}}, \bibinfo {author} {\bibfnamefont {H.}~\bibnamefont {Weyer}}, \
  and\ \bibinfo {author} {\bibfnamefont {M.~E.}\ \bibnamefont {Zeller}}}
  (\bibinfo {year} {1999}),\ \href {\doibase 10.1103/PhysRevLett.83.4482}
  {\bibfield  {journal} {\bibinfo  {journal} {Phys. Rev. Lett.}\ }\textbf
  {\bibinfo {volume} {83}},\ \bibinfo {pages} {4482}}\BibitemShut {NoStop}%
\bibitem [{\citenamefont {Appel}\ \emph {et~al.}(2000)\citenamefont {Appel},
  \citenamefont {Atoyan}, \citenamefont {Bassalleck}, \citenamefont {Bergman},
  \citenamefont {Cheung}, \citenamefont {Dhawan}, \citenamefont {Do},
  \citenamefont {Egger}, \citenamefont {Eilerts}, \citenamefont {Fischer},
  \citenamefont {Herold}, \citenamefont {Issakov}, \citenamefont {Kaspar},
  \citenamefont {Kraus}, \citenamefont {Lazarus}, \citenamefont {Lichard},
  \citenamefont {Lowe}, \citenamefont {Lozano}, \citenamefont {Ma},
  \citenamefont {Majid}, \citenamefont {Menzel}, \citenamefont {Pislak},
  \citenamefont {Poblaguev}, \citenamefont {Rehak}, \citenamefont {Sher},
  \citenamefont {Thompson}, \citenamefont {Tru\"ol},\ and\ \citenamefont
  {Zeller}}]{Appel:2000}%
  \BibitemOpen
  \bibfield  {author} {\bibinfo {author} {\bibnamefont {Appel}, \bibfnamefont
  {R.}}, \bibinfo {author} {\bibfnamefont {G.~S.}\ \bibnamefont {Atoyan}},
  \bibinfo {author} {\bibfnamefont {B.}~\bibnamefont {Bassalleck}}, \bibinfo
  {author} {\bibfnamefont {D.~R.}\ \bibnamefont {Bergman}}, \bibinfo {author}
  {\bibfnamefont {N.}~\bibnamefont {Cheung}}, \bibinfo {author} {\bibfnamefont
  {S.}~\bibnamefont {Dhawan}}, \bibinfo {author} {\bibfnamefont
  {H.}~\bibnamefont {Do}}, \bibinfo {author} {\bibfnamefont {J.}~\bibnamefont
  {Egger}}, \bibinfo {author} {\bibfnamefont {S.}~\bibnamefont {Eilerts}},
  \bibinfo {author} {\bibfnamefont {H.}~\bibnamefont {Fischer}}, \bibinfo
  {author} {\bibfnamefont {W.}~\bibnamefont {Herold}}, \bibinfo {author}
  {\bibfnamefont {V.~V.}\ \bibnamefont {Issakov}}, \bibinfo {author}
  {\bibfnamefont {H.}~\bibnamefont {Kaspar}}, \bibinfo {author} {\bibfnamefont
  {D.~E.}\ \bibnamefont {Kraus}}, \bibinfo {author} {\bibfnamefont {D.~M.}\
  \bibnamefont {Lazarus}}, \bibinfo {author} {\bibfnamefont {P.}~\bibnamefont
  {Lichard}}, \bibinfo {author} {\bibfnamefont {J.}~\bibnamefont {Lowe}},
  \bibinfo {author} {\bibfnamefont {J.}~\bibnamefont {Lozano}}, \bibinfo
  {author} {\bibfnamefont {H.}~\bibnamefont {Ma}}, \bibinfo {author}
  {\bibfnamefont {W.}~\bibnamefont {Majid}}, \bibinfo {author} {\bibfnamefont
  {W.}~\bibnamefont {Menzel}}, \bibinfo {author} {\bibfnamefont
  {S.}~\bibnamefont {Pislak}}, \bibinfo {author} {\bibfnamefont {A.~A.}\
  \bibnamefont {Poblaguev}}, \bibinfo {author} {\bibfnamefont {P.}~\bibnamefont
  {Rehak}}, \bibinfo {author} {\bibfnamefont {A.}~\bibnamefont {Sher}},
  \bibinfo {author} {\bibfnamefont {J.~A.}\ \bibnamefont {Thompson}}, \bibinfo
  {author} {\bibfnamefont {P.}~\bibnamefont {Tru\"ol}}, \ and\ \bibinfo
  {author} {\bibfnamefont {M.~E.}\ \bibnamefont {Zeller}}} (\bibinfo {year}
  {2000}),\ \href {\doibase 10.1103/PhysRevLett.85.2877} {\bibfield  {journal}
  {\bibinfo  {journal} {Phys. Rev. Lett.}\ }\textbf {\bibinfo {volume} {85}},\
  \bibinfo {pages} {2877}}\BibitemShut {NoStop}%
\bibitem [{\citenamefont {Arisaka}\ \emph {et~al.}(1993)\citenamefont
  {Arisaka}, \citenamefont {Auerbach}, \citenamefont {Axelrod}, \citenamefont
  {Belz}, \citenamefont {Biery}, \citenamefont {Buchholz}, \citenamefont
  {Chapman}, \citenamefont {Cousins}, \citenamefont {Diwan}, \citenamefont
  {Eckhause}, \citenamefont {Ginkel}, \citenamefont {Guss}, \citenamefont
  {Hancock}, \citenamefont {Heinson}, \citenamefont {Highland}, \citenamefont
  {Hoffmann}, \citenamefont {Horvath}, \citenamefont {Irwin}, \citenamefont
  {Joyce}, \citenamefont {Kaarsberg}, \citenamefont {Kane}, \citenamefont
  {Kenney}, \citenamefont {Kettell}, \citenamefont {Kinnison}, \citenamefont
  {Knibbe}, \citenamefont {Konigsberg}, \citenamefont {Kuang}, \citenamefont
  {Lang}, \citenamefont {Lee}, \citenamefont {Margulies}, \citenamefont
  {Mathiazhagan}, \citenamefont {McFarlane}, \citenamefont {McKee},
  \citenamefont {Melese}, \citenamefont {Milner}, \citenamefont {Molzon},
  \citenamefont {Ouimette}, \citenamefont {Riley}, \citenamefont {Ritchie},
  \citenamefont {Rubin}, \citenamefont {Sanders}, \citenamefont {Schwartz},
  \citenamefont {Sivertz}, \citenamefont {Slater}, \citenamefont {Urheim},
  \citenamefont {Vulcan}, \citenamefont {Wagner}, \citenamefont {Welsh},
  \citenamefont {Whyley}, \citenamefont {Winter}, \citenamefont {Witkowski},
  \citenamefont {Wojcicki}, \citenamefont {Yamashita},\ and\ \citenamefont
  {Ziock}}]{Arisaka:1993}%
  \BibitemOpen
  \bibfield  {author} {\bibinfo {author} {\bibnamefont {Arisaka}, \bibfnamefont
  {K.}}, \bibinfo {author} {\bibfnamefont {L.~B.}\ \bibnamefont {Auerbach}},
  \bibinfo {author} {\bibfnamefont {S.}~\bibnamefont {Axelrod}}, \bibinfo
  {author} {\bibfnamefont {J.}~\bibnamefont {Belz}}, \bibinfo {author}
  {\bibfnamefont {K.~A.}\ \bibnamefont {Biery}}, \bibinfo {author}
  {\bibfnamefont {P.}~\bibnamefont {Buchholz}}, \bibinfo {author}
  {\bibfnamefont {M.~D.}\ \bibnamefont {Chapman}}, \bibinfo {author}
  {\bibfnamefont {R.~D.}\ \bibnamefont {Cousins}}, \bibinfo {author}
  {\bibfnamefont {M.~V.}\ \bibnamefont {Diwan}}, \bibinfo {author}
  {\bibfnamefont {M.}~\bibnamefont {Eckhause}}, \bibinfo {author}
  {\bibfnamefont {J.~F.}\ \bibnamefont {Ginkel}}, \bibinfo {author}
  {\bibfnamefont {C.}~\bibnamefont {Guss}}, \bibinfo {author} {\bibfnamefont
  {A.~D.}\ \bibnamefont {Hancock}}, \bibinfo {author} {\bibfnamefont {A.~P.}\
  \bibnamefont {Heinson}}, \bibinfo {author} {\bibfnamefont {V.~L.}\
  \bibnamefont {Highland}}, \bibinfo {author} {\bibfnamefont {G.~W.}\
  \bibnamefont {Hoffmann}}, \bibinfo {author} {\bibfnamefont {J.}~\bibnamefont
  {Horvath}}, \bibinfo {author} {\bibfnamefont {G.~M.}\ \bibnamefont {Irwin}},
  \bibinfo {author} {\bibfnamefont {D.}~\bibnamefont {Joyce}}, \bibinfo
  {author} {\bibfnamefont {T.}~\bibnamefont {Kaarsberg}}, \bibinfo {author}
  {\bibfnamefont {J.~R.}\ \bibnamefont {Kane}}, \bibinfo {author}
  {\bibfnamefont {C.~J.}\ \bibnamefont {Kenney}}, \bibinfo {author}
  {\bibfnamefont {S.~H.}\ \bibnamefont {Kettell}}, \bibinfo {author}
  {\bibfnamefont {W.~W.}\ \bibnamefont {Kinnison}}, \bibinfo {author}
  {\bibfnamefont {P.}~\bibnamefont {Knibbe}}, \bibinfo {author} {\bibfnamefont
  {J.}~\bibnamefont {Konigsberg}}, \bibinfo {author} {\bibfnamefont
  {Y.}~\bibnamefont {Kuang}}, \bibinfo {author} {\bibfnamefont
  {K.}~\bibnamefont {Lang}}, \bibinfo {author} {\bibfnamefont {D.~M.}\
  \bibnamefont {Lee}}, \bibinfo {author} {\bibfnamefont {J.}~\bibnamefont
  {Margulies}}, \bibinfo {author} {\bibfnamefont {C.}~\bibnamefont
  {Mathiazhagan}}, \bibinfo {author} {\bibfnamefont {W.~K.}\ \bibnamefont
  {McFarlane}}, \bibinfo {author} {\bibfnamefont {R.~J.}\ \bibnamefont
  {McKee}}, \bibinfo {author} {\bibfnamefont {P.}~\bibnamefont {Melese}},
  \bibinfo {author} {\bibfnamefont {E.~C.}\ \bibnamefont {Milner}}, \bibinfo
  {author} {\bibfnamefont {W.~R.}\ \bibnamefont {Molzon}}, \bibinfo {author}
  {\bibfnamefont {D.~A.}\ \bibnamefont {Ouimette}}, \bibinfo {author}
  {\bibfnamefont {P.~J.}\ \bibnamefont {Riley}}, \bibinfo {author}
  {\bibfnamefont {J.~L.}\ \bibnamefont {Ritchie}}, \bibinfo {author}
  {\bibfnamefont {P.}~\bibnamefont {Rubin}}, \bibinfo {author} {\bibfnamefont
  {G.~H.}\ \bibnamefont {Sanders}}, \bibinfo {author} {\bibfnamefont {A.~J.}\
  \bibnamefont {Schwartz}}, \bibinfo {author} {\bibfnamefont {M.}~\bibnamefont
  {Sivertz}}, \bibinfo {author} {\bibfnamefont {W.~E.}\ \bibnamefont {Slater}},
  \bibinfo {author} {\bibfnamefont {J.}~\bibnamefont {Urheim}}, \bibinfo
  {author} {\bibfnamefont {W.~F.}\ \bibnamefont {Vulcan}}, \bibinfo {author}
  {\bibfnamefont {D.~L.}\ \bibnamefont {Wagner}}, \bibinfo {author}
  {\bibfnamefont {R.~E.}\ \bibnamefont {Welsh}}, \bibinfo {author}
  {\bibfnamefont {R.~J.}\ \bibnamefont {Whyley}}, \bibinfo {author}
  {\bibfnamefont {R.~G.}\ \bibnamefont {Winter}}, \bibinfo {author}
  {\bibfnamefont {M.~T.}\ \bibnamefont {Witkowski}}, \bibinfo {author}
  {\bibfnamefont {S.~G.}\ \bibnamefont {Wojcicki}}, \bibinfo {author}
  {\bibfnamefont {A.}~\bibnamefont {Yamashita}}, \ and\ \bibinfo {author}
  {\bibfnamefont {H.~J.}\ \bibnamefont {Ziock}}} (\bibinfo {year} {1993}),\
  \href {\doibase 10.1103/PhysRevLett.70.1049} {\bibfield  {journal} {\bibinfo
  {journal} {Phys. Rev. Lett.}\ }\textbf {\bibinfo {volume} {70}},\ \bibinfo
  {pages} {1049}}\BibitemShut {NoStop}%
\bibitem [{\citenamefont {Ashkin}\ \emph {et~al.}(1959)\citenamefont {Ashkin},
  \citenamefont {Fazzini}, \citenamefont {Fidecaro}, \citenamefont {Lipman},\
  and\ \citenamefont {Merrison}}]{Ashkin:1959}%
  \BibitemOpen
  \bibfield  {author} {\bibinfo {author} {\bibnamefont {Ashkin}, \bibfnamefont
  {J.}}, \bibinfo {author} {\bibfnamefont {T.}~\bibnamefont {Fazzini}},
  \bibinfo {author} {\bibfnamefont {G.}~\bibnamefont {Fidecaro}}, \bibinfo
  {author} {\bibfnamefont {N.}~\bibnamefont {Lipman}}, \ and\ \bibinfo {author}
  {\bibfnamefont {A.}~\bibnamefont {Merrison}}} (\bibinfo {year} {1959}),\
  \href@noop {} {\bibfield  {journal} {\bibinfo  {journal} {Nuovo Cim.}\
  }\textbf {\bibinfo {volume} {14}},\ \bibinfo {pages} {1266}}\BibitemShut
  {NoStop}%
\bibitem [{\citenamefont {Atoyan}\ \emph {et~al.}(1992)\citenamefont {Atoyan},
  \citenamefont {Gladyshev}, \citenamefont {Gninenko}, \citenamefont {Isakov},
  \citenamefont {Kovzelev}, \citenamefont {Monich}, \citenamefont {Poblaguev},
  \citenamefont {Proskuryakov}, \citenamefont {Semenyuk}, \citenamefont
  {Lapshin}, \citenamefont {Protopopov}, \citenamefont {Rykalin},\ and\
  \citenamefont {Semenov}}]{atoyan:1992}%
  \BibitemOpen
  \bibfield  {author} {\bibinfo {author} {\bibnamefont {Atoyan}, \bibfnamefont
  {G.}}, \bibinfo {author} {\bibfnamefont {V.}~\bibnamefont {Gladyshev}},
  \bibinfo {author} {\bibfnamefont {S.}~\bibnamefont {Gninenko}}, \bibinfo
  {author} {\bibfnamefont {V.}~\bibnamefont {Isakov}}, \bibinfo {author}
  {\bibfnamefont {A.}~\bibnamefont {Kovzelev}}, \bibinfo {author}
  {\bibfnamefont {E.}~\bibnamefont {Monich}}, \bibinfo {author} {\bibfnamefont
  {A.}~\bibnamefont {Poblaguev}}, \bibinfo {author} {\bibfnamefont
  {A.}~\bibnamefont {Proskuryakov}}, \bibinfo {author} {\bibfnamefont
  {I.}~\bibnamefont {Semenyuk}}, \bibinfo {author} {\bibfnamefont
  {V.}~\bibnamefont {Lapshin}}, \bibinfo {author} {\bibfnamefont
  {Y.}~\bibnamefont {Protopopov}}, \bibinfo {author} {\bibfnamefont
  {V.}~\bibnamefont {Rykalin}}, \ and\ \bibinfo {author} {\bibfnamefont
  {V.}~\bibnamefont {Semenov}}} (\bibinfo {year} {1992}),\ \href {\doibase
  10.1016/0168-9002(92)90773-W} {\bibfield  {journal} {\bibinfo  {journal}
  {Nuclear Instruments and Methods in Physics Research Section A: Accelerators,
  Spectrometers, Detectors and Associated Equipment}\ }\textbf {\bibinfo
  {volume} {320}}~(\bibinfo {number} {1‰ÛÒ2}),\ \bibinfo {pages} {144
  }}\BibitemShut {NoStop}%
\bibitem [{\citenamefont {Aubert}\ \emph {et~al.}(2010)\citenamefont {Aubert}
  \emph {et~al.}}]{Aubert:2010}%
  \BibitemOpen
  \bibfield  {author} {\bibinfo {author} {\bibnamefont {Aubert}, \bibfnamefont
  {B.}},  \emph {et~al.} (\bibinfo {collaboration} {BABAR Collaboration})}
  (\bibinfo {year} {2010}),\ \href {\doibase 10.1103/PhysRevLett.104.021802}
  {\bibfield  {journal} {\bibinfo  {journal} {Phys. Rev. Lett.}\ }\textbf
  {\bibinfo {volume} {104}},\ \bibinfo {pages} {021802}}\BibitemShut {NoStop}%
\bibitem [{\citenamefont {Babu}\ and\ \citenamefont {Julio}(2010)}]{Babu:2010}%
  \BibitemOpen
  \bibfield  {author} {\bibinfo {author} {\bibnamefont {Babu}, \bibfnamefont
  {K.}}, \ and\ \bibinfo {author} {\bibfnamefont {J.}~\bibnamefont {Julio}}}
  (\bibinfo {year} {2010}),\ \href {\doibase 10.1016/j.nuclphysb.2010.07.022}
  {\bibfield  {journal} {\bibinfo  {journal} {Nuclear Physics B}\ }\textbf
  {\bibinfo {volume} {841}}~(\bibinfo {number} {1--2}),\ \bibinfo {pages} {130
  }},\ \Eprint {http://arxiv.org/abs/arXiv:1006.1092 [hep-ph]} {arXiv:1006.1092
  [hep-ph]} \BibitemShut {NoStop}%
\bibitem [{\citenamefont {Babu}\ and\ \citenamefont {Kolda}(2002)}]{Babu:2002}%
  \BibitemOpen
  \bibfield  {author} {\bibinfo {author} {\bibnamefont {Babu}, \bibfnamefont
  {K.~S.}}, \ and\ \bibinfo {author} {\bibfnamefont {C.}~\bibnamefont {Kolda}}}
  (\bibinfo {year} {2002}),\ \href {\doibase 10.1103/PhysRevLett.89.241802}
  {\bibfield  {journal} {\bibinfo  {journal} {Phys. Rev. Lett.}\ }\textbf
  {\bibinfo {volume} {89}},\ \bibinfo {pages} {241802}}\BibitemShut {NoStop}%
\bibitem [{\citenamefont {Bachman}\ \emph {et~al.}(1997)\citenamefont {Bachman}
  \emph {et~al.}}]{Bachman:1997}%
  \BibitemOpen
  \bibfield  {author} {\bibinfo {author} {\bibnamefont {Bachman}, \bibfnamefont
  {M.}},  \emph {et~al.} (\bibinfo {collaboration} {MECO})} (\bibinfo {year}
  {1997}),\ \href@noop {} {\enquote {\bibinfo {title} {{BNL} {P}roposal
  {E}-940},}\ }\bibinfo {note}
  {\url{http://mu2e-docdb.fnal.gov/cgi-bin/ShowDocument?docid=284
  }}\BibitemShut {NoStop}%
\bibitem [{\citenamefont {Badertscher}\ \emph {et~al.}(1978)\citenamefont
  {Badertscher}, \citenamefont {Borer}, \citenamefont {Czapek}, \citenamefont
  {Fluckiger}, \citenamefont {Hanni} \emph {et~al.}}]{Badertscher:1978}%
  \BibitemOpen
  \bibfield  {author} {\bibinfo {author} {\bibnamefont {Badertscher},
  \bibfnamefont {A.}}, \bibinfo {author} {\bibfnamefont {K.}~\bibnamefont
  {Borer}}, \bibinfo {author} {\bibfnamefont {G.}~\bibnamefont {Czapek}},
  \bibinfo {author} {\bibfnamefont {A.}~\bibnamefont {Fluckiger}}, \bibinfo
  {author} {\bibfnamefont {H.}~\bibnamefont {Hanni}},  \emph {et~al.}}
  (\bibinfo {year} {1978}),\ \href {\doibase 10.1016/0370-2693(78)90385-4}
  {\bibfield  {journal} {\bibinfo  {journal} {Phys.Lett.}\ }\textbf {\bibinfo
  {volume} {B79}},\ \bibinfo {pages} {371}}\BibitemShut {NoStop}%
\bibitem [{\citenamefont {Badertscher}\ \emph {et~al.}(1980)\citenamefont
  {Badertscher}, \citenamefont {Borer}, \citenamefont {Czapek}, \citenamefont
  {Fluckiger}, \citenamefont {Hanni} \emph {et~al.}}]{Badertscher:1980}%
  \BibitemOpen
  \bibfield  {author} {\bibinfo {author} {\bibnamefont {Badertscher},
  \bibfnamefont {A.}}, \bibinfo {author} {\bibfnamefont {K.}~\bibnamefont
  {Borer}}, \bibinfo {author} {\bibfnamefont {G.}~\bibnamefont {Czapek}},
  \bibinfo {author} {\bibfnamefont {A.}~\bibnamefont {Fluckiger}}, \bibinfo
  {author} {\bibfnamefont {H.}~\bibnamefont {Hanni}},  \emph {et~al.}}
  (\bibinfo {year} {1980}),\ \href {\doibase 10.1007/BF02776193} {\bibfield
  {journal} {\bibinfo  {journal} {Lett.Nuovo Cim.}\ }\textbf {\bibinfo {volume}
  {28}},\ \bibinfo {pages} {401}}\BibitemShut {NoStop}%
\bibitem [{\citenamefont {Badertscher}\ \emph {et~al.}(1977)\citenamefont
  {Badertscher} \emph {et~al.}}]{Badertscher:1977}%
  \BibitemOpen
  \bibfield  {author} {\bibinfo {author} {\bibnamefont {Badertscher},
  \bibfnamefont {A.}},  \emph {et~al.}} (\bibinfo {year} {1977}),\ \href
  {\doibase 10.1103/PhysRevLett.39.1385} {\bibfield  {journal} {\bibinfo
  {journal} {Phys. Rev. Lett.}\ }\textbf {\bibinfo {volume} {39}},\ \bibinfo
  {pages} {1385}}\BibitemShut {NoStop}%
\bibitem [{\citenamefont {Badertscher}\ \emph {et~al.}(1982)\citenamefont
  {Badertscher} \emph {et~al.}}]{Badertscher:1982}%
  \BibitemOpen
  \bibfield  {author} {\bibinfo {author} {\bibnamefont {Badertscher},
  \bibfnamefont {A.}},  \emph {et~al.}} (\bibinfo {year} {1982}),\ \href
  {\doibase 10.1016/0375-9474(82)90049-5} {\bibfield  {journal} {\bibinfo
  {journal} {Nuclear Physics A}\ }\textbf {\bibinfo {volume} {377}}~(\bibinfo
  {number} {2--3}),\ \bibinfo {pages} {406 }}\BibitemShut {NoStop}%
\bibitem [{\citenamefont {Baldini}(2012)}]{Baldini:2012}%
  \BibitemOpen
  \bibfield  {author} {\bibinfo {author} {\bibnamefont {Baldini}, \bibfnamefont
  {A.}}} (\bibinfo {year} {2012}),\ \href@noop {} {}\bibinfo {note} {(MEG
  co-spokesperson), priv.\ comm.}\BibitemShut {Stop}%
\bibitem [{\citenamefont {Baldini}\ \emph {et~al.}(2013)\citenamefont
  {Baldini}, \citenamefont {Cei}, \citenamefont {Cerri}, \citenamefont
  {Dussoni}, \citenamefont {Galli} \emph {et~al.}}]{Baldini:2013}%
  \BibitemOpen
  \bibfield  {author} {\bibinfo {author} {\bibnamefont {Baldini}, \bibfnamefont
  {A.}}, \bibinfo {author} {\bibfnamefont {F.}~\bibnamefont {Cei}}, \bibinfo
  {author} {\bibfnamefont {C.}~\bibnamefont {Cerri}}, \bibinfo {author}
  {\bibfnamefont {S.}~\bibnamefont {Dussoni}}, \bibinfo {author} {\bibfnamefont
  {L.}~\bibnamefont {Galli}},  \emph {et~al.}} (\bibinfo {year} {2013}),\
  \href@noop {} {\ }\Eprint {http://arxiv.org/abs/1301.7225} {arXiv:1301.7225
  [physics.ins-det]} \BibitemShut {NoStop}%
\bibitem [{\citenamefont {Baranov}\ \emph {et~al.}(1991)\citenamefont {Baranov}
  \emph {et~al.}}]{Baranov:1991}%
  \BibitemOpen
  \bibfield  {author} {\bibinfo {author} {\bibnamefont {Baranov}, \bibfnamefont
  {V.~A.}},  \emph {et~al.}} (\bibinfo {year} {1991}),\ \href@noop {}
  {\bibfield  {journal} {\bibinfo  {journal} {Sov.\ J.\ Nucl.\ Phys.}\ }\textbf
  {\bibinfo {volume} {53}},\ \bibinfo {pages} {802}}\BibitemShut {NoStop}%
\bibitem [{\citenamefont {Bartlett}\ \emph {et~al.}(1962)\citenamefont
  {Bartlett}, \citenamefont {Devons},\ and\ \citenamefont
  {Sachs}}]{Bartlett:1962}%
  \BibitemOpen
  \bibfield  {author} {\bibinfo {author} {\bibnamefont {Bartlett},
  \bibfnamefont {D.}}, \bibinfo {author} {\bibfnamefont {S.}~\bibnamefont
  {Devons}}, \ and\ \bibinfo {author} {\bibfnamefont {A.~M.}\ \bibnamefont
  {Sachs}}} (\bibinfo {year} {1962}),\ \href {\doibase
  10.1103/PhysRevLett.8.120} {\bibfield  {journal} {\bibinfo  {journal} {Phys.
  Rev. Lett.}\ }\textbf {\bibinfo {volume} {8}},\ \bibinfo {pages}
  {120}}\BibitemShut {NoStop}%
\bibitem [{\citenamefont {Bartley}\ \emph {et~al.}(1964)\citenamefont
  {Bartley}, \citenamefont {Davies}, \citenamefont {Muirhead},\ and\
  \citenamefont {Woodhead}}]{Bartley:1964}%
  \BibitemOpen
  \bibfield  {author} {\bibinfo {author} {\bibnamefont {Bartley}, \bibfnamefont
  {J.}}, \bibinfo {author} {\bibfnamefont {H.}~\bibnamefont {Davies}}, \bibinfo
  {author} {\bibfnamefont {H.}~\bibnamefont {Muirhead}}, \ and\ \bibinfo
  {author} {\bibfnamefont {T.}~\bibnamefont {Woodhead}}} (\bibinfo {year}
  {1964}),\ \href {\doibase 10.1016/0031-9163(64)90479-2} {\bibfield  {journal}
  {\bibinfo  {journal} {Physics Letters}\ }\textbf {\bibinfo {volume}
  {13}}~(\bibinfo {number} {3}),\ \bibinfo {pages} {258 }}\BibitemShut
  {NoStop}%
\bibitem [{\citenamefont {Batley}\ \emph {et~al.}(2009)\citenamefont {Batley}
  \emph {et~al.}}]{Batley:2009}%
  \BibitemOpen
  \bibfield  {author} {\bibinfo {author} {\bibnamefont {Batley}, \bibfnamefont
  {J.}},  \emph {et~al.}} (\bibinfo {year} {2009}),\ \href {\doibase
  10.1016/j.physletb.2009.05.040} {\bibfield  {journal} {\bibinfo  {journal}
  {Physics Letters B}\ }\textbf {\bibinfo {volume} {677}}~(\bibinfo {number}
  {5}),\ \bibinfo {pages} {246 }}\BibitemShut {NoStop}%
\bibitem [{\citenamefont {Beer}\ \emph {et~al.}(1986)\citenamefont {Beer} \emph
  {et~al.}}]{Beer:1986}%
  \BibitemOpen
  \bibfield  {author} {\bibinfo {author} {\bibnamefont {Beer}, \bibfnamefont
  {G.~A.}},  \emph {et~al.}} (\bibinfo {year} {1986}),\ \href {\doibase
  10.1103/PhysRevLett.57.671} {\bibfield  {journal} {\bibinfo  {journal} {Phys.
  Rev. Lett.}\ }\textbf {\bibinfo {volume} {57}},\ \bibinfo {pages}
  {671}}\BibitemShut {NoStop}%
\bibitem [{\citenamefont {Bellgardt}\ \emph {et~al.}(1988)\citenamefont
  {Bellgardt} \emph {et~al.}}]{Bellgardt:1988}%
  \BibitemOpen
  \bibfield  {author} {\bibinfo {author} {\bibnamefont {Bellgardt},
  \bibfnamefont {U.}},  \emph {et~al.}} (\bibinfo {year} {1988}),\ \href
  {\doibase 10.1016/0550-3213(88)90462-2} {\bibfield  {journal} {\bibinfo
  {journal} {Nuclear Physics B}\ }\textbf {\bibinfo {volume} {299}}~(\bibinfo
  {number} {1}),\ \bibinfo {pages} {1 }}\BibitemShut {NoStop}%
\bibitem [{\citenamefont {Belz}\ \emph {et~al.}(1999)\citenamefont {Belz} \emph
  {et~al.}}]{Belz:1999}%
  \BibitemOpen
  \bibfield  {author} {\bibinfo {author} {\bibnamefont {Belz}, \bibfnamefont
  {J.}},  \emph {et~al.}} (\bibinfo {year} {1999}),\ \href {\doibase
  10.1016/S0168-9002(98)01381-3} {\bibfield  {journal} {\bibinfo  {journal}
  {Nuclear Instruments and Methods in Physics Research Section A: Accelerators,
  Spectrometers, Detectors and Associated Equipment}\ }\textbf {\bibinfo
  {volume} {428}},\ \bibinfo {pages} {239 }}\BibitemShut {NoStop}%
\bibitem [{\citenamefont {Bergbusch}\ \emph {et~al.}(1999)\citenamefont
  {Bergbusch} \emph {et~al.}}]{Bergsbusch:1999}%
  \BibitemOpen
  \bibfield  {author} {\bibinfo {author} {\bibnamefont {Bergbusch},
  \bibfnamefont {P.~C.}},  \emph {et~al.}} (\bibinfo {year} {1999}),\ \href
  {\doibase 10.1103/PhysRevC.59.2853} {\bibfield  {journal} {\bibinfo
  {journal} {Phys. Rev. C}\ }\textbf {\bibinfo {volume} {59}},\ \bibinfo
  {pages} {2853}}\BibitemShut {NoStop}%
\bibitem [{\citenamefont {Berger}\ \emph {et~al.}(2011)\citenamefont {Berger}
  \emph {et~al.}}]{Berger:2011}%
  \BibitemOpen
  \bibfield  {author} {\bibinfo {author} {\bibnamefont {Berger}, \bibfnamefont
  {N.}},  \emph {et~al.}} (\bibinfo {year} {2011}),\ \href@noop {} {}\bibinfo
  {note} {\url{http://arxiv.org/pdf/1110.1504}}\BibitemShut {NoStop}%
\bibitem [{\citenamefont {Beringer}\ \emph {et~al.}(2012)\citenamefont
  {Beringer} \emph {et~al.}}]{RPP}%
  \BibitemOpen
  \bibfield  {author} {\bibinfo {author} {\bibnamefont {Beringer},
  \bibfnamefont {J.}},  \emph {et~al.} (\bibinfo {collaboration} {Particle Data
  Group})} (\bibinfo {year} {2012}),\ \href {\doibase
  10.1103/PhysRevD.86.010001} {\bibfield  {journal} {\bibinfo  {journal}
  {Phys.Rev.}\ }\textbf {\bibinfo {volume} {D86}},\ \bibinfo {pages}
  {010001}}\BibitemShut {NoStop}%
\bibitem [{\citenamefont {Berley}\ \emph {et~al.}(1959)\citenamefont {Berley},
  \citenamefont {Lee},\ and\ \citenamefont {Bardon}}]{Berley:1959}%
  \BibitemOpen
  \bibfield  {author} {\bibinfo {author} {\bibnamefont {Berley}, \bibfnamefont
  {D.}}, \bibinfo {author} {\bibfnamefont {J.}~\bibnamefont {Lee}}, \ and\
  \bibinfo {author} {\bibfnamefont {M.}~\bibnamefont {Bardon}}} (\bibinfo
  {year} {1959}),\ \href {\doibase 10.1103/PhysRevLett.2.357} {\bibfield
  {journal} {\bibinfo  {journal} {Phys. Rev. Lett.}\ }\textbf {\bibinfo
  {volume} {2}},\ \bibinfo {pages} {357}}\BibitemShut {NoStop}%
\bibitem [{\citenamefont {Bertl}\ \emph {et~al.}(1984)\citenamefont {Bertl}
  \emph {et~al.}}]{Bertl:1984}%
  \BibitemOpen
  \bibfield  {author} {\bibinfo {author} {\bibnamefont {Bertl}, \bibfnamefont
  {W.}},  \emph {et~al.}} (\bibinfo {year} {1984}),\ \href {\doibase
  10.1016/0370-2693(84)90757-3} {\bibfield  {journal} {\bibinfo  {journal}
  {Physics Letters B}\ }\textbf {\bibinfo {volume} {140}},\ \bibinfo {pages}
  {299 }}\BibitemShut {NoStop}%
\bibitem [{\citenamefont {Bertl}\ \emph {et~al.}(1985)\citenamefont {Bertl}
  \emph {et~al.}}]{Bertl:1985}%
  \BibitemOpen
  \bibfield  {author} {\bibinfo {author} {\bibnamefont {Bertl}, \bibfnamefont
  {W.}},  \emph {et~al.}} (\bibinfo {year} {1985}),\ \href {\doibase
  10.1016/0550-3213(85)90308-6} {\bibfield  {journal} {\bibinfo  {journal}
  {Nuclear Physics B}\ }\textbf {\bibinfo {volume} {260}}~(\bibinfo {number}
  {1}),\ \bibinfo {pages} {1 }}\BibitemShut {NoStop}%
\bibitem [{\citenamefont {Bertl}\ \emph {et~al.}(2006)\citenamefont {Bertl}
  \emph {et~al.}}]{Bertl:2006}%
  \BibitemOpen
  \bibfield  {author} {\bibinfo {author} {\bibnamefont {Bertl}, \bibfnamefont
  {W.}},  \emph {et~al.}} (\bibinfo {year} {2006}),\ \href {\doibase
  10.1140/epjc/s2006-02582-x} {\bibfield  {journal} {\bibinfo  {journal} {The
  European Physical Journal C - Particles and Fields}\ }\textbf {\bibinfo
  {volume} {47}},\ \bibinfo {pages} {337}}\BibitemShut {NoStop}%
\bibitem [{\citenamefont {Bistirlich}\ \emph {et~al.}(1972)\citenamefont
  {Bistirlich}, \citenamefont {Crowe}, \citenamefont {Parsons}, \citenamefont
  {Skarek},\ and\ \citenamefont {Tru\"ol}}]{Bistirlich:1972}%
  \BibitemOpen
  \bibfield  {author} {\bibinfo {author} {\bibnamefont {Bistirlich},
  \bibfnamefont {J.~A.}}, \bibinfo {author} {\bibfnamefont {K.~M.}\
  \bibnamefont {Crowe}}, \bibinfo {author} {\bibfnamefont {A.~S.~L.}\
  \bibnamefont {Parsons}}, \bibinfo {author} {\bibfnamefont {P.}~\bibnamefont
  {Skarek}}, \ and\ \bibinfo {author} {\bibfnamefont {P.}~\bibnamefont
  {Tru\"ol}}} (\bibinfo {year} {1972}),\ \href {\doibase
  10.1103/PhysRevC.5.1867} {\bibfield  {journal} {\bibinfo  {journal} {Phys.
  Rev. C}\ }\textbf {\bibinfo {volume} {5}},\ \bibinfo {pages}
  {1867}}\BibitemShut {NoStop}%
\bibitem [{\citenamefont {Blanke}\ \emph {et~al.}(2007)\citenamefont {Blanke},
  \citenamefont {Buras}, \citenamefont {Duling}, \citenamefont
  {Poschenrieder},\ and\ \citenamefont {Tarantino}}]{Blanke:2007db}%
  \BibitemOpen
  \bibfield  {author} {\bibinfo {author} {\bibnamefont {Blanke}, \bibfnamefont
  {M.}}, \bibinfo {author} {\bibfnamefont {A.~J.}\ \bibnamefont {Buras}},
  \bibinfo {author} {\bibfnamefont {B.}~\bibnamefont {Duling}}, \bibinfo
  {author} {\bibfnamefont {A.}~\bibnamefont {Poschenrieder}}, \ and\ \bibinfo
  {author} {\bibfnamefont {C.}~\bibnamefont {Tarantino}}} (\bibinfo {year}
  {2007}),\ \href {\doibase 10.1088/1126-6708/2007/05/013} {\bibfield
  {journal} {\bibinfo  {journal} {JHEP}\ }\textbf {\bibinfo {volume} {0705}},\
  \bibinfo {pages} {013}},\ \Eprint {http://arxiv.org/abs/hep-ph/0702136}
  {arXiv:hep-ph/0702136 [hep-ph]} \BibitemShut {NoStop}%
\bibitem [{\citenamefont {Blondel}\ \emph {et~al.}(2012)\citenamefont {Blondel}
  \emph {et~al.}}]{Blondel:2012}%
  \BibitemOpen
  \bibfield  {author} {\bibinfo {author} {\bibnamefont {Blondel}, \bibfnamefont
  {A.}},  \emph {et~al.} (\bibinfo {collaboration} {$\mu$3e})} (\bibinfo {year}
  {2012}),\ \href@noop {} {\enquote {\bibinfo {title} {Letter of intent for an
  experiment to search for the decay $\mu \rightarrow eee $},}\ }\bibinfo
  {note}
  {\url{http://www.physi.uni-heidelberg.de/Forschung/he/mu3e/documents/LOI_Mu3e_PSI.pdf}}\BibitemShut
  {NoStop}%
\bibitem [{\citenamefont {Boer}\ \emph {et~al.}(2011)\citenamefont {Boer} \emph
  {et~al.}}]{Boer:2011}%
  \BibitemOpen
  \bibfield  {author} {\bibinfo {author} {\bibnamefont {Boer}, \bibfnamefont
  {D.}},  \emph {et~al.}} (\bibinfo {year} {2011}),\ \href@noop {} {}\Eprint
  {http://arxiv.org/abs/arXiv:1108.1713v2 [nucl-th]} {arXiv:1108.1713v2
  [nucl-th]} \BibitemShut {NoStop}%
\bibitem [{\citenamefont {Bolton}\ \emph {et~al.}(1986)\citenamefont {Bolton},
  \citenamefont {Bowman}, \citenamefont {Cooper}, \citenamefont {Frank},
  \citenamefont {Hallin}, \citenamefont {Heusi}, \citenamefont {Hoffman},
  \citenamefont {Hogan}, \citenamefont {Mariam}, \citenamefont {Matis},
  \citenamefont {Mischke}, \citenamefont {Nagle}, \citenamefont {Piilonen},
  \citenamefont {Sandberg}, \citenamefont {Sanders}, \citenamefont
  {Sennhauser}, \citenamefont {Werbeck}, \citenamefont {Williams},
  \citenamefont {Wilson}, \citenamefont {Hofstadter}, \citenamefont {Hughes},
  \citenamefont {Ritter}, \citenamefont {Grosnick}, \citenamefont {Wright},
  \citenamefont {Highland},\ and\ \citenamefont {McDonough}}]{Bolton:1986}%
  \BibitemOpen
  \bibfield  {author} {\bibinfo {author} {\bibnamefont {Bolton}, \bibfnamefont
  {R.~D.}}, \bibinfo {author} {\bibfnamefont {J.~D.}\ \bibnamefont {Bowman}},
  \bibinfo {author} {\bibfnamefont {M.~D.}\ \bibnamefont {Cooper}}, \bibinfo
  {author} {\bibfnamefont {J.~S.}\ \bibnamefont {Frank}}, \bibinfo {author}
  {\bibfnamefont {A.~L.}\ \bibnamefont {Hallin}}, \bibinfo {author}
  {\bibfnamefont {P.~A.}\ \bibnamefont {Heusi}}, \bibinfo {author}
  {\bibfnamefont {C.~M.}\ \bibnamefont {Hoffman}}, \bibinfo {author}
  {\bibfnamefont {G.~E.}\ \bibnamefont {Hogan}}, \bibinfo {author}
  {\bibfnamefont {F.~G.}\ \bibnamefont {Mariam}}, \bibinfo {author}
  {\bibfnamefont {H.~S.}\ \bibnamefont {Matis}}, \bibinfo {author}
  {\bibfnamefont {R.~E.}\ \bibnamefont {Mischke}}, \bibinfo {author}
  {\bibfnamefont {D.~E.}\ \bibnamefont {Nagle}}, \bibinfo {author}
  {\bibfnamefont {L.~E.}\ \bibnamefont {Piilonen}}, \bibinfo {author}
  {\bibfnamefont {V.~D.}\ \bibnamefont {Sandberg}}, \bibinfo {author}
  {\bibfnamefont {G.~H.}\ \bibnamefont {Sanders}}, \bibinfo {author}
  {\bibfnamefont {U.}~\bibnamefont {Sennhauser}}, \bibinfo {author}
  {\bibfnamefont {R.}~\bibnamefont {Werbeck}}, \bibinfo {author} {\bibfnamefont
  {R.~A.}\ \bibnamefont {Williams}}, \bibinfo {author} {\bibfnamefont {S.~L.}\
  \bibnamefont {Wilson}}, \bibinfo {author} {\bibfnamefont {R.}~\bibnamefont
  {Hofstadter}}, \bibinfo {author} {\bibfnamefont {E.~B.}\ \bibnamefont
  {Hughes}}, \bibinfo {author} {\bibfnamefont {M.~W.}\ \bibnamefont {Ritter}},
  \bibinfo {author} {\bibfnamefont {D.}~\bibnamefont {Grosnick}}, \bibinfo
  {author} {\bibfnamefont {S.~C.}\ \bibnamefont {Wright}}, \bibinfo {author}
  {\bibfnamefont {V.~L.}\ \bibnamefont {Highland}}, \ and\ \bibinfo {author}
  {\bibfnamefont {J.}~\bibnamefont {McDonough}}} (\bibinfo {year} {1986}),\
  \href {\doibase 10.1103/PhysRevLett.56.2461} {\bibfield  {journal} {\bibinfo
  {journal} {Phys. Rev. Lett.}\ }\textbf {\bibinfo {volume} {56}},\ \bibinfo
  {pages} {2461}}\BibitemShut {NoStop}%
\bibitem [{\citenamefont {Bolton}\ \emph {et~al.}(1984)\citenamefont {Bolton}
  \emph {et~al.}}]{Bolton:1984}%
  \BibitemOpen
  \bibfield  {author} {\bibinfo {author} {\bibnamefont {Bolton}, \bibfnamefont
  {R.~D.}},  \emph {et~al.}} (\bibinfo {year} {1984}),\ \href {\doibase
  10.1103/PhysRevLett.53.1415} {\bibfield  {journal} {\bibinfo  {journal}
  {Phys. Rev. Lett.}\ }\textbf {\bibinfo {volume} {53}},\ \bibinfo {pages}
  {1415}}\BibitemShut {NoStop}%
\bibitem [{\citenamefont {Bolton}\ \emph {et~al.}(1988)\citenamefont {Bolton}
  \emph {et~al.}}]{Bolton:1988}%
  \BibitemOpen
  \bibfield  {author} {\bibinfo {author} {\bibnamefont {Bolton}, \bibfnamefont
  {R.~D.}},  \emph {et~al.}} (\bibinfo {year} {1988}),\ \href {\doibase
  10.1103/PhysRevD.38.2077} {\bibfield  {journal} {\bibinfo  {journal} {Phys.
  Rev. D}\ }\textbf {\bibinfo {volume} {38}},\ \bibinfo {pages}
  {2077}}\BibitemShut {NoStop}%
\bibitem [{\citenamefont {Bott-Bodenhausen}\ \emph {et~al.}(1967)\citenamefont
  {Bott-Bodenhausen}, \citenamefont {Bouard}, \citenamefont {Cassel},
  \citenamefont {Dekkers}, \citenamefont {Felst}, \citenamefont {Mermod},
  \citenamefont {Savin}, \citenamefont {Scharff}, \citenamefont {Vivargent},
  \citenamefont {Willitts},\ and\ \citenamefont {Winter}}]{Bott:1967}%
  \BibitemOpen
  \bibfield  {author} {\bibinfo {author} {\bibnamefont {Bott-Bodenhausen},
  \bibfnamefont {M.}}, \bibinfo {author} {\bibfnamefont {X.~D.}\ \bibnamefont
  {Bouard}}, \bibinfo {author} {\bibfnamefont {D.}~\bibnamefont {Cassel}},
  \bibinfo {author} {\bibfnamefont {D.}~\bibnamefont {Dekkers}}, \bibinfo
  {author} {\bibfnamefont {R.}~\bibnamefont {Felst}}, \bibinfo {author}
  {\bibfnamefont {R.}~\bibnamefont {Mermod}}, \bibinfo {author} {\bibfnamefont
  {I.}~\bibnamefont {Savin}}, \bibinfo {author} {\bibfnamefont
  {P.}~\bibnamefont {Scharff}}, \bibinfo {author} {\bibfnamefont
  {M.}~\bibnamefont {Vivargent}}, \bibinfo {author} {\bibfnamefont
  {T.}~\bibnamefont {Willitts}}, \ and\ \bibinfo {author} {\bibfnamefont
  {K.}~\bibnamefont {Winter}}} (\bibinfo {year} {1967}),\ \href {\doibase
  10.1016/0370-2693(67)90492-3} {\bibfield  {journal} {\bibinfo  {journal}
  {Physics Letters B}\ }\textbf {\bibinfo {volume} {24}}~(\bibinfo {number}
  {4}),\ \bibinfo {pages} {194 }}\BibitemShut {NoStop}%
\bibitem [{\citenamefont {Bowman}\ \emph {et~al.}(1979)\citenamefont {Bowman}
  \emph {et~al.}}]{Bowman:1979}%
  \BibitemOpen
  \bibfield  {author} {\bibinfo {author} {\bibnamefont {Bowman}, \bibfnamefont
  {J.~D.}},  \emph {et~al.}} (\bibinfo {year} {1979}),\ \href {\doibase
  10.1103/PhysRevLett.42.556} {\bibfield  {journal} {\bibinfo  {journal} {Phys.
  Rev. Lett.}\ }\textbf {\bibinfo {volume} {42}},\ \bibinfo {pages}
  {556}}\BibitemShut {NoStop}%
\bibitem [{\citenamefont {Brooks}\ \emph {et~al.}(1999)\citenamefont {Brooks}
  \emph {et~al.}}]{Brooks:1999}%
  \BibitemOpen
  \bibfield  {author} {\bibinfo {author} {\bibnamefont {Brooks}, \bibfnamefont
  {M.~L.}},  \emph {et~al.} (\bibinfo {collaboration} {MEGA})} (\bibinfo {year}
  {1999}),\ \href {\doibase 10.1103/PhysRevLett.83.1521} {\bibfield  {journal}
  {\bibinfo  {journal} {Physical Review Letters}\ }\textbf {\bibinfo {volume}
  {83}}~(\bibinfo {number} {23}),\ \bibinfo {pages} {1521}},\ \Eprint
  {http://arxiv.org/abs/arXiv:hep-ex/9905013v1} {arXiv:hep-ex/9905013v1}
  \BibitemShut {NoStop}%
\bibitem [{\citenamefont {Bryman}\ and\ \citenamefont
  {Tschirhart}(2010)}]{tschirhart:2010}%
  \BibitemOpen
  \bibfield  {author} {\bibinfo {author} {\bibnamefont {Bryman}, \bibfnamefont
  {D.}}, \ and\ \bibinfo {author} {\bibfnamefont {R.}~\bibnamefont
  {Tschirhart}}} (\bibinfo {year} {2010}),\ \href
  {https://indico.fnal.gov/getFile.py/access?resId=1&materialId=0&confId=3579}
  {}\BibitemShut {NoStop}%
\bibitem [{\citenamefont {Bryman}\ \emph {et~al.}(2006)\citenamefont {Bryman}
  \emph {et~al.}}]{Bryman:2007}%
  \BibitemOpen
  \bibfield  {author} {\bibinfo {author} {\bibnamefont {Bryman}, \bibfnamefont
  {D.}},  \emph {et~al.} (\bibinfo {collaboration} {COMET/PRISM/PRIME})}
  (\bibinfo {year} {2006}),\ \href@noop {} {\enquote {\bibinfo {title}
  {{J-PARC} {L}etter of {I}ntent},}\ }\bibinfo {note}
  {\url{http://mu2e-docdb.fnal.gov/cgi-bin/RetrieveFile?docid=2202;filename=PRISM-PRIME.pdf;version=2}}\BibitemShut
  {NoStop}%
\bibitem [{\citenamefont {Bryman}\ \emph {et~al.}(1972)\citenamefont {Bryman},
  \citenamefont {Blecher}, \citenamefont {Gotow},\ and\ \citenamefont
  {Powers}}]{Bryman:1972}%
  \BibitemOpen
  \bibfield  {author} {\bibinfo {author} {\bibnamefont {Bryman}, \bibfnamefont
  {D.~A.}}, \bibinfo {author} {\bibfnamefont {M.}~\bibnamefont {Blecher}},
  \bibinfo {author} {\bibfnamefont {K.}~\bibnamefont {Gotow}}, \ and\ \bibinfo
  {author} {\bibfnamefont {R.~J.}\ \bibnamefont {Powers}}} (\bibinfo {year}
  {1972}),\ \href {\doibase 10.1103/PhysRevLett.28.1469} {\bibfield  {journal}
  {\bibinfo  {journal} {Phys. Rev. Lett.}\ }\textbf {\bibinfo {volume} {28}},\
  \bibinfo {pages} {1469}}\BibitemShut {NoStop}%
\bibitem [{\citenamefont {Carey}\ \emph {et~al.}(2008)\citenamefont {Carey}
  \emph {et~al.}}]{Carey:2008}%
  \BibitemOpen
  \bibfield  {author} {\bibinfo {author} {\bibnamefont {Carey}, \bibfnamefont
  {R.}},  \emph {et~al.} (\bibinfo {collaboration} {Mu2e})} (\bibinfo {year}
  {2008}),\ \href@noop {} {\enquote {\bibinfo {title} {Proposal to search for
  $\mu^- n \rightarrow e^-n$ with a single event sensitivity below $10^{-16}
  $},}\ }\bibinfo {note}
  {\url{http://mu2e-docdb.fnal.gov/cgi-bin/RetrieveFile?docid=388}}\BibitemShut
  {NoStop}%
\bibitem [{\citenamefont {Carpenter}\ \emph {et~al.}(1966)\citenamefont
  {Carpenter}, \citenamefont {Abashian}, \citenamefont {Abrams}, \citenamefont
  {Fisher}, \citenamefont {Nefkens},\ and\ \citenamefont
  {Smith}}]{Carpenter:1966}%
  \BibitemOpen
  \bibfield  {author} {\bibinfo {author} {\bibnamefont {Carpenter},
  \bibfnamefont {D.~W.}}, \bibinfo {author} {\bibfnamefont {A.}~\bibnamefont
  {Abashian}}, \bibinfo {author} {\bibfnamefont {R.~J.}\ \bibnamefont
  {Abrams}}, \bibinfo {author} {\bibfnamefont {G.~P.}\ \bibnamefont {Fisher}},
  \bibinfo {author} {\bibfnamefont {B.~M.~K.}\ \bibnamefont {Nefkens}}, \ and\
  \bibinfo {author} {\bibfnamefont {J.~H.}\ \bibnamefont {Smith}}} (\bibinfo
  {year} {1966}),\ \href {\doibase 10.1103/PhysRev.142.871} {\bibfield
  {journal} {\bibinfo  {journal} {Phys. Rev.}\ }\textbf {\bibinfo {volume}
  {142}},\ \bibinfo {pages} {871}}\BibitemShut {NoStop}%
\bibitem [{\citenamefont {Chekanov}\ \emph {et~al.}(2005)\citenamefont
  {Chekanov} \emph {et~al.}}]{Chekanov:2005au}%
  \BibitemOpen
  \bibfield  {author} {\bibinfo {author} {\bibnamefont {Chekanov},
  \bibfnamefont {S.}},  \emph {et~al.} (\bibinfo {collaboration} {ZEUS
  Collaboration})} (\bibinfo {year} {2005}),\ \href {\doibase
  10.1140/epjc/s2005-02399-1} {\bibfield  {journal} {\bibinfo  {journal}
  {Eur.Phys.J.}\ }\textbf {\bibinfo {volume} {C44}},\ \bibinfo {pages} {463}},\
  \bibinfo {note} {37 pages, 10 figures, Accepted by EPJC. References and 1
  figure (Fig. 6) added Report-no: DESY-05-016},\ \Eprint
  {http://arxiv.org/abs/hep-ex/0501070} {arXiv:hep-ex/0501070 [hep-ex]}
  \BibitemShut {NoStop}%
\bibitem [{\citenamefont {Cirigliano}\ \emph {et~al.}(2009)\citenamefont
  {Cirigliano}, \citenamefont {Kitano}, \citenamefont {Okada},\ and\
  \citenamefont {Tuzon}}]{Okada:2009}%
  \BibitemOpen
  \bibfield  {author} {\bibinfo {author} {\bibnamefont {Cirigliano},
  \bibfnamefont {V.}}, \bibinfo {author} {\bibfnamefont {R.}~\bibnamefont
  {Kitano}}, \bibinfo {author} {\bibfnamefont {Y.}~\bibnamefont {Okada}}, \
  and\ \bibinfo {author} {\bibfnamefont {P.}~\bibnamefont {Tuzon}}} (\bibinfo
  {year} {2009}),\ \href {\doibase 10.1103/PhysRevD.80.013002} {\bibfield
  {journal} {\bibinfo  {journal} {Phys. Rev. D}\ }\textbf {\bibinfo {volume}
  {80}},\ \bibinfo {pages} {013002}},\ \Eprint
  {http://arxiv.org/abs/arXiv:0904.0957 [hep-ph]} {arXiv:0904.0957 [hep-ph]}
  \BibitemShut {NoStop}%
\bibitem [{\citenamefont {Commins}(1973)}]{Commins:1973}%
  \BibitemOpen
  \bibfield  {author} {\bibinfo {author} {\bibnamefont {Commins}, \bibfnamefont
  {E.~D.}}} (\bibinfo {year} {1973}),\ \href@noop {} {\emph {\bibinfo {title}
  {Weak Interactions}}}\ (\bibinfo  {publisher} {Mc-Graw Hill, Inc.})\
  Chap.~\bibinfo {chapter} {2}\BibitemShut {NoStop}%
\bibitem [{\citenamefont {Conforto}\ \emph {et~al.}(1962)\citenamefont
  {Conforto}, \citenamefont {Conversi},\ and\ \citenamefont
  {Lella}}]{Conforto:1962}%
  \BibitemOpen
  \bibfield  {author} {\bibinfo {author} {\bibnamefont {Conforto},
  \bibfnamefont {G.}}, \bibinfo {author} {\bibfnamefont {M.}~\bibnamefont
  {Conversi}}, \ and\ \bibinfo {author} {\bibfnamefont {L.}~\bibnamefont
  {Lella}}} (\bibinfo {year} {1962}),\ \href@noop {} {\bibfield  {journal}
  {\bibinfo  {journal} {Nuovo Cimento}\ }\textbf {\bibinfo {volume} {26}},\
  \bibinfo {pages} {261}}\BibitemShut {NoStop}%
\bibitem [{\citenamefont {Conversi}\ \emph {et~al.}(1961)\citenamefont
  {Conversi}, \citenamefont {di~Lella}, \citenamefont {Egidi}, \citenamefont
  {Rubbia},\ and\ \citenamefont {Toller}}]{Conversi:1961}%
  \BibitemOpen
  \bibfield  {author} {\bibinfo {author} {\bibnamefont {Conversi},
  \bibfnamefont {M.}}, \bibinfo {author} {\bibfnamefont {L.}~\bibnamefont
  {di~Lella}}, \bibinfo {author} {\bibfnamefont {A.}~\bibnamefont {Egidi}},
  \bibinfo {author} {\bibfnamefont {C.}~\bibnamefont {Rubbia}}, \ and\ \bibinfo
  {author} {\bibfnamefont {M.}~\bibnamefont {Toller}}} (\bibinfo {year}
  {1961}),\ \href {\doibase 10.1103/PhysRev.122.687} {\bibfield  {journal}
  {\bibinfo  {journal} {Phys. Rev.}\ }\textbf {\bibinfo {volume} {122}},\
  \bibinfo {pages} {687}}\BibitemShut {NoStop}%
\bibitem [{\citenamefont {Conversi}\ \emph {et~al.}(1947)\citenamefont
  {Conversi}, \citenamefont {Pancini},\ and\ \citenamefont
  {Piccioni}}]{Conversi:1947}%
  \BibitemOpen
  \bibfield  {author} {\bibinfo {author} {\bibnamefont {Conversi},
  \bibfnamefont {M.}}, \bibinfo {author} {\bibfnamefont {E.}~\bibnamefont
  {Pancini}}, \ and\ \bibinfo {author} {\bibfnamefont {O.}~\bibnamefont
  {Piccioni}}} (\bibinfo {year} {1947}),\ \href {\doibase
  10.1103/PhysRev.71.209} {\bibfield  {journal} {\bibinfo  {journal} {Phys.
  Rev.}\ }\textbf {\bibinfo {volume} {71}},\ \bibinfo {pages}
  {209}}\BibitemShut {NoStop}%
\bibitem [{\citenamefont {Cousins}\ \emph {et~al.}(1988)\citenamefont
  {Cousins}, \citenamefont {Konigsberg}, \citenamefont {Kubic}, \citenamefont
  {Melese}, \citenamefont {Rubin}, \citenamefont {Slater}, \citenamefont
  {Frank}, \citenamefont {Hart}, \citenamefont {Kinnison}, \citenamefont {Lee},
  \citenamefont {Milner}, \citenamefont {Sanders}, \citenamefont {Ziock},
  \citenamefont {Arisaka}, \citenamefont {Knibbe}, \citenamefont {Molzon},
  \citenamefont {Urheim}, \citenamefont {Wales}, \citenamefont {Axelrod},
  \citenamefont {Biery}, \citenamefont {Bonneaud}, \citenamefont {Irwin},
  \citenamefont {Lang}, \citenamefont {Martoff}, \citenamefont {Ouimette},
  \citenamefont {Ritchie}, \citenamefont {Trang}, \citenamefont {Wojcicki},
  \citenamefont {Auerbach}, \citenamefont {Buchholz}, \citenamefont {Highland},
  \citenamefont {McFarlane}, \citenamefont {Sivertz}, \citenamefont {Chapman},
  \citenamefont {Eckhause}, \citenamefont {Ginkel}, \citenamefont {Guss},
  \citenamefont {Joyce}, \citenamefont {Kane}, \citenamefont {Kenney},
  \citenamefont {Vulcan}, \citenamefont {Welsh}, \citenamefont {Whyley},\ and\
  \citenamefont {Winter}}]{Cousins:1988}%
  \BibitemOpen
  \bibfield  {author} {\bibinfo {author} {\bibnamefont {Cousins}, \bibfnamefont
  {R.~D.}}, \bibinfo {author} {\bibfnamefont {J.}~\bibnamefont {Konigsberg}},
  \bibinfo {author} {\bibfnamefont {J.}~\bibnamefont {Kubic}}, \bibinfo
  {author} {\bibfnamefont {P.}~\bibnamefont {Melese}}, \bibinfo {author}
  {\bibfnamefont {P.}~\bibnamefont {Rubin}}, \bibinfo {author} {\bibfnamefont
  {W.~E.}\ \bibnamefont {Slater}}, \bibinfo {author} {\bibfnamefont {J.~S.}\
  \bibnamefont {Frank}}, \bibinfo {author} {\bibfnamefont {G.~W.}\ \bibnamefont
  {Hart}}, \bibinfo {author} {\bibfnamefont {W.~W.}\ \bibnamefont {Kinnison}},
  \bibinfo {author} {\bibfnamefont {D.~M.}\ \bibnamefont {Lee}}, \bibinfo
  {author} {\bibfnamefont {E.~C.}\ \bibnamefont {Milner}}, \bibinfo {author}
  {\bibfnamefont {G.~H.}\ \bibnamefont {Sanders}}, \bibinfo {author}
  {\bibfnamefont {H.~J.}\ \bibnamefont {Ziock}}, \bibinfo {author}
  {\bibfnamefont {K.}~\bibnamefont {Arisaka}}, \bibinfo {author} {\bibfnamefont
  {P.}~\bibnamefont {Knibbe}}, \bibinfo {author} {\bibfnamefont {W.~R.}\
  \bibnamefont {Molzon}}, \bibinfo {author} {\bibfnamefont {J.}~\bibnamefont
  {Urheim}}, \bibinfo {author} {\bibfnamefont {W.~D.}\ \bibnamefont {Wales}},
  \bibinfo {author} {\bibfnamefont {S.}~\bibnamefont {Axelrod}}, \bibinfo
  {author} {\bibfnamefont {K.~A.}\ \bibnamefont {Biery}}, \bibinfo {author}
  {\bibfnamefont {G.}~\bibnamefont {Bonneaud}}, \bibinfo {author}
  {\bibfnamefont {G.~M.}\ \bibnamefont {Irwin}}, \bibinfo {author}
  {\bibfnamefont {K.}~\bibnamefont {Lang}}, \bibinfo {author} {\bibfnamefont
  {C.~J.}\ \bibnamefont {Martoff}}, \bibinfo {author} {\bibfnamefont {D.~A.}\
  \bibnamefont {Ouimette}}, \bibinfo {author} {\bibfnamefont {J.~L.}\
  \bibnamefont {Ritchie}}, \bibinfo {author} {\bibfnamefont {Q.~H.}\
  \bibnamefont {Trang}}, \bibinfo {author} {\bibfnamefont {S.~G.}\ \bibnamefont
  {Wojcicki}}, \bibinfo {author} {\bibfnamefont {L.~B.}\ \bibnamefont
  {Auerbach}}, \bibinfo {author} {\bibfnamefont {P.}~\bibnamefont {Buchholz}},
  \bibinfo {author} {\bibfnamefont {V.~L.}\ \bibnamefont {Highland}}, \bibinfo
  {author} {\bibfnamefont {W.~K.}\ \bibnamefont {McFarlane}}, \bibinfo {author}
  {\bibfnamefont {M.~B.}\ \bibnamefont {Sivertz}}, \bibinfo {author}
  {\bibfnamefont {M.~D.}\ \bibnamefont {Chapman}}, \bibinfo {author}
  {\bibfnamefont {M.}~\bibnamefont {Eckhause}}, \bibinfo {author}
  {\bibfnamefont {J.~F.}\ \bibnamefont {Ginkel}}, \bibinfo {author}
  {\bibfnamefont {P.~P.}\ \bibnamefont {Guss}}, \bibinfo {author}
  {\bibfnamefont {D.}~\bibnamefont {Joyce}}, \bibinfo {author} {\bibfnamefont
  {J.~R.}\ \bibnamefont {Kane}}, \bibinfo {author} {\bibfnamefont {C.~J.}\
  \bibnamefont {Kenney}}, \bibinfo {author} {\bibfnamefont {W.~F.}\
  \bibnamefont {Vulcan}}, \bibinfo {author} {\bibfnamefont {R.~E.}\
  \bibnamefont {Welsh}}, \bibinfo {author} {\bibfnamefont {R.~J.}\ \bibnamefont
  {Whyley}}, \ and\ \bibinfo {author} {\bibfnamefont {R.~G.}\ \bibnamefont
  {Winter}}} (\bibinfo {year} {1988}),\ \href {\doibase
  10.1103/PhysRevD.38.2914} {\bibfield  {journal} {\bibinfo  {journal} {Phys.
  Rev. D}\ }\textbf {\bibinfo {volume} {38}},\ \bibinfo {pages}
  {2914}}\BibitemShut {NoStop}%
\bibitem [{\citenamefont {Crittenden}\ \emph {et~al.}(1961)\citenamefont
  {Crittenden}, \citenamefont {Walker},\ and\ \citenamefont
  {Ballam}}]{Crittenden:1961}%
  \BibitemOpen
  \bibfield  {author} {\bibinfo {author} {\bibnamefont {Crittenden},
  \bibfnamefont {R.~R.}}, \bibinfo {author} {\bibfnamefont {W.~D.}\
  \bibnamefont {Walker}}, \ and\ \bibinfo {author} {\bibfnamefont
  {J.}~\bibnamefont {Ballam}}} (\bibinfo {year} {1961}),\ \href {\doibase
  10.1103/PhysRev.121.1823} {\bibfield  {journal} {\bibinfo  {journal} {Phys.
  Rev.}\ }\textbf {\bibinfo {volume} {121}},\ \bibinfo {pages}
  {1823}}\BibitemShut {NoStop}%
\bibitem [{\citenamefont {Cvetic}\ \emph {et~al.}(2002)\citenamefont {Cvetic},
  \citenamefont {Dib}, \citenamefont {Kim},\ and\ \citenamefont
  {Kim}}]{Cvetic:2002}%
  \BibitemOpen
  \bibfield  {author} {\bibinfo {author} {\bibnamefont {Cvetic}, \bibfnamefont
  {G.}}, \bibinfo {author} {\bibfnamefont {C.}~\bibnamefont {Dib}}, \bibinfo
  {author} {\bibfnamefont {C.}~\bibnamefont {Kim}}, \ and\ \bibinfo {author}
  {\bibfnamefont {J.}~\bibnamefont {Kim}}} (\bibinfo {year} {2002}),\ \href
  {\doibase 10.1103/PhysRevD.66.034008, 10.1103/PhysRevD.68.059901} {\bibfield
  {journal} {\bibinfo  {journal} {Phys.Rev.}\ }\textbf {\bibinfo {volume}
  {D66}},\ \bibinfo {pages} {034008}},\ \Eprint
  {http://arxiv.org/abs/hep-ph/0202212} {arXiv:hep-ph/0202212 [hep-ph]}
  \BibitemShut {NoStop}%
\bibitem [{\citenamefont {Czarnecki}\ \emph {et~al.}(2011)\citenamefont
  {Czarnecki}, \citenamefont {Garcia~i Tormo},\ and\ \citenamefont
  {Marciano}}]{Czarnecki:2011}%
  \BibitemOpen
  \bibfield  {author} {\bibinfo {author} {\bibnamefont {Czarnecki},
  \bibfnamefont {A.}}, \bibinfo {author} {\bibfnamefont {X.}~\bibnamefont
  {Garcia~i Tormo}}, \ and\ \bibinfo {author} {\bibfnamefont {W.~J.}\
  \bibnamefont {Marciano}}} (\bibinfo {year} {2011}),\ \href {\doibase
  10.1103/PhysRevD.84.013006} {\bibfield  {journal} {\bibinfo  {journal} {Phys.
  Rev. D}\ }\textbf {\bibinfo {volume} {84}},\ \bibinfo {pages} {013006}},\
  \Eprint {http://arxiv.org/abs/arXiv:1106:4756v1[hep-ph]}
  {arXiv:1106:4756v1[hep-ph]} \BibitemShut {NoStop}%
\bibitem [{\citenamefont {Danby}\ \emph {et~al.}(1962)\citenamefont {Danby},
  \citenamefont {Gaillard}, \citenamefont {Goulianos}, \citenamefont
  {Lederman}, \citenamefont {Mistry}, \citenamefont {Schwartz},\ and\
  \citenamefont {Steinberger}}]{Danby:1962}%
  \BibitemOpen
  \bibfield  {author} {\bibinfo {author} {\bibnamefont {Danby}, \bibfnamefont
  {G.}}, \bibinfo {author} {\bibfnamefont {J.-M.}\ \bibnamefont {Gaillard}},
  \bibinfo {author} {\bibfnamefont {K.}~\bibnamefont {Goulianos}}, \bibinfo
  {author} {\bibfnamefont {L.~M.}\ \bibnamefont {Lederman}}, \bibinfo {author}
  {\bibfnamefont {N.}~\bibnamefont {Mistry}}, \bibinfo {author} {\bibfnamefont
  {M.}~\bibnamefont {Schwartz}}, \ and\ \bibinfo {author} {\bibfnamefont
  {J.}~\bibnamefont {Steinberger}}} (\bibinfo {year} {1962}),\ \href {\doibase
  10.1103/PhysRevLett.9.36} {\bibfield  {journal} {\bibinfo  {journal} {Phys.
  Rev. Lett.}\ }\textbf {\bibinfo {volume} {9}},\ \bibinfo {pages}
  {36}}\BibitemShut {NoStop}%
\bibitem [{\citenamefont {Davis}\ \emph {et~al.}(1959)\citenamefont {Davis},
  \citenamefont {Roberts},\ and\ \citenamefont {Zipf}}]{Davis:1959}%
  \BibitemOpen
  \bibfield  {author} {\bibinfo {author} {\bibnamefont {Davis}, \bibfnamefont
  {H.~F.}}, \bibinfo {author} {\bibfnamefont {A.}~\bibnamefont {Roberts}}, \
  and\ \bibinfo {author} {\bibfnamefont {T.~F.}\ \bibnamefont {Zipf}}}
  (\bibinfo {year} {1959}),\ \href {\doibase 10.1103/PhysRevLett.2.211}
  {\bibfield  {journal} {\bibinfo  {journal} {Phys. Rev. Lett.}\ }\textbf
  {\bibinfo {volume} {2}},\ \bibinfo {pages} {211}}\BibitemShut {NoStop}%
\bibitem [{\citenamefont {Depommier}\ \emph {et~al.}(1977)\citenamefont
  {Depommier} \emph {et~al.}}]{Depommier:1977}%
  \BibitemOpen
  \bibfield  {author} {\bibinfo {author} {\bibnamefont {Depommier},
  \bibfnamefont {P.}},  \emph {et~al.}} (\bibinfo {year} {1977}),\ \href
  {\doibase 10.1103/PhysRevLett.39.1113} {\bibfield  {journal} {\bibinfo
  {journal} {Phys. Rev. Lett.}\ }\textbf {\bibinfo {volume} {39}},\ \bibinfo
  {pages} {1113}}\BibitemShut {NoStop}%
\bibitem [{\citenamefont {Derbenev}\ and\ \citenamefont
  {Johnson}(2005)}]{Johnson:2005}%
  \BibitemOpen
  \bibfield  {author} {\bibinfo {author} {\bibnamefont {Derbenev},
  \bibfnamefont {Y.}}, \ and\ \bibinfo {author} {\bibfnamefont {R.~P.}\
  \bibnamefont {Johnson}}} (\bibinfo {year} {2005}),\ \href {\doibase
  10.1103/PhysRevSTAB.8.041002} {\bibfield  {journal} {\bibinfo  {journal}
  {Phys. Rev. ST Accel. Beams}\ }\textbf {\bibinfo {volume} {8}},\ \bibinfo
  {pages} {041002}}\BibitemShut {NoStop}%
\bibitem [{\citenamefont {{Deshpande}}\ \emph {et~al.}(2012)\citenamefont
  {{Deshpande}}, \citenamefont {{Meziani}}, \citenamefont {{Qiu}},
  \citenamefont {{McKeown}}, \citenamefont {{Vigdor}}, \citenamefont
  {{Aschenauer}}, \citenamefont {{Brooks}}, \citenamefont {{Diehl}},
  \citenamefont {{Ent}}, \citenamefont {{Gao}}, \citenamefont {{Holt}},
  \citenamefont {{Ludlam}}, \citenamefont {{Horn}}, \citenamefont {{Hutton}},
  \citenamefont {{Kovchegov}}, \citenamefont {{Kumar}}, \citenamefont
  {{Mueller}}, \citenamefont {{Ramsey-Musolf}}, \citenamefont {{Roser}},
  \citenamefont {{Sabatie}}, \citenamefont {{Sichtermann}}, \citenamefont
  {{Ullrich}}, \citenamefont {{Vogelsang}},\ and\ \citenamefont
  {{Yuan}}}]{Deshpande:2012}%
  \BibitemOpen
  \bibfield  {author} {\bibinfo {author} {\bibnamefont {{Deshpande}},
  \bibfnamefont {A.}}, \bibinfo {author} {\bibfnamefont {Z.-E.}\ \bibnamefont
  {{Meziani}}}, \bibinfo {author} {\bibfnamefont {J.-W.}\ \bibnamefont
  {{Qiu}}}, \bibinfo {author} {\bibfnamefont {R.}~\bibnamefont {{McKeown}}},
  \bibinfo {author} {\bibfnamefont {S.}~\bibnamefont {{Vigdor}}}, \bibinfo
  {author} {\bibfnamefont {E.~C.}\ \bibnamefont {{Aschenauer}}}, \bibinfo
  {author} {\bibfnamefont {W.}~\bibnamefont {{Brooks}}}, \bibinfo {author}
  {\bibfnamefont {M.}~\bibnamefont {{Diehl}}}, \bibinfo {author} {\bibfnamefont
  {R.}~\bibnamefont {{Ent}}}, \bibinfo {author} {\bibfnamefont
  {H.}~\bibnamefont {{Gao}}}, \bibinfo {author} {\bibfnamefont
  {R.}~\bibnamefont {{Holt}}}, \bibinfo {author} {\bibfnamefont
  {T.}~\bibnamefont {{Ludlam}}}, \bibinfo {author} {\bibfnamefont
  {T.}~\bibnamefont {{Horn}}}, \bibinfo {author} {\bibfnamefont
  {A.}~\bibnamefont {{Hutton}}}, \bibinfo {author} {\bibfnamefont
  {Y.}~\bibnamefont {{Kovchegov}}}, \bibinfo {author} {\bibfnamefont
  {K.}~\bibnamefont {{Kumar}}}, \bibinfo {author} {\bibfnamefont {A.~H.}\
  \bibnamefont {{Mueller}}}, \bibinfo {author} {\bibfnamefont {M.}~\bibnamefont
  {{Ramsey-Musolf}}}, \bibinfo {author} {\bibfnamefont {T.}~\bibnamefont
  {{Roser}}}, \bibinfo {author} {\bibfnamefont {F.}~\bibnamefont {{Sabatie}}},
  \bibinfo {author} {\bibfnamefont {E.}~\bibnamefont {{Sichtermann}}}, \bibinfo
  {author} {\bibfnamefont {T.}~\bibnamefont {{Ullrich}}}, \bibinfo {author}
  {\bibfnamefont {W.}~\bibnamefont {{Vogelsang}}}, \ and\ \bibinfo {author}
  {\bibfnamefont {F.}~\bibnamefont {{Yuan}}}} (\bibinfo {year} {2012}),\
  \href@noop {} {\bibfield  {journal} {\bibinfo  {journal} {ArXiv e-prints}\
  }}\Eprint {http://arxiv.org/abs/1212.1701} {arXiv:1212.1701 [nucl-ex]}
  \BibitemShut {NoStop}%
\bibitem [{\citenamefont {Djilkibaev}\ and\ \citenamefont
  {Lobashev}(2010)}]{Rashid:2010}%
  \BibitemOpen
  \bibfield  {author} {\bibinfo {author} {\bibnamefont {Djilkibaev},
  \bibfnamefont {R.}}, \ and\ \bibinfo {author} {\bibfnamefont
  {V.}~\bibnamefont {Lobashev}}} (\bibinfo {year} {2010}),\ \href {\doibase
  10.1134/S1063778810120057} {\bibfield  {journal} {\bibinfo  {journal}
  {Physics of Atomic Nuclei}\ }\textbf {\bibinfo {volume} {73}},\ \bibinfo
  {pages} {2012}}\BibitemShut {NoStop}%
\bibitem [{\citenamefont {Djilkibaev}\ and\ \citenamefont
  {Konoplich}(2009)}]{Djilkibaev:2009}%
  \BibitemOpen
  \bibfield  {author} {\bibinfo {author} {\bibnamefont {Djilkibaev},
  \bibfnamefont {R.~M.}}, \ and\ \bibinfo {author} {\bibfnamefont {R.~V.}\
  \bibnamefont {Konoplich}}} (\bibinfo {year} {2009}),\ \href {\doibase
  10.1103/PhysRevD.79.073004} {\bibfield  {journal} {\bibinfo  {journal} {Phys.
  Rev. D}\ }\textbf {\bibinfo {volume} {79}},\ \bibinfo {pages}
  {073004}}\BibitemShut {NoStop}%
\bibitem [{\citenamefont {Dohmen}\ \emph {et~al.}(1993)\citenamefont {Dohmen}
  \emph {et~al.}}]{Dohmen:1993}%
  \BibitemOpen
  \bibfield  {author} {\bibinfo {author} {\bibnamefont {Dohmen}, \bibfnamefont
  {C.}},  \emph {et~al.} (\bibinfo {collaboration} {SINDRUM II
  Collaboration.})} (\bibinfo {year} {1993}),\ \href {\doibase
  10.1016/0370-2693(93)91383-X} {\bibfield  {journal} {\bibinfo  {journal}
  {Phys.Lett.}\ }\textbf {\bibinfo {volume} {B317}},\ \bibinfo {pages}
  {631}}\BibitemShut {NoStop}%
\bibitem [{\citenamefont {Dzhilkibaev}\ and\ \citenamefont
  {Lobashev}(1989)}]{Dzhilkibaev:1989zb}%
  \BibitemOpen
  \bibfield  {author} {\bibinfo {author} {\bibnamefont {Dzhilkibaev},
  \bibfnamefont {R.}}, \ and\ \bibinfo {author} {\bibfnamefont
  {V.}~\bibnamefont {Lobashev}}} (\bibinfo {year} {1989}),\ \href@noop {}
  {\bibfield  {journal} {\bibinfo  {journal} {Sov.J.Nucl.Phys.}\ }\textbf
  {\bibinfo {volume} {49}},\ \bibinfo {pages} {384}}\BibitemShut {NoStop}%
\bibitem [{\citenamefont {Feinberg}\ and\ \citenamefont
  {Weinberg}(1961)}]{Feinberg:1961}%
  \BibitemOpen
  \bibfield  {author} {\bibinfo {author} {\bibnamefont {Feinberg},
  \bibfnamefont {G.}}, \ and\ \bibinfo {author} {\bibfnamefont
  {S.}~\bibnamefont {Weinberg}}} (\bibinfo {year} {1961}),\ \href {\doibase
  10.1103/PhysRev.123.1439} {\bibfield  {journal} {\bibinfo  {journal} {Phys.
  Rev.}\ }\textbf {\bibinfo {volume} {123}},\ \bibinfo {pages}
  {1439}}\BibitemShut {NoStop}%
\bibitem [{\citenamefont {Fermi}\ \emph {et~al.}(1947)\citenamefont {Fermi},
  \citenamefont {Teller},\ and\ \citenamefont {Weisskopf}}]{Fermi:1947}%
  \BibitemOpen
  \bibfield  {author} {\bibinfo {author} {\bibnamefont {Fermi}, \bibfnamefont
  {E.}}, \bibinfo {author} {\bibfnamefont {E.}~\bibnamefont {Teller}}, \ and\
  \bibinfo {author} {\bibfnamefont {V.}~\bibnamefont {Weisskopf}}} (\bibinfo
  {year} {1947}),\ \href {\doibase 10.1103/PhysRev.71.314} {\bibfield
  {journal} {\bibinfo  {journal} {Phys. Rev.}\ }\textbf {\bibinfo {volume}
  {71}},\ \bibinfo {pages} {314}}\BibitemShut {NoStop}%
\bibitem [{\citenamefont {Fitch}\ \emph {et~al.}(1967)\citenamefont {Fitch},
  \citenamefont {Roth}, \citenamefont {Russ},\ and\ \citenamefont
  {Vernon}}]{Fitch:1967}%
  \BibitemOpen
  \bibfield  {author} {\bibinfo {author} {\bibnamefont {Fitch}, \bibfnamefont
  {V.~L.}}, \bibinfo {author} {\bibfnamefont {R.~F.}\ \bibnamefont {Roth}},
  \bibinfo {author} {\bibfnamefont {J.}~\bibnamefont {Russ}}, \ and\ \bibinfo
  {author} {\bibfnamefont {W.}~\bibnamefont {Vernon}}} (\bibinfo {year}
  {1967}),\ \href {\doibase 10.1103/PhysRev.164.1711} {\bibfield  {journal}
  {\bibinfo  {journal} {Phys. Rev.}\ }\textbf {\bibinfo {volume} {164}},\
  \bibinfo {pages} {1711}}\BibitemShut {NoStop}%
\bibitem [{\citenamefont {Frankel}\ \emph {et~al.}(1963)\citenamefont
  {Frankel}, \citenamefont {Frati}, \citenamefont {Halpern}, \citenamefont
  {Holloway}, \citenamefont {Wales} \emph {et~al.}}]{Frankel:1963}%
  \BibitemOpen
  \bibfield  {author} {\bibinfo {author} {\bibnamefont {Frankel}, \bibfnamefont
  {S.}}, \bibinfo {author} {\bibfnamefont {W.}~\bibnamefont {Frati}}, \bibinfo
  {author} {\bibfnamefont {J.}~\bibnamefont {Halpern}}, \bibinfo {author}
  {\bibfnamefont {L.}~\bibnamefont {Holloway}}, \bibinfo {author}
  {\bibfnamefont {W.}~\bibnamefont {Wales}},  \emph {et~al.}} (\bibinfo {year}
  {1963}),\ \href {\doibase 10.1007/BF02783278} {\bibfield  {journal} {\bibinfo
   {journal} {Nuovo Cim.}\ }\textbf {\bibinfo {volume} {27}},\ \bibinfo {pages}
  {894}}\BibitemShut {NoStop}%
\bibitem [{\citenamefont {Frankel}\ \emph {et~al.}(1960)\citenamefont
  {Frankel}, \citenamefont {Hagopian}, \citenamefont {Halpern},\ and\
  \citenamefont {Whetstone}}]{Frankel:1960}%
  \BibitemOpen
  \bibfield  {author} {\bibinfo {author} {\bibnamefont {Frankel}, \bibfnamefont
  {S.}}, \bibinfo {author} {\bibfnamefont {V.}~\bibnamefont {Hagopian}},
  \bibinfo {author} {\bibfnamefont {J.}~\bibnamefont {Halpern}}, \ and\
  \bibinfo {author} {\bibfnamefont {A.~L.}\ \bibnamefont {Whetstone}}}
  (\bibinfo {year} {1960}),\ \href {\doibase 10.1103/PhysRev.118.589}
  {\bibfield  {journal} {\bibinfo  {journal} {Phys. Rev.}\ }\textbf {\bibinfo
  {volume} {118}},\ \bibinfo {pages} {589}}\BibitemShut {NoStop}%
\bibitem [{\citenamefont {Frankel}\ \emph {et~al.}(1962)\citenamefont
  {Frankel}, \citenamefont {Halpern}, \citenamefont {Holloway}, \citenamefont
  {Wales}, \citenamefont {Yearian}, \citenamefont {Chamberlain}, \citenamefont
  {Lemonick},\ and\ \citenamefont {Pipkin}}]{Frankel:1962}%
  \BibitemOpen
  \bibfield  {author} {\bibinfo {author} {\bibnamefont {Frankel}, \bibfnamefont
  {S.}}, \bibinfo {author} {\bibfnamefont {J.}~\bibnamefont {Halpern}},
  \bibinfo {author} {\bibfnamefont {L.}~\bibnamefont {Holloway}}, \bibinfo
  {author} {\bibfnamefont {W.}~\bibnamefont {Wales}}, \bibinfo {author}
  {\bibfnamefont {M.}~\bibnamefont {Yearian}}, \bibinfo {author} {\bibfnamefont
  {O.}~\bibnamefont {Chamberlain}}, \bibinfo {author} {\bibfnamefont
  {A.}~\bibnamefont {Lemonick}}, \ and\ \bibinfo {author} {\bibfnamefont
  {F.~M.}\ \bibnamefont {Pipkin}}} (\bibinfo {year} {1962}),\ \href {\doibase
  10.1103/PhysRevLett.8.123} {\bibfield  {journal} {\bibinfo  {journal} {Phys.
  Rev. Lett.}\ }\textbf {\bibinfo {volume} {8}},\ \bibinfo {pages}
  {123}}\BibitemShut {NoStop}%
\bibitem [{\citenamefont {Glashow}(2013)}]{Glashow:2013via}%
  \BibitemOpen
  \bibfield  {author} {\bibinfo {author} {\bibnamefont {Glashow}, \bibfnamefont
  {S.~L.}}} (\bibinfo {year} {2013}),\ \href@noop {} {\ }\Eprint
  {http://arxiv.org/abs/1305.5482} {arXiv:1305.5482 [hep-ph]} \BibitemShut
  {NoStop}%
\bibitem [{\citenamefont {Gonderinger}\ and\ \citenamefont
  {Ramsey-Musolf}(2010)}]{Gonderinger:2010yn}%
  \BibitemOpen
  \bibfield  {author} {\bibinfo {author} {\bibnamefont {Gonderinger},
  \bibfnamefont {M.}}, \ and\ \bibinfo {author} {\bibfnamefont {M.~J.}\
  \bibnamefont {Ramsey-Musolf}}} (\bibinfo {year} {2010}),\ \href {\doibase
  10.1007/JHEP11(2010)045} {\bibfield  {journal} {\bibinfo  {journal} {JHEP}\
  }\textbf {\bibinfo {volume} {1011}},\ \bibinfo {pages} {045}},\ \Eprint
  {http://arxiv.org/abs/1006.5063} {arXiv:1006.5063 [hep-ph]} \BibitemShut
  {NoStop}%
\bibitem [{\citenamefont {de~Gouv\^{e}a}\ and\ \citenamefont
  {Saoulidou}(2010)}]{deGouvea:2010}%
  \BibitemOpen
  \bibfield  {author} {\bibinfo {author} {\bibnamefont {de~Gouv\^{e}a},
  \bibfnamefont {A.}}, \ and\ \bibinfo {author} {\bibfnamefont
  {N.}~\bibnamefont {Saoulidou}}} (\bibinfo {year} {2010}),\ \href {\doibase
  10.1146/annurev-nucl-100809-131949} {\bibfield  {journal} {\bibinfo
  {journal} {Annual Review of Nuclear and Particle Science}\ }\textbf {\bibinfo
  {volume} {60}}~(\bibinfo {number} {1}),\ \bibinfo {pages} {513}},\ \bibinfo
  {note}
  {\url{http://www.annualreviews.org/doi/abs/10.1146/annurev-nucl-100809-131949}}\BibitemShut
  {NoStop}%
\bibitem [{\citenamefont {de~Gouv\^{e}a}\ and\ \citenamefont
  {Vogel}(2013)}]{deGouvea:2013zba}%
  \BibitemOpen
  \bibfield  {author} {\bibinfo {author} {\bibnamefont {de~Gouv\^{e}a},
  \bibfnamefont {A.}}, \ and\ \bibinfo {author} {\bibfnamefont
  {P.}~\bibnamefont {Vogel}}} (\bibinfo {year} {2013}),\ \href@noop {} {\
  }\Eprint {http://arxiv.org/abs/1303.4097} {arXiv:1303.4097 [hep-ph]}
  \BibitemShut {NoStop}%
\bibitem [{\citenamefont {Greenlee}\ \emph {et~al.}(1988)\citenamefont
  {Greenlee}, \citenamefont {Kasha}, \citenamefont {Mannelli}, \citenamefont
  {Mannelli}, \citenamefont {Schaffner}, \citenamefont {Schmidt}, \citenamefont
  {Schwarz}, \citenamefont {Jastrzembski}, \citenamefont {Larsen},
  \citenamefont {Leipuner}, \citenamefont {Morse},\ and\ \citenamefont
  {Adair}}]{Greenlee:1988}%
  \BibitemOpen
  \bibfield  {author} {\bibinfo {author} {\bibnamefont {Greenlee},
  \bibfnamefont {H.~B.}}, \bibinfo {author} {\bibfnamefont {H.}~\bibnamefont
  {Kasha}}, \bibinfo {author} {\bibfnamefont {E.~B.}\ \bibnamefont {Mannelli}},
  \bibinfo {author} {\bibfnamefont {M.}~\bibnamefont {Mannelli}}, \bibinfo
  {author} {\bibfnamefont {S.~F.}\ \bibnamefont {Schaffner}}, \bibinfo {author}
  {\bibfnamefont {M.~P.}\ \bibnamefont {Schmidt}}, \bibinfo {author}
  {\bibfnamefont {C.~B.}\ \bibnamefont {Schwarz}}, \bibinfo {author}
  {\bibfnamefont {E.}~\bibnamefont {Jastrzembski}}, \bibinfo {author}
  {\bibfnamefont {R.~C.}\ \bibnamefont {Larsen}}, \bibinfo {author}
  {\bibfnamefont {L.~B.}\ \bibnamefont {Leipuner}}, \bibinfo {author}
  {\bibfnamefont {W.~M.}\ \bibnamefont {Morse}}, \ and\ \bibinfo {author}
  {\bibfnamefont {R.~K.}\ \bibnamefont {Adair}}} (\bibinfo {year} {1988}),\
  \href {\doibase 10.1103/PhysRevLett.60.893} {\bibfield  {journal} {\bibinfo
  {journal} {Phys. Rev. Lett.}\ }\textbf {\bibinfo {volume} {60}},\ \bibinfo
  {pages} {893}}\BibitemShut {NoStop}%
\bibitem [{\citenamefont {Harrison}\ \emph {et~al.}(2012)\citenamefont
  {Harrison} \emph {et~al.}}]{Harrison:2012}%
  \BibitemOpen
  \bibfield  {author} {\bibinfo {author} {\bibnamefont {Harrison},
  \bibfnamefont {J.}},  \emph {et~al.}} (\bibinfo {year} {2012}),\ \href
  {http://cds.cern.ch/record/1498727/files/LHCb-PROC-2012-064.pdf} {\bibfield
  {journal} {\bibinfo  {journal} {Nuclear Physics B Supplement}\ }\textbf
  {\bibinfo {volume} {00}},\ \bibinfo {pages} {1}}\BibitemShut {NoStop}%
\bibitem [{\citenamefont {Hincks}\ and\ \citenamefont
  {Pontecorvo}(1948)}]{Pontecorvo:1947}%
  \BibitemOpen
  \bibfield  {author} {\bibinfo {author} {\bibnamefont {Hincks}, \bibfnamefont
  {E.~P.}}, \ and\ \bibinfo {author} {\bibfnamefont {B.}~\bibnamefont
  {Pontecorvo}}} (\bibinfo {year} {1948}),\ \href {\doibase
  10.1103/PhysRev.73.257} {\bibfield  {journal} {\bibinfo  {journal} {Phys.
  Rev.}\ }\textbf {\bibinfo {volume} {73}},\ \bibinfo {pages}
  {257}}\BibitemShut {NoStop}%
\bibitem [{\citenamefont {Honecker}\ \emph {et~al.}(1996)\citenamefont
  {Honecker}, \citenamefont {Dohmen}, \citenamefont {Haan}, \citenamefont
  {Junker}, \citenamefont {Otter}, \citenamefont {Starlinger}, \citenamefont
  {Wintz}, \citenamefont {Hofmann}, \citenamefont {Bertl}, \citenamefont
  {Egger}, \citenamefont {Krause}, \citenamefont {Eggli}, \citenamefont
  {Engfer}, \citenamefont {Findeisen}, \citenamefont {Hermes}, \citenamefont
  {Kozlowski}, \citenamefont {Niebuhr}, \citenamefont {Pruys},\ and\
  \citenamefont {van~der Schaaf}}]{Honecker:1996}%
  \BibitemOpen
  \bibfield  {author} {\bibinfo {author} {\bibnamefont {Honecker},
  \bibfnamefont {W.}}, \bibinfo {author} {\bibfnamefont {C.}~\bibnamefont
  {Dohmen}}, \bibinfo {author} {\bibfnamefont {H.}~\bibnamefont {Haan}},
  \bibinfo {author} {\bibfnamefont {D.}~\bibnamefont {Junker}}, \bibinfo
  {author} {\bibfnamefont {G.}~\bibnamefont {Otter}}, \bibinfo {author}
  {\bibfnamefont {M.}~\bibnamefont {Starlinger}}, \bibinfo {author}
  {\bibfnamefont {P.}~\bibnamefont {Wintz}}, \bibinfo {author} {\bibfnamefont
  {J.}~\bibnamefont {Hofmann}}, \bibinfo {author} {\bibfnamefont
  {W.}~\bibnamefont {Bertl}}, \bibinfo {author} {\bibfnamefont
  {J.}~\bibnamefont {Egger}}, \bibinfo {author} {\bibfnamefont
  {B.}~\bibnamefont {Krause}}, \bibinfo {author} {\bibfnamefont
  {S.}~\bibnamefont {Eggli}}, \bibinfo {author} {\bibfnamefont
  {R.}~\bibnamefont {Engfer}}, \bibinfo {author} {\bibfnamefont
  {C.}~\bibnamefont {Findeisen}}, \bibinfo {author} {\bibfnamefont {E.~A.}\
  \bibnamefont {Hermes}}, \bibinfo {author} {\bibfnamefont {T.}~\bibnamefont
  {Kozlowski}}, \bibinfo {author} {\bibfnamefont {C.~B.}\ \bibnamefont
  {Niebuhr}}, \bibinfo {author} {\bibfnamefont {H.~S.}\ \bibnamefont {Pruys}},
  \ and\ \bibinfo {author} {\bibfnamefont {A.}~\bibnamefont {van~der Schaaf}}
  (\bibinfo {collaboration} {SINDRUM II Collaboration})} (\bibinfo {year}
  {1996}),\ \href {\doibase 10.1103/PhysRevLett.76.200} {\bibfield  {journal}
  {\bibinfo  {journal} {Phys. Rev. Lett.}\ }\textbf {\bibinfo {volume} {76}},\
  \bibinfo {pages} {200}}\BibitemShut {NoStop}%
\bibitem [{\citenamefont {Horikawa}\ and\ \citenamefont
  {Sasaki}(1996)}]{Horikawa:1996}%
  \BibitemOpen
  \bibfield  {author} {\bibinfo {author} {\bibnamefont {Horikawa},
  \bibfnamefont {K.}}, \ and\ \bibinfo {author} {\bibfnamefont
  {K.}~\bibnamefont {Sasaki}}} (\bibinfo {year} {1996}),\ \href {\doibase
  10.1103/PhysRevD.53.560} {\bibfield  {journal} {\bibinfo  {journal} {Phys.
  Rev. D}\ }\textbf {\bibinfo {volume} {53}},\ \bibinfo {pages}
  {560}}\BibitemShut {NoStop}%
\bibitem [{\citenamefont {Hou}(1996)}]{Hou:1996}%
  \BibitemOpen
  \bibfield  {author} {\bibinfo {author} {\bibnamefont {Hou}, \bibfnamefont
  {G.~W.-S.}}} (\bibinfo {year} {1996}),\ \href@noop {} {\bibfield  {journal}
  {\bibinfo  {journal} {Nucl.Phys.Proc.Suppl.}\ }\textbf {\bibinfo {volume}
  {51A}},\ \bibinfo {pages} {40}},\ \Eprint
  {http://arxiv.org/abs/hep-ph/9605204} {arXiv:hep-ph/9605204 [hep-ph]}
  \BibitemShut {NoStop}%
\bibitem [{\citenamefont {Hou}\ and\ \citenamefont {Wong}(1995)}]{Hou:1995}%
  \BibitemOpen
  \bibfield  {author} {\bibinfo {author} {\bibnamefont {Hou}, \bibfnamefont
  {W.-S.}}, \ and\ \bibinfo {author} {\bibfnamefont {G.-G.}\ \bibnamefont
  {Wong}}} (\bibinfo {year} {1995}),\ \href {\doibase
  10.1016/0370-2693(95)00893-P} {\bibfield  {journal} {\bibinfo  {journal}
  {Physics Letters B}\ }\textbf {\bibinfo {volume} {357}},\ \bibinfo {pages}
  {145 }},\ \Eprint
  {http://arxiv.org/abs/http://arXiv.org/pdf/hep-ph/9505300v1}
  {http://arXiv.org/pdf/hep-ph/9505300v1} \BibitemShut {NoStop}%
\bibitem [{\citenamefont {Huber}\ \emph {et~al.}(1990)\citenamefont {Huber},
  \citenamefont {Kunselman}, \citenamefont {Janissen}, \citenamefont {Beer},
  \citenamefont {Mason} \emph {et~al.}}]{Huber:1989}%
  \BibitemOpen
  \bibfield  {author} {\bibinfo {author} {\bibnamefont {Huber}, \bibfnamefont
  {T.}}, \bibinfo {author} {\bibfnamefont {A.}~\bibnamefont {Kunselman}},
  \bibinfo {author} {\bibfnamefont {A.}~\bibnamefont {Janissen}}, \bibinfo
  {author} {\bibfnamefont {G.}~\bibnamefont {Beer}}, \bibinfo {author}
  {\bibfnamefont {G.}~\bibnamefont {Mason}},  \emph {et~al.}} (\bibinfo {year}
  {1990}),\ \href {\doibase 10.1103/PhysRevD.41.2709} {\bibfield  {journal}
  {\bibinfo  {journal} {Phys.Rev.}\ }\textbf {\bibinfo {volume} {D41}},\
  \bibinfo {pages} {2709}}\BibitemShut {NoStop}%
\bibitem [{\citenamefont {Hughes}\ \emph {et~al.}(1960)\citenamefont {Hughes},
  \citenamefont {McColm}, \citenamefont {Ziock},\ and\ \citenamefont
  {Prepost}}]{Hughes:1960}%
  \BibitemOpen
  \bibfield  {author} {\bibinfo {author} {\bibnamefont {Hughes}, \bibfnamefont
  {V.~W.}}, \bibinfo {author} {\bibfnamefont {D.~W.}\ \bibnamefont {McColm}},
  \bibinfo {author} {\bibfnamefont {K.}~\bibnamefont {Ziock}}, \ and\ \bibinfo
  {author} {\bibfnamefont {R.}~\bibnamefont {Prepost}}} (\bibinfo {year}
  {1960}),\ \href {\doibase 10.1103/PhysRevLett.5.63} {\bibfield  {journal}
  {\bibinfo  {journal} {Phys. Rev. Lett.}\ }\textbf {\bibinfo {volume} {5}},\
  \bibinfo {pages} {63}}\BibitemShut {NoStop}%
\bibitem [{\citenamefont {Inagaki}\ \emph {et~al.}(1989)\citenamefont
  {Inagaki}, \citenamefont {Kobayashi}, \citenamefont {Sato}, \citenamefont
  {Shinkawa}, \citenamefont {Suekane}, \citenamefont {Takamatsu}, \citenamefont
  {Yoshimura}, \citenamefont {Ishikawa}, \citenamefont {Kishida}, \citenamefont
  {Komatsubara}, \citenamefont {Kuze}, \citenamefont {Sai}, \citenamefont
  {Toyoura}, \citenamefont {Yamamoto},\ and\ \citenamefont
  {Hemmi}}]{Inagaki:1989}%
  \BibitemOpen
  \bibfield  {author} {\bibinfo {author} {\bibnamefont {Inagaki}, \bibfnamefont
  {T.}}, \bibinfo {author} {\bibfnamefont {M.}~\bibnamefont {Kobayashi}},
  \bibinfo {author} {\bibfnamefont {T.}~\bibnamefont {Sato}}, \bibinfo {author}
  {\bibfnamefont {T.}~\bibnamefont {Shinkawa}}, \bibinfo {author}
  {\bibfnamefont {F.}~\bibnamefont {Suekane}}, \bibinfo {author} {\bibfnamefont
  {K.}~\bibnamefont {Takamatsu}}, \bibinfo {author} {\bibfnamefont
  {Y.}~\bibnamefont {Yoshimura}}, \bibinfo {author} {\bibfnamefont
  {K.}~\bibnamefont {Ishikawa}}, \bibinfo {author} {\bibfnamefont
  {T.}~\bibnamefont {Kishida}}, \bibinfo {author} {\bibfnamefont {T.~K.}\
  \bibnamefont {Komatsubara}}, \bibinfo {author} {\bibfnamefont
  {M.}~\bibnamefont {Kuze}}, \bibinfo {author} {\bibfnamefont {F.}~\bibnamefont
  {Sai}}, \bibinfo {author} {\bibfnamefont {J.}~\bibnamefont {Toyoura}},
  \bibinfo {author} {\bibfnamefont {S.~S.}\ \bibnamefont {Yamamoto}}, \ and\
  \bibinfo {author} {\bibfnamefont {Y.}~\bibnamefont {Hemmi}}} (\bibinfo {year}
  {1989}),\ \href {\doibase 10.1103/PhysRevD.40.1712} {\bibfield  {journal}
  {\bibinfo  {journal} {Phys. Rev. D}\ }\textbf {\bibinfo {volume} {40}},\
  \bibinfo {pages} {1712}}\BibitemShut {NoStop}%
\bibitem [{\citenamefont {Jackson}(1975)}]{Jackson:1975}%
  \BibitemOpen
  \bibfield  {author} {\bibinfo {author} {\bibnamefont {Jackson}, \bibfnamefont
  {J.~D.}}} (\bibinfo {year} {1975}),\ \href@noop {} {\emph {\bibinfo {title}
  {Classical Electrodynamics}}}\ (\bibinfo  {publisher} {John Wiley \& Sons})\
  Chap.\ \bibinfo {chapter} {12.5}\BibitemShut {NoStop}%
\bibitem [{\citenamefont {Kaulard}\ \emph {et~al.}(1998)\citenamefont
  {Kaulard}, \citenamefont {Dohmen}, \citenamefont {Haan}, \citenamefont
  {Honecker}, \citenamefont {Junker}, \citenamefont {Otter}, \citenamefont
  {Starlinger}, \citenamefont {Wintz}, \citenamefont {Hofmann}, \citenamefont
  {Bertl}, \citenamefont {Egger}, \citenamefont {Krause}, \citenamefont
  {Eggli}, \citenamefont {Engfer}, \citenamefont {Findeisen}, \citenamefont
  {Hermes}, \citenamefont {Kozlowski}, \citenamefont {Niebuhr}, \citenamefont
  {Rutsche}, \citenamefont {Pruys},\ and\ \citenamefont {van~der
  Schaaf}}]{Kaulard:1998}%
  \BibitemOpen
  \bibfield  {author} {\bibinfo {author} {\bibnamefont {Kaulard}, \bibfnamefont
  {J.}}, \bibinfo {author} {\bibfnamefont {C.}~\bibnamefont {Dohmen}}, \bibinfo
  {author} {\bibfnamefont {H.}~\bibnamefont {Haan}}, \bibinfo {author}
  {\bibfnamefont {W.}~\bibnamefont {Honecker}}, \bibinfo {author}
  {\bibfnamefont {D.}~\bibnamefont {Junker}}, \bibinfo {author} {\bibfnamefont
  {G.}~\bibnamefont {Otter}}, \bibinfo {author} {\bibfnamefont
  {M.}~\bibnamefont {Starlinger}}, \bibinfo {author} {\bibfnamefont
  {P.}~\bibnamefont {Wintz}}, \bibinfo {author} {\bibfnamefont
  {J.}~\bibnamefont {Hofmann}}, \bibinfo {author} {\bibfnamefont
  {W.}~\bibnamefont {Bertl}}, \bibinfo {author} {\bibfnamefont
  {J.}~\bibnamefont {Egger}}, \bibinfo {author} {\bibfnamefont
  {B.}~\bibnamefont {Krause}}, \bibinfo {author} {\bibfnamefont
  {S.}~\bibnamefont {Eggli}}, \bibinfo {author} {\bibfnamefont
  {R.}~\bibnamefont {Engfer}}, \bibinfo {author} {\bibfnamefont
  {C.}~\bibnamefont {Findeisen}}, \bibinfo {author} {\bibfnamefont
  {E.}~\bibnamefont {Hermes}}, \bibinfo {author} {\bibfnamefont
  {T.}~\bibnamefont {Kozlowski}}, \bibinfo {author} {\bibfnamefont
  {C.}~\bibnamefont {Niebuhr}}, \bibinfo {author} {\bibfnamefont
  {M.}~\bibnamefont {Rutsche}}, \bibinfo {author} {\bibfnamefont
  {H.}~\bibnamefont {Pruys}}, \ and\ \bibinfo {author} {\bibfnamefont
  {A.}~\bibnamefont {van~der Schaaf}}} (\bibinfo {year} {1998}),\ \href
  {\doibase 10.1016/S0370-2693(97)01423-8} {\bibfield  {journal} {\bibinfo
  {journal} {Physics Letters B}\ }\textbf {\bibinfo {volume} {422}}~(\bibinfo
  {number} {1-4}),\ \bibinfo {pages} {334 }}\BibitemShut {NoStop}%
\bibitem [{\citenamefont {Kinnison}\ \emph {et~al.}(1982)\citenamefont
  {Kinnison} \emph {et~al.}}]{Kinnison:1982}%
  \BibitemOpen
  \bibfield  {author} {\bibinfo {author} {\bibnamefont {Kinnison},
  \bibfnamefont {W.~W.}},  \emph {et~al.}} (\bibinfo {year} {1982}),\ \href
  {\doibase 10.1103/PhysRevD.25.2846} {\bibfield  {journal} {\bibinfo
  {journal} {Phys. Rev. D}\ }\textbf {\bibinfo {volume} {25}},\ \bibinfo
  {pages} {2846}}\BibitemShut {NoStop}%
\bibitem [{\citenamefont {Kinoshita}\ and\ \citenamefont
  {Sirlin}(1957)}]{Kinoshita:1957}%
  \BibitemOpen
  \bibfield  {author} {\bibinfo {author} {\bibnamefont {Kinoshita},
  \bibfnamefont {T.}}, \ and\ \bibinfo {author} {\bibfnamefont
  {A.}~\bibnamefont {Sirlin}}} (\bibinfo {year} {1957}),\ \href {\doibase
  10.1103/PhysRev.107.593} {\bibfield  {journal} {\bibinfo  {journal} {Phys.
  Rev.}\ }\textbf {\bibinfo {volume} {107}},\ \bibinfo {pages}
  {593}}\BibitemShut {NoStop}%
\bibitem [{\citenamefont {Kitano}\ \emph {et~al.}(2002)\citenamefont {Kitano},
  \citenamefont {Koike},\ and\ \citenamefont {Okada}}]{Okada:2002}%
  \BibitemOpen
  \bibfield  {author} {\bibinfo {author} {\bibnamefont {Kitano}, \bibfnamefont
  {R.}}, \bibinfo {author} {\bibfnamefont {M.}~\bibnamefont {Koike}}, \ and\
  \bibinfo {author} {\bibfnamefont {Y.}~\bibnamefont {Okada}}} (\bibinfo {year}
  {2002}),\ \href {\doibase 10.1103/PhysRevD.66.096002} {\bibfield  {journal}
  {\bibinfo  {journal} {Phys. Rev. D}\ }\textbf {\bibinfo {volume} {66}},\
  \bibinfo {pages} {096002}}\BibitemShut {NoStop}%
\bibitem [{\citenamefont {Kitano}\ \emph {et~al.}(2007)\citenamefont {Kitano},
  \citenamefont {Koike},\ and\ \citenamefont {Okada}}]{Okada:2002e}%
  \BibitemOpen
  \bibfield  {author} {\bibinfo {author} {\bibnamefont {Kitano}, \bibfnamefont
  {R.}}, \bibinfo {author} {\bibfnamefont {M.}~\bibnamefont {Koike}}, \ and\
  \bibinfo {author} {\bibfnamefont {Y.}~\bibnamefont {Okada}}} (\bibinfo {year}
  {2007}),\ \href {\doibase 10.1103/PhysRevD.76.059902} {\bibfield  {journal}
  {\bibinfo  {journal} {Phys. Rev. D}\ }\textbf {\bibinfo {volume} {76}},\
  \bibinfo {pages} {059902}}\BibitemShut {NoStop}%
\bibitem [{\citenamefont {Knoepfel}\ \emph {et~al.}(2013)\citenamefont
  {Knoepfel} \emph {et~al.}}]{Knoepfel:2013ouy}%
  \BibitemOpen
  \bibfield  {author} {\bibinfo {author} {\bibnamefont {Knoepfel},
  \bibfnamefont {K.}},  \emph {et~al.} (\bibinfo {collaboration} {mu2e
  Collaboration})} (\bibinfo {year} {2013}),\ \href@noop {} {\ }\Eprint
  {http://arxiv.org/abs/1307.1168} {arXiv:1307.1168 [physics.ins-det]}
  \BibitemShut {NoStop}%
\bibitem [{\citenamefont {Korenchenko}\ \emph {et~al.}(1971)\citenamefont
  {Korenchenko}, \citenamefont {Kostin}, \citenamefont {Mitselmakher},
  \citenamefont {Nekrasov},\ and\ \citenamefont {Smirnov}}]{Korenchenko:1971}%
  \BibitemOpen
  \bibfield  {author} {\bibinfo {author} {\bibnamefont {Korenchenko},
  \bibfnamefont {S.}}, \bibinfo {author} {\bibfnamefont {B.}~\bibnamefont
  {Kostin}}, \bibinfo {author} {\bibfnamefont {G.}~\bibnamefont
  {Mitselmakher}}, \bibinfo {author} {\bibfnamefont {K.}~\bibnamefont
  {Nekrasov}}, \ and\ \bibinfo {author} {\bibfnamefont {V.}~\bibnamefont
  {Smirnov}}} (\bibinfo {year} {1971}),\ \href@noop {} {\bibfield  {journal}
  {\bibinfo  {journal} {Yad. Fiz.}\ }\textbf {\bibinfo {volume} {13}},\
  \bibinfo {pages} {341}}\BibitemShut {NoStop}%
\bibitem [{\citenamefont {Korenchenko}\ \emph {et~al.}(1976)\citenamefont
  {Korenchenko}, \citenamefont {Kostin}, \citenamefont {Mitselmakher},
  \citenamefont {Nekrasov},\ and\ \citenamefont {Smirnov}}]{Korenchenko:1976}%
  \BibitemOpen
  \bibfield  {author} {\bibinfo {author} {\bibnamefont {Korenchenko},
  \bibfnamefont {S.}}, \bibinfo {author} {\bibfnamefont {B.}~\bibnamefont
  {Kostin}}, \bibinfo {author} {\bibfnamefont {G.}~\bibnamefont
  {Mitselmakher}}, \bibinfo {author} {\bibfnamefont {K.}~\bibnamefont
  {Nekrasov}}, \ and\ \bibinfo {author} {\bibfnamefont {V.}~\bibnamefont
  {Smirnov}}} (\bibinfo {year} {1976}),\ \href@noop {} {\bibfield  {journal}
  {\bibinfo  {journal} {Sov.\ Phys.\ JETP}\ }\textbf {\bibinfo {volume} {43}},\
  \bibinfo {pages} {1}}\BibitemShut {NoStop}%
\bibitem [{\citenamefont {Kosmas}\ and\ \citenamefont
  {Vergados}(1990)}]{Kosmas:1990}%
  \BibitemOpen
  \bibfield  {author} {\bibinfo {author} {\bibnamefont {Kosmas}, \bibfnamefont
  {T.~S.}}, \ and\ \bibinfo {author} {\bibfnamefont {J.~D.}\ \bibnamefont
  {Vergados}}} (\bibinfo {year} {1990}),\ \href {\doibase
  10.1016/0375-9474(90)90353-N} {\bibfield  {journal} {\bibinfo  {journal}
  {Nucl. Phys.}\ }\textbf {\bibinfo {volume} {A510}},\ \bibinfo {pages}
  {641}},\ \bibinfo {note} {and references therein.}\BibitemShut {Stop}%
\bibitem [{\citenamefont {Kosmas}\ \emph {et~al.}(1998)\citenamefont {Kosmas}
  \emph {et~al.}}]{Kosmas:1998}%
  \BibitemOpen
  \bibfield  {author} {\bibinfo {author} {\bibnamefont {Kosmas}, \bibfnamefont
  {T.~S.}},  \emph {et~al.}} (\bibinfo {year} {1998}),\ \href@noop {}
  {\bibfield  {journal} {\bibinfo  {journal} {Phys.\ Atom.\ Nucl.}\ }\textbf
  {\bibinfo {volume} {61}},\ \bibinfo {pages} {1161}},\ \Eprint
  {http://arxiv.org/abs/arXiv:nucl-th/9712016v1} {arXiv:nucl-th/9712016v1}
  \BibitemShut {NoStop}%
\bibitem [{\citenamefont {Kroll}\ and\ \citenamefont
  {Wada}(1955)}]{Kroll:1955}%
  \BibitemOpen
  \bibfield  {author} {\bibinfo {author} {\bibnamefont {Kroll}, \bibfnamefont
  {N.~M.}}, \ and\ \bibinfo {author} {\bibfnamefont {W.}~\bibnamefont {Wada}}}
  (\bibinfo {year} {1955}),\ \href {\doibase 10.1103/PhysRev.98.1355}
  {\bibfield  {journal} {\bibinfo  {journal} {Phys. Rev.}\ }\textbf {\bibinfo
  {volume} {98}},\ \bibinfo {pages} {1355}}\BibitemShut {NoStop}%
\bibitem [{\citenamefont {Kronfeld}\ \emph {et~al.}(2013)\citenamefont
  {Kronfeld}, \citenamefont {Tschirhart}, \citenamefont {Al-Binni},
  \citenamefont {Altmannshofer}, \citenamefont {Ankenbrandt} \emph
  {et~al.}}]{Kronfeld:2013uoa}%
  \BibitemOpen
  \bibfield  {author} {\bibinfo {author} {\bibnamefont {Kronfeld},
  \bibfnamefont {A.~S.}}, \bibinfo {author} {\bibfnamefont {R.~S.}\
  \bibnamefont {Tschirhart}}, \bibinfo {author} {\bibfnamefont
  {U.}~\bibnamefont {Al-Binni}}, \bibinfo {author} {\bibfnamefont
  {W.}~\bibnamefont {Altmannshofer}}, \bibinfo {author} {\bibfnamefont
  {C.}~\bibnamefont {Ankenbrandt}},  \emph {et~al.}} (\bibinfo {year} {2013}),\
  \href@noop {} {\ }\Eprint {http://arxiv.org/abs/1306.5009} {arXiv:1306.5009
  [hep-ex]} \BibitemShut {NoStop}%
\bibitem [{\citenamefont {Kuno}(2008)}]{Kuno:2008zz}%
  \BibitemOpen
  \bibfield  {author} {\bibinfo {author} {\bibnamefont {Kuno}, \bibfnamefont
  {Y.}}} (\bibinfo {year} {2008}),\ \href@noop {} {\bibfield  {journal}
  {\bibinfo  {journal} {PoS}\ }\textbf {\bibinfo {volume} {NUFACT08}},\
  \bibinfo {pages} {111}}\BibitemShut {NoStop}%
\bibitem [{\citenamefont {Kuno}(2010)}]{Kuno:2010}%
  \BibitemOpen
  \bibfield  {author} {\bibinfo {author} {\bibnamefont {Kuno}, \bibfnamefont
  {Y.}}} (\bibinfo {year} {2010}),\ \href@noop {} {\bibfield  {journal}
  {\bibinfo  {journal} {PoS}\ }\textbf {\bibinfo {volume} {FPCP2010}},\
  \bibinfo {pages} {049}},\ \bibinfo {note}
  {\url{http://pos.sissa.it//archive/conferences/116/049/FPCP\%202010\_049.pdf}}\BibitemShut
  {NoStop}%
\bibitem [{\citenamefont {Kuno}\ and\ \citenamefont {Okada}(2001)}]{Kuno:1991}%
  \BibitemOpen
  \bibfield  {author} {\bibinfo {author} {\bibnamefont {Kuno}, \bibfnamefont
  {Y.}}, \ and\ \bibinfo {author} {\bibfnamefont {Y.}~\bibnamefont {Okada}}}
  (\bibinfo {year} {2001}),\ \href {\doibase 10.1103/RevModPhys.73.151}
  {\bibfield  {journal} {\bibinfo  {journal} {Rev.Mod.Phys.}\ }\textbf
  {\bibinfo {volume} {73}},\ \bibinfo {pages} {151}},\ \Eprint
  {http://arxiv.org/abs/hep-ph/9909265} {arXiv:hep-ph/9909265 [hep-ph]}
  \BibitemShut {NoStop}%
\bibitem [{\citenamefont {Lagarrigue}\ and\ \citenamefont
  {Peyrou}(1952)}]{Lagarrigue:1952}%
  \BibitemOpen
  \bibfield  {author} {\bibinfo {author} {\bibnamefont {Lagarrigue},
  \bibfnamefont {A.}}, \ and\ \bibinfo {author} {\bibfnamefont
  {C.}~\bibnamefont {Peyrou}}} (\bibinfo {year} {1952}),\ \href@noop {}
  {\bibfield  {journal} {\bibinfo  {journal} {Comptes Rendus Acad. Sci. Paris}\
  }\textbf {\bibinfo {volume} {234}},\ \bibinfo {pages} {1873}}\BibitemShut
  {NoStop}%
\bibitem [{\citenamefont {Landsberg}(2005)}]{Landsberg:2005}%
  \BibitemOpen
  \bibfield  {author} {\bibinfo {author} {\bibnamefont {Landsberg},
  \bibfnamefont {L.}}} (\bibinfo {year} {2005}),\ \href {\doibase
  10.1134/1.1992575, 10.1134/1.1992575} {\bibfield  {journal} {\bibinfo
  {journal} {Phys.Atom.Nucl.}\ }\textbf {\bibinfo {volume} {68}},\ \bibinfo
  {pages} {1190}},\ \bibinfo {note} {extended version of the talk given at the
  Chicago Flavor Seminar, 27 FEb 2004},\ \Eprint
  {http://arxiv.org/abs/hep-ph/0410261} {arXiv:hep-ph/0410261 [hep-ph]}
  \BibitemShut {NoStop}%
\bibitem [{\citenamefont {Lazzeroni}\ \emph {et~al.}(2011)\citenamefont
  {Lazzeroni} \emph {et~al.}}]{na62:2012}%
  \BibitemOpen
  \bibfield  {author} {\bibinfo {author} {\bibnamefont {Lazzeroni},
  \bibfnamefont {C.}},  \emph {et~al.}} (\bibinfo {year} {2011}),\ \href
  {\doibase http://dx.doi.org/10.1016/j.physletb.2011.02.064} {\bibfield
  {journal} {\bibinfo  {journal} {Physics Letters B}\ }\textbf {\bibinfo
  {volume} {698}}~(\bibinfo {number} {2}),\ \bibinfo {pages} {105}}\BibitemShut
  {NoStop}%
\bibitem [{\citenamefont {Lee}\ \emph {et~al.}(1990)\citenamefont {Lee},
  \citenamefont {Alliegro}, \citenamefont {Campagnari}, \citenamefont
  {Chaloupka}, \citenamefont {Cooper}, \citenamefont {Egger}, \citenamefont
  {Gordon}, \citenamefont {Hadley}, \citenamefont {Herold}, \citenamefont
  {Jagel}, \citenamefont {Kaspar}, \citenamefont {Lazarus}, \citenamefont
  {Lubatti}, \citenamefont {Rehak}, \citenamefont {Zeller},\ and\ \citenamefont
  {Zhao}}]{Lee:1990}%
  \BibitemOpen
  \bibfield  {author} {\bibinfo {author} {\bibnamefont {Lee}, \bibfnamefont
  {A.~M.}}, \bibinfo {author} {\bibfnamefont {C.}~\bibnamefont {Alliegro}},
  \bibinfo {author} {\bibfnamefont {C.}~\bibnamefont {Campagnari}}, \bibinfo
  {author} {\bibfnamefont {V.}~\bibnamefont {Chaloupka}}, \bibinfo {author}
  {\bibfnamefont {P.~S.}\ \bibnamefont {Cooper}}, \bibinfo {author}
  {\bibfnamefont {J.}~\bibnamefont {Egger}}, \bibinfo {author} {\bibfnamefont
  {H.~A.}\ \bibnamefont {Gordon}}, \bibinfo {author} {\bibfnamefont {N.~J.}\
  \bibnamefont {Hadley}}, \bibinfo {author} {\bibfnamefont {W.~D.}\
  \bibnamefont {Herold}}, \bibinfo {author} {\bibfnamefont {E.~A.}\
  \bibnamefont {Jagel}}, \bibinfo {author} {\bibfnamefont {H.}~\bibnamefont
  {Kaspar}}, \bibinfo {author} {\bibfnamefont {D.~M.}\ \bibnamefont {Lazarus}},
  \bibinfo {author} {\bibfnamefont {H.~J.}\ \bibnamefont {Lubatti}}, \bibinfo
  {author} {\bibfnamefont {P.}~\bibnamefont {Rehak}}, \bibinfo {author}
  {\bibfnamefont {M.~E.}\ \bibnamefont {Zeller}}, \ and\ \bibinfo {author}
  {\bibfnamefont {T.}~\bibnamefont {Zhao}}} (\bibinfo {year} {1990}),\ \href
  {\doibase 10.1103/PhysRevLett.64.165} {\bibfield  {journal} {\bibinfo
  {journal} {Phys. Rev. Lett.}\ }\textbf {\bibinfo {volume} {64}},\ \bibinfo
  {pages} {165}}\BibitemShut {NoStop}%
\bibitem [{\citenamefont {Lee}\ and\ \citenamefont {Shrock}(1977)}]{Lee:1977}%
  \BibitemOpen
  \bibfield  {author} {\bibinfo {author} {\bibnamefont {Lee}, \bibfnamefont
  {B.~W.}}, \ and\ \bibinfo {author} {\bibfnamefont {R.~E.}\ \bibnamefont
  {Shrock}}} (\bibinfo {year} {1977}),\ \href {\doibase
  10.1103/PhysRevD.16.1444} {\bibfield  {journal} {\bibinfo  {journal} {Phys.
  Rev. D}\ }\textbf {\bibinfo {volume} {16}},\ \bibinfo {pages}
  {1444}}\BibitemShut {NoStop}%
\bibitem [{\citenamefont {Lee}\ and\ \citenamefont {Samios}(1959)}]{Lee:1959}%
  \BibitemOpen
  \bibfield  {author} {\bibinfo {author} {\bibnamefont {Lee}, \bibfnamefont
  {J.}}, \ and\ \bibinfo {author} {\bibfnamefont {N.~P.}\ \bibnamefont
  {Samios}}} (\bibinfo {year} {1959}),\ \href {\doibase
  10.1103/PhysRevLett.3.55} {\bibfield  {journal} {\bibinfo  {journal} {Phys.
  Rev. Lett.}\ }\textbf {\bibinfo {volume} {3}},\ \bibinfo {pages}
  {55}}\BibitemShut {NoStop}%
\bibitem [{\citenamefont {Lees}\ \emph {et~al.}(2012)\citenamefont {Lees} \emph
  {et~al.}}]{Lees:2012}%
  \BibitemOpen
  \bibfield  {author} {\bibinfo {author} {\bibnamefont {Lees}, \bibfnamefont
  {J.}},  \emph {et~al.} (\bibinfo {collaboration} {BABAR Collaboration})}
  (\bibinfo {year} {2012}),\ \href@noop {} {\bibfield  {journal} {\bibinfo
  {journal} {submitted to Phys.Rev.D}\ }}\Eprint
  {http://arxiv.org/abs/1204.2852} {arXiv:1204.2852 [hep-ex]} \BibitemShut
  {NoStop}%
\bibitem [{\citenamefont {Lees}\ \emph {et~al.}(2010)\citenamefont {Lees} \emph
  {et~al.}}]{Aubert:2010a}%
  \BibitemOpen
  \bibfield  {author} {\bibinfo {author} {\bibnamefont {Lees}, \bibfnamefont
  {J.~P.}},  \emph {et~al.} (\bibinfo {collaboration} {BABAR Collaboration})}
  (\bibinfo {year} {2010}),\ \href {\doibase 10.1103/PhysRevD.81.111101}
  {\bibfield  {journal} {\bibinfo  {journal} {Phys. Rev. D}\ }\textbf {\bibinfo
  {volume} {81}},\ \bibinfo {pages} {111101}}\BibitemShut {NoStop}%
\bibitem [{\citenamefont {Littenberg}\ and\ \citenamefont
  {Shrock}(2000)}]{Littenberg:2000}%
  \BibitemOpen
  \bibfield  {author} {\bibinfo {author} {\bibnamefont {Littenberg},
  \bibfnamefont {L.~S.}}, \ and\ \bibinfo {author} {\bibfnamefont
  {R.}~\bibnamefont {Shrock}}} (\bibinfo {year} {2000}),\ \href {\doibase
  10.1016/S0370-2693(00)01041-8} {\bibfield  {journal} {\bibinfo  {journal}
  {Phys.Lett.}\ }\textbf {\bibinfo {volume} {B491}},\ \bibinfo {pages} {285}},\
  \Eprint {http://arxiv.org/abs/hep-ph/0005285} {arXiv:hep-ph/0005285 [hep-ph]}
  \BibitemShut {NoStop}%
\bibitem [{\citenamefont {Liu}(2009)}]{Liu:2009}%
  \BibitemOpen
  \bibfield  {author} {\bibinfo {author} {\bibnamefont {Liu}, \bibfnamefont
  {B.}}} (\bibinfo {year} {2009}),\ \href {\doibase 10.1103/PhysRevD.79.015001}
  {\bibfield  {journal} {\bibinfo  {journal} {Phys. Rev. D}\ }\textbf {\bibinfo
  {volume} {79}},\ \bibinfo {pages} {015001}}\BibitemShut {NoStop}%
\bibitem [{\citenamefont {Lynch}\ \emph {et~al.}(1958)\citenamefont {Lynch},
  \citenamefont {Orear},\ and\ \citenamefont {Rosendorff}}]{Lynch:1958}%
  \BibitemOpen
  \bibfield  {author} {\bibinfo {author} {\bibnamefont {Lynch}, \bibfnamefont
  {G.}}, \bibinfo {author} {\bibfnamefont {J.}~\bibnamefont {Orear}}, \ and\
  \bibinfo {author} {\bibfnamefont {S.}~\bibnamefont {Rosendorff}}} (\bibinfo
  {year} {1958}),\ \href {\doibase 10.1103/PhysRevLett.1.471} {\bibfield
  {journal} {\bibinfo  {journal} {Phys. Rev. Lett.}\ }\textbf {\bibinfo
  {volume} {1}},\ \bibinfo {pages} {471}},\ \bibinfo {note} {the paper set a
  limit on the sum of several modes, and said ``probably less than" rather than
  setting a 90\% CL.}\BibitemShut {Stop}%
\bibitem [{\citenamefont {Marciano}\ and\ \citenamefont
  {Sanda}(1977)}]{Marciano:1977}%
  \BibitemOpen
  \bibfield  {author} {\bibinfo {author} {\bibnamefont {Marciano},
  \bibfnamefont {W.}}, \ and\ \bibinfo {author} {\bibfnamefont
  {A.}~\bibnamefont {Sanda}}} (\bibinfo {year} {1977}),\ \href {\doibase
  10.1016/0370-2693(77)90377-X} {\bibfield  {journal} {\bibinfo  {journal}
  {Physics Letters B}\ }\textbf {\bibinfo {volume} {67}}~(\bibinfo {number}
  {3}),\ \bibinfo {pages} {303 }}\BibitemShut {NoStop}%
\bibitem [{\citenamefont {Marciano}\ \emph {et~al.}(2008)\citenamefont
  {Marciano}, \citenamefont {Mori},\ and\ \citenamefont
  {Roney}}]{Marciano:2008}%
  \BibitemOpen
  \bibfield  {author} {\bibinfo {author} {\bibnamefont {Marciano},
  \bibfnamefont {W.~J.}}, \bibinfo {author} {\bibfnamefont {T.}~\bibnamefont
  {Mori}}, \ and\ \bibinfo {author} {\bibfnamefont {J.~M.}\ \bibnamefont
  {Roney}}} (\bibinfo {year} {2008}),\ \href {\doibase
  10.1146/annurev.nucl.58.110707.171126} {\bibfield  {journal} {\bibinfo
  {journal} {Annual Review of Nuclear and Particle Science}\ }\textbf {\bibinfo
  {volume} {58}}~(\bibinfo {number} {1}),\ \bibinfo {pages} {315}},\ \bibinfo
  {note}
  {\url{http://www.annualreviews.org/doi/abs/10.1146/annurev.nucl.58.110707.171126}}\BibitemShut
  {NoStop}%
\bibitem [{\citenamefont {Marshall}(2012)}]{Marshall:2012}%
  \BibitemOpen
  \bibfield  {author} {\bibinfo {author} {\bibnamefont {Marshall},
  \bibfnamefont {G.}}} (\bibinfo {year} {2012}),\ \href@noop {} {}\bibinfo
  {note} {Priv.\ comm.}\BibitemShut {Stop}%
\bibitem [{\citenamefont {Marshall}\ \emph {et~al.}(1982)\citenamefont
  {Marshall}, \citenamefont {Warren}, \citenamefont {Oram},\ and\ \citenamefont
  {Kiefl}}]{Marshall:1982}%
  \BibitemOpen
  \bibfield  {author} {\bibinfo {author} {\bibnamefont {Marshall},
  \bibfnamefont {G.~M.}}, \bibinfo {author} {\bibfnamefont {J.~B.}\
  \bibnamefont {Warren}}, \bibinfo {author} {\bibfnamefont {C.~J.}\
  \bibnamefont {Oram}}, \ and\ \bibinfo {author} {\bibfnamefont {R.~F.}\
  \bibnamefont {Kiefl}}} (\bibinfo {year} {1982}),\ \href {\doibase
  10.1103/PhysRevD.25.1174} {\bibfield  {journal} {\bibinfo  {journal} {Phys.
  Rev. D}\ }\textbf {\bibinfo {volume} {25}},\ \bibinfo {pages}
  {1174}}\BibitemShut {NoStop}%
\bibitem [{\citenamefont {Mathiazhagan}\ \emph {et~al.}(1989)\citenamefont
  {Mathiazhagan}, \citenamefont {Molzon}, \citenamefont {Cousins},
  \citenamefont {Konigsberg}, \citenamefont {Kubic}, \citenamefont {Melese},
  \citenamefont {Rubin}, \citenamefont {Slater}, \citenamefont {Wagner},
  \citenamefont {Hart}, \citenamefont {Kinnison}, \citenamefont {Lee},
  \citenamefont {McKee}, \citenamefont {Milner}, \citenamefont {Sanders},
  \citenamefont {Ziock}, \citenamefont {Arisaka}, \citenamefont {Knibbe},
  \citenamefont {Urheim}, \citenamefont {Axelrod}, \citenamefont {Biery},
  \citenamefont {Irwin}, \citenamefont {Lang}, \citenamefont {Margulies},
  \citenamefont {Ouimette}, \citenamefont {Ritchie}, \citenamefont {Trang},
  \citenamefont {Wojcicki}, \citenamefont {Auerbach}, \citenamefont {Buchholz},
  \citenamefont {Highland}, \citenamefont {McFarlane}, \citenamefont {Sivertz},
  \citenamefont {Chapman}, \citenamefont {Eckhause}, \citenamefont {Ginkel},
  \citenamefont {Hancock}, \citenamefont {Joyce}, \citenamefont {Kane},
  \citenamefont {Kenney}, \citenamefont {Vulcan}, \citenamefont {Welsh},
  \citenamefont {Whyley},\ and\ \citenamefont {Winter}}]{Mathiazhagan:1989}%
  \BibitemOpen
  \bibfield  {author} {\bibinfo {author} {\bibnamefont {Mathiazhagan},
  \bibfnamefont {C.}}, \bibinfo {author} {\bibfnamefont {W.~R.}\ \bibnamefont
  {Molzon}}, \bibinfo {author} {\bibfnamefont {R.~D.}\ \bibnamefont {Cousins}},
  \bibinfo {author} {\bibfnamefont {J.}~\bibnamefont {Konigsberg}}, \bibinfo
  {author} {\bibfnamefont {J.}~\bibnamefont {Kubic}}, \bibinfo {author}
  {\bibfnamefont {P.}~\bibnamefont {Melese}}, \bibinfo {author} {\bibfnamefont
  {P.}~\bibnamefont {Rubin}}, \bibinfo {author} {\bibfnamefont {W.~E.}\
  \bibnamefont {Slater}}, \bibinfo {author} {\bibfnamefont {D.}~\bibnamefont
  {Wagner}}, \bibinfo {author} {\bibfnamefont {G.~W.}\ \bibnamefont {Hart}},
  \bibinfo {author} {\bibfnamefont {W.~W.}\ \bibnamefont {Kinnison}}, \bibinfo
  {author} {\bibfnamefont {D.~M.}\ \bibnamefont {Lee}}, \bibinfo {author}
  {\bibfnamefont {R.~J.}\ \bibnamefont {McKee}}, \bibinfo {author}
  {\bibfnamefont {E.~C.}\ \bibnamefont {Milner}}, \bibinfo {author}
  {\bibfnamefont {G.~H.}\ \bibnamefont {Sanders}}, \bibinfo {author}
  {\bibfnamefont {H.~J.}\ \bibnamefont {Ziock}}, \bibinfo {author}
  {\bibfnamefont {K.}~\bibnamefont {Arisaka}}, \bibinfo {author} {\bibfnamefont
  {P.}~\bibnamefont {Knibbe}}, \bibinfo {author} {\bibfnamefont
  {J.}~\bibnamefont {Urheim}}, \bibinfo {author} {\bibfnamefont
  {S.}~\bibnamefont {Axelrod}}, \bibinfo {author} {\bibfnamefont {K.~A.}\
  \bibnamefont {Biery}}, \bibinfo {author} {\bibfnamefont {G.~M.}\ \bibnamefont
  {Irwin}}, \bibinfo {author} {\bibfnamefont {K.}~\bibnamefont {Lang}},
  \bibinfo {author} {\bibfnamefont {J.}~\bibnamefont {Margulies}}, \bibinfo
  {author} {\bibfnamefont {D.~A.}\ \bibnamefont {Ouimette}}, \bibinfo {author}
  {\bibfnamefont {J.~L.}\ \bibnamefont {Ritchie}}, \bibinfo {author}
  {\bibfnamefont {Q.~H.}\ \bibnamefont {Trang}}, \bibinfo {author}
  {\bibfnamefont {S.~G.}\ \bibnamefont {Wojcicki}}, \bibinfo {author}
  {\bibfnamefont {L.~B.}\ \bibnamefont {Auerbach}}, \bibinfo {author}
  {\bibfnamefont {P.}~\bibnamefont {Buchholz}}, \bibinfo {author}
  {\bibfnamefont {V.~L.}\ \bibnamefont {Highland}}, \bibinfo {author}
  {\bibfnamefont {W.~K.}\ \bibnamefont {McFarlane}}, \bibinfo {author}
  {\bibfnamefont {M.}~\bibnamefont {Sivertz}}, \bibinfo {author} {\bibfnamefont
  {M.~D.}\ \bibnamefont {Chapman}}, \bibinfo {author} {\bibfnamefont
  {M.}~\bibnamefont {Eckhause}}, \bibinfo {author} {\bibfnamefont {J.~F.}\
  \bibnamefont {Ginkel}}, \bibinfo {author} {\bibfnamefont {A.~D.}\
  \bibnamefont {Hancock}}, \bibinfo {author} {\bibfnamefont {D.}~\bibnamefont
  {Joyce}}, \bibinfo {author} {\bibfnamefont {J.~R.}\ \bibnamefont {Kane}},
  \bibinfo {author} {\bibfnamefont {C.~J.}\ \bibnamefont {Kenney}}, \bibinfo
  {author} {\bibfnamefont {W.~F.}\ \bibnamefont {Vulcan}}, \bibinfo {author}
  {\bibfnamefont {R.~E.}\ \bibnamefont {Welsh}}, \bibinfo {author}
  {\bibfnamefont {R.~J.}\ \bibnamefont {Whyley}}, \ and\ \bibinfo {author}
  {\bibfnamefont {R.~G.}\ \bibnamefont {Winter}}} (\bibinfo {year} {1989}),\
  \href {\doibase 10.1103/PhysRevLett.63.2181} {\bibfield  {journal} {\bibinfo
  {journal} {Phys. Rev. Lett.}\ }\textbf {\bibinfo {volume} {63}},\ \bibinfo
  {pages} {2181}}\BibitemShut {NoStop}%
\bibitem [{\citenamefont {Matsushita}\ and\ \citenamefont
  {Nagamine}(1996)}]{Tungsten:2012}%
  \BibitemOpen
  \bibfield  {author} {\bibinfo {author} {\bibnamefont {Matsushita},
  \bibfnamefont {A.}}, \ and\ \bibinfo {author} {\bibfnamefont
  {K.}~\bibnamefont {Nagamine}}} (\bibinfo {year} {1996}),\ \href {\doibase
  10.1016/0039-6028(96)00299-3} {\bibfield  {journal} {\bibinfo  {journal}
  {Surface Science}\ }\textbf {\bibinfo {volume} {357}}~(\bibinfo {number}
  {0}),\ \bibinfo {pages} {961 }}\BibitemShut {NoStop}%
\bibitem [{\citenamefont {Matthias}\ \emph {et~al.}(1991)\citenamefont
  {Matthias} \emph {et~al.}}]{Matthias:1991}%
  \BibitemOpen
  \bibfield  {author} {\bibinfo {author} {\bibnamefont {Matthias},
  \bibfnamefont {B.~E.}},  \emph {et~al.}} (\bibinfo {year} {1991}),\ \href
  {\doibase 10.1103/PhysRevLett.66.2716} {\bibfield  {journal} {\bibinfo
  {journal} {Phys. Rev. Lett.}\ }\textbf {\bibinfo {volume} {66}},\ \bibinfo
  {pages} {2716}}\BibitemShut {NoStop}%
\bibitem [{\citenamefont {Measday}(2001)}]{Measday:2001}%
  \BibitemOpen
  \bibfield  {author} {\bibinfo {author} {\bibnamefont {Measday}, \bibfnamefont
  {D.}}} (\bibinfo {year} {2001}),\ \href {\doibase
  10.1016/S0370-1573(01)00012-6} {\bibfield  {journal} {\bibinfo  {journal}
  {Physics Reports}\ }\textbf {\bibinfo {volume} {354}}~(\bibinfo {number}
  {4--5}),\ \bibinfo {pages} {243 }}\BibitemShut {NoStop}%
\bibitem [{\citenamefont {Michel}(1950)}]{Michel:1950}%
  \BibitemOpen
  \bibfield  {author} {\bibinfo {author} {\bibnamefont {Michel}, \bibfnamefont
  {L.}}} (\bibinfo {year} {1950}),\ \href@noop {} {\bibfield  {journal}
  {\bibinfo  {journal} {Proc.\ Phys.\ Soc.\ (London)}\ }\textbf {\bibinfo
  {volume} {A}}~(\bibinfo {number} {63}),\ \bibinfo {pages} {514}}\BibitemShut
  {NoStop}%
\bibitem [{\citenamefont {Morrissey}\ \emph {et~al.}(2012)\citenamefont
  {Morrissey}, \citenamefont {Plehn},\ and\ \citenamefont
  {Tait}}]{Morrissey:2012}%
  \BibitemOpen
  \bibfield  {author} {\bibinfo {author} {\bibnamefont {Morrissey},
  \bibfnamefont {D.~E.}}, \bibinfo {author} {\bibfnamefont {T.}~\bibnamefont
  {Plehn}}, \ and\ \bibinfo {author} {\bibfnamefont {T.~M.}\ \bibnamefont
  {Tait}}} (\bibinfo {year} {2012}),\ \href {\doibase
  10.1016/j.physrep.2012.02.007} {\bibfield  {journal} {\bibinfo  {journal}
  {Physics Reports}\ }\textbf {\bibinfo {volume} {515}}~(\bibinfo {number}
  {1}),\ \bibinfo {pages} {1 }},\ \bibinfo {note} {physics searches at the
  LHC}\BibitemShut {NoStop}%
\bibitem [{\citenamefont {Mukhopadhyay}(1977)}]{Mukhopadhyay:1977}%
  \BibitemOpen
  \bibfield  {author} {\bibinfo {author} {\bibnamefont {Mukhopadhyay},
  \bibfnamefont {N.~C.}}} (\bibinfo {year} {1977}),\ \href {\doibase
  10.1016/0370-1573(77)90073-4} {\bibfield  {journal} {\bibinfo  {journal}
  {Physics Reports}\ }\textbf {\bibinfo {volume} {30}}~(\bibinfo {number}
  {1}),\ \bibinfo {pages} {1 }}\BibitemShut {NoStop}%
\bibitem [{\citenamefont {Nakamura}\ \emph {et~al.}(2010)\citenamefont
  {Nakamura} \emph {et~al.}}]{Nakamura:2010}%
  \BibitemOpen
  \bibfield  {author} {\bibinfo {author} {\bibnamefont {Nakamura},
  \bibfnamefont {K.}},  \emph {et~al.} (\bibinfo {collaboration} {Particle Data
  Group})} (\bibinfo {year} {2010}),\ \href {\doibase
  10.1088/0954-3899/37/7A/075021} {\bibfield  {journal} {\bibinfo  {journal}
  {J.Phys.G}\ }\textbf {\bibinfo {volume} {G37}},\ \bibinfo {pages}
  {075021}}\BibitemShut {NoStop}%
\bibitem [{\citenamefont {Nishiguchi}(2008)}]{Nishiguchi:2008}%
  \BibitemOpen
  \bibfield  {author} {\bibinfo {author} {\bibnamefont {Nishiguchi},
  \bibfnamefont {H.}}} (\bibinfo {year} {2008}),\ \emph {\bibinfo {title} {An
  Innovative Positron Spectrometer to Search for the Lepton Flavour Violating
  Muon Decay with a Sensitivity of $10^{-13}$}},\ \href@noop {} {Ph.D. thesis}\
  (\bibinfo  {school} {University of Tokyo}),\ \bibinfo {note}
  {\url{http://meg.web.psi.ch/docs/theses/nishiguchi\_phd.pdf}}\BibitemShut
  {NoStop}%
\bibitem [{\citenamefont {Nishimura}(2010)}]{Nishimura:2010}%
  \BibitemOpen
  \bibfield  {author} {\bibinfo {author} {\bibnamefont {Nishimura},
  \bibfnamefont {Y.}}} (\bibinfo {year} {2010}),\ \emph {\bibinfo {title} {A
  Search for the Decay $\mu^+ \rightarrow e^+ \gamma$ Using a High-Resolution
  Liquid Xenon Gamma-Ray Detector}},\ \href@noop {} {Ph.D. thesis}\ (\bibinfo
  {school} {University of Tokyo}),\ \bibinfo {note}
  {\url{http://meg.web.psi.ch/docs/theses/nishimura\_phd.pdf}}\BibitemShut
  {NoStop}%
\bibitem [{\citenamefont {Okada}\ \emph {et~al.}(1998)\citenamefont {Okada},
  \citenamefont {Okumura},\ and\ \citenamefont {Shimizu}}]{Okada:1997fz}%
  \BibitemOpen
  \bibfield  {author} {\bibinfo {author} {\bibnamefont {Okada}, \bibfnamefont
  {Y.}}, \bibinfo {author} {\bibfnamefont {K.-i.}\ \bibnamefont {Okumura}}, \
  and\ \bibinfo {author} {\bibfnamefont {Y.}~\bibnamefont {Shimizu}}} (\bibinfo
  {year} {1998}),\ \href {\doibase 10.1103/PhysRevD.58.051901} {\bibfield
  {journal} {\bibinfo  {journal} {Phys.Rev.}\ }\textbf {\bibinfo {volume}
  {D58}},\ \bibinfo {pages} {051901}},\ \Eprint
  {http://arxiv.org/abs/hep-ph/9708446} {arXiv:hep-ph/9708446 [hep-ph]}
  \BibitemShut {NoStop}%
\bibitem [{\citenamefont {O'Keefe}\ \emph {et~al.}(1959)\citenamefont
  {O'Keefe}, \citenamefont {Rigby},\ and\ \citenamefont
  {Wormald}}]{Okeefe:1959}%
  \BibitemOpen
  \bibfield  {author} {\bibinfo {author} {\bibnamefont {O'Keefe}, \bibfnamefont
  {T.}}, \bibinfo {author} {\bibfnamefont {M.}~\bibnamefont {Rigby}}, \ and\
  \bibinfo {author} {\bibfnamefont {J.}~\bibnamefont {Wormald}}} (\bibinfo
  {year} {1959}),\ \href@noop {} {\bibfield  {journal} {\bibinfo  {journal}
  {Proc. Phys. Soc.}\ }\textbf {\bibinfo {volume} {73}},\ \bibinfo {pages}
  {951}}\BibitemShut {NoStop}%
\bibitem [{\citenamefont {Paradisi}(2006)}]{Paradisi:2006}%
  \BibitemOpen
  \bibfield  {author} {\bibinfo {author} {\bibnamefont {Paradisi},
  \bibfnamefont {P.}}} (\bibinfo {year} {2006}),\ \href@noop {} {\bibfield
  {journal} {\bibinfo  {journal} {JHEP}\ }\textbf {\bibinfo {volume} {0602}}},\
  \Eprint {http://arxiv.org/abs/hep-ph/0508054v2} {hep-ph/0508054v2}
  \BibitemShut {NoStop}%
\bibitem [{\citenamefont {Park}\ \emph {et~al.}(2002)\citenamefont {Park},
  \citenamefont {Burnstein}, \citenamefont {Chakravorty}, \citenamefont {Chan},
  \citenamefont {Chen}, \citenamefont {Choong}, \citenamefont {Clark},
  \citenamefont {Dukes}, \citenamefont {Durandet}, \citenamefont {Felix},
  \citenamefont {Gidal}, \citenamefont {Gu}, \citenamefont {Gustafson},
  \citenamefont {Ho}, \citenamefont {Holmstrom}, \citenamefont {Huang},
  \citenamefont {James}, \citenamefont {Jenkins}, \citenamefont {Kaplan},
  \citenamefont {Lederman}, \citenamefont {Leros}, \citenamefont {Longo},
  \citenamefont {Lopez}, \citenamefont {Lu}, \citenamefont {Luebke},
  \citenamefont {Luk}, \citenamefont {Nelson}, \citenamefont {Perroud},
  \citenamefont {Rajaram}, \citenamefont {Rubin}, \citenamefont {Teng},
  \citenamefont {Volk}, \citenamefont {White}, \citenamefont {White},\ and\
  \citenamefont {Zyla}}]{Park:2002}%
  \BibitemOpen
  \bibfield  {author} {\bibinfo {author} {\bibnamefont {Park}, \bibfnamefont
  {H.~K.}}, \bibinfo {author} {\bibfnamefont {R.~A.}\ \bibnamefont
  {Burnstein}}, \bibinfo {author} {\bibfnamefont {A.}~\bibnamefont
  {Chakravorty}}, \bibinfo {author} {\bibfnamefont {A.}~\bibnamefont {Chan}},
  \bibinfo {author} {\bibfnamefont {Y.~C.}\ \bibnamefont {Chen}}, \bibinfo
  {author} {\bibfnamefont {W.~S.}\ \bibnamefont {Choong}}, \bibinfo {author}
  {\bibfnamefont {K.}~\bibnamefont {Clark}}, \bibinfo {author} {\bibfnamefont
  {E.~C.}\ \bibnamefont {Dukes}}, \bibinfo {author} {\bibfnamefont
  {C.}~\bibnamefont {Durandet}}, \bibinfo {author} {\bibfnamefont
  {J.}~\bibnamefont {Felix}}, \bibinfo {author} {\bibfnamefont
  {G.}~\bibnamefont {Gidal}}, \bibinfo {author} {\bibfnamefont
  {P.}~\bibnamefont {Gu}}, \bibinfo {author} {\bibfnamefont {H.~R.}\
  \bibnamefont {Gustafson}}, \bibinfo {author} {\bibfnamefont {C.}~\bibnamefont
  {Ho}}, \bibinfo {author} {\bibfnamefont {T.}~\bibnamefont {Holmstrom}},
  \bibinfo {author} {\bibfnamefont {M.}~\bibnamefont {Huang}}, \bibinfo
  {author} {\bibfnamefont {C.}~\bibnamefont {James}}, \bibinfo {author}
  {\bibfnamefont {C.~M.}\ \bibnamefont {Jenkins}}, \bibinfo {author}
  {\bibfnamefont {D.~M.}\ \bibnamefont {Kaplan}}, \bibinfo {author}
  {\bibfnamefont {L.~M.}\ \bibnamefont {Lederman}}, \bibinfo {author}
  {\bibfnamefont {N.}~\bibnamefont {Leros}}, \bibinfo {author} {\bibfnamefont
  {M.~J.}\ \bibnamefont {Longo}}, \bibinfo {author} {\bibfnamefont
  {F.}~\bibnamefont {Lopez}}, \bibinfo {author} {\bibfnamefont
  {L.}~\bibnamefont {Lu}}, \bibinfo {author} {\bibfnamefont {W.}~\bibnamefont
  {Luebke}}, \bibinfo {author} {\bibfnamefont {K.~B.}\ \bibnamefont {Luk}},
  \bibinfo {author} {\bibfnamefont {K.~S.}\ \bibnamefont {Nelson}}, \bibinfo
  {author} {\bibfnamefont {J.~P.}\ \bibnamefont {Perroud}}, \bibinfo {author}
  {\bibfnamefont {D.}~\bibnamefont {Rajaram}}, \bibinfo {author} {\bibfnamefont
  {H.~A.}\ \bibnamefont {Rubin}}, \bibinfo {author} {\bibfnamefont {P.~K.}\
  \bibnamefont {Teng}}, \bibinfo {author} {\bibfnamefont {J.}~\bibnamefont
  {Volk}}, \bibinfo {author} {\bibfnamefont {C.}~\bibnamefont {White}},
  \bibinfo {author} {\bibfnamefont {S.}~\bibnamefont {White}}, \ and\ \bibinfo
  {author} {\bibfnamefont {P.}~\bibnamefont {Zyla}} (\bibinfo {collaboration}
  {(HyperCP Collaboration)})} (\bibinfo {year} {2002}),\ \href {\doibase
  10.1103/PhysRevLett.88.111801} {\bibfield  {journal} {\bibinfo  {journal}
  {Phys. Rev. Lett.}\ }\textbf {\bibinfo {volume} {88}},\ \bibinfo {pages}
  {111801}}\BibitemShut {NoStop}%
\bibitem [{\citenamefont {Parker}\ \emph {et~al.}(1964)\citenamefont {Parker},
  \citenamefont {Anderson},\ and\ \citenamefont {Rey}}]{Parker:1964}%
  \BibitemOpen
  \bibfield  {author} {\bibinfo {author} {\bibnamefont {Parker}, \bibfnamefont
  {S.}}, \bibinfo {author} {\bibfnamefont {H.~L.}\ \bibnamefont {Anderson}}, \
  and\ \bibinfo {author} {\bibfnamefont {C.}~\bibnamefont {Rey}}} (\bibinfo
  {year} {1964}),\ \href {\doibase 10.1103/PhysRev.133.B768} {\bibfield
  {journal} {\bibinfo  {journal} {Phys. Rev.}\ }\textbf {\bibinfo {volume}
  {133}},\ \bibinfo {pages} {B768}}\BibitemShut {NoStop}%
\bibitem [{\citenamefont {Parker}\ and\ \citenamefont
  {Penman}(1962)}]{Parker:1962}%
  \BibitemOpen
  \bibfield  {author} {\bibinfo {author} {\bibnamefont {Parker}, \bibfnamefont
  {S.}}, \ and\ \bibinfo {author} {\bibfnamefont {S.}~\bibnamefont {Penman}}}
  (\bibinfo {year} {1962}),\ \href@noop {} {\bibfield  {journal} {\bibinfo
  {journal} {Nuovo Cimento}\ }\textbf {\bibinfo {volume} {23}},\ \bibinfo
  {pages} {485}}\BibitemShut {NoStop}%
\bibitem [{\citenamefont {Perkins}(1999)}]{Perkins:1999}%
  \BibitemOpen
  \bibfield  {author} {\bibinfo {author} {\bibnamefont {Perkins}, \bibfnamefont
  {D.~H.}}} (\bibinfo {year} {1999}),\ \href@noop {} {\emph {\bibinfo {title}
  {Introduction to High Energy Physics, $4^{{\rm th}}$ Edition}}}\ (\bibinfo
  {publisher} {Cambridge University Press})\BibitemShut {NoStop}%
\bibitem [{\citenamefont {Pham}(1999)}]{Pham:1998}%
  \BibitemOpen
  \bibfield  {author} {\bibinfo {author} {\bibnamefont {Pham}, \bibfnamefont
  {X.-Y.}}} (\bibinfo {year} {1999}),\ \href {\doibase 10.1007/s100529901088}
  {\bibfield  {journal} {\bibinfo  {journal} {Eur.Phys.J.}\ }\textbf {\bibinfo
  {volume} {C8}},\ \bibinfo {pages} {513}},\ \Eprint
  {http://arxiv.org/abs/hep-ph/9810484} {arXiv:hep-ph/9810484 [hep-ph]}
  \BibitemShut {NoStop}%
\bibitem [{\citenamefont {Pontecorvo}(1958)}]{Pontecorvo:1958}%
  \BibitemOpen
  \bibfield  {author} {\bibinfo {author} {\bibnamefont {Pontecorvo},
  \bibfnamefont {B.}}} (\bibinfo {year} {1958}),\ \href@noop {} {\bibfield
  {journal} {\bibinfo  {journal} {Sov.\ Phys.\ JETP}\ }\textbf {\bibinfo
  {volume} {6}},\ \bibinfo {pages} {429}}\BibitemShut {NoStop}%
\bibitem [{\citenamefont {Povel}\ \emph {et~al.}(1977)\citenamefont {Povel},
  \citenamefont {Dey}, \citenamefont {Walter}, \citenamefont {Pfeiffer},
  \citenamefont {Sennhauser} \emph {et~al.}}]{Povel:1977}%
  \BibitemOpen
  \bibfield  {author} {\bibinfo {author} {\bibnamefont {Povel}, \bibfnamefont
  {H.}}, \bibinfo {author} {\bibfnamefont {W.}~\bibnamefont {Dey}}, \bibinfo
  {author} {\bibfnamefont {H.}~\bibnamefont {Walter}}, \bibinfo {author}
  {\bibfnamefont {H.}~\bibnamefont {Pfeiffer}}, \bibinfo {author}
  {\bibfnamefont {U.}~\bibnamefont {Sennhauser}},  \emph {et~al.}} (\bibinfo
  {year} {1977}),\ \href {\doibase 10.1016/0370-2693(77)90697-9} {\bibfield
  {journal} {\bibinfo  {journal} {Phys.Lett.}\ }\textbf {\bibinfo {volume}
  {B72}},\ \bibinfo {pages} {183}}\BibitemShut {NoStop}%
\bibitem [{\citenamefont {Raidal}\ \emph {et~al.}(2008)\citenamefont {Raidal}
  \emph {et~al.}}]{Raidal:2008}%
  \BibitemOpen
  \bibfield  {author} {\bibinfo {author} {\bibnamefont {Raidal}, \bibfnamefont
  {M.}},  \emph {et~al.}} (\bibinfo {year} {2008}),\ \href@noop {} {\bibfield
  {journal} {\bibinfo  {journal} {Report of Working Group 3 of the CERN
  Workshop on Flavour in the Era of the LHC}\ }}\Eprint
  {http://arxiv.org/abs/arXiv:0801.1826 [hep-ph]} {arXiv:0801.1826 [hep-ph]}
  \BibitemShut {NoStop}%
\bibitem [{\citenamefont {Sard}\ and\ \citenamefont
  {Althaus}(1948)}]{Sard:1948}%
  \BibitemOpen
  \bibfield  {author} {\bibinfo {author} {\bibnamefont {Sard}, \bibfnamefont
  {R.~D.}}, \ and\ \bibinfo {author} {\bibfnamefont {E.~J.}\ \bibnamefont
  {Althaus}}} (\bibinfo {year} {1948}),\ \href {\doibase
  10.1103/PhysRev.74.1364} {\bibfield  {journal} {\bibinfo  {journal} {Phys.
  Rev.}\ }\textbf {\bibinfo {volume} {74}},\ \bibinfo {pages}
  {1364}}\BibitemShut {NoStop}%
\bibitem [{\citenamefont {Sard}\ \emph {et~al.}(1961)\citenamefont {Sard},
  \citenamefont {Crowe},\ and\ \citenamefont {Kruger}}]{Sard:1961}%
  \BibitemOpen
  \bibfield  {author} {\bibinfo {author} {\bibnamefont {Sard}, \bibfnamefont
  {R.~D.}}, \bibinfo {author} {\bibfnamefont {K.~M.}\ \bibnamefont {Crowe}}, \
  and\ \bibinfo {author} {\bibfnamefont {H.}~\bibnamefont {Kruger}}} (\bibinfo
  {year} {1961}),\ \href {\doibase 10.1103/PhysRev.121.619} {\bibfield
  {journal} {\bibinfo  {journal} {Phys. Rev.}\ }\textbf {\bibinfo {volume}
  {121}},\ \bibinfo {pages} {619}}\BibitemShut {NoStop}%
\bibitem [{\citenamefont {Sawada}(2008)}]{Sawada:2008}%
  \BibitemOpen
  \bibfield  {author} {\bibinfo {author} {\bibnamefont {Sawada}, \bibfnamefont
  {R.}}} (\bibinfo {year} {2008}),\ \emph {\bibinfo {title} {A Liquid Xenon
  Scintillation Detector to Search for the Lepton Flavor Violating Muon Decay
  with a Sensitivity of $10^{-13}$}},\ \href@noop {} {Ph.D. thesis}\ (\bibinfo
  {school} {University of Tokyo}),\ \bibinfo {note}
  {\url{http://meg.web.psi.ch/docs/theses/sawada\_phd.pdf}}\BibitemShut
  {NoStop}%
\bibitem [{\citenamefont {Schaffner}\ \emph {et~al.}(1989)\citenamefont
  {Schaffner}, \citenamefont {Greenlee}, \citenamefont {Kasha}, \citenamefont
  {Mannelli}, \citenamefont {Mannelli}, \citenamefont {Ohl}, \citenamefont
  {Schmidt}, \citenamefont {Jastrzembski}, \citenamefont {Larsen},
  \citenamefont {Leipuner},\ and\ \citenamefont {Morse}}]{Schaffner:1989}%
  \BibitemOpen
  \bibfield  {author} {\bibinfo {author} {\bibnamefont {Schaffner},
  \bibfnamefont {S.~F.}}, \bibinfo {author} {\bibfnamefont {H.~B.}\
  \bibnamefont {Greenlee}}, \bibinfo {author} {\bibfnamefont {H.}~\bibnamefont
  {Kasha}}, \bibinfo {author} {\bibfnamefont {E.~B.}\ \bibnamefont {Mannelli}},
  \bibinfo {author} {\bibfnamefont {M.}~\bibnamefont {Mannelli}}, \bibinfo
  {author} {\bibfnamefont {K.~E.}\ \bibnamefont {Ohl}}, \bibinfo {author}
  {\bibfnamefont {M.~P.}\ \bibnamefont {Schmidt}}, \bibinfo {author}
  {\bibfnamefont {E.}~\bibnamefont {Jastrzembski}}, \bibinfo {author}
  {\bibfnamefont {R.~C.}\ \bibnamefont {Larsen}}, \bibinfo {author}
  {\bibfnamefont {L.~B.}\ \bibnamefont {Leipuner}}, \ and\ \bibinfo {author}
  {\bibfnamefont {W.~M.}\ \bibnamefont {Morse}}} (\bibinfo {year} {1989}),\
  \href {\doibase 10.1103/PhysRevD.39.990} {\bibfield  {journal} {\bibinfo
  {journal} {Phys. Rev. D}\ }\textbf {\bibinfo {volume} {39}},\ \bibinfo
  {pages} {990}}\BibitemShut {NoStop}%
\bibitem [{\citenamefont {Shanker}(1982)}]{Shanker:1982}%
  \BibitemOpen
  \bibfield  {author} {\bibinfo {author} {\bibnamefont {Shanker}, \bibfnamefont
  {O.}}} (\bibinfo {year} {1982}),\ \href {\doibase 10.1103/PhysRevD.25.1847}
  {\bibfield  {journal} {\bibinfo  {journal} {Phys. Rev. D}\ }\textbf {\bibinfo
  {volume} {25}},\ \bibinfo {pages} {1847}}\BibitemShut {NoStop}%
\bibitem [{\citenamefont {Shanker}\ and\ \citenamefont
  {Roy}(1997)}]{Shanker:1997}%
  \BibitemOpen
  \bibfield  {author} {\bibinfo {author} {\bibnamefont {Shanker}, \bibfnamefont
  {O.}}, \ and\ \bibinfo {author} {\bibfnamefont {R.}~\bibnamefont {Roy}}}
  (\bibinfo {year} {1997}),\ \href {\doibase 10.1103/PhysRevD.55.7307}
  {\bibfield  {journal} {\bibinfo  {journal} {Phys. Rev. D}\ }\textbf {\bibinfo
  {volume} {55}},\ \bibinfo {pages} {7307}}\BibitemShut {NoStop}%
\bibitem [{\citenamefont {Sher}\ \emph {et~al.}(2005)\citenamefont {Sher} \emph
  {et~al.}}]{Sher:2005}%
  \BibitemOpen
  \bibfield  {author} {\bibinfo {author} {\bibnamefont {Sher}, \bibfnamefont
  {A.}},  \emph {et~al.}} (\bibinfo {year} {2005}),\ \href {\doibase
  10.1103/PhysRevD.72.012005} {\bibfield  {journal} {\bibinfo  {journal} {Phys.
  Rev. D}\ }\textbf {\bibinfo {volume} {72}},\ \bibinfo {pages}
  {012005}}\BibitemShut {NoStop}%
\bibitem [{\citenamefont {Sher}\ and\ \citenamefont {Yuan}(1991)}]{Sher:1991}%
  \BibitemOpen
  \bibfield  {author} {\bibinfo {author} {\bibnamefont {Sher}, \bibfnamefont
  {M.}}, \ and\ \bibinfo {author} {\bibfnamefont {Y.}~\bibnamefont {Yuan}}}
  (\bibinfo {year} {1991}),\ \href {\doibase 10.1103/PhysRevD.44.1461}
  {\bibfield  {journal} {\bibinfo  {journal} {Phys. Rev. D}\ }\textbf {\bibinfo
  {volume} {44}},\ \bibinfo {pages} {1461}}\BibitemShut {NoStop}%
\bibitem [{\citenamefont {Signorelli}(2004)}]{Signorelli:2004}%
  \BibitemOpen
  \bibfield  {author} {\bibinfo {author} {\bibnamefont {Signorelli},
  \bibfnamefont {G.}}} (\bibinfo {year} {2004}),\ \emph {\bibinfo {title} {A
  Sensitive Search for lepton-flavor violation: the MEG experiment and the new
  LXe calorimetry}},\ \href@noop {} {Ph.D. thesis}\ (\bibinfo  {school} {Scuola
  Normale Superiore Di Pisa}),\ \bibinfo {note}
  {\url{http://meg.web.psi.ch/docs/theses/tesi\_signorelli.pdf}}\BibitemShut
  {NoStop}%
\bibitem [{\citenamefont {Steinberger}\ and\ \citenamefont
  {Lokanathan}(1955)}]{Steinberger:1955a}%
  \BibitemOpen
  \bibfield  {author} {\bibinfo {author} {\bibnamefont {Steinberger},
  \bibfnamefont {J.}}, \ and\ \bibinfo {author} {\bibfnamefont
  {S.}~\bibnamefont {Lokanathan}}} (\bibinfo {year} {1955}),\ \href@noop {}
  {\bibfield  {journal} {\bibinfo  {journal} {Phys. Rev.}\ }\textbf {\bibinfo
  {volume} {240}}},\ \bibinfo {note} {presented as a talk in Session Q. This
  result was reported to be published in Nuovo Cimento Supplemento II Serie
  X,151 (1955) but that is a paper titled {\it Search for the $\beta$-Decay of
  the Pion} and does not discuss $\mu \rightarrow e \gamma$. Perhaps the result
  referred to in the Phys.\ Rev.\ talk was never published.}\BibitemShut
  {Stop}%
\bibitem [{\citenamefont {Steinberger}\ and\ \citenamefont
  {Wolfe}(1955)}]{Steinberger:1955}%
  \BibitemOpen
  \bibfield  {author} {\bibinfo {author} {\bibnamefont {Steinberger},
  \bibfnamefont {J.}}, \ and\ \bibinfo {author} {\bibfnamefont {H.~B.}\
  \bibnamefont {Wolfe}}} (\bibinfo {year} {1955}),\ \href {\doibase
  10.1103/PhysRev.100.1490} {\bibfield  {journal} {\bibinfo  {journal} {Phys.
  Rev.}\ }\textbf {\bibinfo {volume} {100}},\ \bibinfo {pages}
  {1490}}\BibitemShut {NoStop}%
\bibitem [{\citenamefont {Symon}\ \emph {et~al.}(1956)\citenamefont {Symon},
  \citenamefont {Kerst}, \citenamefont {Jones}, \citenamefont {Laslett},\ and\
  \citenamefont {Terwilliger}}]{Symon:1956}%
  \BibitemOpen
  \bibfield  {author} {\bibinfo {author} {\bibnamefont {Symon}, \bibfnamefont
  {K.~R.}}, \bibinfo {author} {\bibfnamefont {D.~W.}\ \bibnamefont {Kerst}},
  \bibinfo {author} {\bibfnamefont {L.~W.}\ \bibnamefont {Jones}}, \bibinfo
  {author} {\bibfnamefont {L.~J.}\ \bibnamefont {Laslett}}, \ and\ \bibinfo
  {author} {\bibfnamefont {K.~M.}\ \bibnamefont {Terwilliger}}} (\bibinfo
  {year} {1956}),\ \href {\doibase 10.1103/PhysRev.103.1837} {\bibfield
  {journal} {\bibinfo  {journal} {Phys. Rev.}\ }\textbf {\bibinfo {volume}
  {103}},\ \bibinfo {pages} {1837}}\BibitemShut {NoStop}%
\bibitem [{\citenamefont {Varlamov}\ \emph {et~al.}(1999)\citenamefont
  {Varlamov} \emph {et~al.}}]{Varlamov:1999}%
  \BibitemOpen
  \bibfield  {author} {\bibinfo {author} {\bibnamefont {Varlamov},
  \bibfnamefont {A.}},  \emph {et~al.}} (\bibinfo {year} {1999}),\ \href@noop
  {} {\bibfield  {journal} {\bibinfo  {journal} {{ International Nuclear Data
  Committee}}\ }}\bibinfo {note} {\url
  {http://www-nds.iaea.org/publications/indc/indc-nds-0394.pdf}}\BibitemShut
  {NoStop}%
\bibitem [{\citenamefont {Cveti\ifmmode~\check{c}\else \v{c}\fi{}}\ \emph
  {et~al.}(2005)\citenamefont {Cveti\ifmmode~\check{c}\else \v{c}\fi{}},
  \citenamefont {Dib}, \citenamefont {Kim},\ and\ \citenamefont
  {Kim}}]{Cvetic:2005}%
  \BibitemOpen
  \bibfield  {author} {\bibinfo {author} {\bibnamefont
  {Cveti\ifmmode~\check{c}\else \v{c}\fi{}}, \bibfnamefont {G.}}, \bibinfo
  {author} {\bibfnamefont {C.~O.}\ \bibnamefont {Dib}}, \bibinfo {author}
  {\bibfnamefont {C.~S.}\ \bibnamefont {Kim}}, \ and\ \bibinfo {author}
  {\bibfnamefont {J.~D.}\ \bibnamefont {Kim}}} (\bibinfo {year} {2005}),\ \href
  {\doibase 10.1103/PhysRevD.71.113013} {\bibfield  {journal} {\bibinfo
  {journal} {Phys. Rev. D}\ }\textbf {\bibinfo {volume} {71}},\ \bibinfo
  {pages} {113013}}\BibitemShut {NoStop}%
\bibitem [{\citenamefont {Willmann}\ and\ \citenamefont
  {Jungmann}(1998)}]{Willmann:1998}%
  \BibitemOpen
  \bibfield  {author} {\bibinfo {author} {\bibnamefont {Willmann},
  \bibfnamefont {L.}}, \ and\ \bibinfo {author} {\bibfnamefont
  {K.}~\bibnamefont {Jungmann}}} (\bibinfo {year} {1998}),\ \href@noop {}
  {\bibfield  {journal} {\bibinfo  {journal} {Lect. Notes Phys.}\ }\textbf
  {\bibinfo {volume} {499}},\ \bibinfo {pages} {43}},\ \Eprint
  {http://arxiv.org/abs/hep-ex/9805013} {hep-ex/9805013} \BibitemShut {NoStop}%
\bibitem [{\citenamefont {Willmann}\ \emph {et~al.}(1999)\citenamefont
  {Willmann} \emph {et~al.}}]{Willmann:1999}%
  \BibitemOpen
  \bibfield  {author} {\bibinfo {author} {\bibnamefont {Willmann},
  \bibfnamefont {L.}},  \emph {et~al.}} (\bibinfo {year} {1999}),\ \href
  {\doibase 10.1103/PhysRevLett.82.49} {\bibfield  {journal} {\bibinfo
  {journal} {Phys. Rev. Lett.}\ }\textbf {\bibinfo {volume} {82}},\ \bibinfo
  {pages} {49}}\BibitemShut {NoStop}%
\bibitem [{\citenamefont {Zhu}(2006)}]{Zhu:2006cn}%
  \BibitemOpen
  \bibfield  {author} {\bibinfo {author} {\bibnamefont {Zhu}, \bibfnamefont
  {R.}}} (\bibinfo {year} {2006}),\ \href {\doibase 10.1117/12.796834,
  10.1063/1.2396939} {\bibfield  {journal} {\bibinfo  {journal} {AIP
  Conf.Proc.}\ }\textbf {\bibinfo {volume} {867}},\ \bibinfo {pages}
  {61}}\BibitemShut {NoStop}%
\end{thebibliography}%


%





\end{document}